DOCTORADO EN PLANEACIÓN ESTRATÉGICA Y DIRECCIÓN DE TECNOLOGÍA

Universidad Popular Autónoma del Estado de Puebla

Centro Interdisciplinario de Posgrados

Investigación y Consultoría

Departamento de Ingeniería

Doctorado en Planeación Estratégica

y Dirección de Tecnología

Título de la investigación

# "Propuesta de un estándar para México de Gestión del Conocimiento e Innovación Tecnológica"

Tesis que para obtener el Grado de Doctor
en Planeación Estratégica y Dirección de Tecnología

Presenta

Jorge Alberto Romero Hidalgo

Puebla, México. 2017

Universidad Popular Autónoma del Estado de Puebla
Centro Interdisciplinario de Posgrados
Investigación y Consultoría
Departamento de Ingeniería
Doctorado en Planeación Estratégica
y Dirección de Tecnología

**Se aprueba la Tesis:**
"Propuesta de un Estándar para México de Gestión del Conocimiento e Innovación Tecnológica"

**Nombre del Alumno:**
Jorge Alberto Romero Hidalgo

**Comité Asesor**

**Dr. Ricardo Rodríguez Carvajal**
Director de Tesis

**Dr. Cuauhtémoc Lemus Olalde,**
Centro de Investigación en Matemáticas A. C.
**Asesor**

**Dr. José Arturo Mora Soto,**
Centro de Investigación en Matemáticas A. C.
**Asesor**

Puebla, México.

12 de enero de 2017

# Agradecimientos

A mi familia, Papá, Mamá, Rudy y Rafa que hoy me acompañan en espíritu y a Pila y Genaro que siempre son solidarios y amorosos. A todos mis sobrinos y familias Romero e Hidalgo.

A Claudia y Fer por el apoyo, paciencia y colaboración para este trabajo, pero sobre todo por la motivación que representan en mi vida para seguir adelante. Que Fer siga adelante en el derrotero que se forje en la vida y logre su pleno desarrollo y felicidad. Mientras yo viva, contará conmigo. Gracias a La Güera por darle cariño y sazón a nuestra vida.

Al Maestro Juan Carlos Romero Hicks por haberme brindado siempre su amistad, confianza, apoyo y consejos e impulsarme y encauzarme hacia este reto. Mi eterno agradecimiento, amistad y respeto para él y su apreciada familia.

Al Doctor Luis Felipe Guerrero Agripino por su irrestricto apoyo a lo largo de este proceso y su constante apremio para su culminación. Mi reconocimiento y enorme gratitud.

A mis amigos y cercanos colaboradores que me orientaron en el desarrollo del trabajo con atinadas observaciones, aportaciones y sugerencias, Edgar Vázquez, Cecy Ramos y en especial a Héctor Pérez por su invaluable apoyo para conjuntar diversos elementos de análisis para el desarrollo de la tesis. Mil gracias por su tiempo, talento colaborativo y crítico para la culminación del estudio.

A Ricardo Rodríguez Carvajal por tan acertada y comprometida dirección de tesis. Sin su cercanía, orientación, talento, constante apoyo y apremio, el resultado final estaría aún lejano. Gracias a su distinguida esposa, Paula, por su apoyo.

A mis solidarios y talentosos asesores, Cuauhtémoc Lemus y Arturo Mora, quienes sembraron en mi horizonte la factibilidad para la realización de un estándar en la materia y que espero el que hoy se presenta se encuentre a la altura de sus expectativas. Mi amplio reconocimiento por todo su apoyo.

A mi querida Universidad de Guanajuato por haberme brindado esta invaluable oportunidad de formarme y colaborar en ella. A mis profesores y compañeros de estudio y trabajo. A Claudia Gutiérrez, Domingo Herrera, Héctor E. Rodríguez, Claudia Gómez, J. Iván Vázquez, gracias por el apoyo para alcanzar este reto.

A la UPAEP sus profesores y coordinadores.

A todos mis amigos, hermanos, Pancho, Gus, Javier C., Javier M., Memo, Carlitos, Javier B., Erasmo, Eloy, Adrián J., Mario, Alex L. Víctor, Rodrigo, Rafa, Elmer, Carlos H., Mago,.

# Resumen

El objetivo de esta investigación es elaborar una metodología que permita construir un indicador de gestión del conocimiento e innovación tecnológica aplicable al ámbito de las organizaciones en México cuya utilización redunde en un mejor uso de sus recursos y el mejoramiento de su productividad. A partir de la revisión de la literatura existente, se formula una propuesta que incluye diversos elementos y posibilitan a las organizaciones establecer una referencia específica sobre su posicionamiento con respecto a ambos conceptos.

La propuesta que a continuación se presenta se ha construido a partir de un esfuerzo sistemático por entender e integrar los modelos de gestión del conocimiento e innovación publicados en los últimos años, así como los resultados de las experiencias para proponer estándares de gestión del conocimiento e innovación. Para elaborar la propuesta, se han analizado los factores y sus componentes asociados, mediante una revisión sistemática de la literatura, para a partir de ello construir una propuesta de estándar.

Para validar empíricamente el estándar propuesto se ha construido un modelo de investigación de seis fases. Para ello se ha realizado un estudio exploratorio a profundidad en una organización del sector público, particularmente en un área que permite la replicabilidad del estándar. De esta manera, se han analizado los resultados bajo la perspectiva de interés de este trabajo: construir y validar empíricamente la propuesta del Estándar de Gestión del Conocimiento e Innovación Tecnológica.

Finalmente, luego del análisis estadístico correspondiente, se presentan los resultados obtenidos derivado de la aplicación del instrumento de evaluación, previamente validado, que apoya a la definición del estándar propuesto.

# Abstract

The purpose of this work is to offer a methodology that allows to construct a standard in Knowledge Management and Technological Innovation which may be used in various organizations in México to improve the operation of their resources and productivity. Based on the review of the existing literature, a model is offered including several elements to enable organizations to establish their position in relation to both concepts.

The following proposal is based on a systematic effort to understand and integrate models of Knowledge Management and Innovation published in recent years as well as the results of the experiences to propose standards of Knowledge Management and Technological Innovation. In order to elaborate the proposal, factors and their associated components have been analyzed through a review of the literature in order to build and validate a standard proposal.

To test the research study, a six-stage research model has been constructed. For this purpose, an in-depth exploratory research study has been carried out in a public sector organization, in an area that allows the replicability of the model. The results have been analyzed to construct and empirically validate the Mexican Standard of Knowledge Management an Technological Innovation.

Finally, after the statistical analysis, results obtained from the application of the validated instrument are shown , which supports the definition of the model.

# Índice General

# Índice de figuras

# Índice de tablas

# Lista de acrónimos

**ADIAT:** Asociación Mexicana de Directivos de la Innovación Aplicada y el Desarrollo Tecnológico

**BM:** Banco Mundial

**CMMI**: Capability Maturity Model Integration®

**EMGCIT:** Estándar Mexicano de Gestión del Conocimiento e Innovación Tecnológica

**GC:** Gestión del Conocimiento

**IEEE:** Institute of Electrical and Electronics Engineers

**IM:** Innovation Management

**MPS:** Mejora de Procesos de Software

**OCDE:** Organización para la Cooperación y el Desarrollo Económicos

**ONU:** Organización de las Naciones Unidas

**PGC:** Proceso de Gestión del Conocimiento

**PMI**: Project Management Institute®

**RSL:** Revisión Sistemática de la Literatura

**SGC:** Sistema de Gestión del Conocimiento

**TRL**: Technology Readiness Level

# Introducción

La sociedad actual se caracteriza por sus complejas dinámicas y transformaciones constantes. Somos testigos y actores de una sociedad digital, llamada sociedad del conocimiento, cuyos elementos esenciales son la información, la velocidad y el cambio. En los últimos años se ha producido tal cantidad de información, datos y conocimiento como no se tenía algún precedente en otras épocas. Ello, sin duda ha transformado la manera en cómo interactuamos en nuestro entorno con otros, con nosotros mismos y en la sociedad.

Ante tales circunstancias las organizaciones del presente no están exentas. Cada vez es más común escuchar términos como innovación, desarrollo científico y tecnologías de información al interior de éstas. A medida que nos aproximamos a la primera mitad de este siglo con frecuencia escuchamos hablar de la cuarta revolución industrial (Schwab & World Economic Forum, 2016; World Economic Forum, 2016b). Hoy, por ejemplo, es común escuchar hablar de *big data*, manufactura aditiva, internet de las cosas (IoT), industrias inteligentes, productos inteligentes y conectados, inteligencia artificial (IA), robótica, vehículos autónomos, nanotecnología, computadoras cuánticas, entre otras cosas.

Al transformarse las sociedades del siglo XXI, a sociedades en donde la producción y uso del conocimiento es intensa, también se hace necesario que las organizaciones encausen de manera estratégica el cambio definitivo que les permita desarrollare y transformarse para entregar más valor en sus operaciones (Cornell University, INSEAD, & WIPO, 2016). En la sociedad digital de la cuarta revolución industrial (Baller, Dutta,

& Lanvin, 2016; Michelman, 2016; Weill & Woerner, 2015), la innovación se ha vuelto un imperativo insoslayable (Baller et al., 2016).

En virtud de lo anterior, el presente trabajo plantea la elaboración de una herramienta que permita a las organizaciones –públicas, privadas o de otra naturaleza- establecer su posición relativa con respecto a dos conceptos que tienen un impacto significativo en su desempeño y productividad y que contribuyen en su conjunto a definir el logro de sus objetivos primordiales. La gestión del conocimiento (GC) y la innovación han sido abordados por connotados especialistas a partir de la sexta década del siglo pasado (Carneiro, 2000) y aunque han sido analizados en lo general, desde ópticas independientes, se pretende establecer cuál es el grado de integración entre ambos conceptos, para finalmente realizar una propuesta que, de forma integral, converja en un estándar para la determinación de la gestión del conocimiento y la innovación tecnológica.

En el **capítulo 1,** se aborda la justificación del presente trabajo. En el mismo se analiza el desarrollo de la ciencia, la tecnología y la innovación en los últimos años, para dar pauta a la justificación natural sobre el objeto de estudio. Asimismo, debido a la escasa oferta de herramientas que con rigor técnico y metodológico permitan establecer al interior de las organizaciones su grado de avance con respecto a la aplicación de estrategias y acciones en materia de gestión del conocimiento e innovación tecnológica se hace un análisis del estado del arte en este ámbito.

Para abonar en la justificación del presente trabajo, se estudian las definiciones y se analizan los principales conceptos objeto de estudio; enseguida se establecen los objetivos de esta tesis, las preguntas de investigación, así como la determinación de los alcances y limitaciones.

El **capítulo 2**, profundiza en el análisis del marco teórico, a través de una búsqueda sistematizada de las principales corrientes de pensamiento y contribuciones de expertos en el objeto de estudio. Así se puede dar pauta a la construcción sistémica y metodológica de un instrumento válido que permita la contrastación empírica del objeto de estudio en un entorno específico.

El **capítulo 3** describe la metodología a utilizar, partiendo del diseño de la propia investigación, determinación de variables, así como el diseño de los instrumentos necesarios para obtener la información correspondiente. A partir de la formulación de la hipótesis y objetivos de la investigación planteados en el Capítulo 1, se realizó la recolección de la muestra en la organización que fue definida para probar la hipótesis de trabajo y cumplir los objetivos planteados.

El **capítulo 4**, da cuenta de los resultados obtenidos, el análisis a fondo de los principales hallazgos y la contribución empírica y conceptual que se realizó, así como las dificultades en la aplicación del estándar y limitantes del propio trabajo de investigación.

Las conclusiones y recomendaciones son el corolario al presente trabajo.

Es importante mencionar que este trabajo de tesis se ha presentado de manera parcial (Romero Hidalgo, Pérez López-Portillo, & Rodríguez Carvajal, 2016) en el 8º Congreso de Investigación Científica Multidisciplinaria en el mes de noviembre del año 2016, producto de esta presentación la versión final se ha enriquecido. El avance de investigación está disponible en: https://goo.gl/Ui64Ux

# CAPÍTULO 1. JUSTIFICACIÓN Y ESTRUCTURA DE LA TESIS

Las diversas reconfiguraciones sociales, que han tenido lugar a raíz de los procesos de mundialización detonados por la globalización, el auge en la industrialización, y el desarrollo de nuevas tecnologías de la información y la comunicación, han propiciado, como nunca antes en la historia, transformaciones en múltiples sectores de la sociedad. Un mundo de cambios constantes y de crecimientos exponenciales en distintos ámbitos ha sido un elemento persistente en los últimos años; tan sólo la población mundial pasó de 3.03 billones de habitantes en 1960, a 7.2 billones de habitantes en el año 2014, es decir más de 4 billones de nuevos habitantes en menos de 55 años (Department of Economic and Social Affairs, 2015; United Nations, 2015).

**Figura 1.1.** ***Crecimiento de la población mundial***

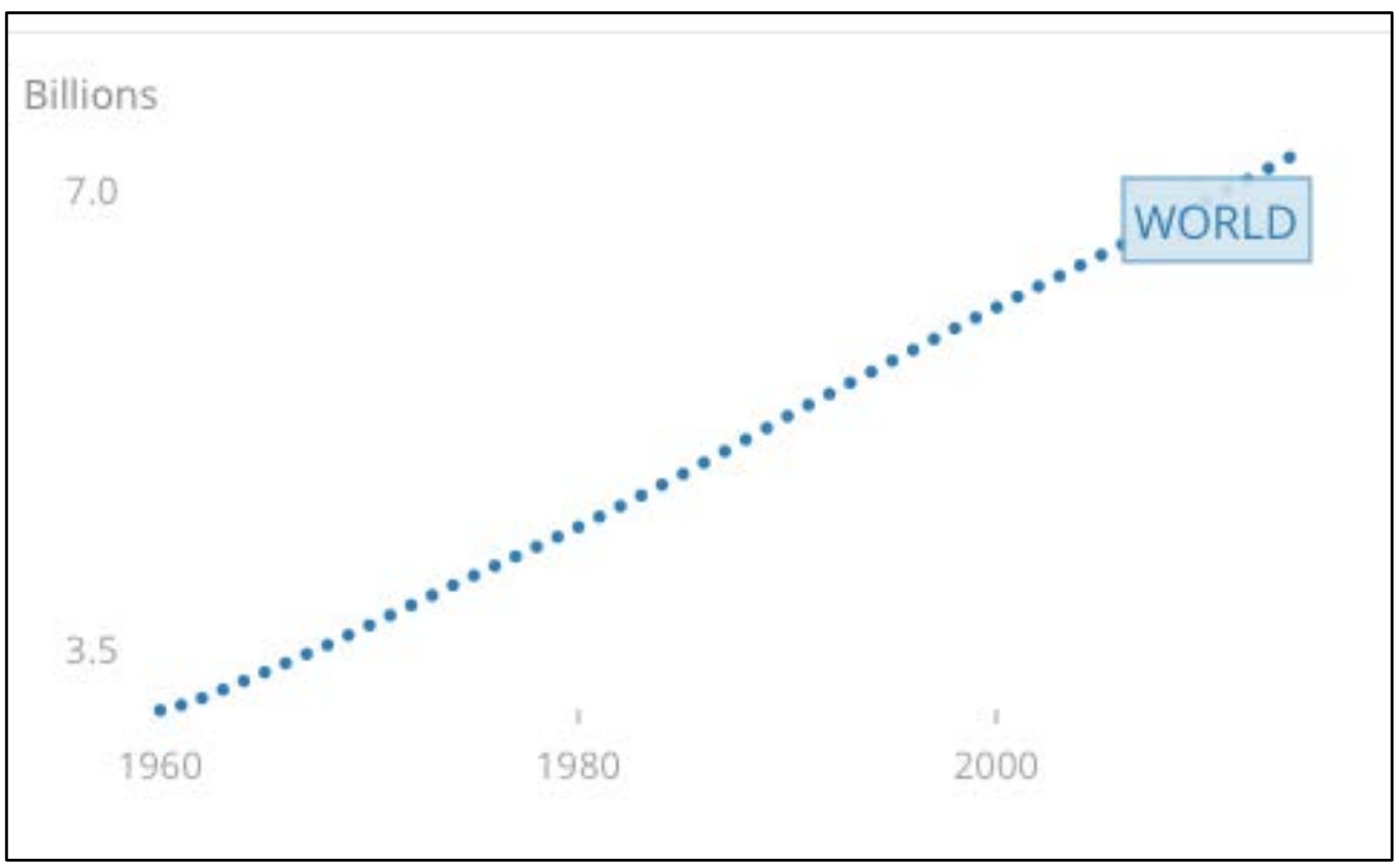

**Fuente**: Banco Mundial (2016)

La acelerada producción de información, trajo consigo a lo que se denominó la Era de la Información, espacio temporal caracterizado por la profusión de la información, los datos y el conocimiento. Consecuentemente, desde comienzos del siglo XX, la producción del conocimiento se ha incrementado sustancialmente en distintos sectores de la sociedad (OECD, 2000). Este incremento ha sido tan sustancial, que desde finales de ese siglo se comenzó a hablar de una economía del conocimiento (Powell & Snellman, 2004), que como su nombre lo indica esta economía usa como su principal insumo, o activo, el conocimiento para la producción de más conocimiento.

Hoy en día, se estima que el 46.4% de la población mundial utiliza internet (Miniwatts Marketing Group, 2016), esto equivale a más de 3.3 billones de usuarios, tendencia que crece año con año. Para el caso de México, se estima que son 60 millones de usuarios, lo que equivale a un 73.2% del total de la población. Innegablemente nos encontramos en la era de la híper-conectividad (World Economic Forum, 2016a). En nuestro país existen 82.2 líneas de teléfono celular por cada 100 habitantes. Los usuarios de internet mexicanos, refieren que lo usan principalmente para obtener información y para consultar sus redes sociales (INEGI, 2015).

**Figura 1. 2.** ***Líneas celulares por cada 100 habitantes, en México***

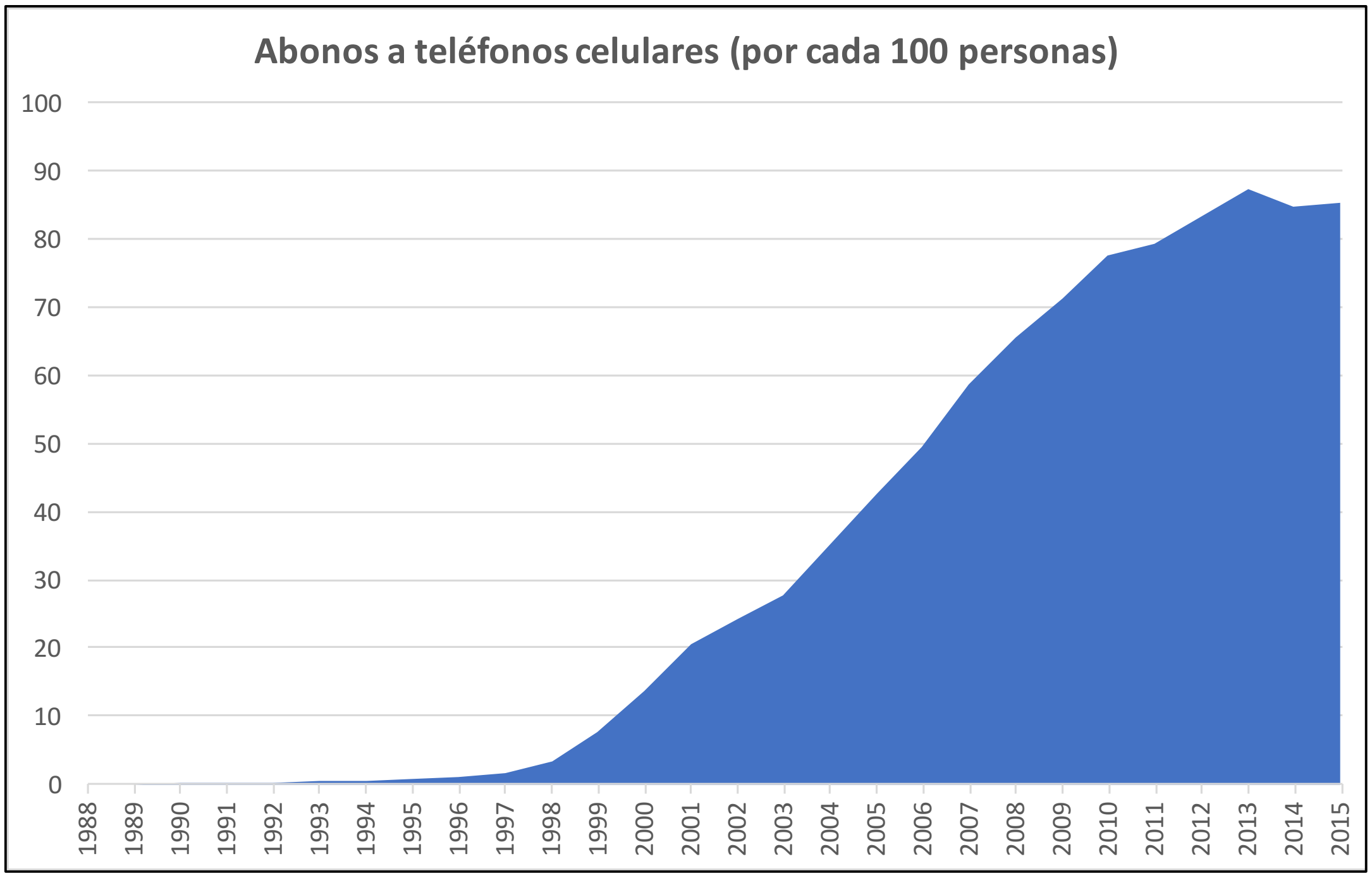

**Fuente:** Banco Mundial (2016)

Entonces, podemos observar como la economía del conocimiento ha ido configurando, lo que algunos autores, denominan sociedades del conocimiento (Bindé & Matsuura, 2005), en las que el conocimiento es el recurso más importante y de mayor valor, y que generalmente se caracterizan por ser economías industrialmente avanzadas (Hislop, 2005).

Consecuentemente, una sociedad del conocimiento es una sociedad que se nutre de sus diversidades y capacidades (UNESCO & Bindé, 2005). Este tipo de sociedades y economías necesitan organizaciones basadas en el conocimiento (David Rooney, Hearn, & Ninan, 2005). Las economías del conocimiento suponen nuevos paradigmas para la innovación y para el avance en el conocimiento con relación al desarrollo económico (OECD, 2004).

El tema que se pretende estudiar es relevante y pertinente a la luz de su, tantas veces reconocido, poder transformador en la sociedad actual (OECD, 2004, 2015e). Nuestra época ha sido escenario de tantas transformaciones y cambios radicales tan considerables que aun cuando se seguía hablando de una tercera revolución industrial (UNESCO & Bindé, 2005, p. 25), hemos transitado hacia una cuarta revolución industrial (Schwab & World Economic Forum, 2016), caracterizada por sistemas ciber-físicos, que desafían lo conocido hasta ahora, la forma en que hacemos lo que hacemos, y nuestros paradigmas para habitar la tierra. Incluso, hay quienes ya comienzan a hablar de la Era, denominada, de la híper-conexión (World Economic Forum, 2016a).

**Figura 1.3. Porcentaje de artículos científicos publicados en el mundo, por área del conocimiento**

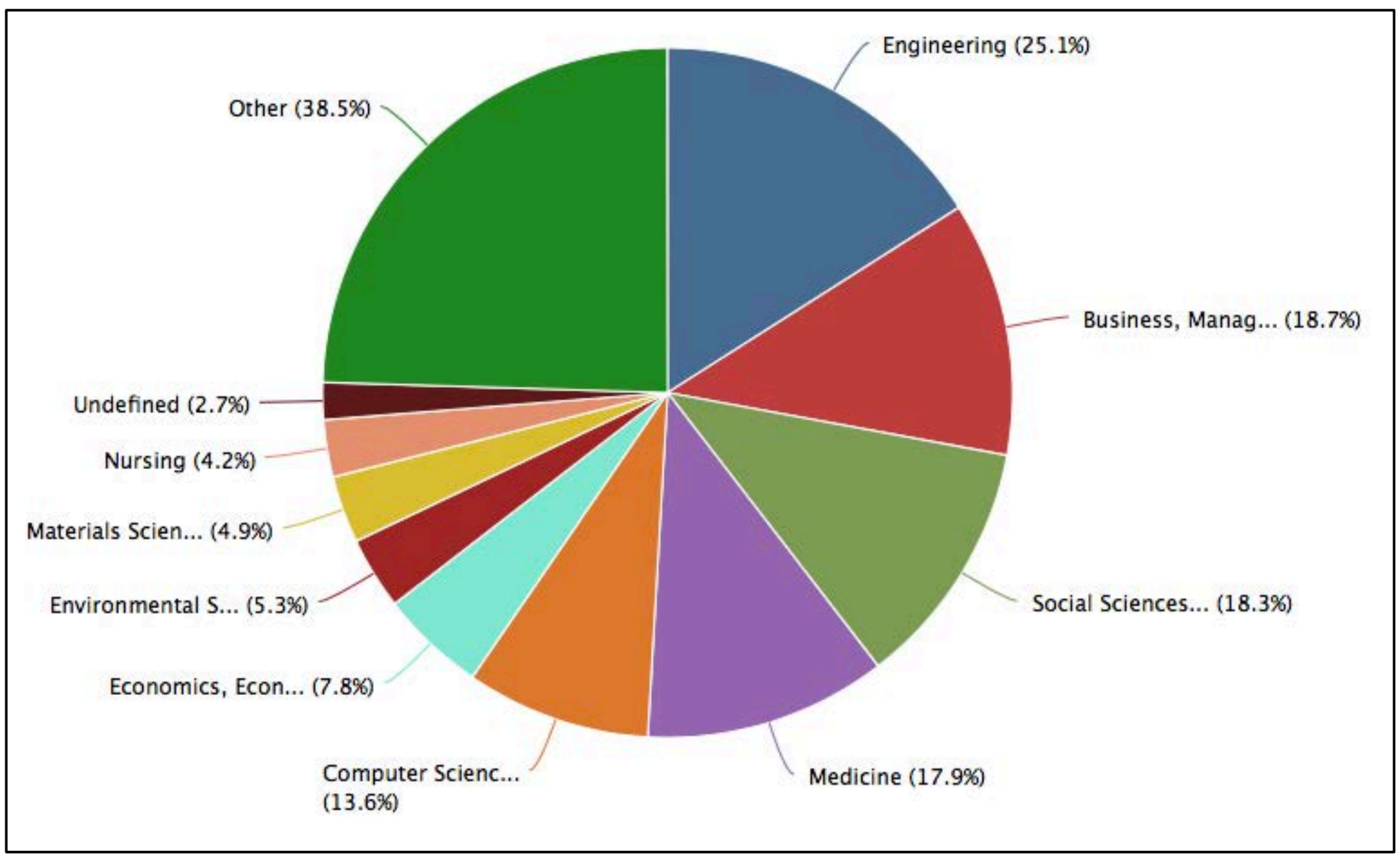

**Fuente**: Scopus® (2016)

En la actualidad, hablar de ciencia, tecnología e innovación resulta común y hasta ordinario. Nos hemos, al menos en los últimos años, acostumbrado a estos conceptos de alta plasticidad semántica, de definición multívoca, que incluso hemos abusado en sus referencias. La producción científica ha crecido considerablemente, la figura 1.3, muestra el número de artículos científicos publicados en las áreas de física, biología, química, matemática, medicina clínica, investigación biomédica, ingeniería y tecnología, y ciencias de la tierra y el espacio.

**Figura 1.4.** ***Olas de desarrollo tecnológico, entre los años 1770–1990***

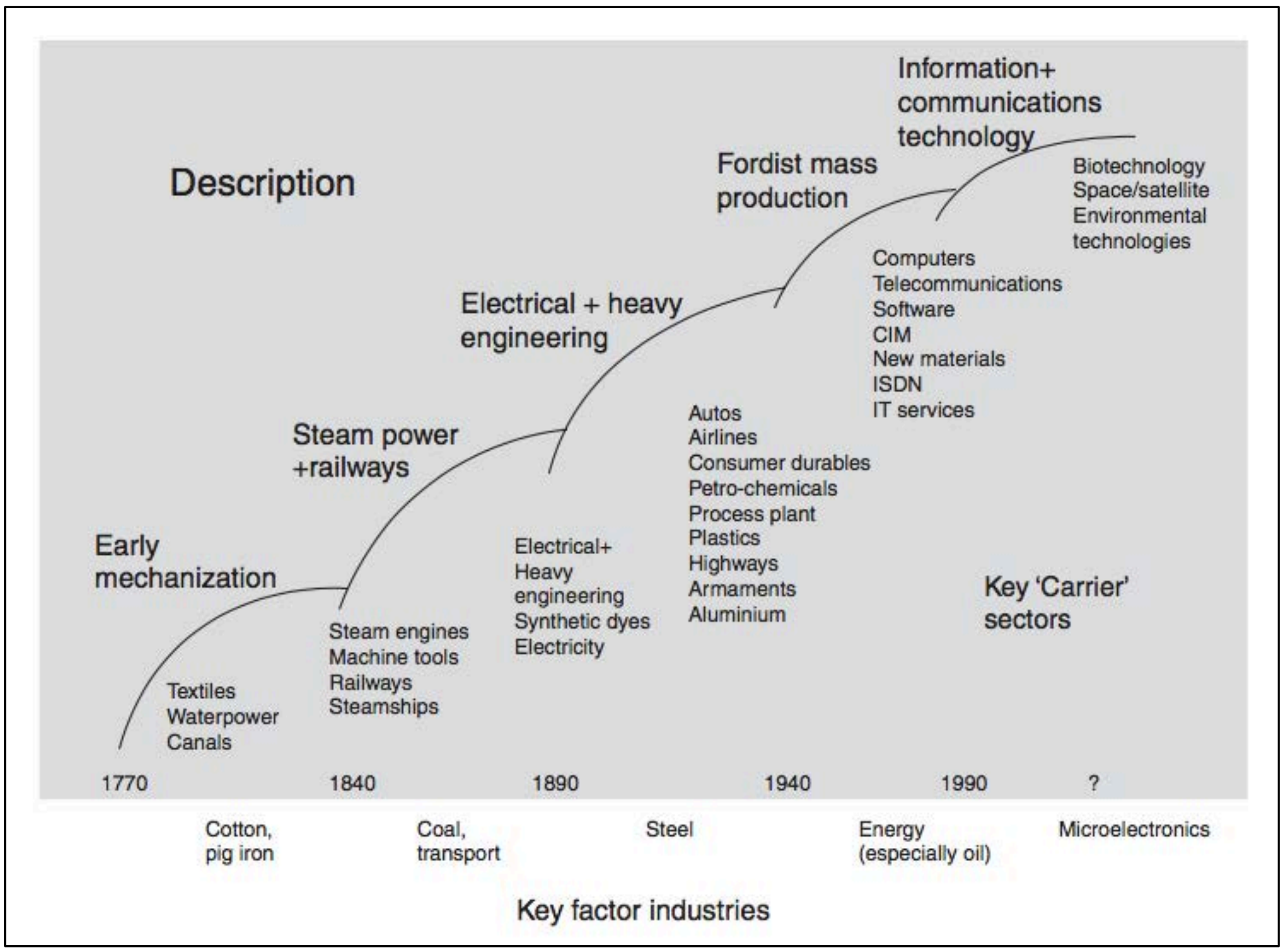

**Fuente**: (Dodgson, Gann, & Salter, 2008, p. 27)

Sin embargo, aún hoy nos encontramos con grandes retos: la inversión en investigación y desarrollo con respecto al Producto Interno Bruto (OECD, 2016c) sigue siendo en nuestro país una de las más bajas entre los países miembros de la OCDE al no alcanzar ni siquiera el medio punto del PIB, tal y como se muestra en el siguiente gráfico.

**Figura 1.5.** ***Gasto en I+D como porcentaje del PIB***

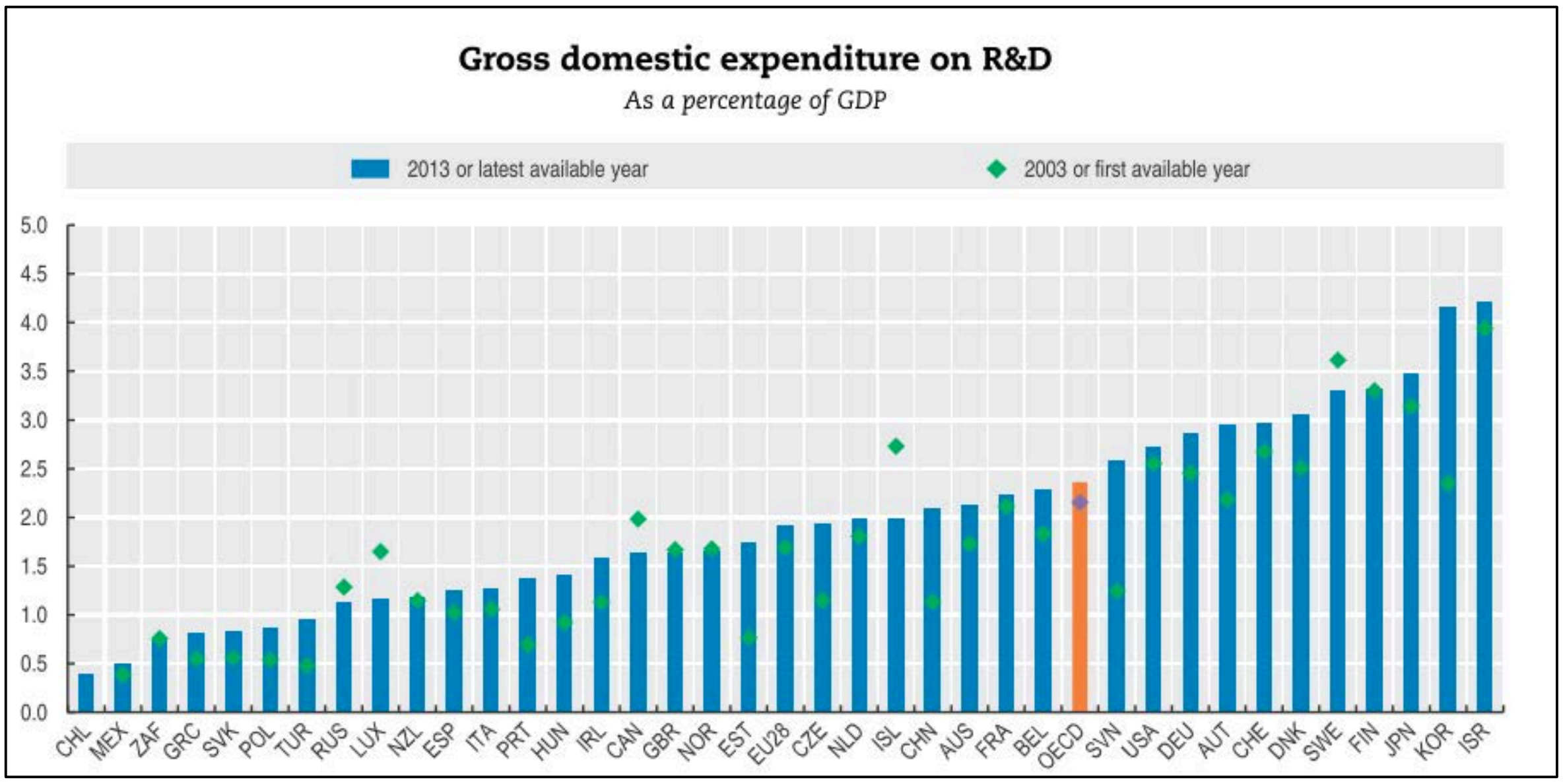

**Fuente**: OCDE Factbook (OECD, 2016c)

El número de investigadores de tiempo completo por cada 10,000 habitantes en México es significativamente bajo pues no llegamos ni a 2. Y para agravar la problemática, un alto porcentaje de ellos se vinculan laboralmente al sector académico y no al sector industrial.

**Figura 1.6. Número de investigadores (en miles)**

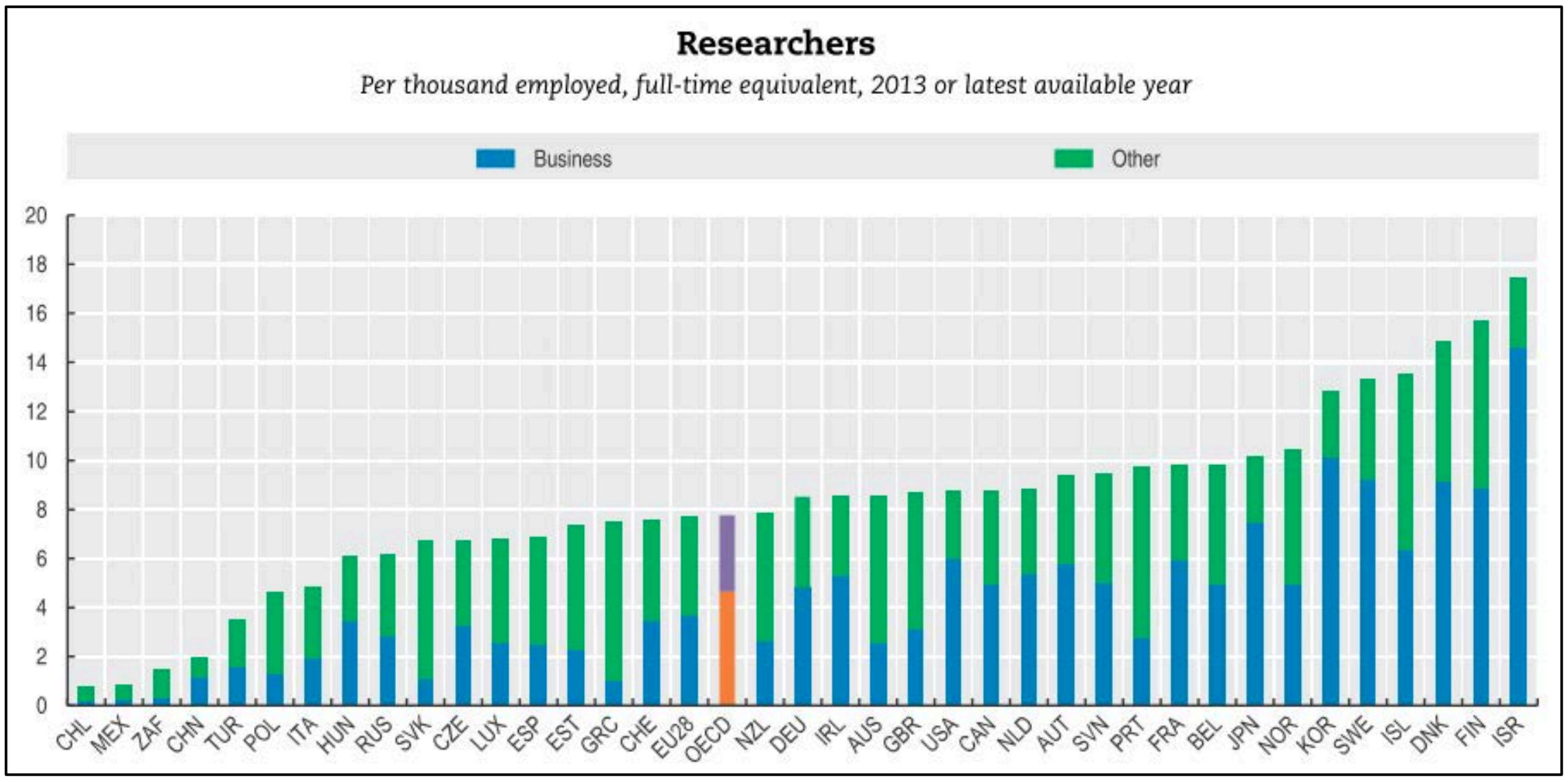

**Fuente**: OCDE Factbook (OECD, 2016c)

Las patentes registradas en las tres principales oficinas de registro de patentes: Europa, Japón y Estados Unidos nos indican que estamos a una gran distancia del promedio de los países de la OCDE que arroja 40 por cada millón de habitantes.

**Figura 1.7.** ***Número de patentes (por millones de habitantes)***

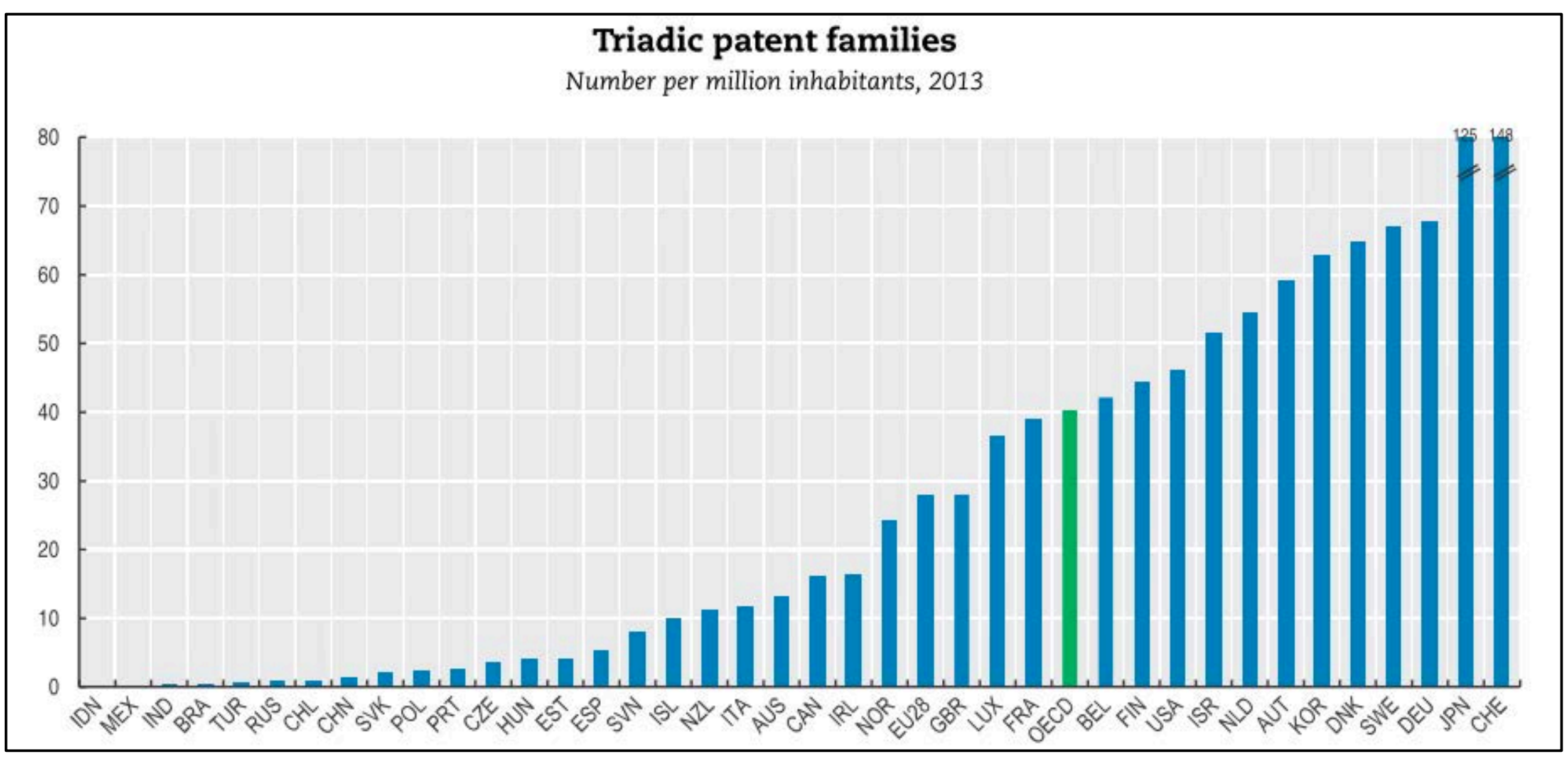

**Fuente**: OCDE Factbook (OECD, 2016c)

Desde mediados de la década de los 80, el gasto de la OCDE en investigación básica ha aumentado más rápido que el gasto en investigación aplicada y desarrollo experimental, reflejo del énfasis que muchos gobiernos dan al financiamiento de la investigación científica. La investigación básica sigue muy concentrada en las universidades y en los centros de investigación públicos. Un porcentaje considerable de la I+D que se realiza en esas instituciones se dedica al desarrollo; en Corea (35%) y en China (43%). En términos generales, en 2013 China invirtió relativamente poco (4%) en investigación básica en comparación con la mayoría de las economías de la OCDE (17%) y su gasto en I+D sigue muy orientado al desarrollo de infraestructura para ciencia y tecnología; es decir, laboratorios y equipos (OECD, 2015d).

Una nueva generación de Tecnologías de la Información y la Comunicación (TIC), como las relacionadas con el Internet de las Cosas, los datos masivos y la computación cuántica, más una ola de invenciones en salud y materiales avanzados están sentando las bases para transformaciones profundas en la forma en que vamos a trabajar y vivir en el futuro. En los años 2010 al 2012, Estados Unidos, Japón y Corea lideraron la invención en esos campos (representando en conjunto más del 65% de las familias de patentes solicitadas en Europa y Estados Unidos) seguidos por Alemania, Francia y China (OECD, 2015d)

**Figura 1. 8. Gasto público y privado en educación como porcentaje del PIB *(2013)***

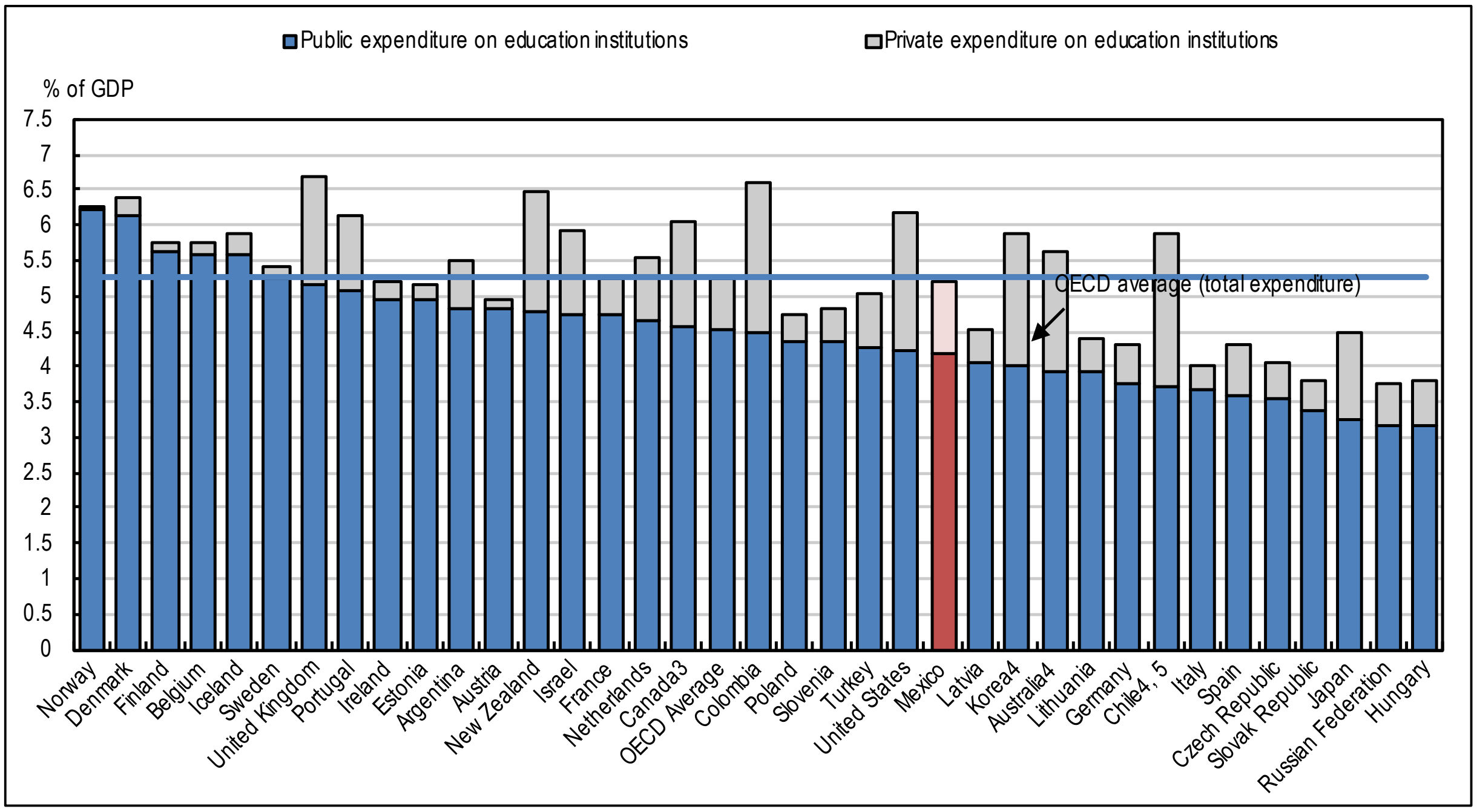

**Fuente**: (OECD, 2016)

## 1.1. Contexto de la innovación y el conocimiento en años recientes

Por otra parte, se decidió hacer uso de la herramienta Scopus Top Cited (2016), *por ser la que  la mayor base referencial de resúmenes y citas que actualmente hay en el mundo* (CONACYT, 2016), para ubicar crecimiento del número de publicaciones en la materia, usando las palabras clave innovación y gestión del conocimiento (inglés), se encontraron los siguientes resultados: 273,928 documentos usando la palabra “innovación”, que muestra claramente una tendencia con incremento exponencial a raíz de la conformación de la web en el año 1993. Una aparente disminución en el número de publicaciones a partir del año 2010 podría deberse al tiempo de evaluación de las publicaciones

**Figura 1.9. Investigadores y gasto en investigación y desarrollo (2014)**

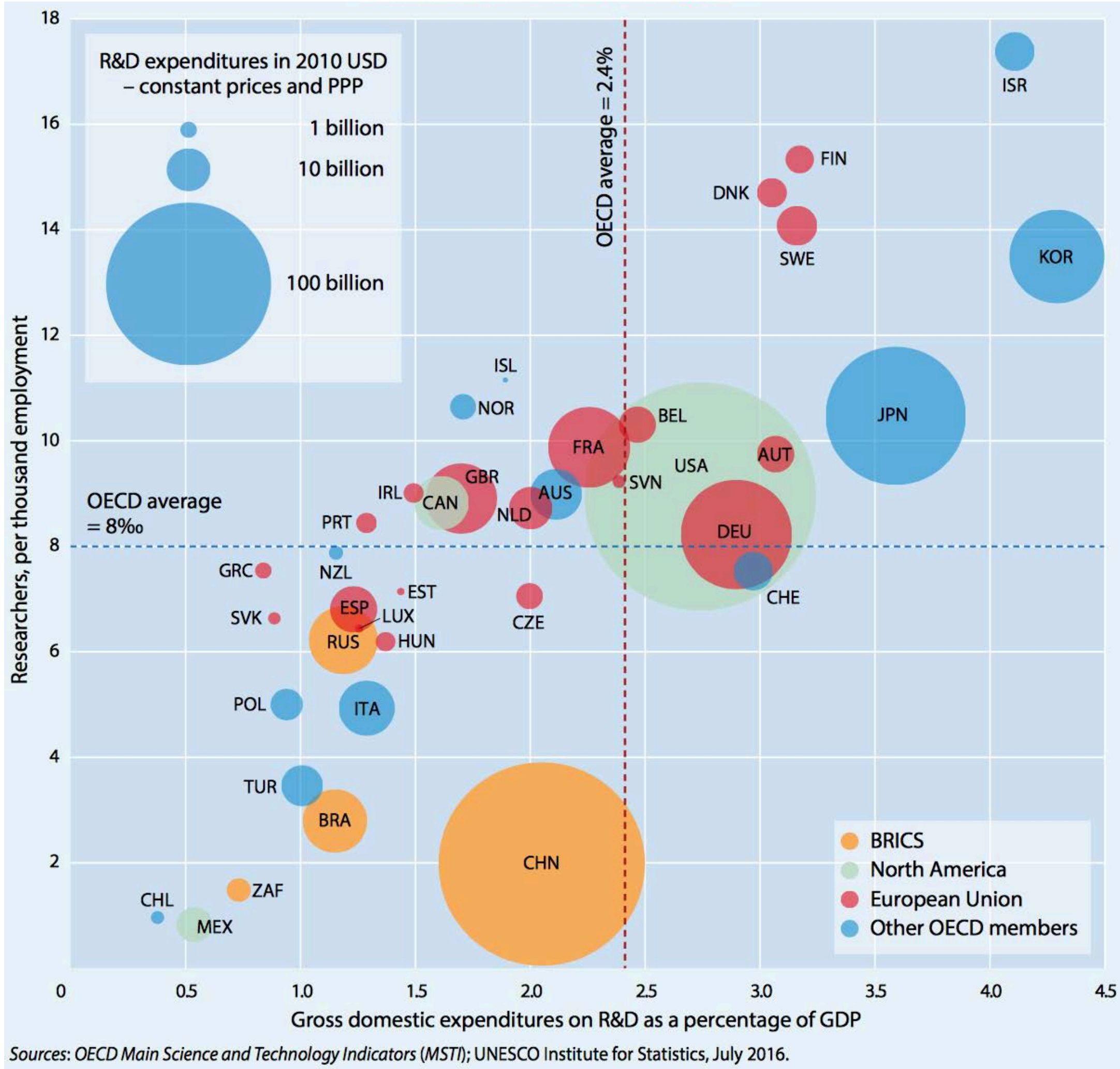

**Fuente**: (OECD, 2015d, 2016d)

**Figura 1.10. Número de artículos publicados sobre innovación en el mundo, por año**

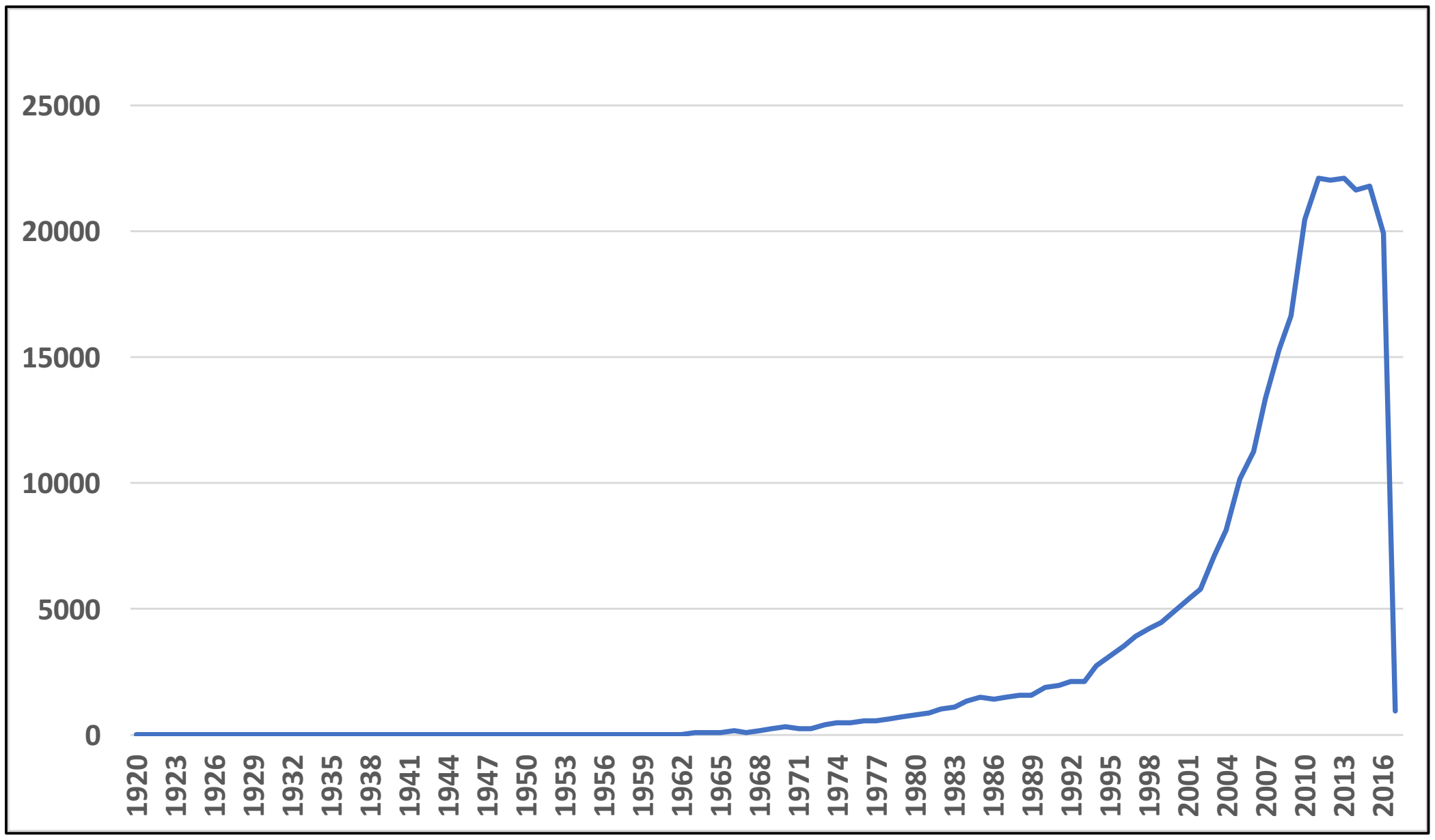

**Fuente:** (Scopus®, 2016)

Una referencia contextual contundente, es la producción de documentos alusivos a innovación y gestión del conocimiento arrojado por la búsqueda en la misma base de datos de Scopus® (2016).

**Figura 1.11.** ***Países con mayor número de artículos publicados sobre innovación***

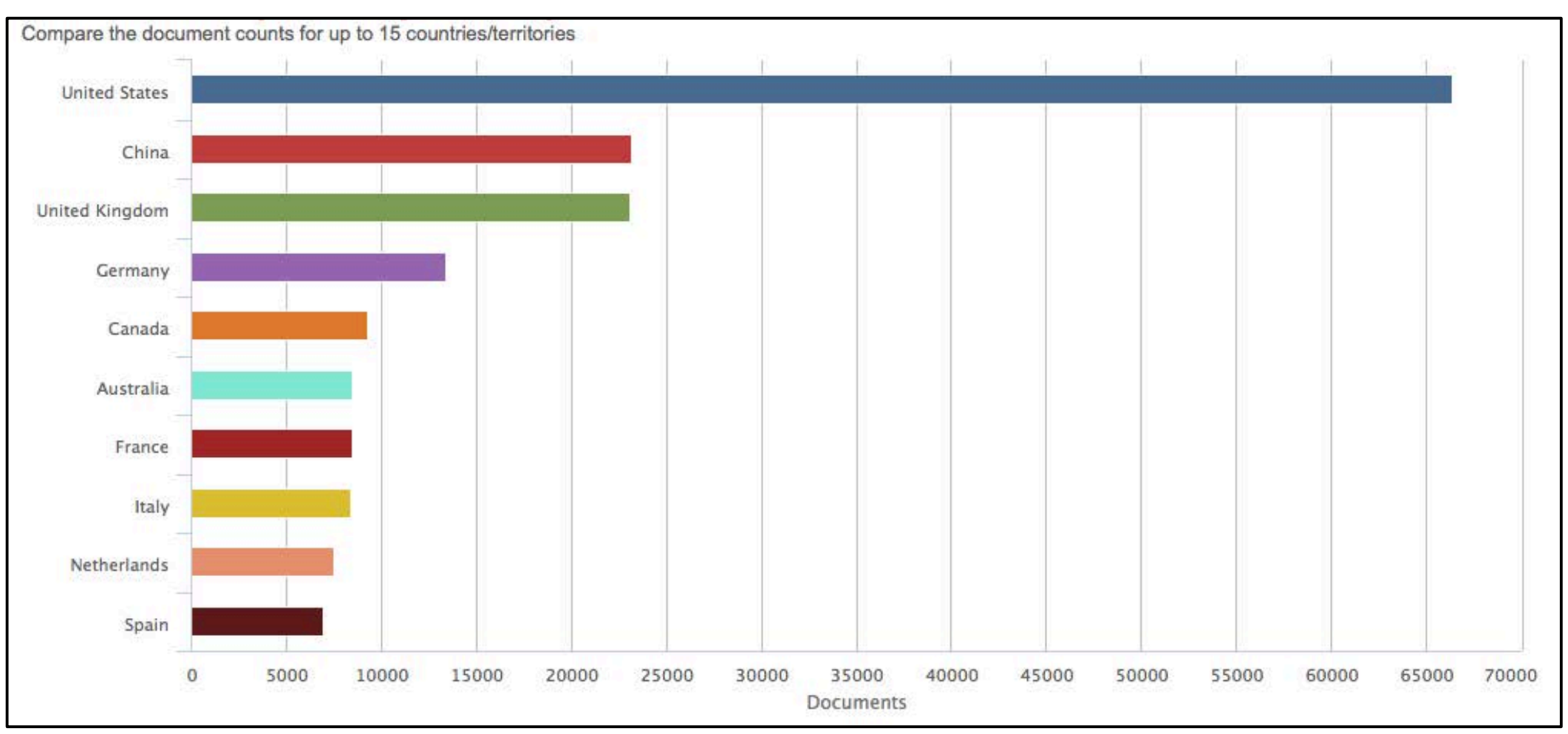

**Fuente:** (Scopus®, 2016)

Utilizando el mismo método de búsqueda descrito líneas arriba, se encontraron 357,962 documentos usando la palabra conocimiento en inglés (*knowledge*) desde 1950 a la fecha, encontrando la misma tendencia de crecimiento a partir de la puesta en marcha de Internet.

**Figura 1.12. Número de artículos publicados sobre conocimiento en el mundo, por año**

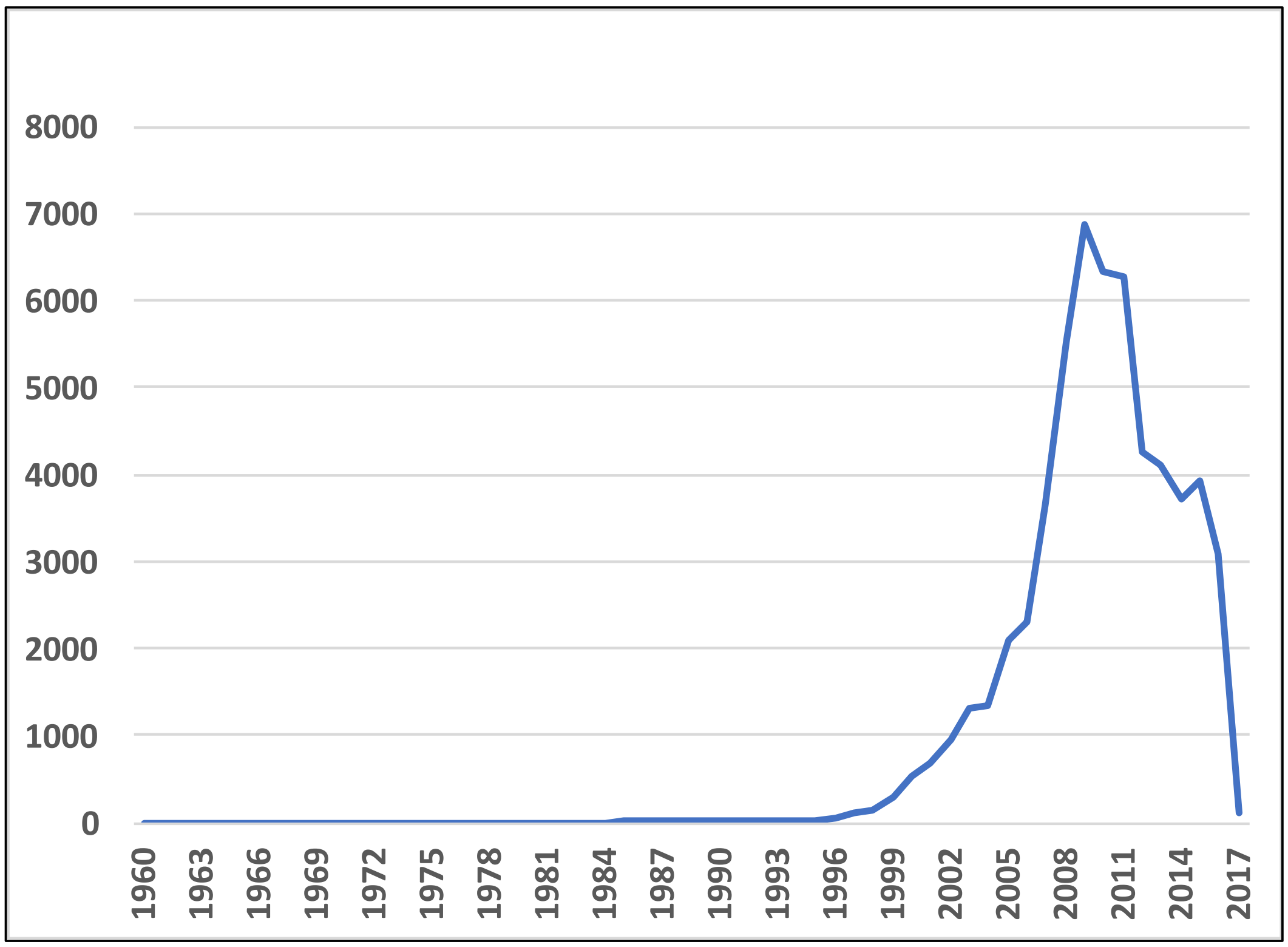

**Fuente**: Scopus® (2016)

En ese sentido y con el objetivo de describir sistemáticamente los elementos que serán objeto de estudio de la presente tesis, tomaremos algunos conceptos que son claves para aterrizar esta propuesta de investigación, por un lado el conocimiento, quizá, como ya se ha expuesto genéricamente, el activo intangible más importante de este siglo (Davenport & Prusak, 1998b; Polyani, 1958; J.-C. J. Spender, 1996), y sus múltiples manifestaciones y definiciones; y por otra parte; la innovación, fuente crucial de competitividad, de desarrollo económico y de transformación de la sociedad moderna (OECD, 2004, p. 7), imperativo para el desarrollo del mundo actual (Cornell University, INSEAD, & WIPO, 2015; Cornell University et al., 2016; OECD, 2015e) y elemento clave del crecimiento económico y del desarrollo, que permite a las naciones alcanzar estándares de vida más altos (OECD, 2015e).

**Figura 1. 13. Estructura general del desarrollo de la tesis**

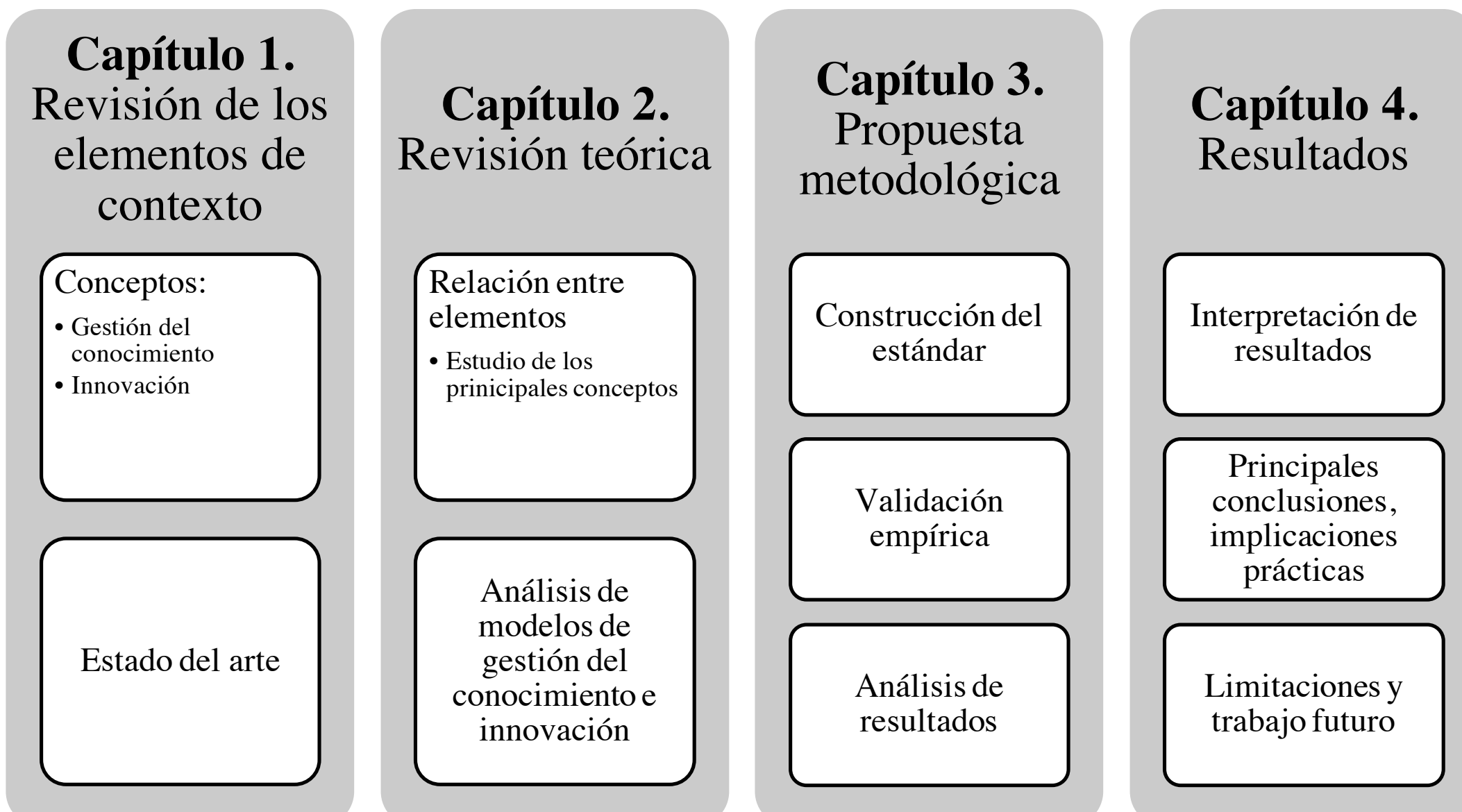

**Fuente**: elaboración propia (2016)

Posteriormente, analizamos la relación entre gestión del conocimiento e innovación, binomio indisoluble de los cimientos de una sociedad moderna, de una sociedad, denominada por antonomasia, del conocimiento, pues se observa una determinada propuesta de valor del rol de la gestión del conocimiento en la innovación (Plessis & du Plessis, 2007), a saber: apoya la creación de herramientas, plataformas y procesos para la creación de conocimiento tácito; facilita la conversión de conocimiento tácito en conocimiento explícito; fomenta la colaboración al interior de las organizaciones (Dugan & Gabriel, 2013; Fateh Rad, Seyedesfahani, & Jalilvand, 2015; Hamel, 2006; Morten T.; Birkinshaw, 2007) y propicia le creación de redes al exterior; incrementa la disponibilidad de conocimiento tácito y explícito necesarios para la innovación pues posibilita tener una perspectiva global sobre el conocimiento disponible (Leal-Millan, Roldan, Leal-Rodriguez, & Ortega-Gutierrez, 2016).

Así, es posible observar desde dónde puede ser obtenido y cuáles son las brechas que existen y que, en consecuencia, no han sido abordadas, pero que se requiere cubrirlas para alentar y materializar la innovación tanto en áreas estratégicas como las de procesos; apoya en la identificación de competencias clave por parte de las personas y propicia la obtención de conocimiento tácito y explicito para fortalecer dichas competencias (Kogut & Zander, 1992; Kotter, 2012); provee un contexto organizacional al cuerpo de conocimiento del grupo; aporta una cultura con enfoque en el conocimiento dentro de la cual la innovación puede ser incubada (Tan, Fischer, Mitchell, & Phan, 2009; B. Thomas, Miller, & Murphy, 2011).

Es un hecho incuestionable que las organizaciones enfrentan cotidianamente retos para los cuales no se encuentran preparadas pues las formas rutinarias para hacerles frente ya no responden ante circunstancias nuevas y complejas (Chick, Huchzermeier, & Netessine, 2014; Johnson, Yip, & Hensmans, 2012). El conocimiento explícito acumulado, y del que pueden echar mano, no es suficiente para dar una respuesta oportuna a nuevos planteamientos e incluso a nuevas oportunidades (Cohen, 1999; Dodgson et al., 2008).

En el peor de los casos, muchas de ellas ni siquiera son conscientes de que tienen ante sí alguno de estos desafíos pues sus mecanismos para inferir el medio ambiente o bien son nulos o en el mejor de los casos, demasiado ineficaces (Van der Panne, Van Beers, & Kleinknecht, 2003). Pocas tienen opción para hacer alguna labor mínima de inteligencia de mercado (Lester, 2014) u observaciones de tendencias tecnológicas que les adviertan sobre riesgos u oportunidades que el mercado les plantea.

Por otra parte, hablar de innovación es hablar de la transformación en la sociedad (den Ouden, 2012; OECD, 2015b). Es, sin duda, algo en lo que todas las naciones se deberían involucrar (Atkinson & Ezell, 2015). Aunque el concepto de innovación es complejo y diverso (Gonzalez, Llopis, & Gasco, 2013, p. 2025); la innovación es un proceso de creación de conocimiento a partir del conocimiento existente, de exploración y explotación de nuevas oportunidades (Martínez Piva, 2008, p. 335), va más allá de una simple actividad de descubrimiento por prueba y error (Siekmann, 2006). En la figura 14, se muestra la lista de los 10 países con mejores calificaciones a nivel global en el ranking de innovación (Global Innovation Index 2016).

**Figura 1. 14. Listado de países con mejores calificaciones en el índice global de innovación 2016**

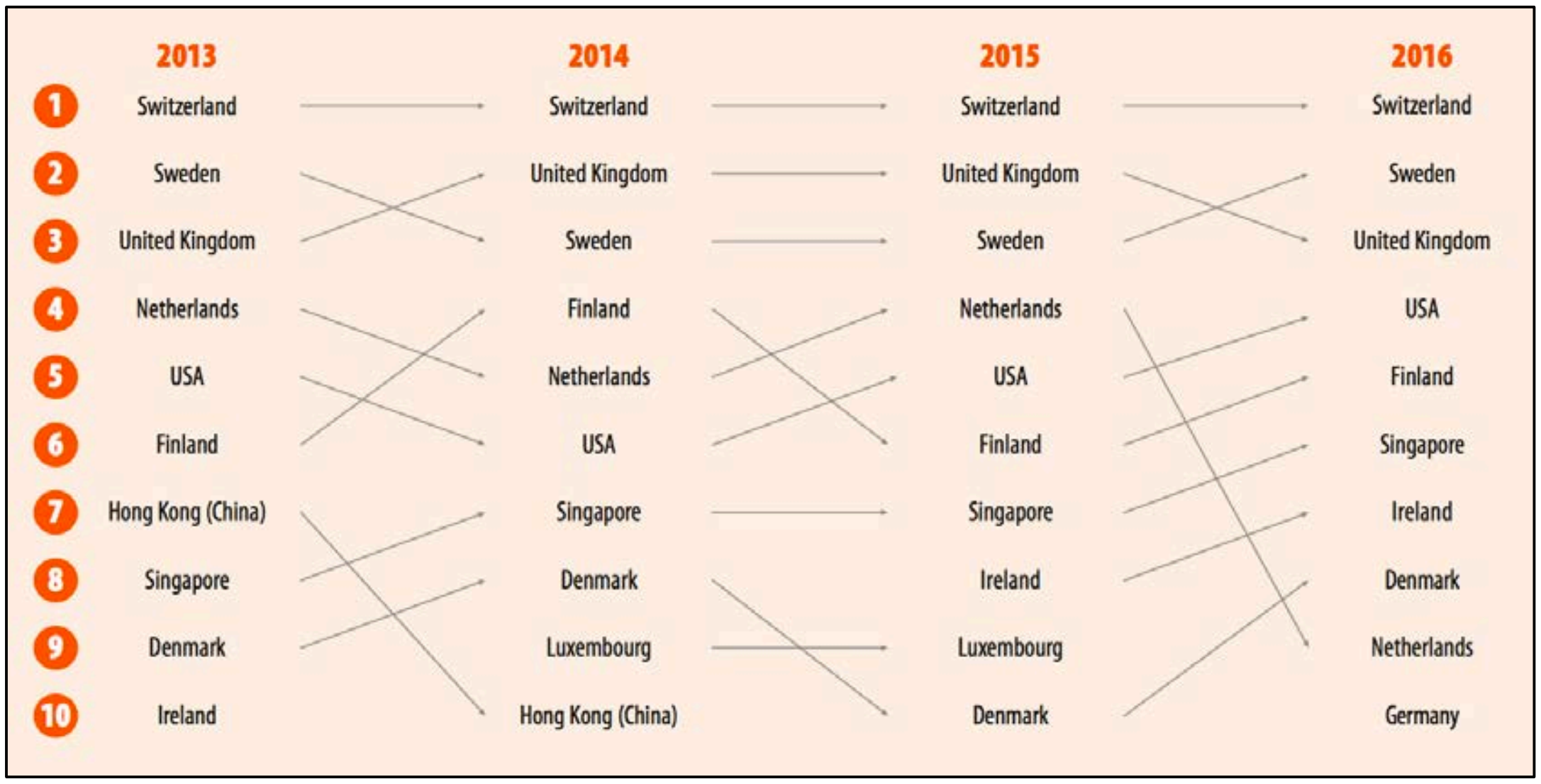

**Fuente**: (Cornell University et al., 2016, p. 15)

Es de destacar que la combinación de políticas que promueven la innovación, acompañadas de otras que tienen que ver con la facilidad para abrir negocios y la de transparencia y anticorrupción, son elementos que propician en su conjunto una mejor calificación en este tipo de rankings (Cornell University et al., 2016; Dutta, 2016).

Se describe a la innovación, generalmente, como una actividad intensiva en producción y uso de conocimiento, que implica el descubrimiento, la experimentación y el desarrollo de nuevas tecnologías, servicios, procesos de producción y las estructuras organizativas (Carneiro, 2000; Murray E Jennex & Smolnik, 2011).

Es entonces, el acto de generar y adoptar nuevas ideas, productos y procesos que son beneficiosos para el cumplimiento de los objetivos organizacionales (Clyde W.

Holsapple & Luo, 1996, p. 16), una innovación es la implementación de un producto nuevo o significativamente mejorado (VV. AA., 2009, p. 196).

De hecho, es la clave en la creación de valor (Cano-Kollmann, Cantwell, Hannigan, Mudambi, & Song, 2016, p. 255) y una fuerza muy disruptiva, que contribuye al proceso de *destrucción creativa en la economía*, y por lo tanto a la pérdida de empleos y a la reasignación de mano de obra y capital en la economía (OECD, 2015e, p. 24)

La innovación puede verse como resultado de la gran producción y análisis de datos (*big data*) (OECD, 2015a), y como síntesis de la implementación de prácticas de gestión del conocimiento (Belkahla & Triki, 2011; H. Inkinen, 2016; Lundvall & Nielsen, 2007; VV.AA., 2005), y de la efectividad de las mismas prácticas (Donate & Guadamillas, 2013; H. T. Inkinen, Kianto, & Vanhala, 2015). Es entonces, un parámetro de competencia (Lundvall & Nielsen, 2007), y con relación a la gestión del conocimiento, la innovación requiere una constante producción de conocimiento (Becerra-Fernandez & Sabherwal, 2015).

Sin duda, la innovación es crucial para la competitividad (VV. AA., 2015) y éxito de las organizaciones modernas (Chutivongse & Gerdsri, 2015); podemos ver su materialización a través de varias maneras. Algunas de ellas son : la reducción de los costes de producción, la mejora de los productos existentes y la creación de otros nuevos, o mediante la presentación y la venta de productos con mayor eficacia (OECD, 2014, p. 108).

En suma, en este trabajo de investigación se abordan ambos conceptos de manera interdependiente para estudiar y describir la relación que tienen como una estrategia de mejora en las organizaciones. Por ello, en el capítulo 3, se plantea la propuesta metodológica que busca integrar esta revisión teórica del estado del arte en un modelo de investigación, que permita (1) conocer la frontera del conocimiento sobre la innovación y la gestión del conocimiento y (2) proponer un estándar mexicano de gestión del conocimiento y la innovación. Finalmente, en el capítulo 4, se presentan los resultados obtenidos del estudio realizado, además se discuten algunas implicaciones prácticas y limitaciones del estudio realizado.

### *1.1. Objetivo general y objetivos particulares*

A partir de lo anterior, se presentan los objetivos de la investigación:

**Objetivo general:** Elaborar un estándar de gestión del conocimiento e innovación tecnológica aplicable a las organizaciones en México.

**Objetivos particulares:**

[OP1] Identificar las variables que integrarán el estándar de referencia.

[OP2] Construir un instrumento válido y confiable que pueda ser aplicado en las organizaciones en México.

[OP3] Aplicar la propuesta de estándar en una organización.

### *1.2. Planteamiento del problema*

Actualmente no existe un estándar mexicano de gestión del conocimiento e innovación tecnológica que permita a las organizaciones tener una guía, no prescriptiva, sobre cómo éstas gestionan lo que saben, transfieren su conocimiento, experiencias, mejores prácticas y resultados de sus innovaciones a otras organizaciones que les permita incrementar su conocimiento, la calidad del mismo y potencializar la contribución de sus innovaciones tecnológicas en las dinámicas organizacionales del presente.

En nuestro país aún nos queda un gran trabajo por hacer, México ha sido calificado como la nación 61 de entre 128 países en el índice global de innovación (Cornell University et al., 2016), que evalúa aspectos como las instituciones, el capital humano y la investigación, la infraestructura, el nivel de sofisticación de los mercados, el nivel de sofisticación de las empresas, los productos del conocimiento y la tecnología, así como los productos de la creatividad.

Aunado a lo anterior, de la revisión de la literatura se desprende que aún existe cierta falta de consenso metodológico entre los factores, y sus componentes, relacionados con la gestión del conocimiento y la innovación tecnológica en las organizaciones mexicanas, que podría aprovechar de mejor manera lo que saben y cómo saben lo que saben si existieran criterios orientadores sobre esta cuestión.

Particularmente sobre la gestión del conocimiento se ha observado que los elementos que deben estar presentes para que ocurra en las organizaciones, aún en la actualidad parecen estar dispersos y/o fragmentados (Garlatti & Massaro, 2016; Garlatti, Massaro, Dumay, & Zanin, 2014; Massaro, Dumay, & Garlatti, 2015). Sobre la innovación, más allá de que se hayan presentado distintos modelos a través del tiempo (Cornell University et al., 2016; Hernandez-Munoz, Torane, Amini, & Vivekanandan-Dhukaram, 2015; İzadi, Zarrabi, & Zarrabi, 2013; Stošić & Išljamović, 2010; Tidd, 2006), aún no existe un estándar que permita generar una estrategia que potencialice sus resultados, productos y servicios (Pisano, 2015).

Finalmente, el doctorado cursado en Planeación Estratégica y Dirección de Tecnología le ha permitido al autor de este trabajo, conjuntar su experiencia con los contenidos desarrollados a través de este programa doctoral, de tal manera que la propuesta es producto de cada una de las clases, seminarios, pláticas y orientaciones recibidas a lo largo de este periodo de formación, producto de lo cual se ha enriquecido.

### *1.3. Supuestos de investigación*

- **S1:** La definición de un indicador de Gestión de Conocimiento e Innovación Tecnológica posibilita a las organizaciones incrementar sus capacidades para crear, adquirir, compartir, almacenar, incrementar y usar racionalmente el conocimiento para mejorar su eficiencia y productividad.
- **S2:** La gestión del conocimiento favorece la innovación en las organizaciones y les permite ofrecer mejores productos y servicios.
- **S3:** En la medida en que se hace mejor gestión del conocimiento se facilita la innovación.

# CAPÍTULO 2. REVISIÓN DE LA LITERATURA Y ESTADO DEL ARTE

En este capítulo se abordan los conceptos que dan sustento al estudio presentando, siendo estos: gestión del conocimiento, innovación y estandarización. Lo anterior, a partir de las propuestas teóricas de los principales autores que abordan estos conceptos.

Se ha observado, a través de la revisión de la literatura, que contar con un estándar, podría apoyar a las organizaciones a identificar las diversas variables que inciden en el mejor aprovechamiento de sus recursos, así como para establecer, a partir de una medición objetiva, la mejor estrategia para hacerse ya sea del conocimiento que requieren o bien la mejor forma de administrarlo, comercializarlo y transferirlo para lograr una mejor innovación en sus productos, procesos o servicios.

**Figura 2.1. Descripción gráfica del contenido del capítulo**

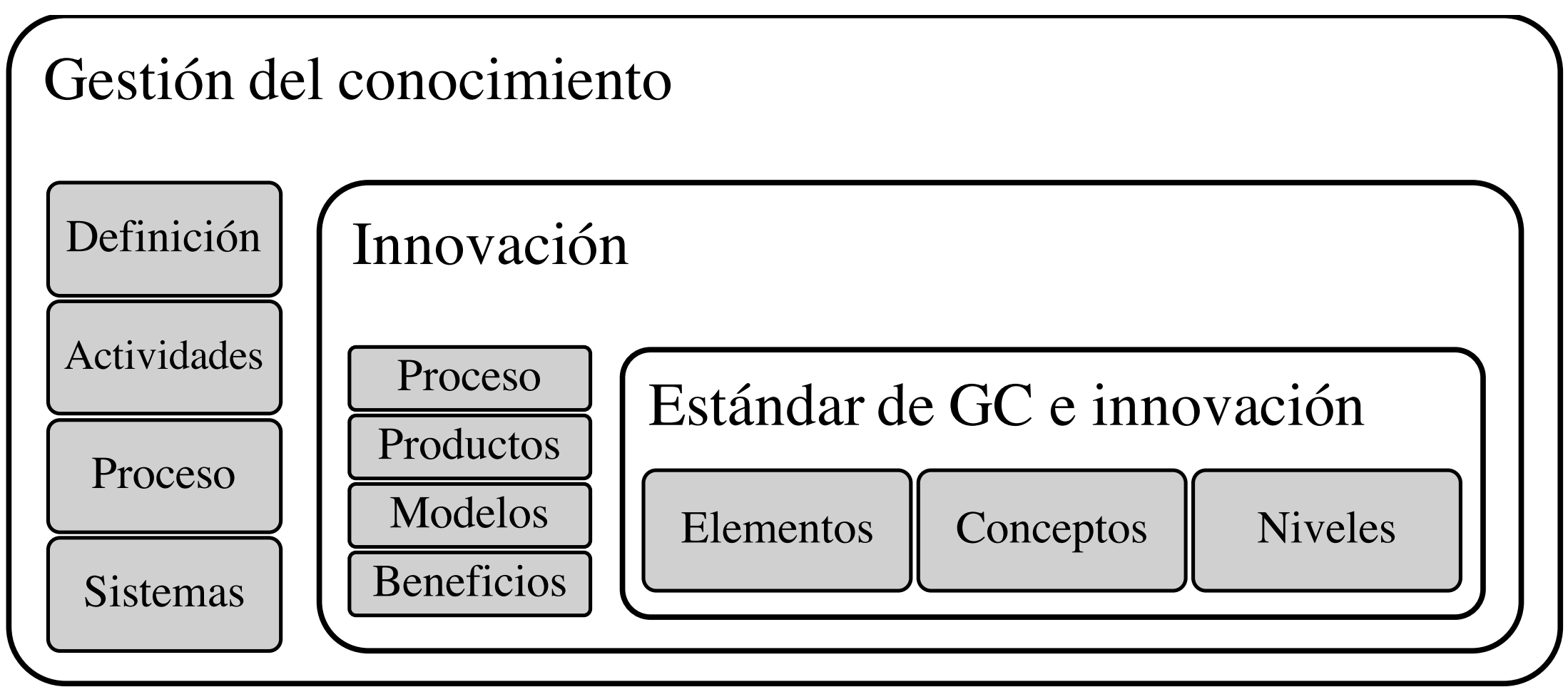

**Fuente:** Elaboración propia (2016)

## 2.1. El conocimiento y la gestión del conocimiento

### *2.1.1. El conocimiento*

La necesidad de aprovechar el sinnúmero de activos intangibles que poseen las organizaciones es un imperativo que podría determinar su éxito o fracaso. Desde finales del siglo XX, la producción mundial de conocimiento se ha incrementado considerablemente. Se puede afirmar que el conocimiento es un recurso fundamental de las sociedades modernas. Es por excelencia, como ya se ha señalado, el activo más importante de la denominada economía del conocimiento (OECD, 2004; Powell & Snellman, 2004; David Rooney et al., 2005).

Incluso, desde hace algunos años, en el siglo XVI, el conocimiento se ha reconocido su importancia, con la ya por excelencia y multicitada frase "*scientia est potentia*", atribuida al célebre filósofo inglés del Trinity College, Sir Francis Bacon, que da sustancia del cómo este recurso esencial se ha ido consumando como un elemento transformador del entorno.

No tan lejos de aquellos años, en nuestro tiempo, el conocimiento es un recurso fundamental y estratégico de las organizaciones (Wilcox King & Zeithaml, 2003), generalmente relacionado con otro recurso de igual importancia: el tiempo (Ragab & Arisha, 2013). Por tal motivo, se refuerza la idea de que es quizá el activo más importante del siglo XXI (Tianyong Zhang, 2010, p. 572).

El conocimiento engloba las ideas guardadas en la mente, realidades, conceptos, datos y técnicas de la memoria humana (Ikujiro Nonaka, 1991; Ikujiro Nonaka &

Takeuchi, 1999). Su fuente es la mente humana, y se basa en la información que se obtiene a través de la experiencia, las creencias y los valores personales. Su transformación ocurre cuando se asocia con decisiones o acciones (M. Allameh, Zamani, & Davoodi, 2011, p. 1227).

Como se ha referido, la producción mundial de conocimiento ha incrementado con gran impulso en los últimos años. Un indicador comúnmente asociado con la creación del conocimiento es el número de patentes concedidas. En la figura 2.2, observamos el número de patentes concedidas en los últimos años, desde los años 60's a la fecha, mismo que ha ido creciendo exponencialmente.

**Figura 2.2. Número de solicitudes de patentes vs. patentes concedidas**

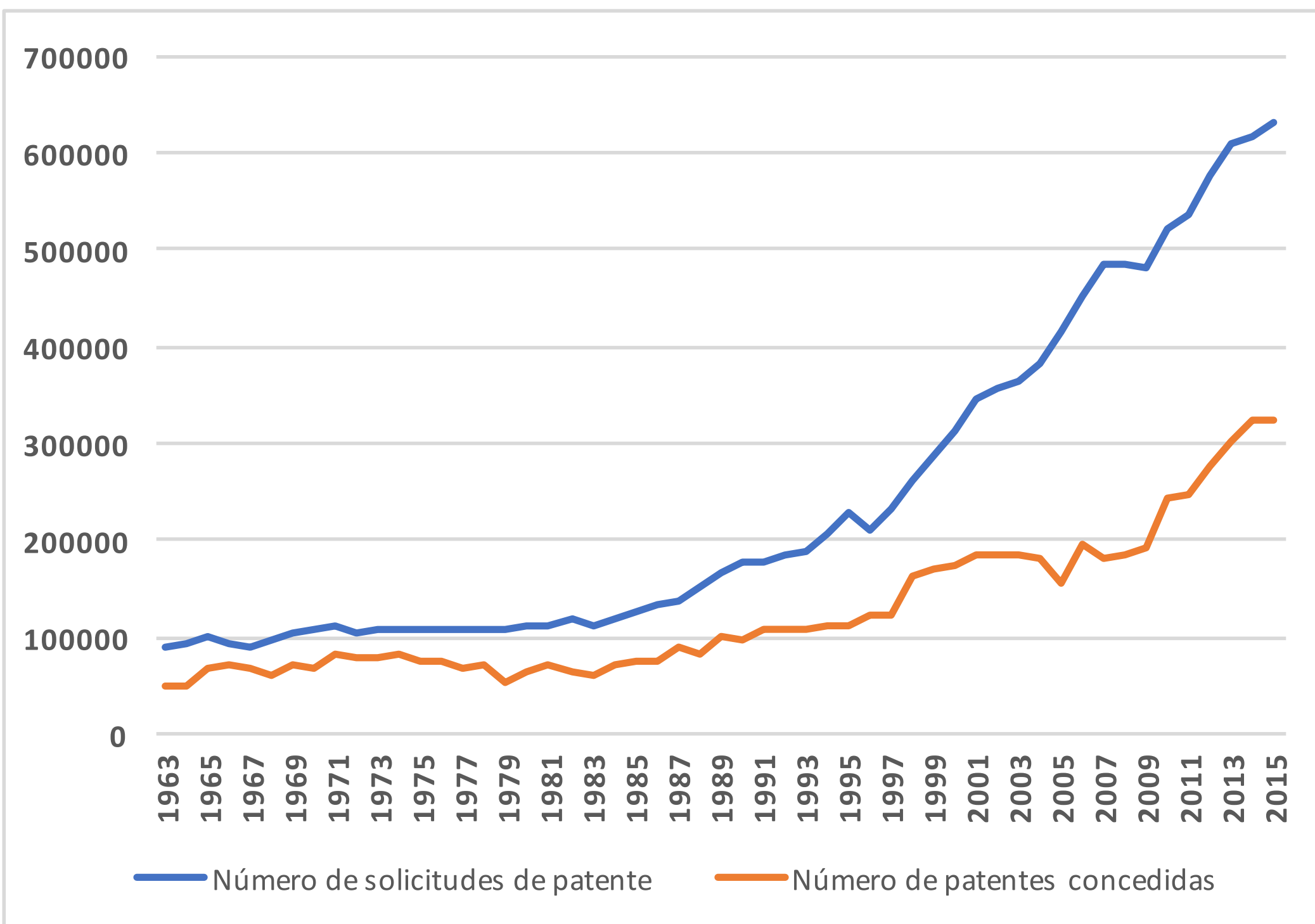

**Fuente**: U.S. patent and trademark office (2016)

Dada la compleja naturaleza del conocimiento, como un activo intangible (Petty & Guthrie, 2000), se ha vuelto necesario tener que administrarlo de alguna manera; por una parte para evitar que se pierda (Shafique, 2015), pero además para obtener el máximo beneficio de este activo (J.-C. Spender & Grant, 1996). En consecuencia, gestionar el conocimiento se ha vuelto cada vez más necesario en las organizaciones, para que éstas "conozcan lo que saben" (Davenport & Prusak, 1998a; Kogut & Zander, 1992) y como una fuente de ventaja competitiva en entornos altamente cambiantes (Coff, 2003).

### *2.1.2. La gestión del conocimiento*

La Gestión del Conocimiento (GC) (Choo, 1998; Ikujiro Nonaka, 1991), aparece como una solución que permite administrar lo que la organización sabe, integra a las personas, la tecnología, los procesos y la estructura de una organización (Tianyong Zhang, 2010), consiste en "*llevar el conocimiento correcto a las personas adecuadas en el momento que lo necesitan con el fin de tomar una acción concertada*" (Murray E Jennex & Smolnik, 2011, p. 75).

Por tanto, puede ser vista como "un esfuerzo sistemático y deliberado para coordinar a las personas, la tecnología y la estructura de una organización y su ambiente a través de la reutilización del conocimiento y la innovación" (Tianyong Zhang, 2010, p. 572). Su éxito se materializa cuando el conocimiento es reutilizado para mejorar el desempeño de la organización (M.E. Jennex & Olfman, 2004).

Por esa razón, aparece la gestión del conocimiento (GC), que puede ser definida, como el arte de crear valor con los activos intangibles de una organización (Sarvary, 1999), y observada como un proceso que busca optimizar la aplicación efectiva del capital intelectual para alcanzar los objetivos organizacionales (Tocan, 2012), a través de la integración de todas las unidades de la organización para identificar y compartir los conocimientos que se producen y se acumulan (Mojibi, Khojasteh, & Khojasteh-Ghamari, 2015), con el objetivo de mejorar el manejo sistemático del conocimiento dentro de la organización (Heisig, 2009).

La conceptualización de la GC se solapa en cierta medida con distintas propuestas en la literatura, incluido el aprendizaje organizacional (Swan, Newell, Scarbrough, & Hislop, 1999) y la innovación. Muchos académicos del área de negocios y economía consideran que el potencial de la gestión del conocimiento crea el capital intelectual como fuente de innovación y de renovación, las estrategias de negocios deben centrarse más en estos temas.

Para Beckman (1999), la GC se refiere a la formalización y el acceso a la experiencia, el conocimiento y la experiencia que crea nuevas capacidades, permite un rendimiento superior, fomenta la innovación y mejora el valor del cliente.

Coleman (1988) define a la GC como un término general para una amplia variedad de funciones interdependientes y entrelazados que consisten en: la creación de conocimiento; valoración del conocimiento y la métrica; mapa del conocimiento y la indexación; el transporte de conocimientos, almacenamiento y distribución; y el intercambio de conocimientos.

La definición de conocimiento como "información, combinada con la experiencia, el contexto, la interpretación y la reflexión" (Davenport & Prusak, 1998a). La GC es un término relativamente nuevo que abarca no sólo las nociones relacionadas con la transferencia de conocimientos y el intercambio de conocimientos (externamente de otras empresas para la pequeña empresa y/o internamente entre los miembros de la firma), sino también el proceso de utilización del conocimiento (Bontis, 2001; Ichijo & Nonaka, 2007; Ikujiro Nonaka, 2006)

En suma, la gestión de la innovación es la disciplina de la gestión de los procesos de innovación. Puede ser utilizado para desarrollar productos, procesos e innovaciones organizacionales. La GC incluye un conjunto de herramientas que permiten a los administradores e ingenieros cooperar con una comprensión común de los objetivos y procesos. El enfoque de gestión de la innovación es permitir a la organización responder a una oportunidad externa o interna, y usar sus esfuerzos creativos para introducir nuevas ideas, procesos o productos (Birkinshaw, Hamel, & Mol, 2008a; Costa & Monteiro, 2016).

### *2.1.3. Modelos de gestión del conocimiento*

Existen múltiples modelos propuestos de gestión del conocimiento, Heisig (2009), tan sólo analizó 160 de ellos, normalmente estos modelos se asocian con el proceso general de gestión del conocimiento, ya explorado por distintos autores (Chang Lee, Lee, & Kang, 2005; Ding, Liang, Tang, & van Vliet, 2014), incluso existen modelos de transferencia del conocimiento y la tecnología (González & Rodríguez, 2016).

Típicamente estos modelos integran las fases de creación, acumulación, recuperación, trasferencia y aplicación del conocimiento.

**Figura 2.3. Proceso de Gestión del Conocimiento (PGC)**

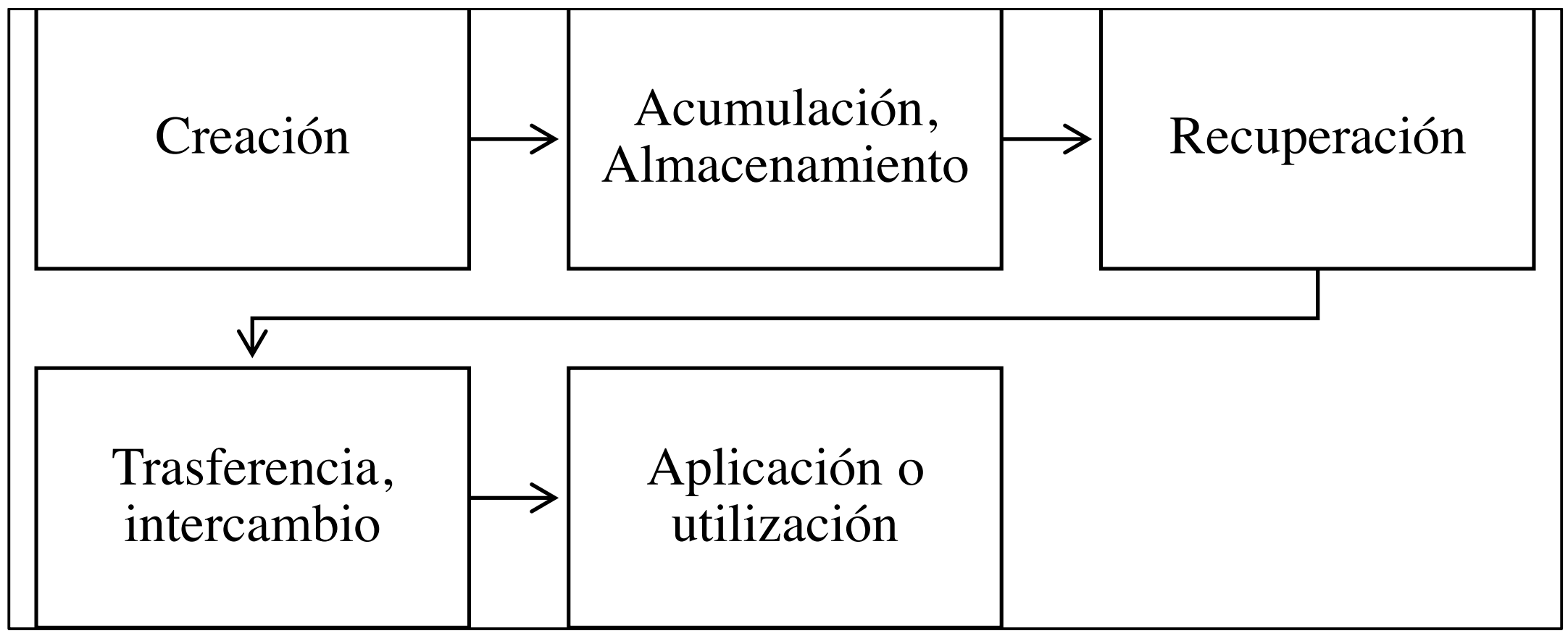

**Fuente**: Elaboración propia, adaptación de la propuesta de Chang Lee, Lee, & Kang (2005)

Más allá de adentrarnos de manera extensa en el concepto de epistemología como objeto de estudio del conocimiento, el alcance del presente trabajo pretende circunscribirse al contexto de las organizaciones y más específicamente al de las que se clasifican en el ámbito empresarial. No por ello algunas de las recomendaciones y conclusiones que se realicen se podrían excluir de ser adoptadas por otras organizaciones de naturaleza diferente.

De esta manera, el conocimiento que se encuentra en las personas y que es difícil de articular y codificar y  es comúnmente el resultado de la experiencia y la continua repetición es el que se denomina conocimiento tácito (Polyani, 1958; J.-C. Spender & Grant, 1996), siendo este el concepto que posteriormente daría origen a lo que varios autores denominaron la gestión del conocimiento y que se entiende como la forma de dar a conocer y administrar las actividades relacionadas con el conocimiento, así como su creación, captura, transformación y uso (Wiig, 1997a). Por la trascendencia del concepto, varios autores han realizado diversas representaciones de cómo ocurre y cuáles son los elementos que interactúan en la gestión del conocimiento (Chang Lee et al., 2005; Chong & Chong, 2009; Ding et al., 2014; O'Brien, Clifford, & Southern, 2010).

En este contexto, Lassi (2010), se refiere al valioso documento generado por Nonaka y Takeuchi denominado "The Knowledge-Creating Company" (1991, 2008) cuyo trabajo seminal parte del libro del mismo nombre escrito en el año de 1995 en el que se refieren a dos tipos de conocimiento; el tácito, que es aquél que se circunscribe al ámbito del individuo, que es difícil de extraer y plasmar en palabras precisas y que las organizaciones pocas veces tienen la capacidad y mecanismos para hacerlo transparente y útil para toda la organización; y el explícito, que puede ser articulado, codificado, almacenado y por lo tanto utilizado de forma estructurada por un mayor número de integrantes de la organización.

Ambos autores, quienes está más inclinados culturalmente hacia valorar más el "conocimiento tácito", que es más individualizado, intuitivo, ambiguo y de difícil expresión o reducción al ámbito de las palabras, que el explícito, generaron el siguiente modelo para explicar ambos tipos de conocimiento y que se muestra en la figura 2.4.

**Figura 2.4. Modelo SECI (Ikujiro Nonaka, 1991)**

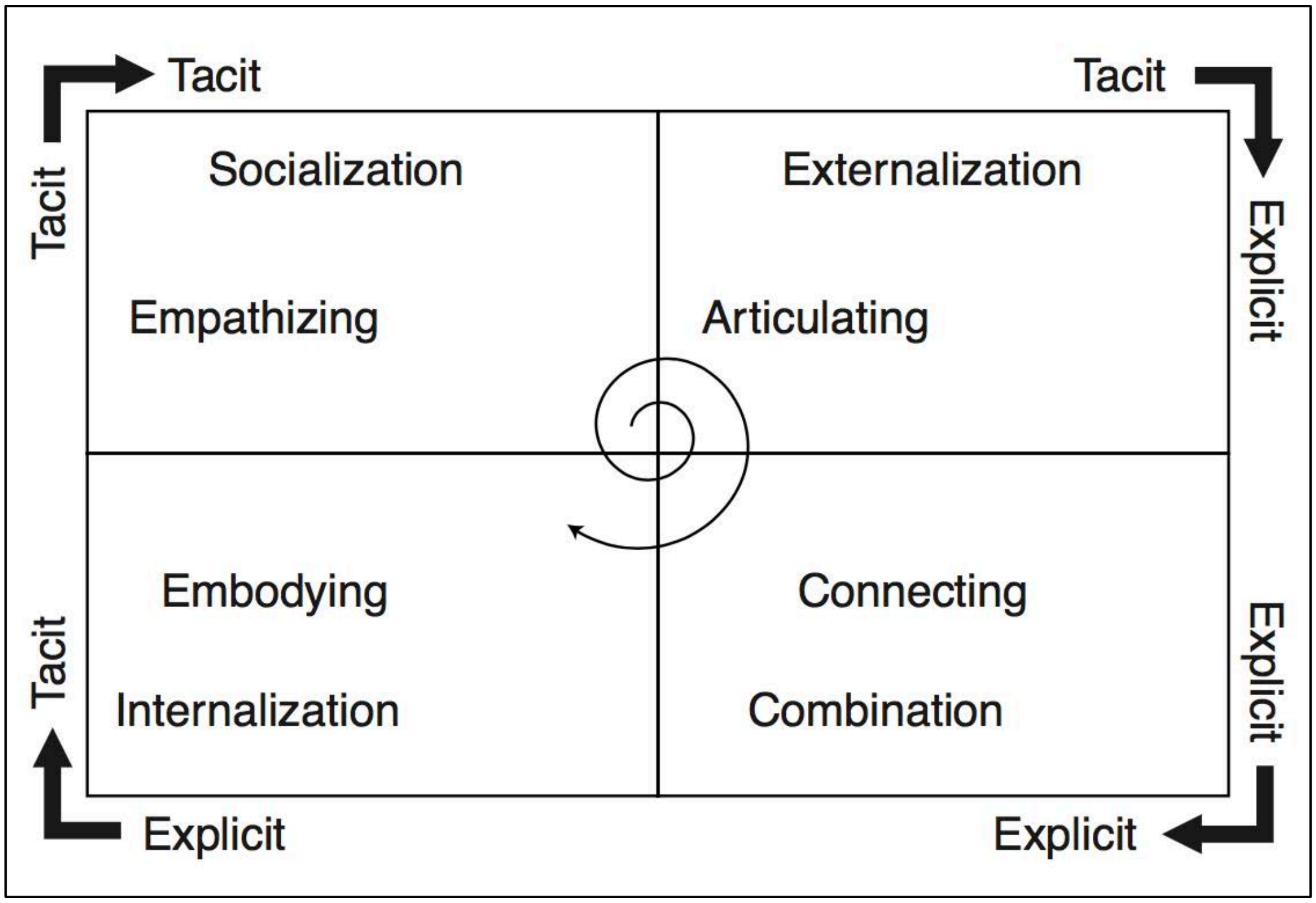

**Fuente**: (I. Nonaka & D. Teece, 2001, p. 20)

El modelo explica que una vez que el conocimiento tácito de un individuo es socializado, se torna en explícito y que a su vez es encarnado o interiorizado por otros individuos que lo hacen nuevamente un conocimiento tácito (individual), pero de un nivel superior y refinado. Es en este punto en donde es posible encontrar la conexión entre la generación del conocimiento y su aplicación.

En la espiral de creación del conocimiento organizacional (Ikujiro Nonaka, 1994, p. 20), se observa el proceso de gestión del conocimiento (PGC) a través del modelo SECI de conversión del conocimiento interno-tácito en conocimiento externo-explícito, mediante representaciones que lo modelan a la gestión del conocimiento (GC) como un proceso en donde las entradas son el conocimiento acumulado en las mentes de las personas (tácito) y las salidas, en este caso son descritas, como las representaciones del conocimiento aplicado o explícito, tal como se muestra en la figura 2.5.

**Figura 2.5. Espiral de creación del conocimiento organizacional**

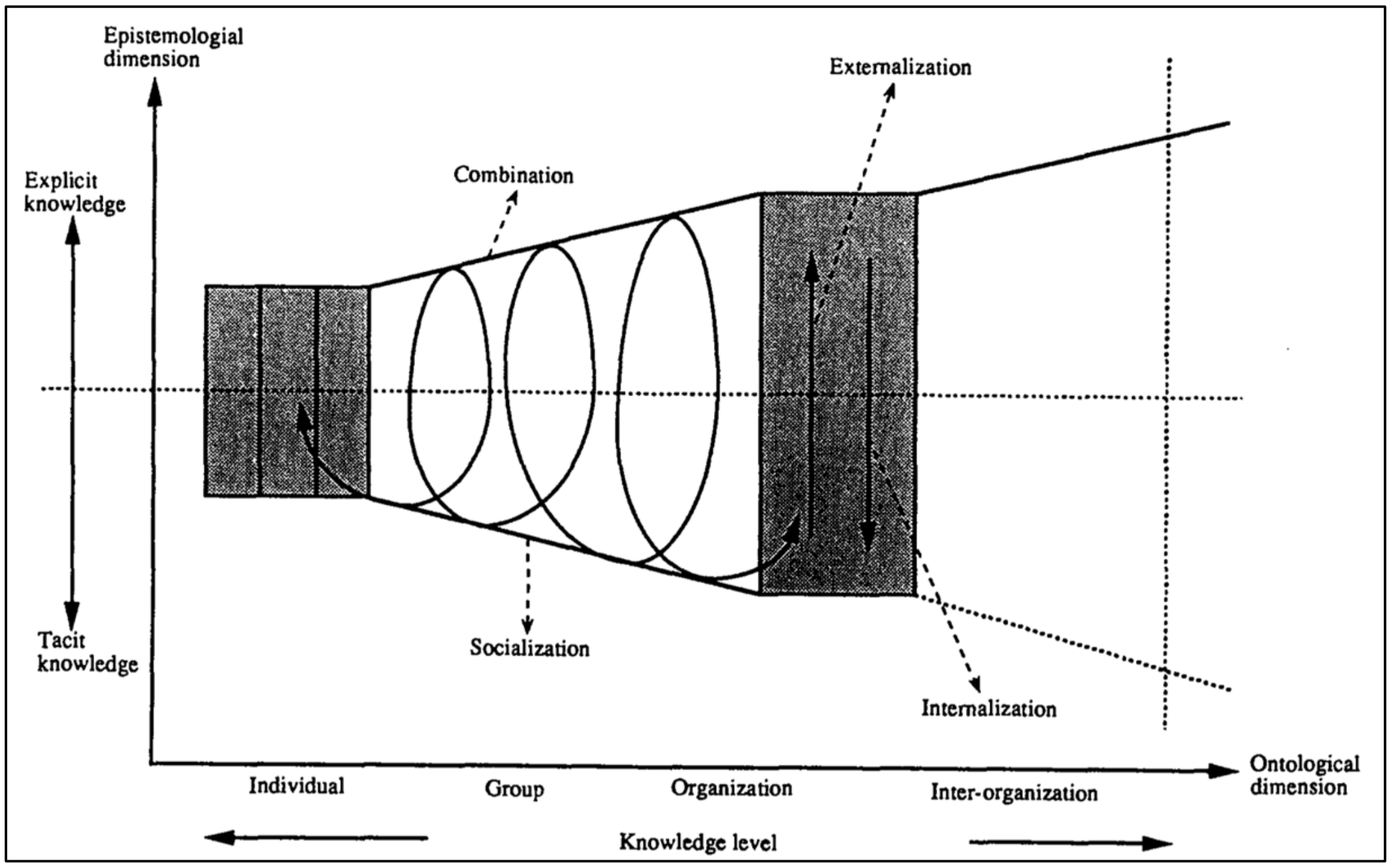

**Fuente**: (Ikujiro Nonaka, 1994, p. 20)

Aunado a lo anterior, diversos autores como Ding. et. al, (2014, p. 548) han descrito el proceso de gestión del conocimiento (PGC) dentro de un proceso general de documentación en el mundo de desarrollo de software, tal como se muestra en la figura 2.6.

**Figura 2.6. Proceso general de documentación de software**

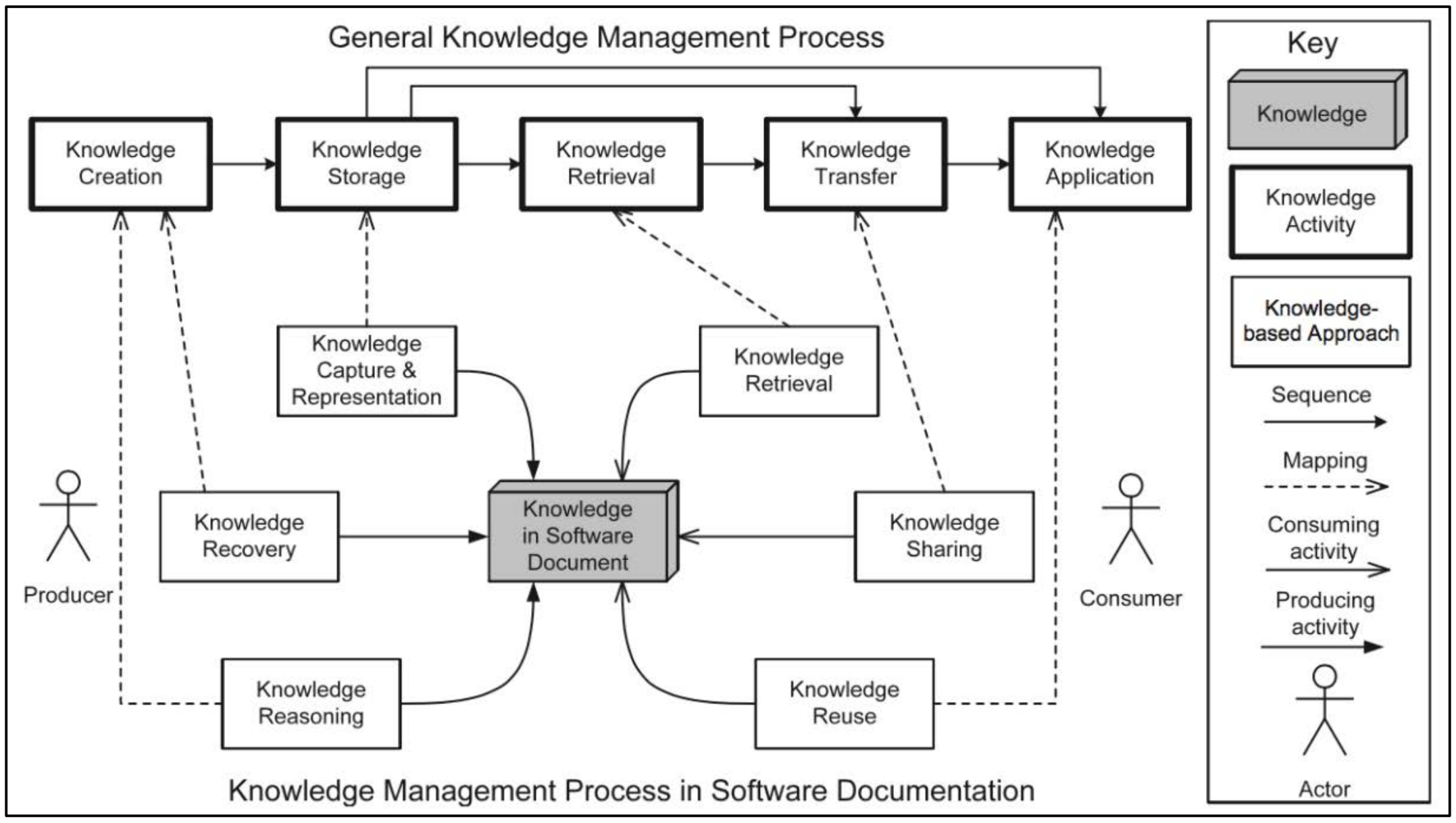

**Fuente**: Ding. et. al, (2014, p. 548)

Por otra parte, el modelo del Ciclo de Vida del desarrollo de software (IEEE, 2004), es otro ejemplo de cómo las representaciones se han modelado como un proceso de desarrollo, en donde la creación de conocimiento se observa de manera incremental, en la figura 2.7, se observa a través de este modelo como el conocimiento puede pasar de una toma de requerimientos a un desarrollo de un producto.

De igual forma se han desarrollado modelos específicos para la gestión del conocimiento organizativo, como el denominado *Promise Framework*, (Mora-Soto, 2011) dirigido a empresas y organizaciones de desarrollo de software para asegurar la valoración, accesibilidad, usabilidad y aprendizaje del conocimiento organizativo.

**Figura 2.7.** Ciclo de vida de desarrollo de software (Muench Dean, 1994)

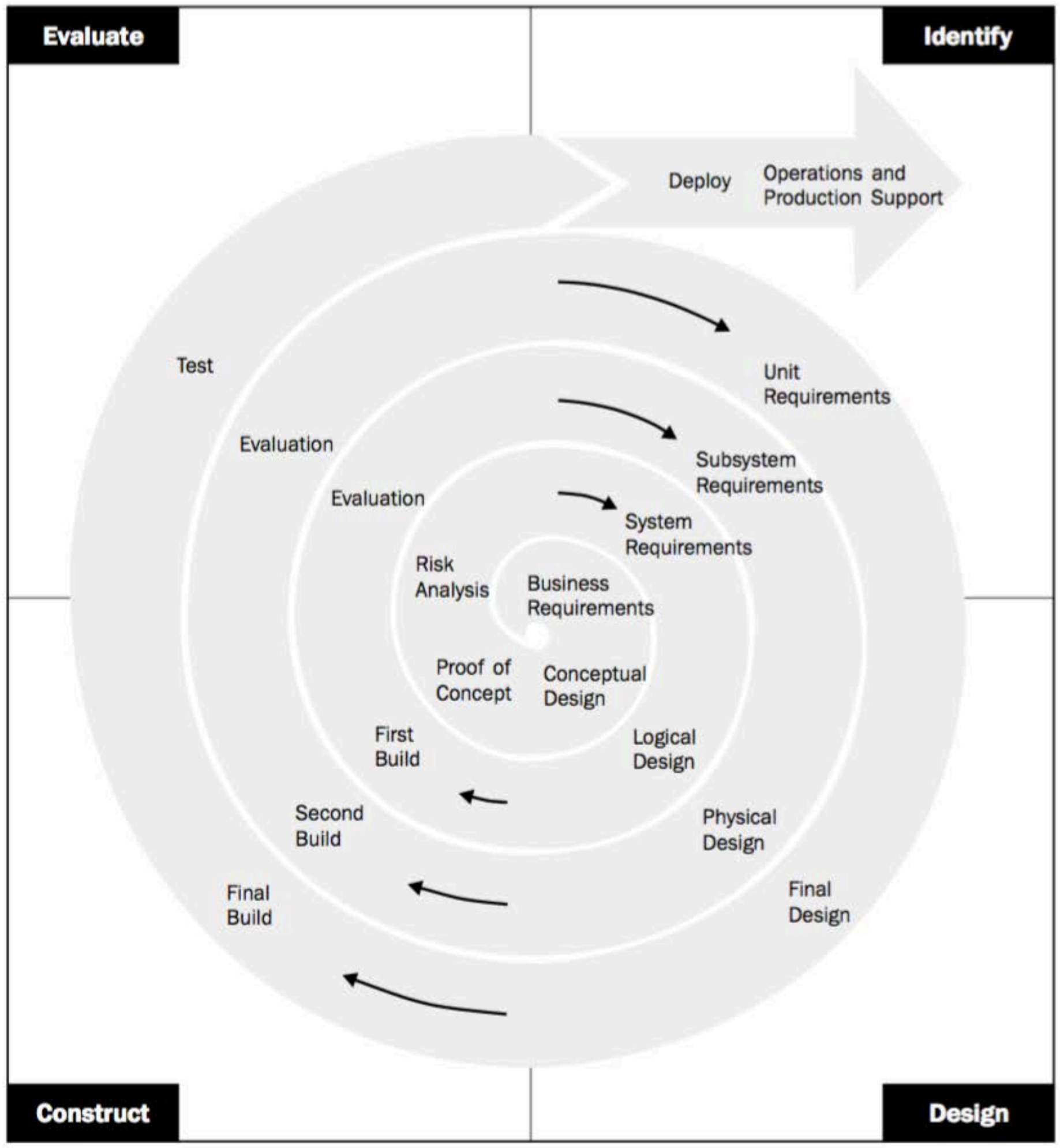

**Fuente**: (IEEE, 2004, p. 17)

Consecuentemente, es posible observar en la literatura académica, la coincidencia sobre la afirmación de que no es la cantidad de conocimiento existente en un momento dado lo que es importante, sino la capacidad de la empresa para aplicar efectivamente los conocimientos existentes para crear nuevos conocimientos (Alavi, Kayworth, & Leidner, 2006; Massa & Testa, 2004, p. 611). De tal manera que, la gestión del conocimiento incentiva la innovación (Quast, 2012; Wiig, 1997b).

Además, como se ha podido observar en los modelos descritos la gestión del conocimiento es un proceso recursivo que inicia con los individuos (Ikujiro Nonaka, 1991), y que constituye una espiral de creación del conocimiento (Ikujiro Nonaka, 1991, p. 164), que es incremental y que puede llegar a externalizarse en toda la organización (Ikujiro Nonaka, 2008, p. 10). Por tanto, toda la organización puede beneficiarse de la creación de conocimiento que es enriquecido a través de la experiencia.

Así, la GC involucra actividades relacionadas con la captura, el uso y el intercambio de conocimiento por parte de la organización y es una parte importante del proceso de innovación (OECD & Eurostat, 2005, p. 24). Desde sus orígenes ha sido un elemento que puede contribuir a alcanzar los objetivos de la organización, al poner al servicio de la misma lo que las personas saben, sus experiencias y sus habilidades (Choo, 2006; Ichijo & Nonaka, 2007; I. Nonaka & D. Teece, 2001; Ikujiro Nonaka, 1991). Sin lugar a dudas, hoy más que nunca es necesario gestionar lo que sabemos (Davenport & Prusak, 1998b) y es importante asegurar un manejo sistemático del conocimiento (Heisig, 2009) en las organizaciones.

Tal como se ha apuntado existe una gran diversidad de modelos de gestión del conocimiento (Haslinda & Sarinah, 2009; Heisig, 2009; Hislop, 2005; Piraquive, García, & Crespo, 2015). Para facilitar la comprensión de los mismos se han agrupado en la tabla 2.1.

**Tabla 2.1.** Características encontradas en los modelos de gestión del conocimiento

| Características | Descripción | Autores |
|---|---|---|
| **Actividades de gestión del conocimiento** | Integra a las actividades del proceso de gestión del conocimiento:<br>1. Creación,<br>2. Acumulación,<br>3. Recuperación,<br>4. Trasferencia Y<br>5. Aplicación | (M. Allameh et al., 2011; S. M. Allameh, Zare, & Davoodi, 2011; Chang Lee et al., 2005; Chong & Chong, 2009; Ding et al., 2014; Kulkarni, Ravindran, & Freeze, 2007) |
| **Factores críticos de éxito** | 1. **Humanos**: Cultura, personas y liderazgo<br>2. **Procesos organizacionales y de estructura**:<br>3. **Tecnología**: infraestructura y aplicaciones<br>4. **Procesos de gestión**: estrategia, metas y medición.<br>Los factores socio-técnicos sin duda apoyan el proceso de gestión del conocimiento (Handzic, 2011). | (Al-Alawi, Al-Marzooqi, & Mohammed, 2007; M. Allameh et al., 2011; S. M. Allameh et al., 2011; Esterhuizen, Schutte, & Du Toit, 2012; Heisig, 2009; Mas-Machuca & Martínez Costa, 2012; McAdam & McCreedy, 1999; Naghavi, Dastaviz, & Nezakati, 2013; Wong, 2005) |
| **Herramientas de gestión del conocimiento** | Estas herramientas apoyan a las actividades del PGC | (P. Massingham, 2014a, 2014b; Young & Organization, 2010) |

**Fuente**: Elaboración propia (2017)

## 2.2. La innovación y sus impactos

El concepto de innovación abarca las nuevas tecnologías de procesos de producción, nuevas estructuras o sistemas administrativos, y los nuevos planes o programas pertenecientes a miembros de la organización (Dalkir, 2005, p. 335). La innovación se asocia a menudo con el aumento de la productividad que reduce la cantidad de trabajo físico necesario para la producción de bienes y servicios. (Audretsch, Coad, & Segarra, 2014, p. 746), pero no se limita a las fronteras internas de la organización, sino que implica procesos interactivos, en los que las organizaciones interactúan con socios externos, incluidos los clientes y usuarios (Winter, 2013, p. 187).

Para efectos de la presente tesis se sostiene lo señalado por el Manual de Oslo, que apunta que la innovación es

> "la implementación de un producto nuevo o significativamente mejorado (bien o servicio), o proceso, un nuevo método de marketing, o un nuevo método organizacional en las prácticas comerciales, la organización del lugar de trabajo o las relaciones externas" (OECD & Eurostat, 2005, p. 46).

A medida que los paradigmas en la creación de valor han ido cambiando, hacia una economía digital (Baller et al., 2016; OECD, 2014) y de la transformación (den Ouden, 2012, p. 11), la sociedad ha tenido que modificar sus esquemas de pensamiento y considerar nuevos elementos para crear valor. La innovación siempre ha estado presente, a través de ella es posible ofrecer mejores productos y servicios, la innovación puede entregar valor al usuario, a la organización, al ecosistema y a la sociedad (den Ouden, 2012, p. 61).

La base de la innovación es el aprendizaje organizacional, porque esta es la manera de aumentar el conocimiento de la empresa (J.-C. J. Spender, 1996; Teece, Pisano, & Shuen, 1997). Cuanto más conocimiento se comparte entre los empleados de una empresa, mayor es la capacidad de innovación será (VV AA, 2009, p. 70). Consecuentemente, el intercambio, y la diseminación (Darroch & McNaughton, 2002) del conocimiento facilita la innovación (Zyngier, Burstein, & McKay, 2006, p. 86).

Es así como la innovación es extremadamente dependiente de la disponibilidad de conocimiento (Yousuf Al-Aama, 2014) y, por tanto, la complejidad generada por la explosión de riqueza y alcance del conocimiento tiene que ser reconocida y manejada para asegurar la innovación exitosa (Plessis & du Plessis, 2007).

**Figura 2.8. Paradigmas en la creación de valor**

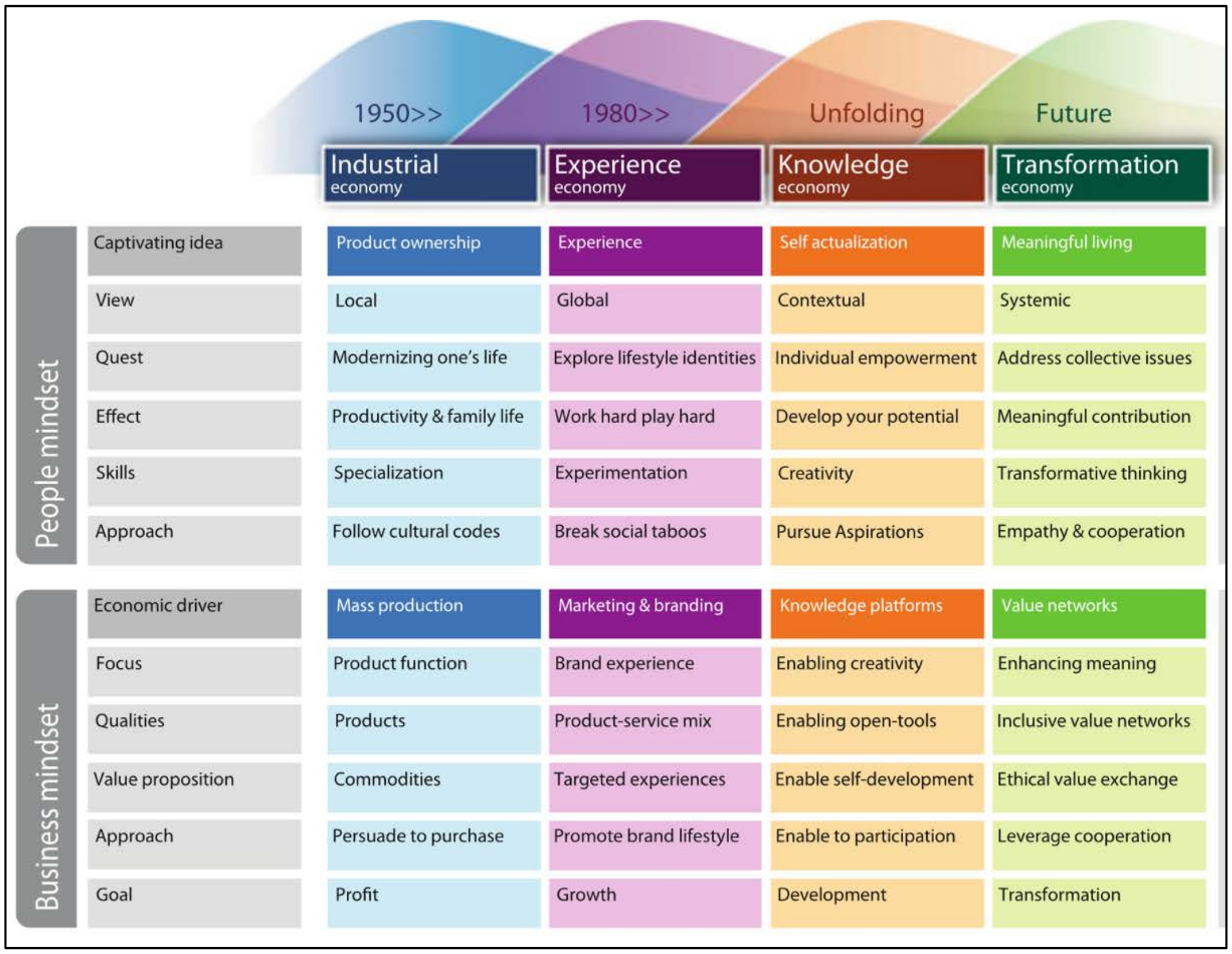

**Fuente**: (den Ouden, 2012, p. 11)

Una innovación se alcanza por completo cuando se integra y se combina con el conocimiento previo (Winter, 2013, p. 286). Sin embargo, la innovación en la economía basada en el conocimiento no puede significar la mera reproducción del conocimiento (Stehr, Adolf, & Mast, 2013, p. 1189). El proceso de innovación se equipara comúnmente con una búsqueda permanente de la utilización de los conocimientos nuevos y únicos (Ikujiro Nonaka, 1991, 1994). Razonablemente, podemos hablar de gestión del conocimiento e innovación; como un resultado de la gestión efectiva del conocimiento organizacional (Leal-Rodríguez, Leal-Millán, Roldán-Salgueiro, & Ortega-Gutiérrez, 2013).

Lo que es una realidad es que la innovación, en nuestros tiempos, es aún más rápida y la competencia es más difícil y cada vez más global (Massa & Testa, 2004, p. 610). Diversos autores han analizado a la innovación como un proceso dinámico en el cual el conocimiento es acumulado a través del aprendizaje y la interacción (OECD & Eurostat, 2005, p. 33).

**Figura 2.9. Modelo simple de la estrategia de innovación**

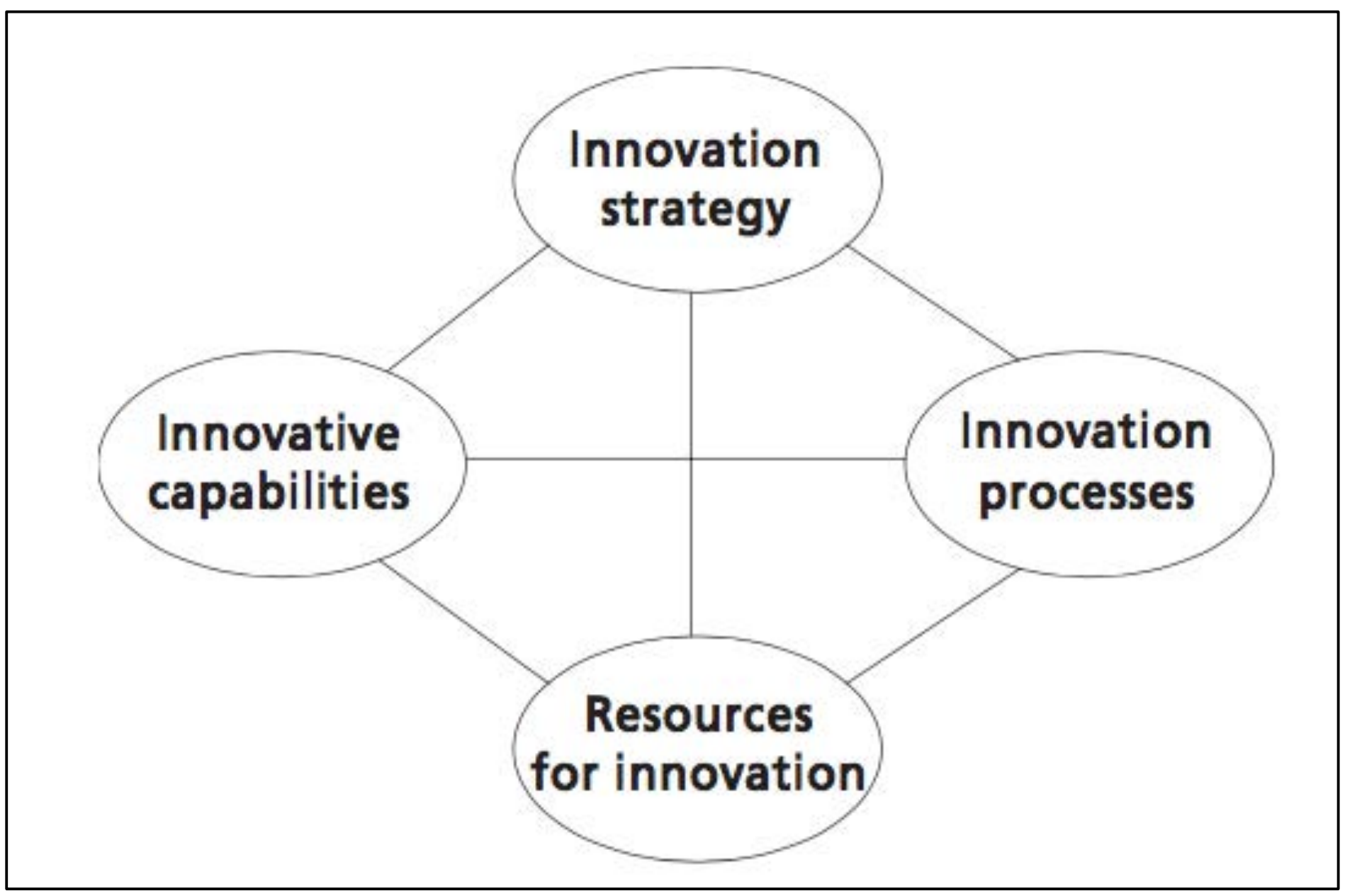

**Fuente**: (Dodgson et al., 2008, p. 96)

Gracias al conocimiento, y a la gestión del mismo, es que podemos detonar innovaciones de mercado, innovaciones tecnológicas e innovaciones organizacionales (Anthony, Johnson, & Sinfield, 2008) o administrativas (Popadiuk & Choo, 2006) que favorecen el logro de los objetivos de la organización. La innovación y la creación de conocimiento (Popadiuk & Choo, 2006), son entonces un binomio indisoluble, dado que la innovación es esencialmente creación de nuevo conocimiento.

Por ello, la relación entre el conocimiento y la innovación (Swan et al., 1999) es amplia y profusa, así existe una alta dependencia entre la innovación y la evolución del conocimiento (Carneiro, 2000), que deberá ser medida (Zheng, Chanaron, You, & Chen, 2009) y analizada para mejorarse sistemáticamente, en aras de lograr los objetivos organizacionales.

**Figura 2.10. Innovación como producto de la gestión del conocimiento**

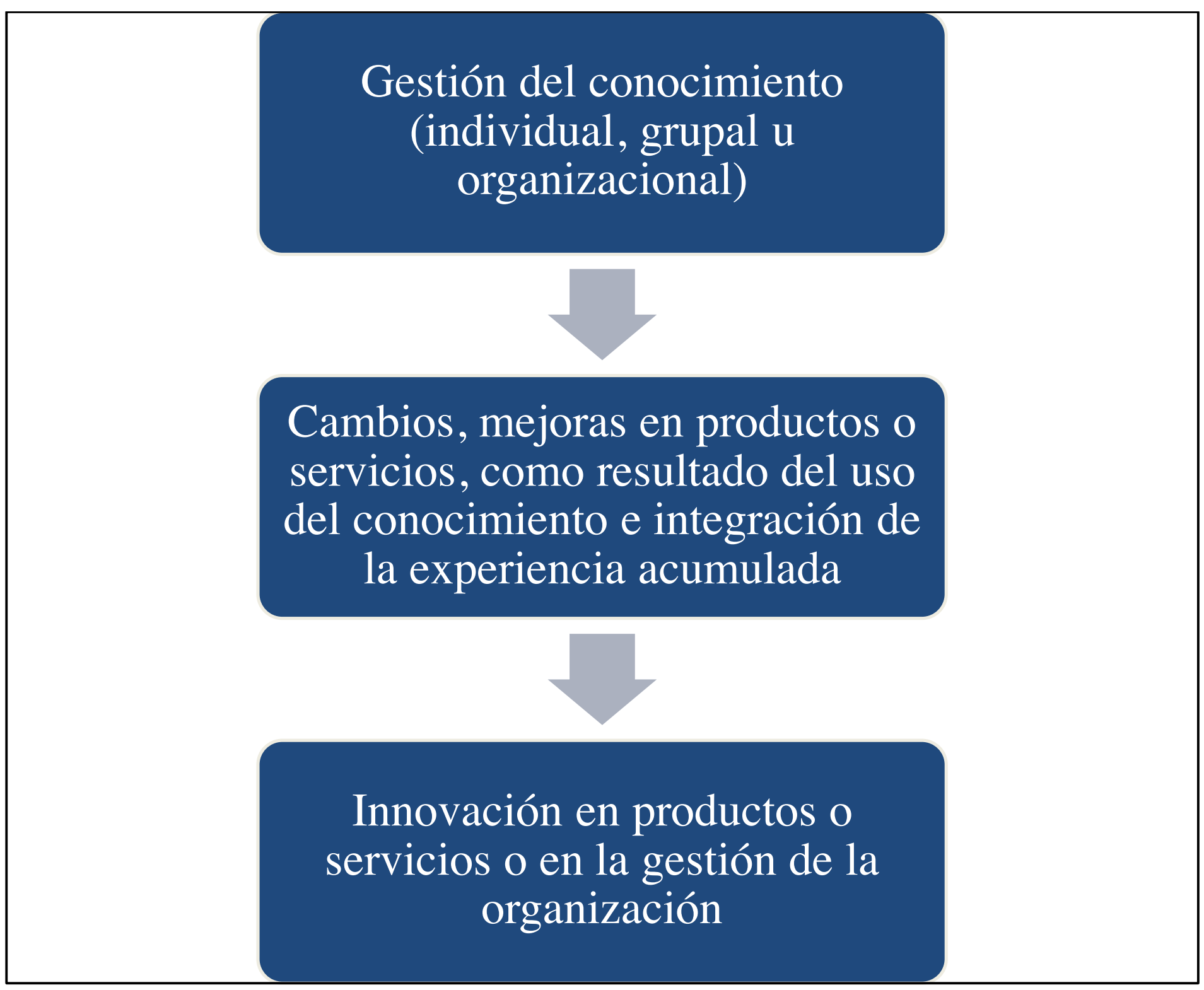

**Fuente**: Elaboración propia, a partir de la perspectiva que sostiene que la innovación es producto de la gestión del conocimiento, véase autores como Inkinen (2016).

Existen distintos tipos de ella, como la innovación abierta (Jenny S. Z. Eriksson Lundström, Wiberg, Hrastinski, Edenius, & Ågerfalk, 2013; Striukova & Rayna, 2015) cerrada (Herzog, 2011), disruptiva (Hongyi Sun, 2012), radical (Khedhaouria & Jamal, 2015), la innovación social (Murray, Caulier-Grice, & Mulgan, 2010), y la innovación educativa (OECD, 2016b).

Al igual que con la gestión del conocimiento, en la literatura académica se pueden observar en gran variedad de modelos sobre innovación (Tidd, 2006), en realidad la diversidad de los mismos, es tan amplia como el campo disciplinar o la procedencia de sus autores (Birkinshaw, Hamel, & Mol, 2008b; den Ouden, 2012; Errasti & Zabaleta, 2011; Hernandez-Munoz et al., 2015; İzadi et al., 2013; Marinova & Phillimore, 2003; Massa & Testa, 2004; Stošić & Išljamović, 2010; Vasconcellos, Bruno, Campanario, & Noffs, 2009; Xia, Zhang, Zhu, & Jia, 2012).

En consecuencia, por ejemplo, Errasti & Zabaleta (2011), han analizado más de 40 modelos de innovación, observando cinco tipos de modelos:

1. De etapa departamental,
2. De etapa de actividad,
3. De etapa de decisión,
4. De proceso de conversión y
5. Modelos de respuesta

De esta manera los autores, observan que los modelos tienen como elementos comunes a la creatividad y el conocimiento, relacionados con el comportamiento humano o la gestión de recursos humanos, y que se define en los modelos de innovación como un factor clave para el éxito del proceso de innovación.

**Figura 2.11. Procesos de innovación de 5ta generación**

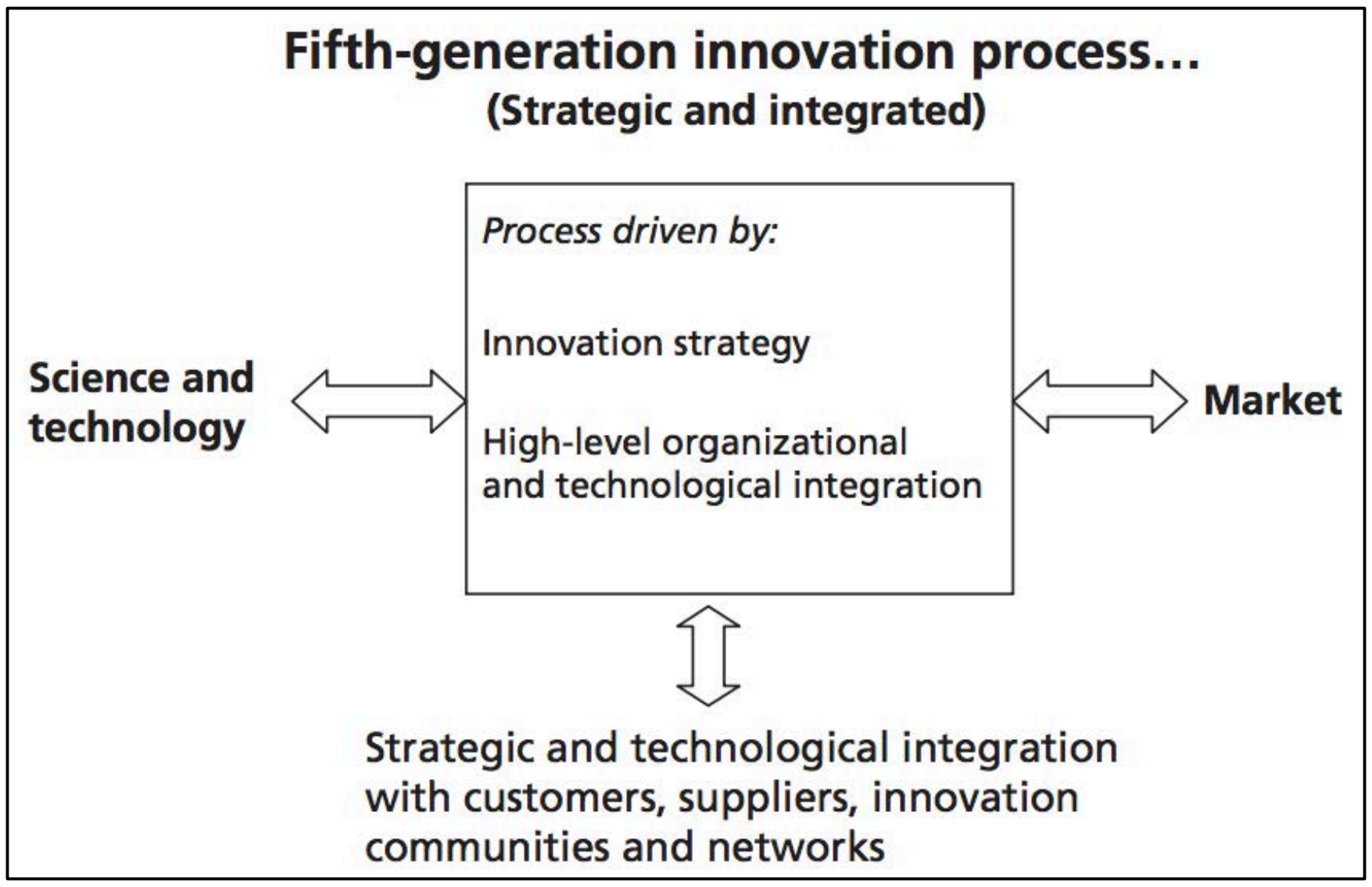

**Fuente**: (Dodgson et al., 2008, p. 63)

Por su parte, Marinova & Phillimore (2003), describen seis generaciones de modelos de innovación y observa como a medida que más se estudia a la innovación, es posible darse cuenta de que es un proceso complejo y difícil de dominar. De hecho, Cooper (1990; 1983) analizó el proceso de desarrollo de nuevos productos y observó que la innovación no es un proceso homogéneo, pero que tiene elementos comunes.

**Figura 2.12. Vista general del sistema Stage-Gate®**

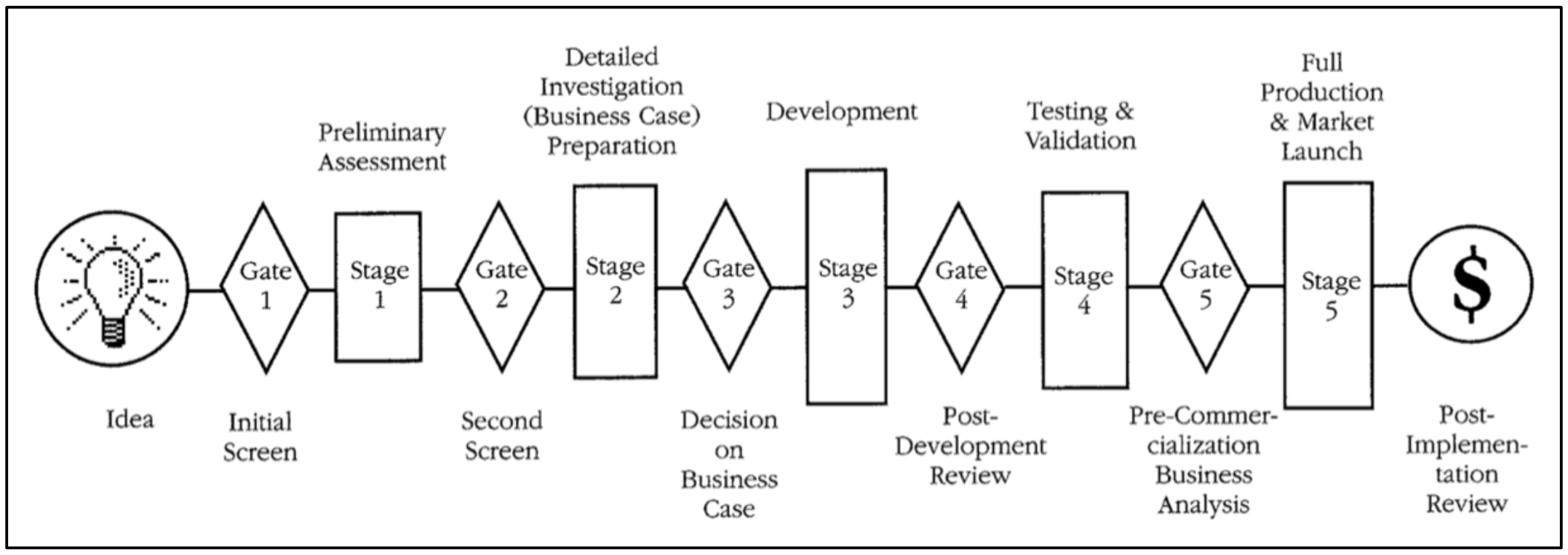

**Fuente**: Cooper (1990, p. 46)

Por ello, cada vez más los modelos de innovación deben considerar factores externos a la organización, como puede ser la cultura, o la estrategia que soporta a la innovación (Pisano, 2015; Stošić & Išljamović, 2010). Cada nueva generación de estos modelos refleja un creciente cuerpo de conocimientos académicos y una visión analítica más profunda del proceso de innovación (İzadi et al., 2013). Desde la innovación cerrada a la innovación abierta que hace posible acelerar el desarrollo de ideas (Xia et al., 2012), hasta la gestión de la innovación que potencializa la gestión tecnológica (Vasconcellos et al., 2009).

**Figura 2.13. Paradigma de innovación abierta**

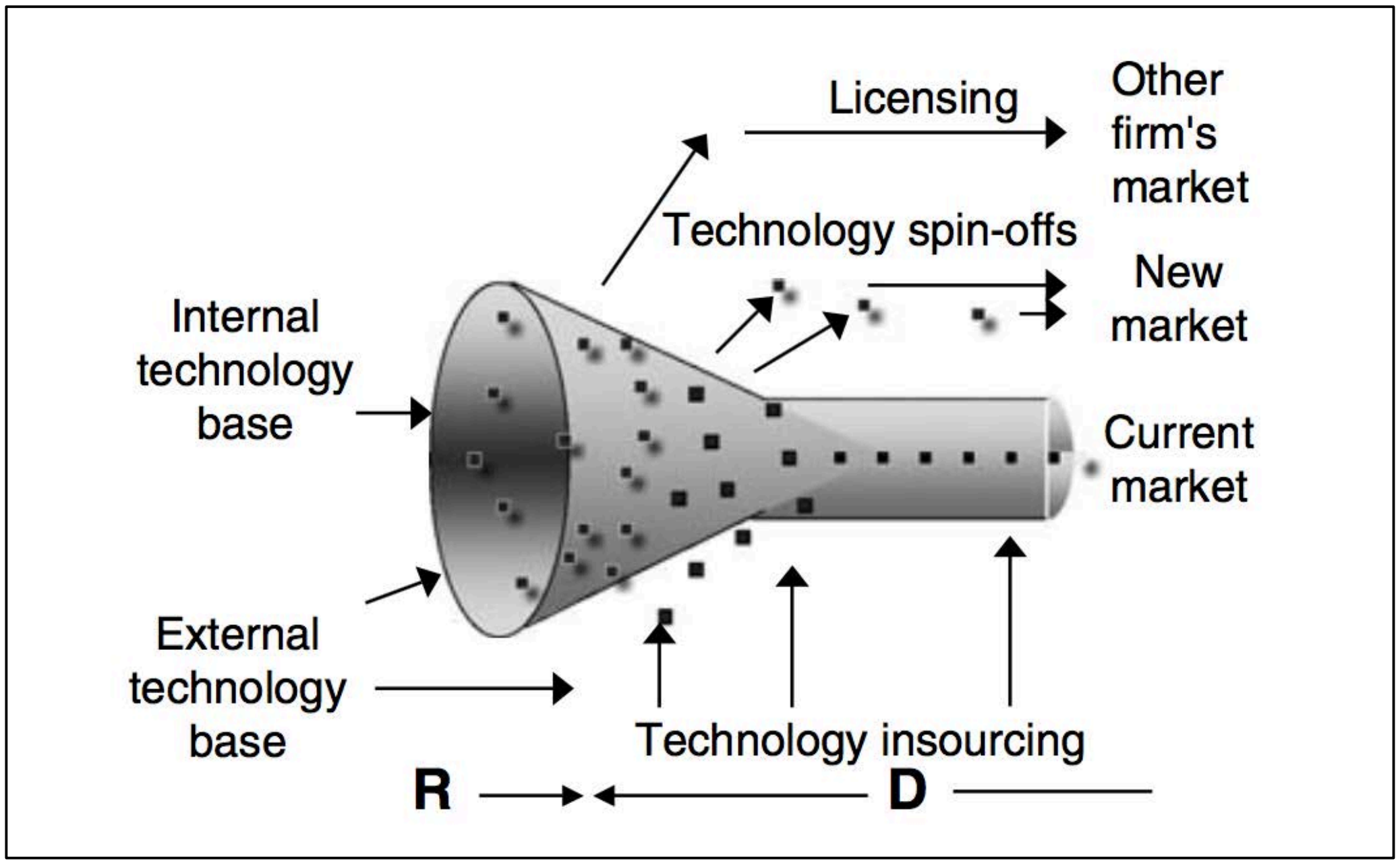

**Fuente**: (Chesbrough, Vanhaverbeke, & West, 2006, p. 3)

Consecuentemente, gestionar la innovación se ha vuelto necesario en las organizaciones para mejorar los resultados de la innovación *per se* (Birkinshaw et al., 2008b). La gestión de la innovación en las organizaciones se refiere a la invención y la implementación de una práctica, proceso y estructura que mejore los resultados de la organización y la acerque más a sus objetivos (Birkinshaw et al., 2008b, p. 829).

**Figura 2.14. Modelo del proceso de gestión de la innovación**

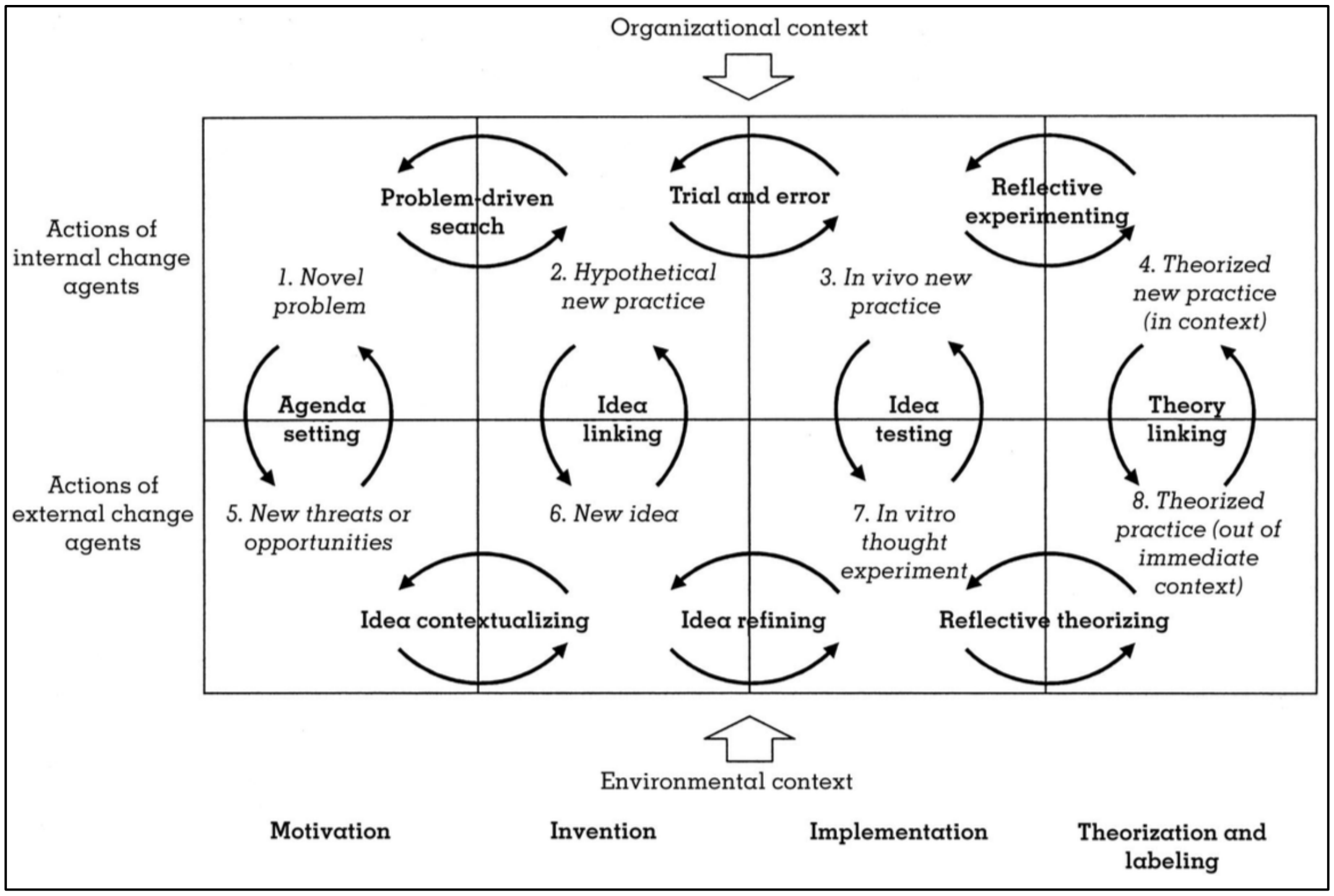

**Fuente**: Management Innovation Process Framework (Birkinshaw et al., 2008b, p. 832)

En la figura 2.15, se aprecia el modelo de comercialización de tecnología propuesto por Vijay (1997) que hace evidente la gradualidad en el desarrollo de un proceso de innovación tecnológica que parte de imaginar el producto, proceso o servicio hasta el sostenimiento y comercialización, pasando por las etapas que se ilustran en el modelo. Este modelo se asemeja a otros modelos a los que se hace referencia en esta tesis, y que tienen distintas etapas, que involucran procesos de desarrollo basados en metodologías.

**Figura 2.15. Modelo de comercialización de tecnología**

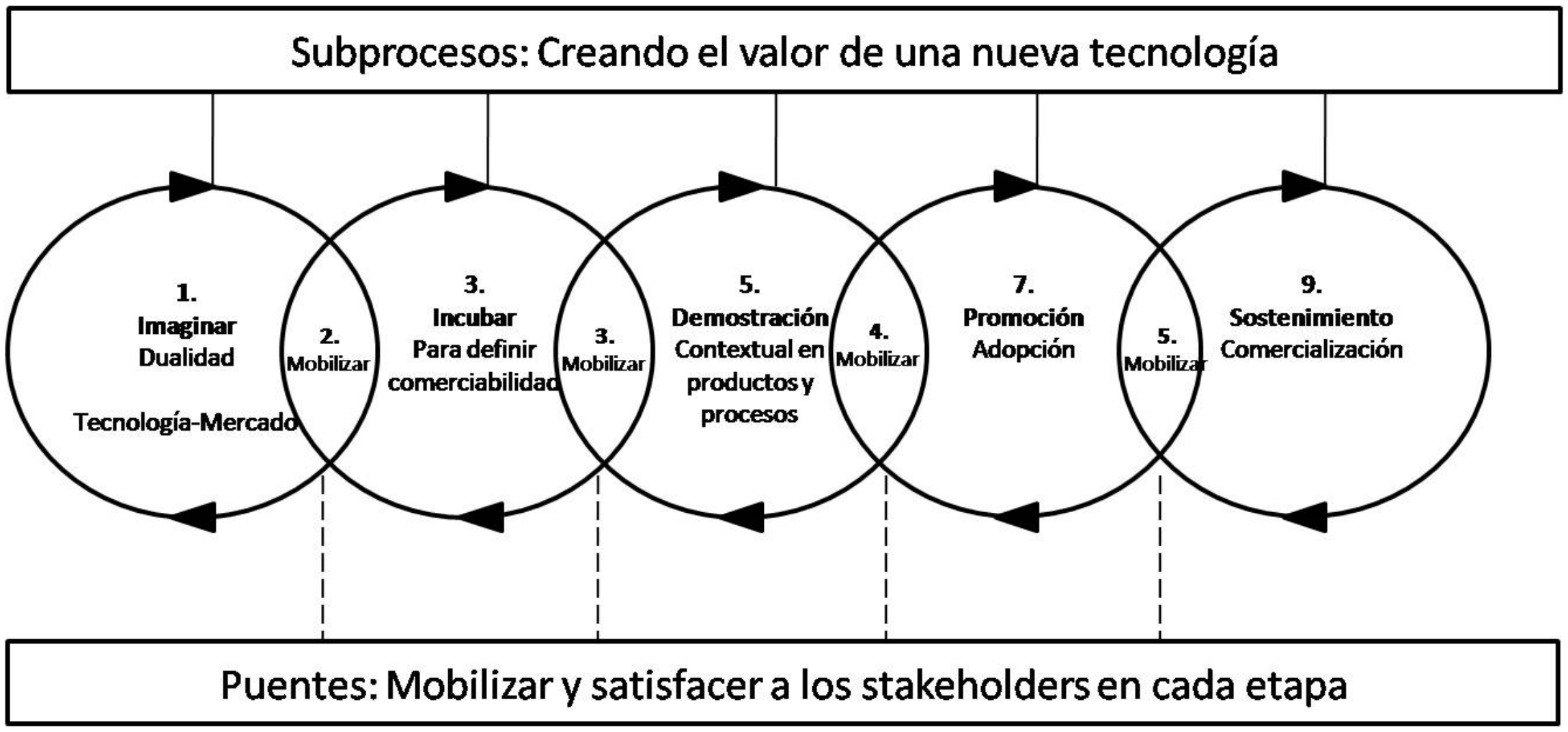

**Fuente:** Adaptación el Modelo, “Commercializing new technologies: getting from mind to market” (V. K. Jolly, 1997).

Finalmente, y al igual que la gestión del conocimiento, en modelos más recientes (Corsi & Neau, 2015) se aprecia como la innovación podría tener distintos niveles en función del grado de madurez de las innovaciones.

**Figura 2.16. Modelo innovación de cinco niveles de propuesto por Corsi & Neau**

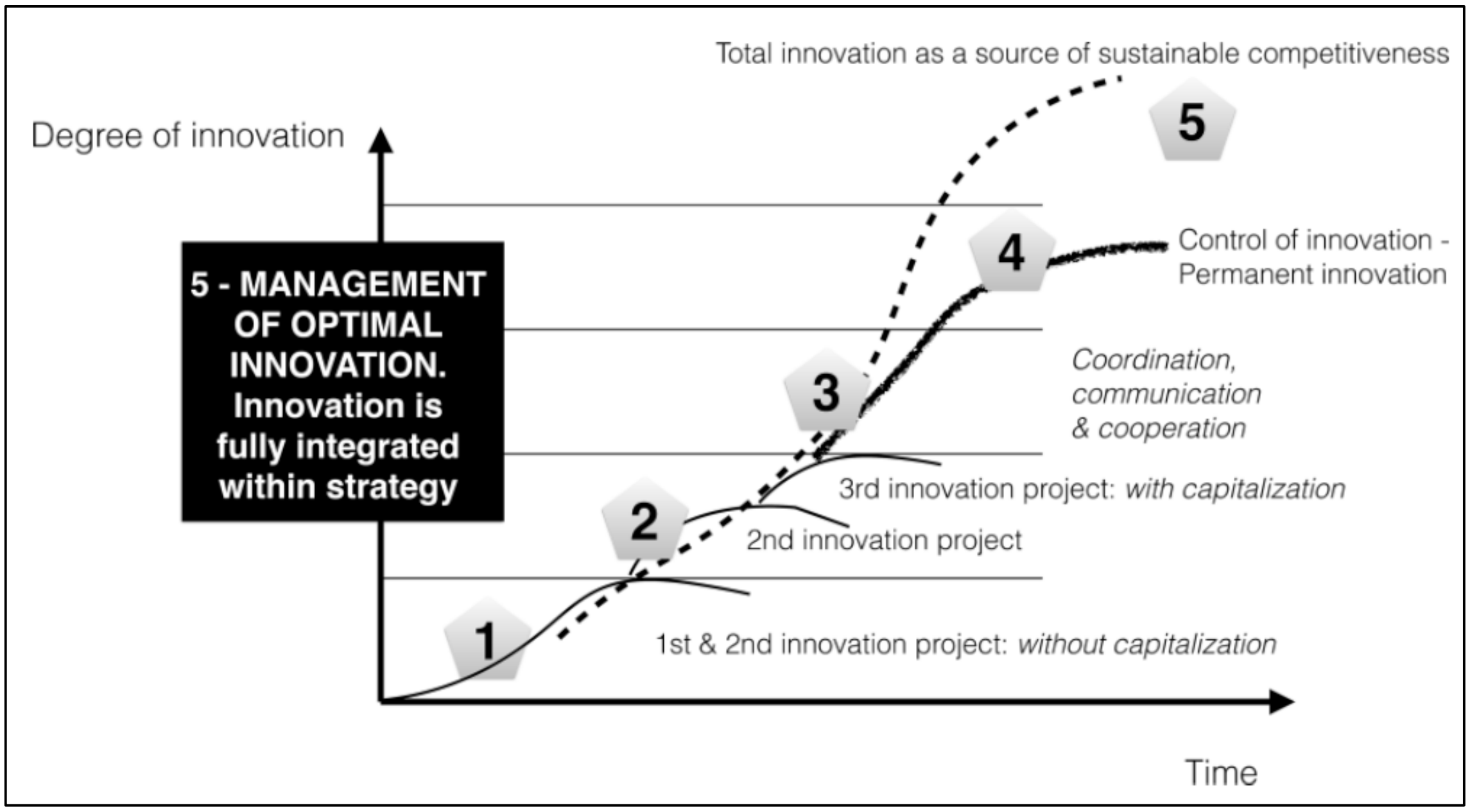

**Fuente**: (Corsi & Neau, 2015, p. 209)

### *2.2.1. Importancia de la innovación*

Todas las organizaciones saludables generan y usan conocimiento. A medida que las organizaciones interactúan con sus entornos, absorben información, la convierten en conocimiento y llevan a cabo acciones sobre la base de la combinación de ese conocimiento y de sus experiencias, valores y normas internas. Sienten y responden. Sin conocimiento, una organización no se podría organizar a sí misma[...] (Davenport & Prusak, 1998a, p. 61)

> [...] la capacidad de una compañía para generar nuevos conocimientos, diseminarlos entre los miembros de la organización y materializarlos en productos, servicios y sistemas. La creación de conocimiento organizacional es la clave del proceso peculiar a través del cual estas firmas innovan. Son especialmente aptas para innovar continuamente, en cantidades cada vez mayores y en espiral [generando ventaja competitiva para la organización (Ikujiro Nonaka & Takeuchi, 1999).

La innovación se ha vuelto un imperativo en nuestros días (Dutta, 2015; Galindo-

Rueda & Millot, 2015; OECD, 2010, 2015a, 2015b, 2015e, 2016d). Su presencia como un componente estratégico (Porter, 1990, 2008) del desarrollo de las organizaciones y del crecimiento económico de las naciones es innegable. Las organizaciones deben ser capaces de proponer nuevos modelos para hacer negocios, modificar sus competencias organizacionales y crear o proveer nuevas capacidades a sus miembros (Prahalad & Mashelkar, 2010).

Como hemos podido dar cuenta, la innovación se caracteriza, a menudo, por ser principalmente un proceso de creación de conocimiento (Hislop, 2005, p. 157), y por tanto potencializa la contribución del conocimiento y, consecuentemente, de la gestión del conocimiento.

No obstante, de la gran diversidad de modelos de innovación, la gran mayoría de éstos sostienen como elemento central la creación y utilización de conocimiento para avanzar hacia los objetivos estratégicos de la organización.

En ese orden de ideas, gestionar el conocimiento y la innovación podría permitirles a las organizaciones mexicanas avanzar en la sociedad del conocimiento e integrarse en cadenas de producción de alto valor que les permitan un crecimiento sostenido y mantener ventajas competitivas, favoreciendo la competitividad regional y la competitividad de las organizaciones mexicanas en el entorno cada vez más global.

### *2.2.2. Relación entre la gestión del conocimiento y la innovación*

La innovación y la creación de conocimiento (Popadiuk & Choo, 2006), son entonces un binomio indisoluble, dado que la innovación es esencialmente creación de nuevo conocimiento. Por tanto, la relación entre el conocimiento y la innovación (Swan et al., 1999) es amplia y profusa, así existe una alta dependencia entre la innovación y la evolución del conocimiento (Carneiro, 2000), que deberá ser medida (Zheng et al., 2009) y analizada para mejorarse sistemáticamente, en aras de lograr los objetivos organizacionales.

Una de las dimensiones de la gestión del conocimiento es la innovación, tal como lo señala (Diakoulakis, Georgopoulos, Koulouriotis, & Emiris, 2004), que sucede mediante la aplicación del conocimiento y la creación de nuevo conocimiento. La relación de la GC con la innovación ha sido vista como una externalidad, como un producto de la gestión del conocimiento e incluso como un beneficio.

**Figura 2.17. Relación entre creación y aplicación del conocimiento e innovación**

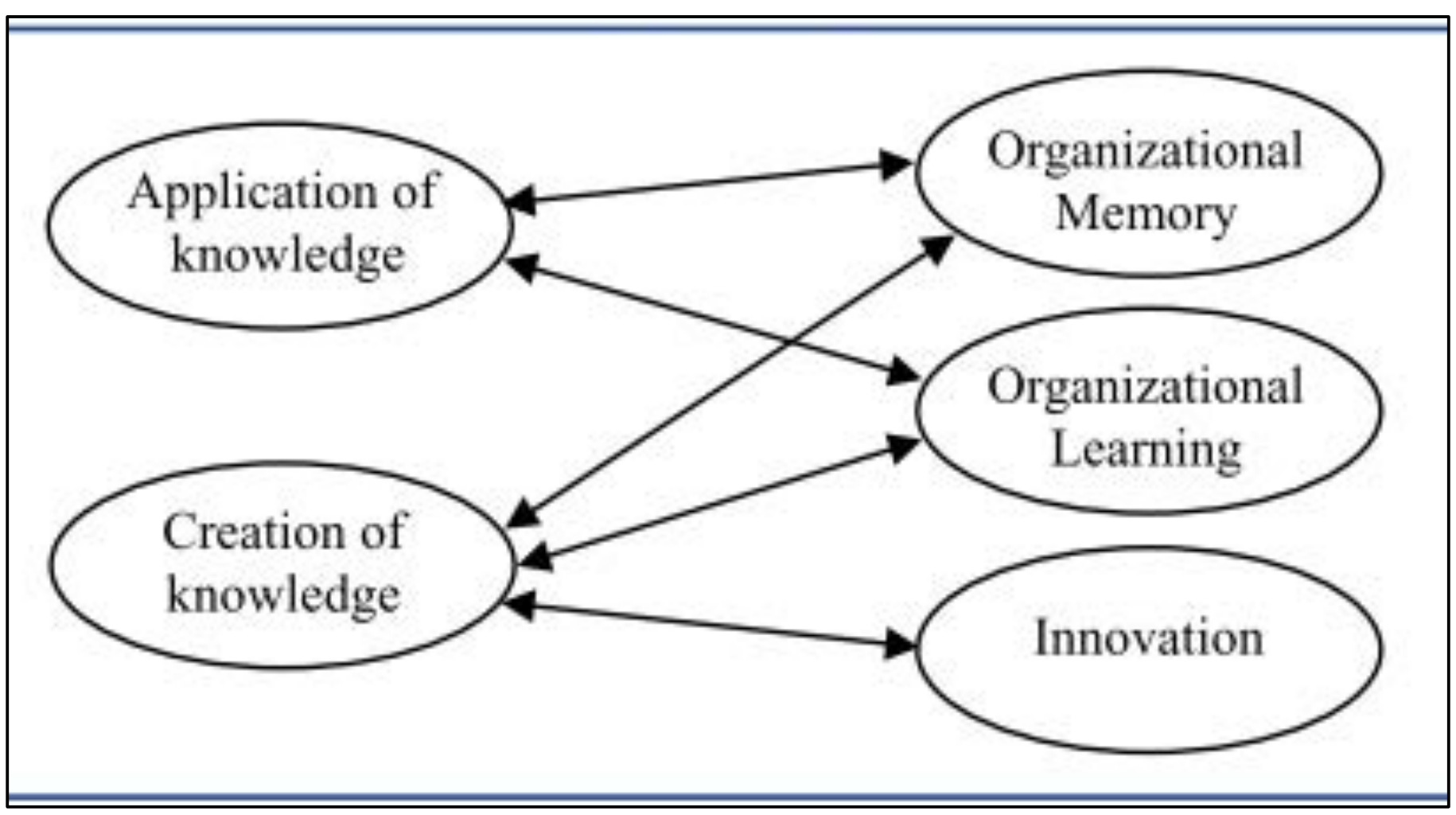

**Fuente**: (Diakoulakis et al., 2004, p. 35)

La GC en el largo plazo sostiene la innovación continua (A. K. Jain & Moreno,

2015). La GC puede verse a través del uso colectivo del conocimiento organizacional, así como la aplicación de conocimiento en la resolución de problemas y el desarrollo de productos y la innovación (Lustri, Miura, & Takahashi, 2007, p. 193). Así, la innovación favorece el desarrollo de nuevos proyectos y la GC es un factor clave para el buen desarrollo de los proyectos (Piraquive et al., 2015).

Sin duda, desde hace algún tiempo se reconoce que la gestión del conocimiento promueve la innovación (Ikujiro Nonaka & Takeuchi, 1999), porque la creación del conocimiento es un proceso dinámico, iniciado en la mente de las personas que se autoactualiza y que, por tanto, es interminable. Entonces, en la actualidad la GC es utilizada por distintas organizaciones para lograr sus objetivos y para producir productos de excelente calidad e innovadores (Piraquive et al., 2015). Uno de los mayores impactos para la organización es que la gestión del conocimiento permite la innovación sostenible y ayuda a resolver los problemas de adaptación y buen funcionamiento organizacional (A. K. Jain & Moreno, 2015).

Además, la innovación ocurre a través de la diseminación y uso del conocimiento organizacional (Pandey, 2016). Por ello, la GC puede derivar en la mejora de la productividad, la mejora del entorno empresarial y el aumento de los niveles de innovación (Pandey, 2016).

Por tanto, la innovación tiene la capacidad de mejorar el rendimiento, resolver problemas, agregar valor y crear una ventaja competitiva para las organizaciones. La innovación puede ser descrita en términos generales como la aplicación de ambos descubrimientos e invenciones y el proceso por el cual los nuevos resultados, sean de productos, sistemas, procesos o formas de organización, vienen a la existencia (Williams, 1999).

El proceso de innovación depende en gran medida en el conocimiento, en particular desde que el conocimiento representa un ámbito mucho más profundo que simplemente el de datos, la información y la lógica convencional.

La gestión de la innovación (Birkinshaw et al., 2008a) y la gestión del conocimiento (Choo, 1998; Ikujiro Nonaka, 2008) van de la mano al proveer un espacio en donde el aprendizaje es sencillo y continuo (Argyris, 1977; March, 1991; Salleh, Chong, Syed Ahmad, & Syed Ikhsan, 2012). El rol de la GC y la innovación es en general, amplificar y acelerar la creación de valor para la organización, incluso más que cualquier estrategia o modelo de negocios.

**Figura 2.18. Sistemas de gestión de la innovación**

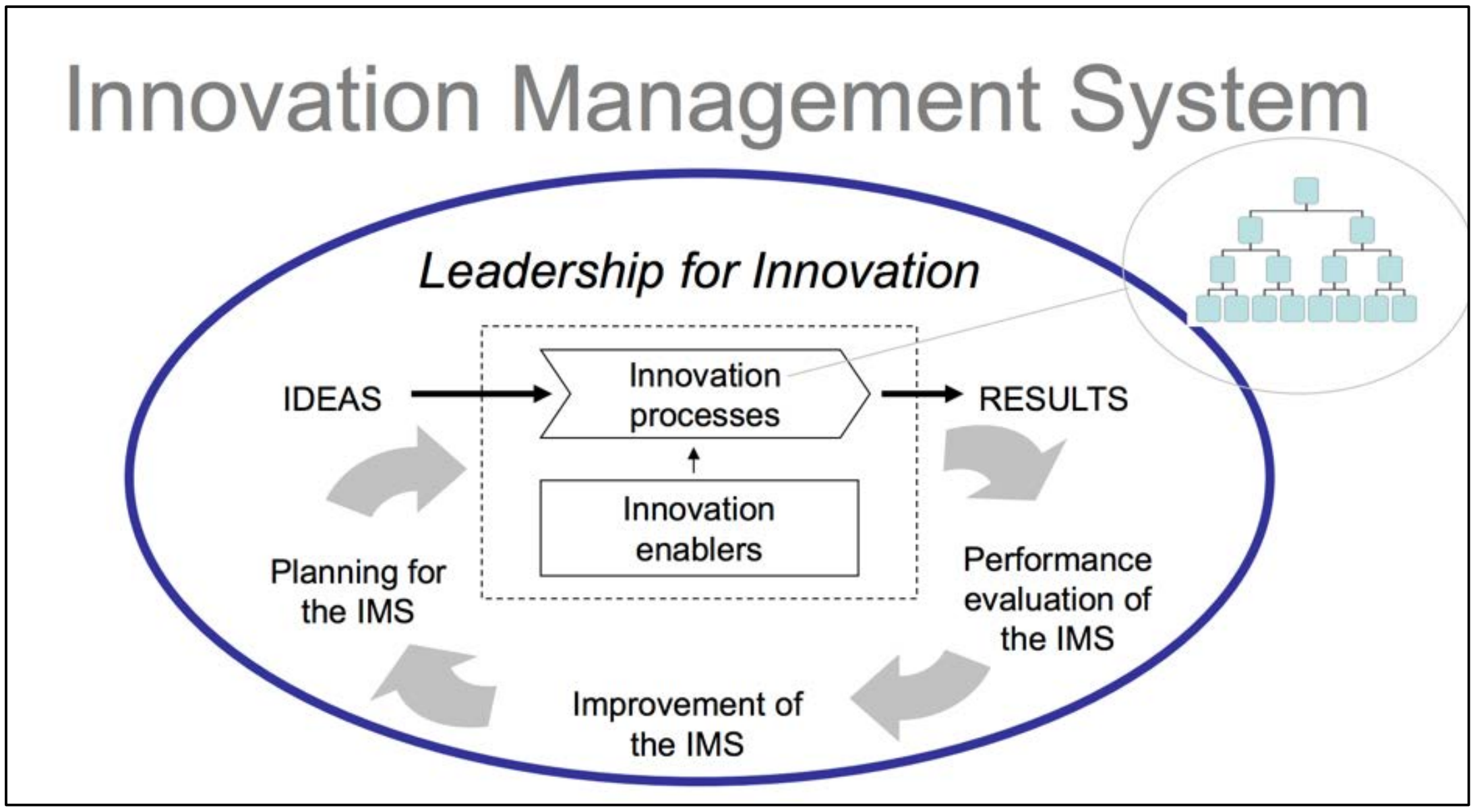

**Fuente**: (Karlsson, 2013, p. 11)

El conocimiento desarrolla el concepto, la innovación es producto del instinto creativo y el pensamiento imaginativo, el aprendizaje organizacional eventualmente crea competencias esenciales para las organizaciones. El conocimiento es esencial para la innovación, la cual crea ventajas competitivas a través del uso de competencias clave de la organización.

El conocimiento ayuda a crear una visión más profunda de los productos, los proceso y la tecnología, y está compuesto por el aprendizaje organizacional que unido a la imaginación permite la innovación. La GC y la gestión de la innovación están interrelacionadas y el aprendizaje organizacional crea las bases para ello. Por tanto, la innovación es producto y medio de la gestión del conocimiento (Paterson, 2013).

Los procesos de GC podrían afectar positivamente la innovación (Darroch & McNaughton, 2002). La innovación se ha convertido en una de la prioridades clave para las organizaciones que quieren alcanzar una ventaja competitiva (Bosilj Vukšić & Pejić Bach, 2015). A través de la codificación del conocimiento adquirido, su reutilización, almacenamiento, refinamiento y mejora se pueden generar innovaciones organizacionales (Ng, Yip, Din, & Bakar, 2012, p. 210).

En suma, derivado del análisis de la relación entre la gestión del conocimiento y la innovación es posible apuntar lo siguiente:

1. Una buena gestión del conocimiento incentiva la innovación (Brand, 1998).
2. El conocimiento necesario para la innovación se distribuye dentro de las organizaciones (en todas las funciones y unidades de negocio geográficamente dislocados) y entre las organizaciones (por ejemplo, a través de los proveedores de TI, consultores y empresas implicadas) (Swan et al., 1999).
3. Una buena GC permite la mejora continua de los productos, servicios y la reducción de costos (Levett & Guenov, 2000).
4. Los procesos de GC podrían afectar positivamente la innovación (Darroch & McNaughton, 2002).
5. A través de la codificación del conocimiento adquirido, su reutilización almacenamiento, refinamiento y mejora se pueden generar innovaciones organizacionales (Ng et al., 2012, p. 210).
6. A través de la gestión del conocimiento y innovación se puede alcanzar una ventaja competitiva sostenible (Bosilj Vukšić & Pejić Bach, 2015; Plessis & du Plessis, 2007).

No obstante, de la importancia en la relación entre GC e innovación, tal como se describe en el problema de investigación de esta tesis, aún no existe un marco común que permita encauzar los esfuerzos en estas dos importantes materias. Este trabajo de investigación pretende proponer y validar un estándar mexicano de gestión del conocimiento e innovación tecnológica. Para dar soporte a lo anterior, se ha realizado una búsqueda mediante las bases de datos de Scopus® (2016), y derivado de ello es posible observar como la combinación entre gestión del conocimiento y estándar; así como la combinación innovación y estándar, aparecen en la literatura académica. Ésta última con mayor presencia que la primera, tal como se muestra en la figura 2.20.

Es importante mencionar que esta búsqueda se ha realizado empleando combinaciones de términos de búsqueda en los títulos, palabras claves y resúmenes de los trabajos y se ha dejado abierto el año, por ello es posible obtener resultados desde la segunda década del siglo pasado.

**Figura 2.19. Documentos publicados con los términos “innovación y estándar”**

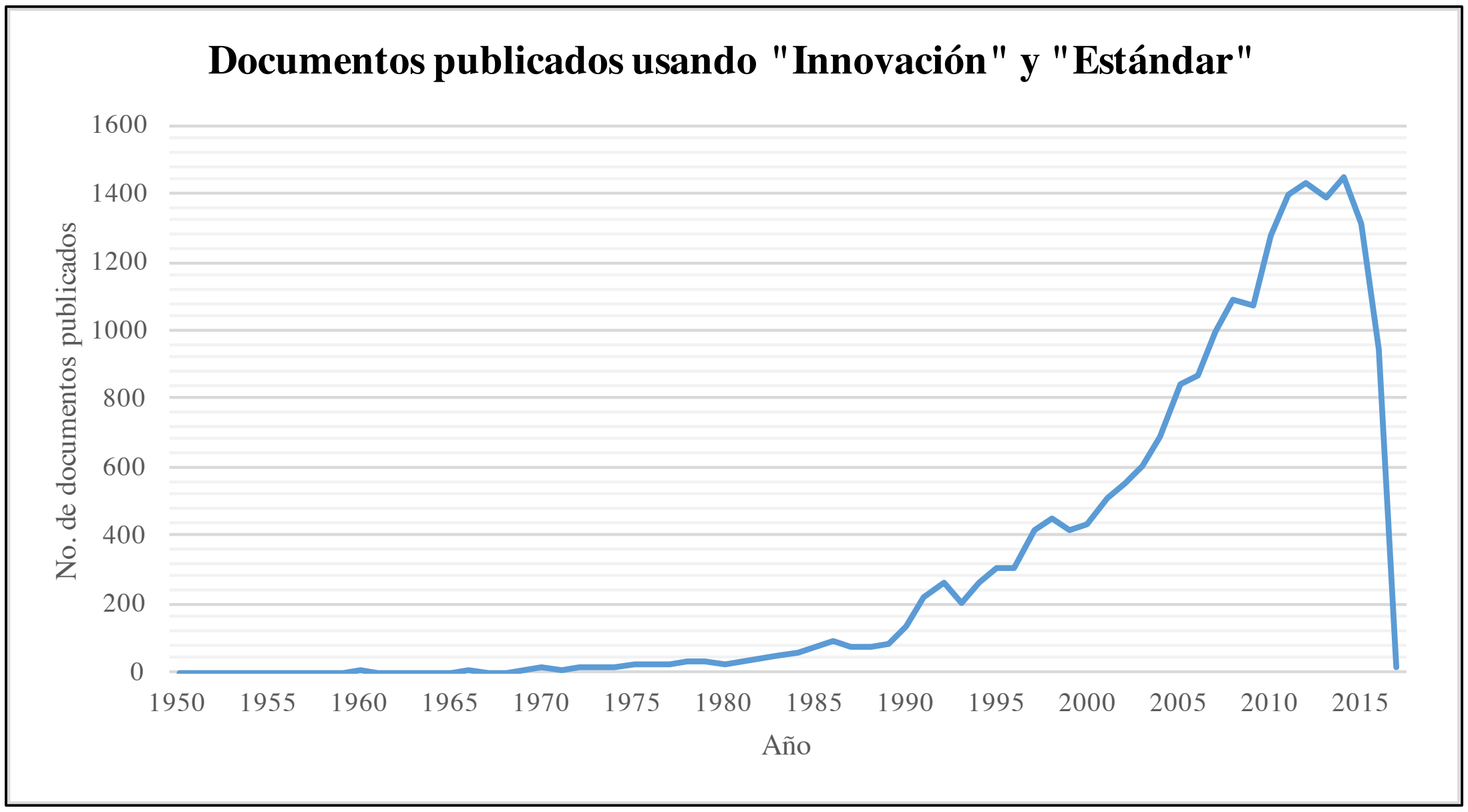

**Fuente**: (Scopus®, 2016)

Tal como se observa en la figura 2.19, existen 20,627 documentos desde el año de 1926 a la fecha usando los términos de “innovación” y “estándar”, principalmente en las áreas de medicina, ingenierías y ciencias de la computación. Por otra parte, usando los términos de búsqueda “gestión del conocimiento” y “estándar” hay un total de 2,829. Se observa cómo, para el segundo ejemplo, fue a partir del año 2000 cuando empieza a proliferar en el mundo académico esta combinación, como se observa en la figura 2.20.

**Figura 2.20. Documentos publicados con los términos “gestión del conocimiento y estándar”**

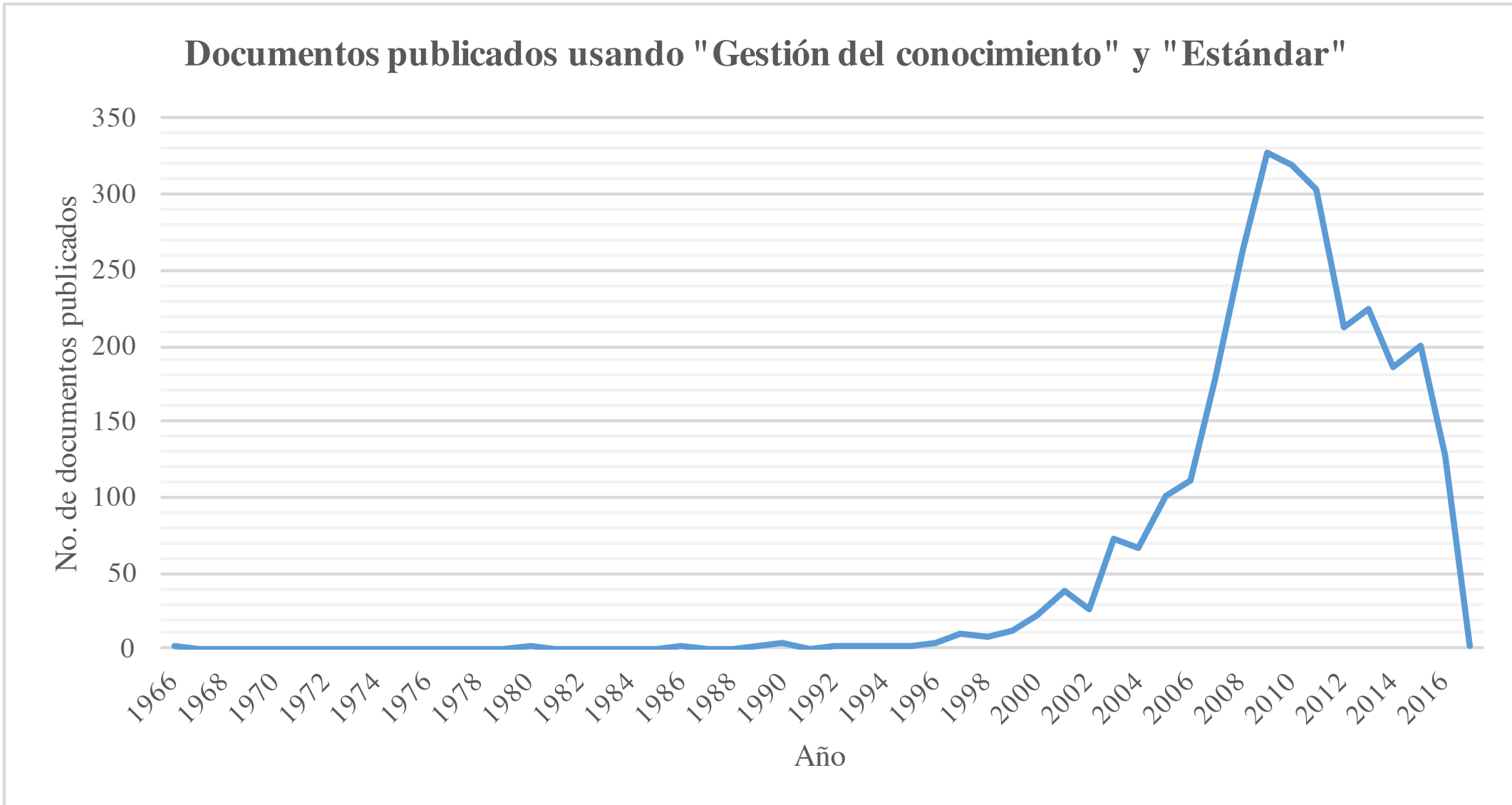

**Fuente**: (Scopus®, 2016)

## 2.3. Hacia un estándar mexicano de gestión del conocimiento e innovación

La estandarización es el proceso que hace que las cosas del mismo tipo tengan las mismas características básicas (Cambridge University Press, 2016). El estándar surge de un enfoque transdisciplinario, para mejorar los resultados de la organización y el aprendizaje, a través de maximizar el uso del conocimiento. Implica el diseño, implementación y revisión de actividades y procesos sociales y tecnológicos para mejorar la creación, el intercambio, la aplicación o el uso del conocimiento (Bouthillier & Shearer, 2002), tal y como se ha referido en diversos trabajos previos que establecen la importancia de construir estándares específicos para la medición de la gestión del conocimiento (Rodríguez, 2013, p. 146).

El punto clave es que la normalización es que es un proceso voluntario para el desarrollo de especificaciones técnicas, pero cada vez más también otras, basadas en el consenso entre las partes interesadas: la industria en primer lugar, pero también una variedad de usuarios, grupos de interés y público autoridades (Blind, 2013). El nivel de consenso de un estándar está en función del tiempo de desarrollo que éste conlleva, tal como se muestra en la figura 2.21.

Normalmente son guías que pueden usarse en diversos tipos de organizaciones con las características mínimas que deben reunir las organizaciones si quieren alcanzar metas u objetivos comunes en algún campo disciplinar. Diversos esfuerzos en materia de estandarización relativos a la innovación y a sus productos, mediciones e indicadores (Blind, 2016). La propuesta que a continuación se presenta se ha enriquecido de las presentaciones realizadas durante el foro: "*OECD Blue Sky Forum on Science and Innovation Indicators*" (OECD, 2016a, 2016d).

**Figura 2.21. Grado de consenso y tiempo de desarrollo de un estándar**

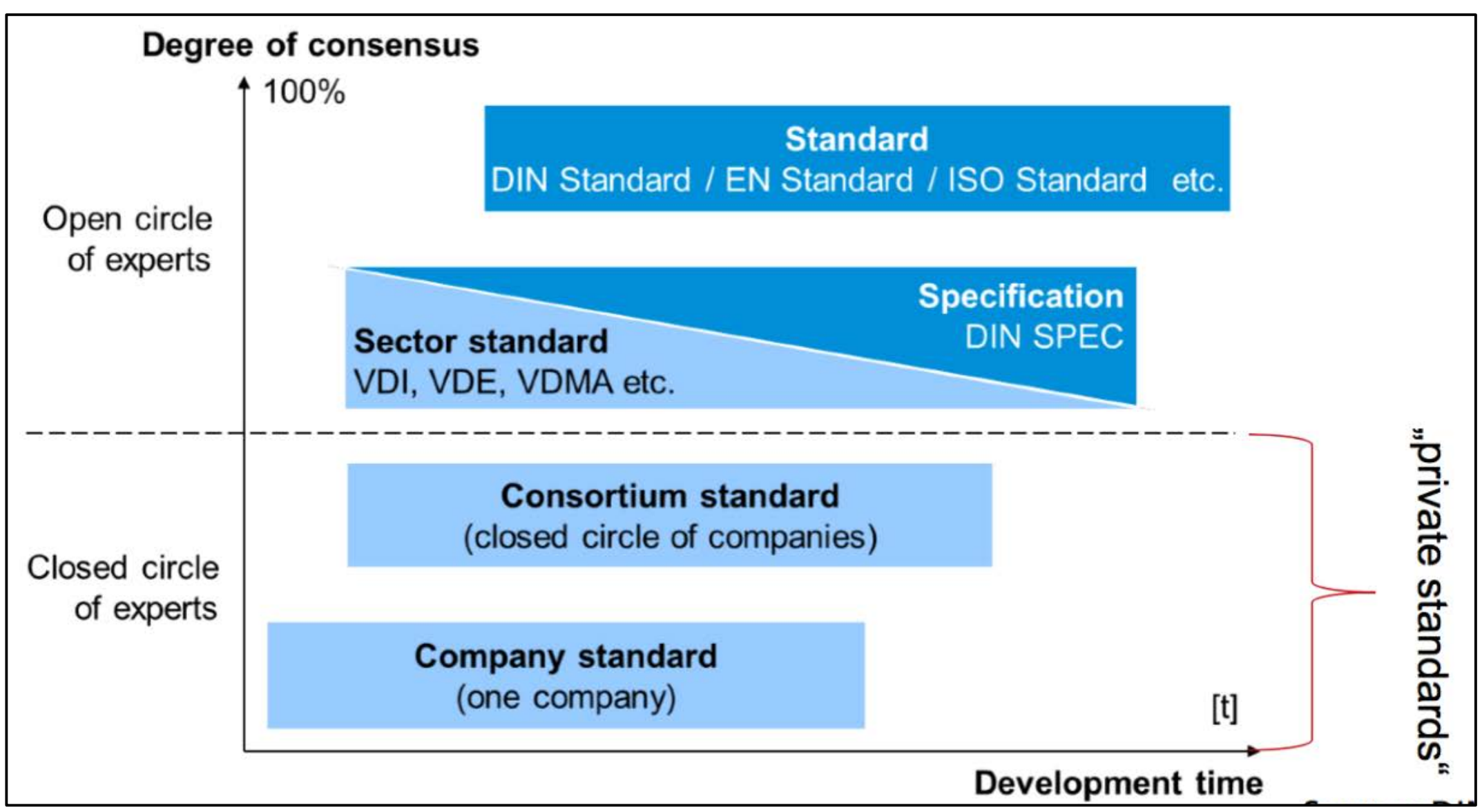

**Fuente:** (Blind, 2016; OECD, 2016a)

### *2.3.1. Iniciativas de estandarización de la gestión del conocimiento y la innovación*

Diversas iniciativas de estandarización de la gestión del conocimiento se han llevado a cabo en Australia (Burford & Ferguson, 2011; Standards Australia, 2005), Inglaterra (Burford & Ferguson, 2011; Chatwin, 2006; Ferguson, 2006), Alemania y a nivel europeo han tratado de alcanzar un entendimiento común de la GC. Estas iniciativas se han dado, principalmente por la necesidades de contar con elementos que permitan la innovación sistemática, la creación de un sistema replicable (Karlsson, 2013), que se materialice a través de marcos comunes u homogéneos de aplicación. Ello, visto a través de esquemas que se puedan comparar.

Las propuestas de estandarización relativas a la gestión del conocimiento surgen de la necesidad de administrar el conocimiento (Burford & Ferguson, 2011). La estandarización relativa a la innovación surge por la necesidad de asegurar el rendimiento, conformidad y seguridad de los procesos y productos, existe una relación intrínseca entre los estándares y la innovación, ya que crean un entorno competitivo (Farrell & Saloner, 1985; Henning, 2013).

Tener una propuesta de estándar mexicano de gestión del conocimiento e innovación se fundamenta, en que como se ha expuesto, México tiene un gran potencial para alcanzar mejores resultados en estos dos temas (Cornell University et al., 2016; OECD, 2016d), importantes en una economía digital y que poco a poco avanza hacia una cuarta revolución industrial. La innovación, como hemos observado se ha vuelto un imperativo en las organizaciones del siglo XXI (OECD, 2015e, 2015f) y nuestro país no puede permanecer ajeno a ello (Bonilla & Agencia Informativa Conacyt, 2016).

### *2.3.2. Estandarización y gestión del conocimiento e innovación*

Cuando inicialmente se planteaba el término estandarización relacionado con la innovación, se analizaba en función de procesos de manufactura, debido a que, con frecuencia, los estándares se derivan de las innovaciones tecnológicas. Así es como los beneficios de las normas sobre la innovación en el diseño y la fabricación superan las posibles limitaciones de la creatividad impuestas por dichas normas (Allen & Sriram, 2000).

Es un hecho que existe cierta influencia en ambos sentidos entre la estandarización y la innovación (Jakobs, 2006). En realidad existe una serie de estándares que apoyan el intercambio de información sobre productos, procesos, operaciones y cadena de suministro, armonización del producto y gestión del ciclo de vida (Rachuri et al., 2008), no así un estándar de gestión del conocimiento e innovación. La existencia de lo anterior, podría favorecer la innovación en las organizaciones.

La estandarización relacionada con la innovación puede ser una herramienta importante para el crecimiento económico (Shin, Kim, & Hwang, 2015; Tidd, 2006), los esfuerzos de estandarización constituyen acciones importantes para encauzar los esfuerzos y lograr determinados objetivos más amplios (Yoo, Lyytinen, & Yang, 2005). La innovación puede ser gestionada a través de sistemas o marcos estandarizados de gestión de la innovación (Mir, Casadesús, & Petnji, 2016, p. 26)

Además, la estandarización requiere la emisión formal, aceptación e implementación de políticas, procesos, sistemas, planes y procedimientos que se utilizan de manera consistente en toda la empresa para autorizar y administrar todas las carteras, programas y proyectos (Bolles & Hubbard, 2007). Uno de sus principales beneficios es crear consenso en los elementos, procesos y actividades mínimas, pero también nivelar las capacidades individuales y organizacionales (Bolles & Hubbard, 2007).

En el mundo académico, hay también quienes sugieren que los estándares relacionados con innovación deberían ser flexibles para permitir a que todos los tipos de empresas puedan innovar (Drahos & Maher, 2004), con el mínimo grado de restricciones, y hay quienes incluso critican que la presencia de estándares, al menos mínimos, por el hecho de que pueden inhibir la capacidad para innovar (Maxwell, 1998). De hecho uno de los problemas clave en la gestión de la innovación es dar sentido a un conjunto de fenómenos complejos, inciertos y altamente arriesgados (Tidd, 2006). Otra crítica comúnmente referida es que la adopción de un estándar suele ser compleja, principalmente por compatibilidad entre las organizaciones (Farrell & Saloner, 1985).

A través de los años recientes se ha observado que los estándares pueden hacer una diferencia significativa en el éxito de la innovación creando un marco compartido para la innovación y estableciendo las reglas del juego (Shin et al., 2015, p. 155). Por ello, el diseño de estándares debería involucrar la participación de distintas áreas (Marsal-Llacuna & Wood-Hill, 2017) organizacionales.

En tal sentido, es posible proponer un estándar de gestión del conocimiento e innovación tecnológica, atendiendo a que ambos elementos se materializan en la organización a través de procesos (Birkinshaw et al., 2008a; Chang Lee et al., 2005;

Ding et al., 2014; Tidd, 2006), y tienen impactos, a partir de sus resultados, en la estrategia de negocios (Eyring & Johnson, 2005; Matzler, 2013).

David Teece (1997) sostiene que la capacidad de la empresa para apropiarse de los beneficios de su inversión en tecnología depende de dos factores: (i) la capacidad de la empresa para traducir su ventaja tecnológica en productos o procesos comercialmente viables; y (ii) la capacidad de la empresa para defender su ventaja frente a los imitadores. Así , la estandarización es un componente importante que permite a una organización beneficiarse de sus desarrollos tecnológicos (Tidd, 2006).

A través de las competencias como medio de estandarización de las conocimientos, habilidades y actitudes que deben tener los trabajadores (de Sordi & Carvalho Azevedo, 2008) es posible contribuir a tener colaboradores mejor calificados en sus capacidades y en el valor que agregan en sus operaciones.

### *2.3.3. Beneficios potenciales de la estandarización de la gestión del conocimiento y la innovación*

En este sentido, la elaboración de un estándar que permita a la organización dimensionar la forma en la que genera, adquiere, almacena, usa y administra su conocimiento puede dar lugar a que determine una serie de acciones estratégicas que le permitan incrementar su productividad y mejorar su desempeño económico (Alsadhan, Zairi, & Keoy, 2008). La definición de un estándar ayudaría a las organizaciones a lo siguiente:

1. Poder replicar las experiencias exitosas en materia de gestión del conocimiento e innovación en otras organizaciones, tal como se hace en la gestión de proyectos (Burford & Ferguson, 2011; IEEE, 2004), al tener una *guía fácil de leer y no prescriptiva sobre la gestión del conocimiento (Standards Australia, 2005);*

2. Mejorar el entendimiento que se tiene sobre los procesos de gestión del conocimiento y sobre la gestión de la innovación en las organizaciones (Jonsson, 2013; Pee & Kankanhalli, 2008), al proponer un piso común que sirva como base para encausar y guiar (Burford & Ferguson, 2011), sin ánimo de ser prescriptivo, de las iniciativas de gestión del conocimiento e innovación en las organizaciones mexicanas;
3. Medir el volumen de conocimiento que poseen (Edison, Bin Ali, & Torkar, 2013; Hubbard, 2014; Phusavat, Anussornnitisarn, Helo, & Dwight, 2009);
4. Evaluar su capacidad para crear, usar, administrar y retener nuevo conocimiento (Yamaguchi, 2014);
5. Medir su capacidad para adquirir y usar nuevo conocimiento proveniente del exterior (Burford & Ferguson, 2011) y su capacidad de innovación (Adams, Bessant, & Phelps, 2006);
6. Utilizar de mejor forma sus recursos humanos y tecnológicos definiendo repositorios de conocimiento (Arora, 2002; Kovačič, 2007; Milner, 2000) y facilitando mecanismos para su accesibilidad (Karr, 2016; Koh, Ryan, & Prybutok, 2005);
7. Valuar sus activos intelectuales como un elemento de apreciación comercial del negocio (Burford & Ferguson, 2011);

8. Asegurar la disponibilidad inmediata del conocimiento para el óptimo desempeño de los procesos clave, insertando nuevo y mejorado conocimiento en su cadena de valor (Plessis & du Plessis, 2007);
9. Guardar y preservar la "memoria organizacional" (Kimble, de Vasconcelos, & Rocha, 2016; Sánchez & Morrison-Saunders, 2011) como una herramienta de valor para futuras tomas de decisiones (Quast, 2012);
10. Apreciar si una adecuada gestión del conocimiento y de la innovación genera una ventaja competitiva para la organización (Hamel, 2006; Johannessen, 2008; Kör & Maden, 2013; Williamson & Yin, 2014);
11. Establecer la relación que existe entre la óptima gestión del conocimiento y la capacidad de innovación tecnológica de la organización (Altunok & Cakmak, 2010; Sauser, Verma, Ramirez-Marquez, & Gove, 2006; Straub, 2015; Swan et al., 1999; van der Heiden, Pohl, Mansor, & van Genderen, 2016);
12. Un estándar puede potencializar los resultados del proceso de gestión del conocimiento (Ferguson, 2006) (Mir et al., 2016);
13. Potencializar las actividades y resultados de los trabajadores del conocimiento (Burford & Ferguson, 2011; Kulkarni et al., 2007);
14. La interoperabilidad de los esfuerzos y de sus resultados entre organizaciones (Henning, 2013);
15. Incrementar la capacidad para innovar de las organizaciones (Blind, 2016; Mir et al., 2016);

16. Tener un estándar podría permitir una mejor gestión del conocimiento y de la innovación en las organizaciones (Blind, 2013, 2016; Mir et al., 2016);
17. Ayudar a las personas ya las organizaciones a profundizar su comprensión de los conceptos de gestión del conocimiento (Burford & Ferguson, 2011; Standards Australia, 2005);
18. Ofrecer un marco escalable y flexible para diseñar, planificar, implementar y evaluar intervenciones de conocimiento e innovación que respondan al entorno de la organización y al estado de preparación conocimiento (Burford & Ferguson, 2011; Standards Australia, 2005);
19. Mejorar la difusión de mejores prácticas de gestión del conocimiento y la innovación y permite tener un marco común de medición entre organizaciones del mismo sector, lo que podría crear competencia y contribuyen al crecimiento impulsado por la innovación. (Blind, 2013, p. 9).

## 2.4. Propuesta estándar mexicano de gestión del conocimiento e innovación tecnológica

### *2.4.1. Factores del estándar propuesto*

Después de analizar distintos modelos sobre gestión del conocimiento e innovación, identificar los elementos que están interrelacionados entre ambos conceptos, es posible observar que, en la literatura académica, hay factores (F) que inciden en la GC: (1) Factor Humano, (2) Factor Organización, (3) Factor Infraestructura y (4) Factor Estrategia (Heisig, 2009), y que asimismo hay etapas comunes en el Proceso de Gestión del Conocimiento (PGC): creación, almacenamiento, obtención, transferencia y aplicación del conocimiento (Chang Lee et al., 2005; Ding et al., 2014).

**Tabla 2.2. Descripción de los factores y sus componentes integrados en la propuesta**

| | Descripción |
|---|---|
| **Factor 1. Humano** | Implica todos los elementos socio-organizacionales que hacen posible la gestión del conocimiento y la innovación, tal como las personas, los estilos de liderazgo, y la cultura de la organización (Esterhuizen et al., 2012; Heisig, 2009; Ho, Hsieh, & Hung, 2014; Mas-Machuca & Martínez Costa, 2012) |
| **Componentes del Factor 1** | **C1. Confianza y colaboración:** La confianza se refiere a la creencia común entre los miembros de una organización que otros hacen esfuerzos de buena fe para comportarse de acuerdo a un compromiso, actúan con honestidad y no toman ventaja de otros aun cuando se presenta una oportunidad de hacerlo (Pee & Kankanhalli, 2008, p. 442). La confianza promueve el intercambio abierto de conocimiento (Pee & Kankanhalli, 2008; Yuen, 2007). Por su parte, la colaboración es el grado de voluntad que los individuos exhiben para apoyarse entre ellos (H. Lee & Choi, 2003), requiere de mecanismos que faciliten la interoperabilidad, para el intercambio de información y de recursos tecnológicos (Luna-Reyes, Gil-Garcia, & Cruz, 2007) entre individuos e instituciones.<br><br>**C2. Cultura de la organización:** La cultura organizacional es considerada uno de los factores críticos más importantes en el éxito de la gestión del conocimiento y la innovación (M. Allameh et al., 2011; Anthony et al., 2008; Armson, 2008; Ho et al., 2014), tiene influencia esencial en la decisión acerca de cuándo, dónde y con quién debe ser intercambiado el conocimiento (Omar Sharifuddin Syed-Ikhsan & Rowland, 2004). La cultura puede ser definida como "los valores de una organización, principios, normas, reglas no escritas y procedimientos de la organización (C.W. Holsapple & Joshi, 2001, p. 46).<br><br>**C3. Formación y desarrollo:** Implica el desarrollo de conocimientos, habilidades y actitudes (competencias) (P. R. Massingham & Massingham, 2014), para ello se deberán identificar las necesidades de formación organizacional e implementar programas de entrenamiento y desarrollo de habilidades (Hafeez & Abdelmeguid, 2003; P. Massingham, 2014a). Este componente involucra la formación continua (March, 1991) de las personas que trabajan en la organización y el esfuerzo institucional para fomentar el deseo, entre los trabajadores, para mantener actualizados sus conocimientos. A nivel de la organización una actitud de aprendizaje es deseable para obtener resultados de la gestión del conocimiento y la innovación (Chang Lee et al., 2005; Salleh et al., 2012). |
| **Factor 2. Organización** | Integra a los elementos de orden lógico de la organización, sus procesos, y procedimientos internos; así como su estructura, documentación y metodologías empleadas para capturar, usar y reutiliza el conocimiento y sus innovaciones (Daft, 2011; Perez Lopez-Portillo, 2016) |
| **Componentes del Factor 2** | **C4. Documentación:** La documentación en las organizaciones es un elemento indispensable que permite reducir las curvas de aprendizaje (Ding et al., 2014), nivelar el conocimiento y preservar la memoria organizacional (Lj Todorović, Č Petrović, Mihić, Lj Obradović, & Bushuyev, 2015; Mao, Liu, Zhang, & Deng, 2016).<br><br>**C5. Metodologías:** Las metodologías son formas de llevar a cabo las actividades (Brockmann & Roztocki, 2015; Chen, Shih, & Yang, 2009; IEEE, 2004), implican cierta alineación con objetivos estratégicos (Batini & Scannapieco, 2016; Dehghani & Ramsin, 2015; Fine & Hax, 1985; Levett & Guenov, 2000), así como proveer marcos de gestión de actividades que estén estandarizados y normalizados en toda la organización, lo que incentiva la gestión del conocimiento y la innovación.<br><br>**C6: Procesos y procedimientos:** implican el uso del método más eficiente para "transformar" el conocimiento implícito, fragmentado y privado de los individuos o grupos, tanto dentro como fuera de la organización en activos intelectuales valiosos para la organización (Ho et al., 2014, p. 736). La claridad en los procesos internos da secuencia y estructura lógica a las actividades de la organización (Savvas & Bassiliades, 2009). Esto contribuye a su medición y mejora continua, y por tanto a la efectividad organizacional de la gestión del conocimiento y la innovación (Chong & Chong, 2009). |
| **Factor 3. Infraestructura** | Agrupa a todos elementos tecnológicos, aplicaciones, sistemas que utiliza la organización para la gestión del conocimiento y la innovación. La infraestructura permite establecer redes de colaboración, como las wikis, foros, redes sociales, al tiempo que facilitan el flujo del conocimiento entre los empleados y hacia la organización (Tabrizi, Ebrahimi, & Delpisheh, 2011, p. 695), proporciona sistemas de almacenamiento eficiente y mecanismos de recuperación y transferencia del conocimiento (Pee & Kankanhalli, 2008; Perez Lopez-Portillo, 2016; Pérez López-Portillo, Romero Hidalgo, & Mora Martínez, 2016) y la innovación (Armson, 2008; Quintane, Mitch Casselman, Sebastian Reiche, & Nylund, 2011). |
| **Componentes del Factor 3** | **C7: Infraestructura física:** se refiere a todas aquellas herramientas tecnológicas producto de tecnologías de la información y sus capacidades de apoyo a la GC (Pee & Kankanhalli, 2008), como sistemas de comunicación, redes, repositorios de conocimiento, programas de |
| | formación a distancia (Wiig, 2002), sistemas de manejo de la información, aplicaciones de inteligencia, software experto, bases de datos, tecnologías específicas, y modelos que se emplean para gestionar el conocimiento de la organización (Ragab & Arisha, 2013, p. 878).<br><br>**C8: Re-uso del conocimiento:** Esta actividad facilita las tareas mediante la aplicación de conocimientos, es sobre el grado de conocimiento utilizado por la organización y el conocimiento (Chang Lee et al., 2005; Ho et al., 2014), un indicador de éxito de la innovación es precisamente la reutilización del conocimiento enriquecido por la experiencia.<br><br>**C9: Sistemas de información:** son soluciones basadas en las TIC que apoyan la implementación de la GC en las organizaciones (Ragab & Arisha, 2013), mediante actividades de captura y representación, recuperación, intercambio, reutilización, razonamiento y recuperación del conocimiento en una organización (Ding et al., 2014). Algunos ejemplos de sistemas son: los de gestión de documentos, motores de recuperación de información, bases de datos relacionales y de objetos, sistemas de trabajo en grupo y flujo de trabajo, tecnologías de empuje y agentes, así como herramientas de minería de datos (Giudice & Peruta, 2016; Little & Deokar, 2012; Wong & Aspinwall, 2005; Zyngier et al., 2006). Además son un elemento de apoyo para la planeación y la toma de decisiones en las instituciones (P. Jain, 2009), porque permiten generar marcos institucionales de innovación y procesos de colaboración a través sistemas de gestión integrados (Puron-Cid, 2014; United Nations, 2014) e interoperables entre diferentes instituciones y/o sectores (Savvas & Bassiliades, 2009; Yang & Maxwell, 2011). Se ha observado que la tecnología puede proveer la red para relacionar a grupos geográficamente dispersos (Swan et al., 1999). |
| **Factor 4. Estrategia** | Integra todos aquellos elementos relacionados con la gestión estratégica de la gestión del conocimiento y la innovación en la organización (Man, 2008; Perez Lopez-Portillo, 2016; Pérez López-Portillo et al., 2016; Pisano, 2015; Woolf, 2010) |
| **Componentes del Factor 4** | C10. Estrategia de gestión del conocimiento: Se refiere a los objetivos de gestión del conocimiento (Pee & Kankanhalli, 2008) y la forma en que se van a medir. La estrategia provee las bases sobre cómo la organización desplegará sus competencias y recursos para alcanzar los objetivos de la GC (Akhavan, Jafari, & Fathian, 2006, p. 108). En suma, la estrategia de gestión del conocimiento da sentido y orienta los esfuerzos organizacionales en materia de GC.<br><br>**C11. Estrategia de innovación:** Se refiere a la estrategia que se implementa para evitar las fallas comunes en las iniciativas de innovación, es necesaria para orientar los recursos y objetivos de la organización (Allen & Sriram, 2000, p. 30; Dodgson et al., 2008; OECD, 2015c; Pisano, 2015). |

**Fuente**: Elaboración propia (2016)

### *2.4.2. Gradualidad del estándar*

Por otra parte, asumimos que el estándar tendría que ser gradual como el desarrollo mismo de proyectos (IEEE, 2004; Khedhaouria & Jamal, 2015; P. R. Massingham & Massingham, 2014). En tal sentido, creemos que el estándar debería tener al menos cuatro niveles: Básico, Intermedio, Avanzado y Experto en relación a la gestión del conocimiento y la innovación, tal como los niveles que se observan en el TRL Technology Readiness Level ((Fast-Berglund, Bligård, Åkerman, & Karlsson, 2014; Sauser et al., 2006; Straub, 2015).

El estándar asume el supuesto de que la innovación ocurre cuando se implementan las ideas generadas (Popadiuk & Choo, 2006). En la tabla 2.3., se describe a grandes rasgos los niveles propuestos:

**Tabla 2.3. Descripción de los niveles propuestos en el estándar**

| Nivel | Descripción |
|---|---|
| **Nivel básico** | Caracterizado por que los procesos están parcialmente documentados, se observa un bajo nivel de formación de las personas para la gestión del conocimiento y la innovación; falta de instrumentos o mecanismos que permitan la creación o utilización del conocimiento organizacional. Este nivel se relaciona con la creación del conocimiento (Arora, 2002; Diakoulakis et al., 2004; Farzin, Kahreh, Hesan, & Khalouei, 2014; Gasik, 2011; Hunter, Webster, & Wyatt, 2005; C. S. Lee & Wong, 2015; Plessis & du Plessis, 2007). |
| **Nivel intermedio** | En este nivel la organización comienza a conocer su conocimiento, es capaz de manejar sus actividades mediante procesos definidos y establecidos. Utiliza el conocimiento que obtiene y produce para tomar decisiones e involucra a sus miembros en procesos de mejora continua en sus actividades y procesos clave. En este nivel la organización conoce cómo acumula conocimiento y es capaz de inventariarlos para aplicarlos. Este nivel de relaciona con la acumulación del conocimiento (Carneiro, 2000; Costa & Monteiro, 2016; Edenius & Borgerson, 2003). |
| **Nivel avanzado** | La organización utiliza el conocimiento que obtiene y produce como un elemento esencial de sus operaciones cotidianas, transfiere el conocimiento entre sus miembros y en otras áreas de la organización para mejorar la toma de decisiones. Permite la auto actualización de sus procesos clave mediante la gestión del conocimiento y la innovación. Produce conocimiento de gran valor y transfiere el resultado de sus innovaciones al entorno. Este nivel guarda relación con la transferencia del conocimiento (Bate & Robert, 2002; Boland et al., 2001; Edenius & Borgerson, 2003; González & Rodríguez, 2016; Klaus-Dieter, Frank, & Carsten, 2000; Zapata Cantu & Mondragon, 2016). |
| **Nivel experto** | La gestión del conocimiento y la innovación son actividades esenciales de la organización. Se caracteriza por su intensa producción de conocimiento, que reutiliza y mejora sistemáticamente mediante la producción de nuevo conocimiento. Es capaz de evaluar la calidad del conocimiento que genera. Otras organizaciones buscan a esta organización por sus productos de gestión del conocimiento e innovación, que transfiere mediante distintos mecanismos (Fateh Rad et al., 2015). El conocimiento de la organización se encuentra disponible para todos los miembros. Este nivel se relaciona con la transferencia-mejora del conocimiento a través de la innovación (Chan & Chow, 2007; Donate & Sánchez de Pablo, 2014; Kogut & Zander, 1992; Plessis & du Plessis, 2007). |

**Fuente**: Elaboración propia (2016)

Consecuentemente, en la figura 2.22, se muestra la propuesta conceptual del estándar con los elementos que debería integrar. Se ha decidido representar con un espiral, tal como el espiral de creación del conocimiento del Modelo SECI (Ikujiro Nonaka, 2008), que da el sentido de gradualidad en la creación, utilización e implementación del conocimiento y la innovación.

En este sentido, el proponer una herramienta que adopta la forma de un estándar, podría apoyar a las organizaciones a identificar las diversas variables que inciden en el mejor aprovechamiento de sus recursos, así como para establecer, a partir de una medición objetiva, la mejor estrategia para hacerse ya sea del conocimiento que requieren o bien la mejor forma de administrar el que podrían transferir para lograr una mejor innovación en sus productos, procesos o servicios.

**Figura 2.22. Modelo del estándar propuesto**

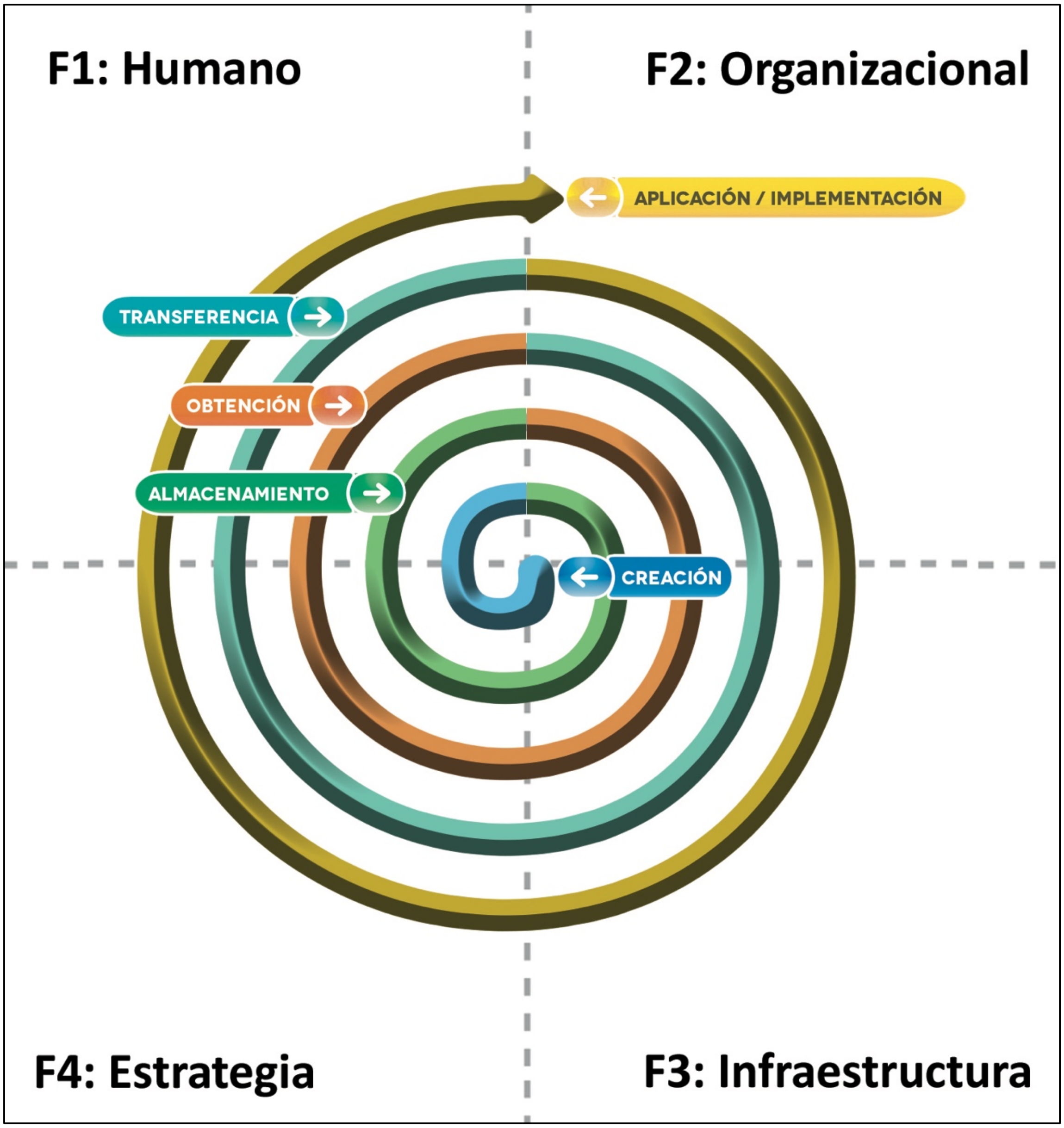

**Fuente:** Elaboración propia (2016).

Los niveles que se han definido para el estándar propuesto se han derivado a partir de trabajos previos realizados en el mundo académico (Gasik, 2011; Nasa, 2014; Shareef, Kumar, Kumar, & Dwivedi, 2011), y a partir de la gradualidad que normalmente se observa en la gestión del conocimiento y en la innovación. Así como a la aproximación que comúnmente se asocia en otros modelos, por ejemplo: valoraciones de los niveles del *Technology Readiness Level* (TRL) (Altunok & Cakmak, 2010; Fast-Berglund et al., 2014; Mankins, 1995; Nasa, 2014; Sauser et al., 2006; Straub, 2015); los modelos de generación o producción del conocimiento, originalmente propuestos (Ikujiro Nonaka, 1994; Polyani, 1958; J.-C. Spender & Grant, 1996).

**Figura 2.23. Technology Readiness Level**

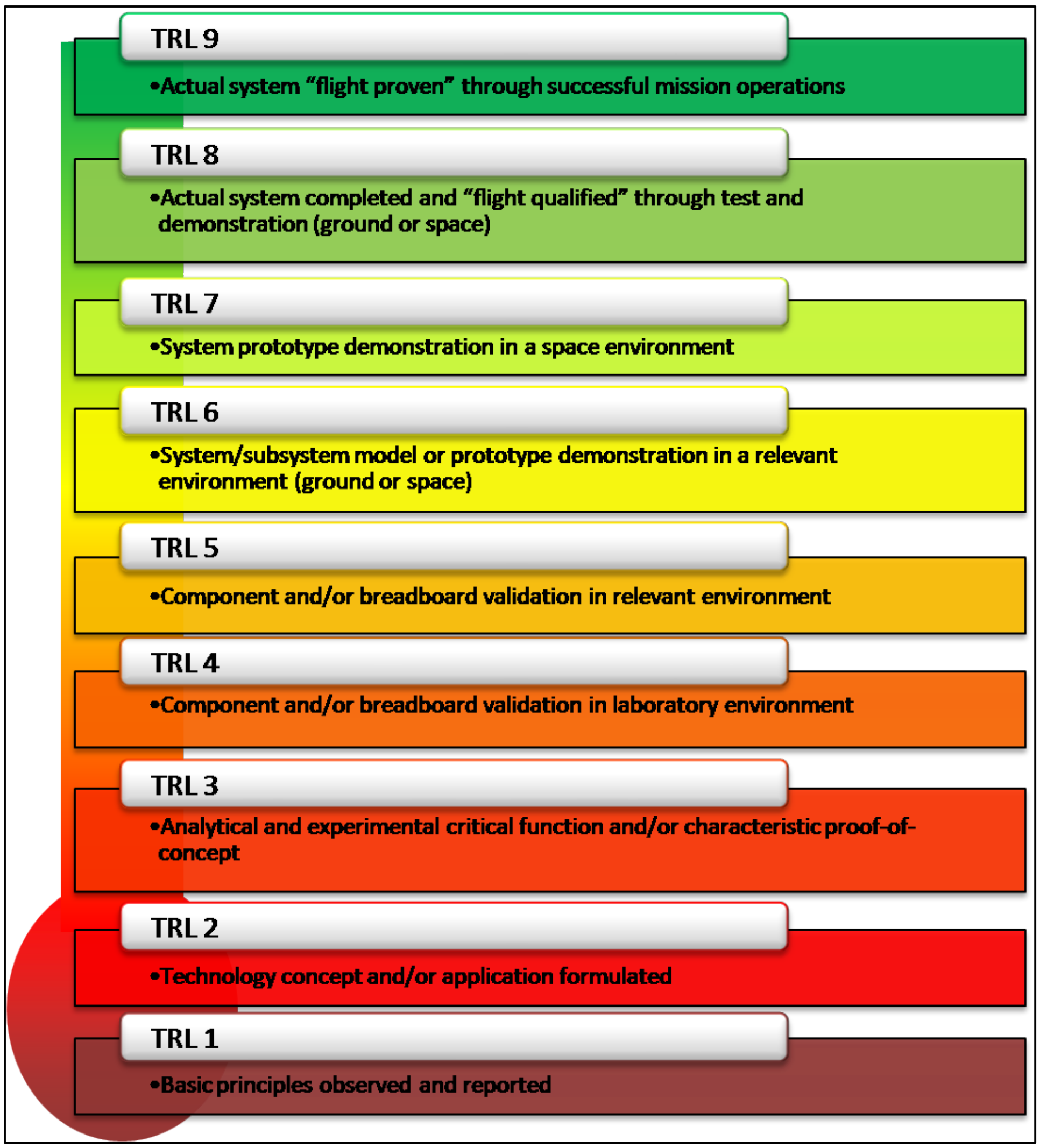

**Fuente**: (Mankins, 1995; NASA, 2012).

Así como las etapas del Proceso de Gestión del Conocimiento, *Knowledge Management Process* (KMP) (Chang Lee et al., 2005; Ding et al., 2014; Maier & Schmidt, 2014; N. Thomas & Vohra, 2015); el funcionamiento de los Sistemas de Gestión del Conocimiento, *Knowledge Management Systems*, (KMS) (Balaid, Abd Rozan, Hikmi, & Memon, 2016; Dorasamy, Raman, & Kaliannan, 2013; Kulkarni et al., 2007; Un Jan & Contreras, 2016); el proceso de gestión de la innovación (Innovation Management) (Birkinshaw et al., 2008b; Carneiro, 2000; R G Cooper, 1990; Darroch, 2005; Vijay K Jolly, 1997; Moore, 1999; Razmerita, Phillips-Wren, & Jain, 2016); el Integración de modelos de madurez de capacidades, *Capability Maturity Model Integration* (CMMI) (Dayan & Evans, 2006; Software Engineering Institute, 2010; Strutt, Sharp, Terry, & Miles, 2006); así como la teoría general de desarrollo y gestión de proyectos (IEEE, 2004; Kerzner, 2013; Project Management Institute, 2013; Sokhanvar, Matthews, & Yarlagadda, 2014).

**Figura 2.24. Ciclo de vida de adopción de tecnología (Moore, 1999)**

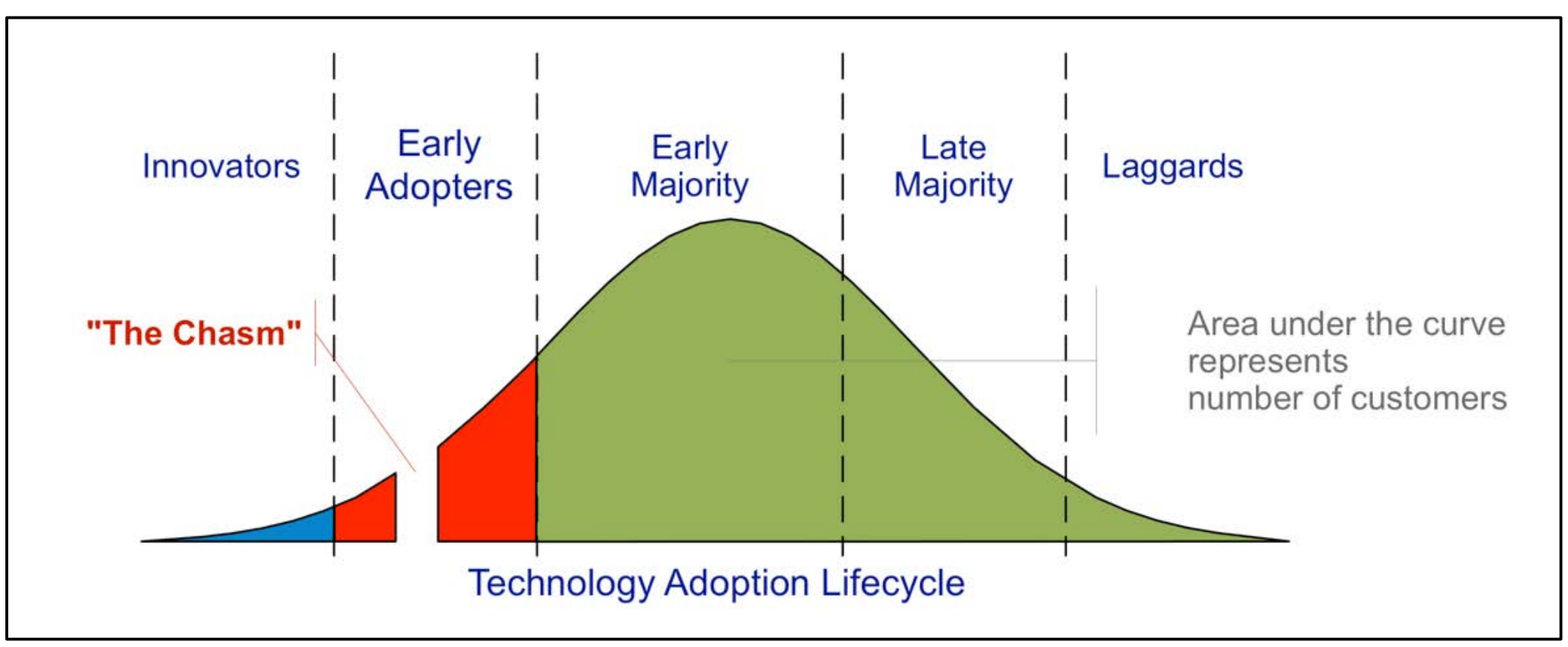

**Fuente:** (Moore, 1999)

Otros elementos que también se consideraron fueron las normas IEEE de desarrollo de software y proyectos de tecnología (Augustine, 2008; Carmo & Albuquerque, 2014; Engineering & Committee, 2013; Gasik, 2011; IEEE, 2004; Imnc & Projects-, 2008; Society, 2007; Standard, 2009), comúnmente utilizadas en el área de ingeniería de software.

A razón de lo anterior, los niveles propuestos para el estándar se deberán observar en función de los del progreso lógico de la organización en cuanto a la creación, almacenamiento, obtención, transferencia y aplicación del conocimiento desarrollado en la organización y en la implementación de innovaciones (Carneiro, 2000; Darroch, 2005; Greenhalgh, Robert, Macfarlane, Bate, & Kyriakidou, 2004; Johannessen, 2008; Mageswari, Sivasubramanian, & Dath, 2015; Razmerita et al., 2016), lo que subyace sobre la relación entre la gestión del conocimiento y la innovación.

### *2.4.3. Descripción del modelo propuesto*

En el modelo se puede apreciar un espiral como el que se propuso hace algunos años para la creación del conocimiento (Ikujiro Nonaka, 1994; Ikujiro Nonaka & Takeuchi, 1999), y más recientemente en el mundo de la gestión de proyectos y el desarrollo de software (IEEE, 2004). En el último giro de la espiral se representa la implementación del conocimiento (innovación).

El modelo propuesto tiene cuatro cuadrantes que consideran a los cuatro factores objeto de estudio: humano, infraestructura, organización y estrategia, estos factores han sido observados y validados por distintos autores (M. Allameh et al., 2011; S. M. Allameh et al., 2011; Heisig, 2009).

Se ha pensado que cada uno de los factores tenga una integración con las etapas del proceso de gestión del conocimiento (Chang Lee et al., 2005; Ding et al., 2014).

### 2.5. Validación del Estándar mexicano de gestión del conocimiento e innovación

Para realizar una validación empírica de los elementos que conforman a la propuesta descrita anteriormente se ha desarrollado una propuesta metodológica, mediante un modelo de investigación, que permita por un lado validar los elementos del modelo y, por otra parte, comprobar el nivel de gestión del conocimiento e innovación en una organización mexicana. Todo ello se presenta en el capítulo 3, Propuesta Metodológica.

# CAPÍTULO 3. PROPUESTA METODOLÓGICA

En esta sección se describe el tipo de estudio realizado, así como las características teóricas y metodológicas del mismo. Además, se describen los elementos utilizados para definir, construir, validar e implementar el instrumento de evaluación del presente trabajo de investigación, **véase Apéndice 1**. Finalmente, se relata la propuesta de análisis que ha permitido observar el fenómeno objeto de estudio de esta tesis.

**Figura 3.1. Propuesta metodológica**

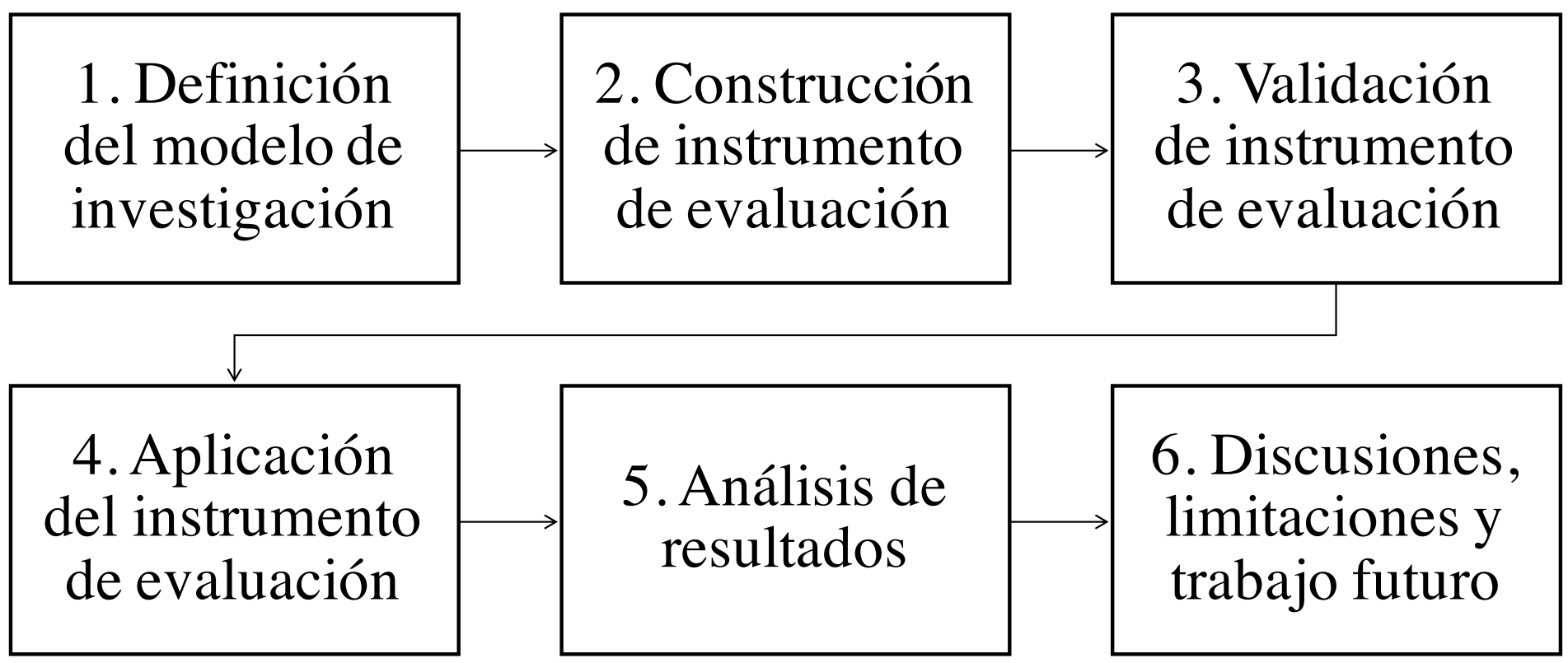

**Fuente:** Elaboración propia (2016)

## 3.1. Propuesta metodológica

El objetivo del presente trabajo es conocer la influencia que tiene la gestión del conocimiento y la innovación en las organizaciones, así como la gradualidad con la cual estos conceptos se conocen, implementan y se usan desde un enfoque estratégico. Para ello, se ha realizado un abordaje metodológico de **seis fases**: en la primera se realizó una revisión de la literatura más relevante (Taylor & Procter, 2008), así como el análisis de los modelos de innovación y gestión del conocimiento que tienen mayor número de citas en el mundo académico.

En la fase 2, se diseñó y construyó el instrumento de evaluación para observar la presencia de los fenómenos objeto de estudio (Corbetta, 2007), en este caso, la gestión del conocimiento y la innovación en las organizaciones. En la fase 3, se validó el instrumento con expertos, se aplicó una prueba piloto y se constató su consistencia interna realizando los ajustes necesarios previos a su aplicación en extenso (Hamann, Schiemann, Bellora, & Guenther, 2013; Henseler, Ringle, & Sarstedt, 2014).

Posteriormente, en la fase 4, se realizó la aplicación del instrumento de evaluación, alineado con el modelo de investigación propuesto, ver figura 3.0.1. Consecuentemente, en la fase 5, se realizó el análisis de resultados utilizando software de tratamiento estadístico; y finalmente en la fase 6 se presentan discusiones sobre los resultados, así como implicaciones prácticas del presente estudio, limitaciones y futuras líneas de investigación a partir de la propuesta.

### *3.1.1. Enfoque o aproximación metodológica*

Esta investigación se ha realizado a partir del paradigma neopositivista. Las principales construcciones conceptuales y metodológicas han sido abordadas desde el enfoque de la administración (*management*), con particular énfasis en las estrategias organizacionales. Así mismo, por la propia naturaleza del objeto de estudio y los estudios previos del autor del presente trabajo, se han adoptado referentes conceptuales del área de ciencias de la computación y el desarrollo de software (Society, 2007; Standard, 2009).

Esta investigación es de tipo transeccional, porque se ha realizado la recolección de datos en un único punto en el tiempo, de tipo exploratorio-descriptivo (Corbetta, 2007; Zorrilla, 2009) y no experimental porque se ha definido, construido, validado e implementado un único instrumento de evaluación que permite la aproximación directa con el objeto de estudio (Corbetta, 2007; Hernández Sampieri, Fernández-Collado, & Baptista Lucio, 2010), la aproximación metodológica de esta tesis se trata de un estudio cuantitativo a profundidad, en este caso en función del propósito de la investigación.

Así, tomando como premisa inicial el objetivo del presente trabajo de investigación: elaborar una propuesta de estándar de gestión del conocimiento e innovación tecnológica aplicable a las organizaciones en México, se planteó un modelo de investigación de seis fases que permita, asimismo cumplir con los objetivos particulares de esta tesis:

**OP1.** Identificar las variables que integrarán el estándar de referencia.

**OP2.** Construir un instrumento válido y confiable que pueda ser aplicado en las organizaciones en México

**OP3.** Aplicar la propuesta de estándar en una organización.

### *3.1.2. Desarrollo de las fases de la propuesta metodológica*

En esta sección se describen detalladamente las fases desarrolladas en esta propuesta metodológica. Se determinó conveniente a partir de los estudios previos (M. Allameh et al., 2011; S. M. Allameh et al., 2011; Liberona & Ruiz, 2013; Mas-Machuca & Martínez Costa, 2012), con la intención de garantizar la mejor definición, construcción y validación del instrumento de evaluación, así como la correspondencia con el modelo de investigación propuesto, que ayude a explicar, observar y analizar el objeto de estudio.

**Figura 3.2. Fases de la propuesta metodológica llevadas a cabo previo a la implementación del instrumento de evaluación**

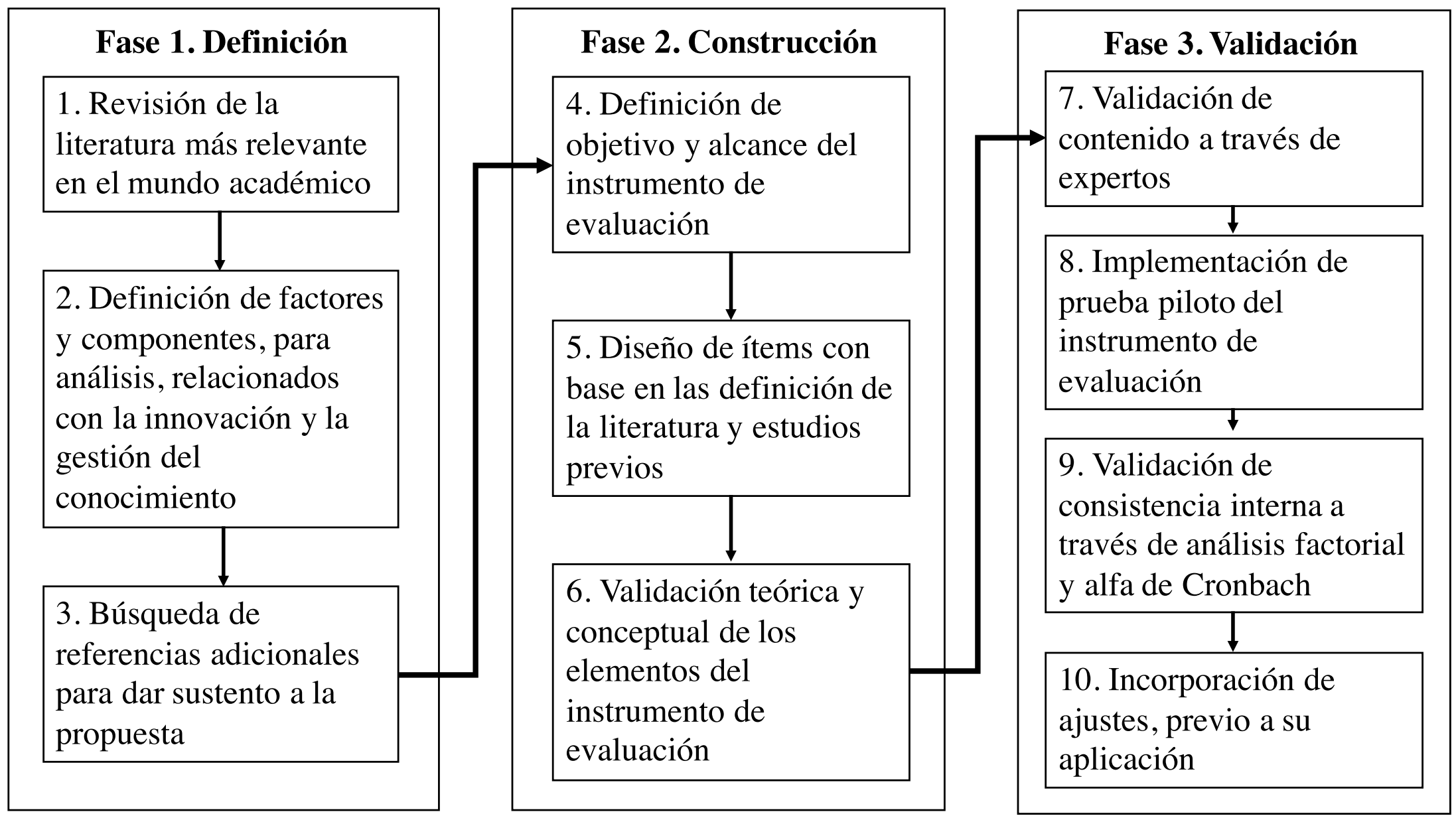

**Fuente:** Elaboración propia (2016)

Dentro de la **Fase 1. Definición**, del modelo de investigación, se han determinado los elementos objetos de análisis y su definición conceptual, para facilitar la aproximación en términos de factores y componentes de exploración, partiendo de (1) una revisión exhaustiva de la literatura más relevante en el mundo académico, sobre el estado del arte (Mitre-Hernández, Mora-Soto, López-Portillo, & Lara-Alvarez, 2015).

Como resultado de lo anterior, a partir de las ideas de (Chesbrough et al., 2006; Ikujiro Nonaka, 1991; Polyani, 1958; Porter, 1990; Schumpeter, 1954), (2) se definieron los factores que conforman la propuesta del estándar mexicano de gestión del conocimiento e innovación, motivo de este trabajo y finalmente se (3) se integraron algunas referencias adicionales que sustentarán la propuesta.

El modelo se desarrolló, como ya se ha mencionado, con base en la revisión de la literatura (Anthony et al., 2008; Balaid et al., 2016; Costa & Monteiro, 2016; Ding et al., 2014; Dorasamy et al., 2013; Greenhalgh et al., 2004; H. Inkinen, 2016)**,** quedando definidos cuatro factores:

(i) **humano** (Al-Tit, 2015; Alreemy, Chang, Walters, & Wills, 2016; Anantatmula & Kanungo, 2007; Heisig, 2009),

(ii) **organizacional** (Alsadhan et al., 2008; Mas-Machuca & Martínez Costa, 2012; Moffett & McAdam, 2009),

(iii) **Infraestructura** (Board & Chen, 2013; Mojibi et al., 2015)y

(iv) **estrategia** (Carneiro, 2000; OECD, 2015c; Ping, 2008).

La madurez de estos factores nos lleva a que en la organización se genere la gestión del conocimiento y la innovación, es decir que en la medida en que la organización avanza en estos factores se materializa la gestión del conocimiento y la innovación (Carneiro, 2000; Darroch, 2005; Greenhalgh et al., 2004; Johannessen, 2008; Matei & Savulescu, 2014; Razmerita et al., 2016)**.** La propuesta de medición de este modelo se desarrolló para cuatro niveles: (a) Básico, (b) Intermedio, (c) Avanzado y (d) Experto.

**Figura 3.3. Modelo de investigación propuesto**

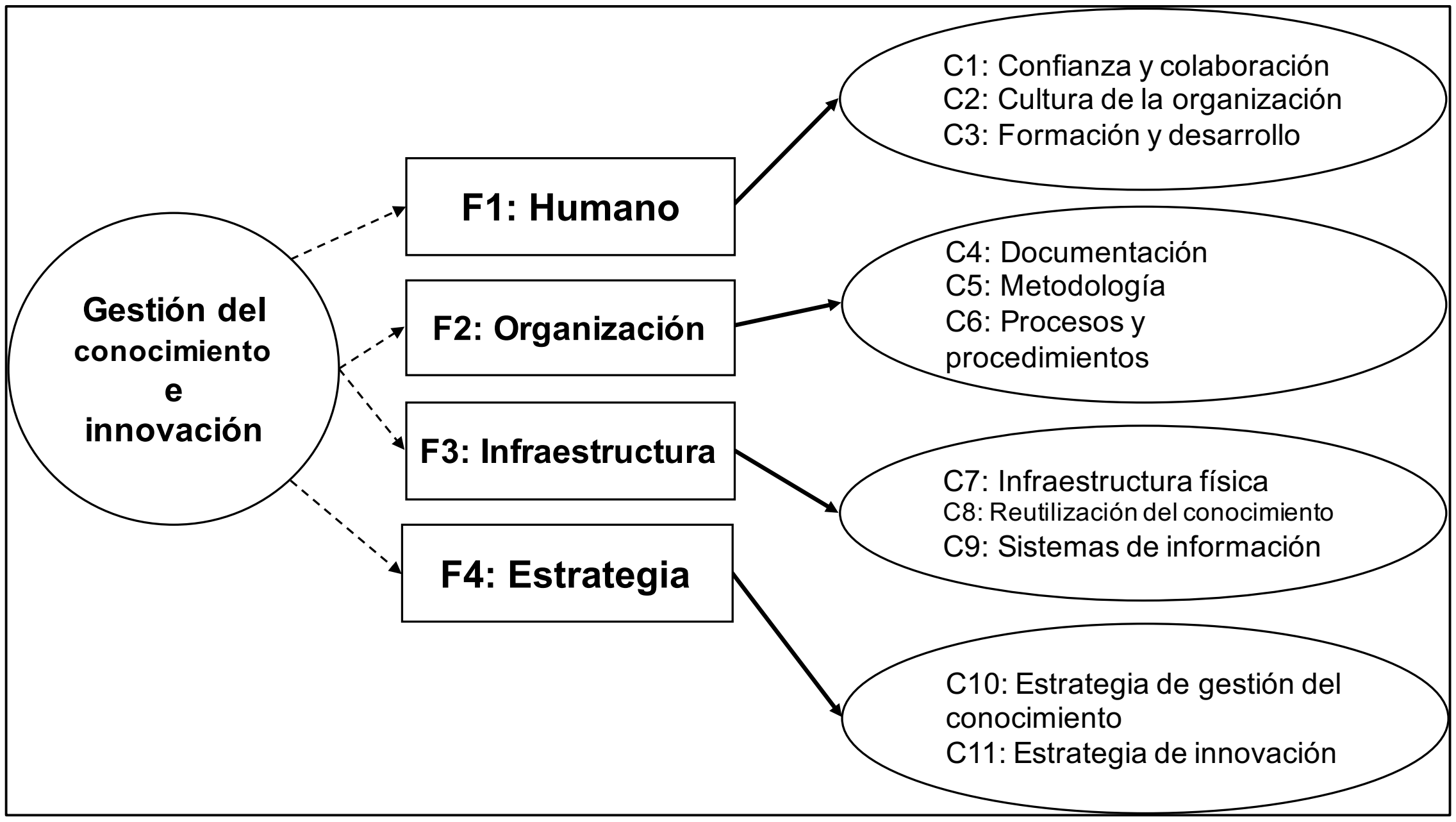

**Fuente:** Elaboración propia (2016)

A partir del modelo de investigación propuesto, en la siguiente fase alusiva a la construcción del instrumento de evaluación, para dar soporte al desarrollo teórico y conceptual de la presente tesis y, al mismo tiempo, permitir la validación empírica de los supuestos de investigación, así como del Estándar de gestión del conocimiento e innovación tecnológica, descrito y propuesto en esta tesis.

Por tal motivo, en esta fase se (4) definieron tanto el objetivo como el alcance del objetivo del instrumento de evaluación; (5) se construyeron los ítems a partir de la literatura existente más relevante en el mundo académico, y (6) se realizó una validación teórica y conceptual de los elementos (Aranda & Molina-Fernández, 2002; Badewi & Shehab, 2016), así como sus componentes asociados, propuestos en el modelo y que integran al instrumento de evaluación. Ello, también de acuerdo con comentarios y sugerencias de expertos en la materia (Heisig et al., 2016; Sánchez & Morrison-Saunders, 2011).

Finalmente, se definió que el instrumento de evaluación que está conformado por 83 ítems, fue dividido en 3 secciones: en la primera se realiza la presentación del instrumento de evaluación y se declara que a las respuestas se les dará un tratamiento confidencial; así mismo se agradece al participante por sus respuestas; enseguida, en la segunda sección, se pregunta sobre información general al participante, tales como su escolaridad, tipo de puesto, adscripción, antigüedad, entre otras cuestiones.

En la última sección, que también es la más extensa del instrumento de evaluación, se cuestionan los factores objeto de análisis descritos en el modelo de investigación propuesto, a través de una escala Likert de siete puntos, por ser la que más se recomienda en este tipo de estudios (Malhotra, 2008) , en donde 1 representa: “totalmente en desacuerdo” y 7 “totalmente de acuerdo”. En esta sección se exploran los 4 factores y los 11 componentes que los conforman. De tal manera que esta sección tiene una sub-sección por cada factor, lo que facilita su análisis.

Enseguida, en la **fase 3. Validación del instrumento de evaluación propuesto**, se validó el instrumento a través de distintas técnicas, tal como (7) la validación de contenido mediante expertos (Hamann et al., 2013; Hernández Sampieri et al., 2010); asimismo se implementó una prueba piloto del instrumento evaluación, con el objetivo de revisar la construcción semántica de los ítems para evitar el error (Malhotra, 2008) que producen las interpretaciones subjetivas de los individuos que participaron en el estudio (Hardy & Ford, 2014).

Es importante mencionar que además se ha realizado un análisis preliminar del instrumento de evaluación, con cinco expertos (Hernández Sampieri et al., 2010), en el mundo académico, en la elaboración de instrumentos de evaluación y en la gestión del conocimiento y la innovación, provenientes de distintas instituciones de educación públicas y privadas, y centros de investigación. Posteriormente se realizaron los ajustes en la redacción de cinco ítems, derivados de las observaciones y recomendaciones de los expertos (Merrill, Keeling, & Gebbie, 2009)

Esta versión inicial o prueba piloto del instrumento de evaluación, denominada "Versión Piloto – Instrumento de Evaluación", fue aplicada de manera aleatoria e intencionada a un grupo de 11 participantes provenientes de distintas organizaciones, para verificar si la construcción semántica de los ítems era correcta y entendible para todas los participantes.

La prueba piloto consistió en un instrumento de evaluación, integrado por 89 ítems organizados en tres secciones:

1. Información general sobre el objetivo del estudio y declaración de confidencialidad;
2. Información general del participante;
3. Evaluación de factores
    a. Factor Humano (F1)
    b. Factor Organización (F2)
    c. Factor Infraestructura (F3)

d. Factor Estrategia (F4).

La aplicación de la prueba piloto se llevó de a cabo mediante instrumentos de evaluación impresos en hojas de papel, directamente con los participantes seleccionados de manera intencional.

Es importante señalar que la investigación se ha realizado entre los meses de enero del año 2015 al mes de noviembre del año 2016, particularmente la construcción, validación e implementación del instrumento de evaluación se ha realizado durante los meses de agosto a diciembre del año 2016, de conformidad con el calendario definido previamente con los asesores para el desarrollo de la tesis.

Además para cuidar integridad del instrumento se realizó una (8) validación de consistencia interna a través del análisis factorial (Hamann et al., 2013; Henseler et al., 2014) utilizando el software SPSS®, versión 23, y el cálculo del alfa de Cronbach (Cho & Kim, 2015; Cronbach, 1951; Cronbach & Meehl, 1955a), obtenida inicialmente para los resultados obtenidos de la prueba piloto. Todo ello, permitió (10) incorporar ajustes semánticos principalmente al instrumento de evaluación, previo a su aplicación.

> La medición válida es la condición sine qua non de la ciencia. En un sentido general, la validez se refiere al grado en que los instrumentos miden realmente los constructos que se pretende medir. Si las medidas utilizadas en una disciplina no han demostrado tener un alto grado de validez, esa disciplina no es una ciencia (Peter, 1979, p. 6).

La validez de los constructos es una elemento esencial en la confiabilidad y validez de los instrumentos de evaluación (Shepard, 1993a). Estos constructos tienen características especiales, por tanto, su medición para establecer relaciones entre los elementos de análisis de la organización (Venkatraman & Grant, 1986). La correcta definición y validación de los constructos que conforman a los instrumentos de evaluación, sustentados en modelos de investigación, es un tema fundamental en la investigación, por tal motivo se ha puesto especial énfasis en esta actividad.

En sentido de lo anterior, la validez de constructo (Cronbach & Meehl, 1955b; Hamann et al., 2013; Herrero, 2010; Ledesma, Ibañez, & Mora, 2002; Shepard, 1993b), a su vez garantiza que el instrumento de evaluación es confiable y que los resultados permiten interpretar el fenómeno que se intenta medir (Hernández Sampieri et al., 2010; Kerlinger, 1975; Ruiz Olabuénaga, 2012; Zorrilla, 2009).

**Figura 3.4. Fases llevadas a cabo, a partir de la aplicación del instrumento de evaluación del presente estudio**

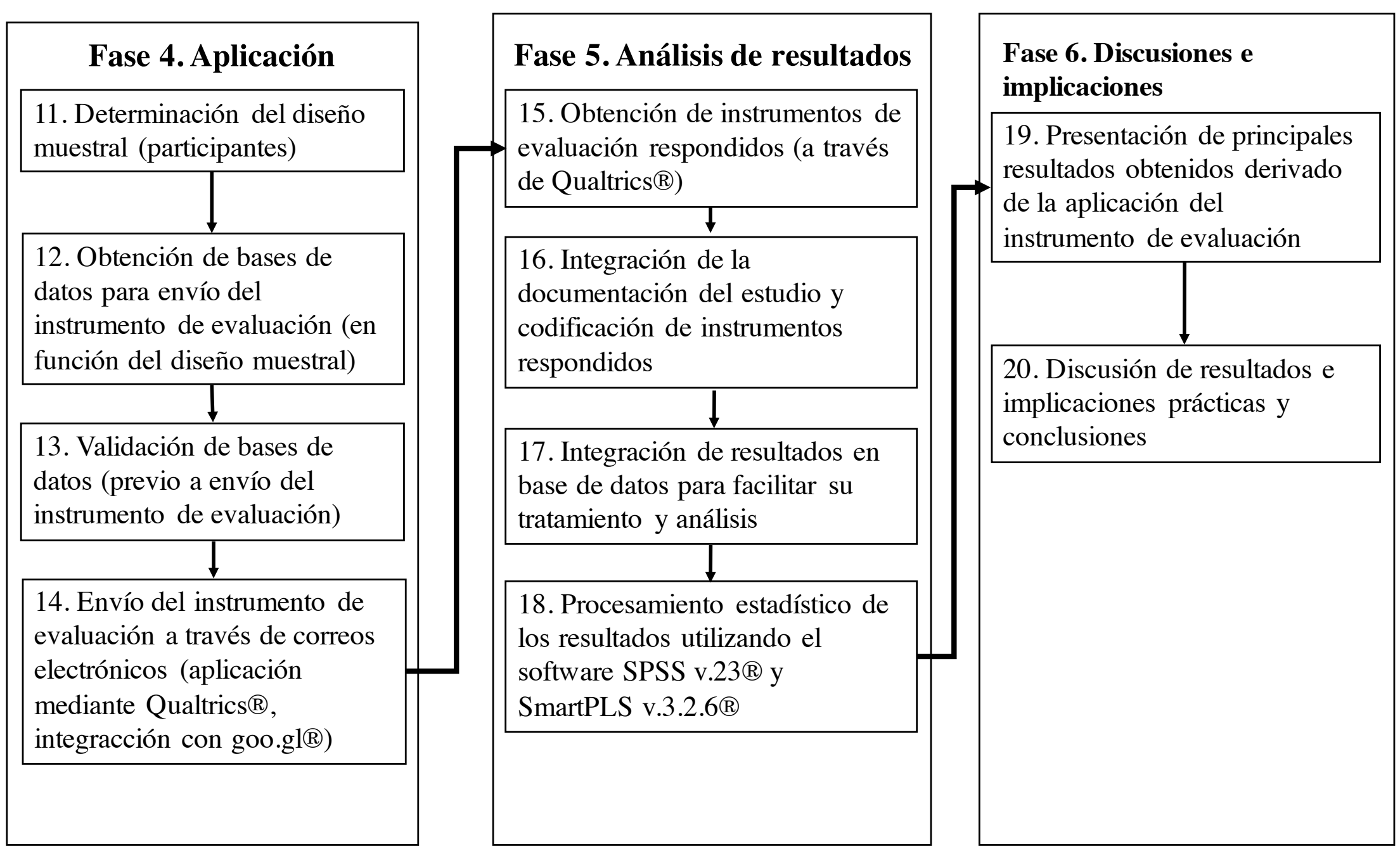

**Fuente:** Elaboración propia (2016)

Seguidamente, se llevó a cabo la **fase 4. Aplicación del instrumento de evaluación,** para llegar a esta fase, toda vez que se había definido el instrumento de evaluación, a partir de un modelo de investigación, que integra a los principales referentes conceptuales sobre la gestión del conocimiento y la innovación. Se procedió a la aplicación del instrumento de evaluación, para ello se siguieron los siguientes pasos:

(11) Determinación del diseño muestral (participantes): Para llevar a cabo la implementación del estudio propuesto, bajo el modelo de investigación previamente definido se diseñó la muestra a la cual se le aplicaría el instrumento de evaluación. Por tanto:

- Se definió el tipo de institución. En tal virtud, se consideraron distintas opciones de organizaciones relacionadas con la gestión del conocimiento y la innovación (Dutta, 2015; OECD, 2004; "OECD Multilingual Summaries OECD Science , Technology and Industry Outlook Perspectivas de Ciencia , tecnología e industria de la OCDE 2012 Innovación en tiempos de crisis," 2012; Soto-Acosta & Cegarra-Navarro, 2016).
- Se definió que la organización sería una universidad pública. Las Universidades son típicamente centros intensivos de producción conocimiento, razón por la cual se decidió aplicar el instrumento de evaluación en un sector específico de esta institución relacionado con su administración.
    - Este sector incluye áreas alusivas a Recursos Humanos, Recursos Financieros, Adquisiciones, Recursos Materiales y Servicios de Apoyo, incluida el área que provee servicios médicos a toda la

población de trabajadores universitarios, Tecnologías de la Información y Biblioteca.

Consecuentemente, el estudio fue realizado en una universidad pública, por ser considerada un centro en donde se produce conocimiento de manera intensiva (Den Hertog, 2000; Makani & Marche, 2010), y porque por su propia naturaleza generadora y regeneradora del conocimiento, se producen innovaciones (Alkhuraiji, Liu, Oderanti, & Megicks, 2015; Andrés, Vallé, Peralta, Farioli, & Giacosa, 2016; Galbraith, 1982; Vrakking, 1990). Se trata de una organización de más de 5,000 empleados, del sector público.

Se definió el área específica en donde se realizaría la implementación del instrumento de evaluación, atendiendo al objetivo de la tesis, se pretende proponer un estándar que sea útil para las organizaciones relacionadas con la producción de conocimiento e innovación de manera intensa (ReinhardtKN, Schmidt, Sloep, & Drachsler, 2011). Por ello, respecto al área en donde se aplicaría el estudio, se definió que sería en un área que típicamente estuviera presente en la mayoría de organizaciones, sin importar si se trataba de una organización pública o privada. Resultando el área de alta gestión de la administración de los recursos institucionales de la universidad, que tiene como responsabilidad la gestión de los recursos financieros, humanos, tecnológicos, y de infraestructura de la organización.

Finalmente, se (12) obtuvieron de bases de datos, para el envío del instrumento de evaluación en función de la sección anterior (11), en este paso se realizaron algunas gestiones para obtener los permisos correspondientes para realizar el presente estudio y aplicar el instrumento de evaluación. Es importante mencionar que agradecemos la disposición mostrada por la institución, que por razones de confidencialidad de la información no se mencionaran sus particularidades.

**Tabla 3.1. Etapas de depuración de bases de datos en función del diseño muestral**

| Etapa de depuración de bases de datos, para obtener muestra | Total |
|---|---|
| (A) Universo de participantes disponibles (personal de la organización) | *5,300* |
| (B) Total de integrantes del área definida dentro de la organización (se incluyen jubilados y pensionados) | *788* |
| (C) Primera depuración, después de aplicar filtro de criterios de elegibilidad de los participantes (personal que no estuviera comisionado, con nombramiento activo y adscrito a las áreas que integran la dirección definida para el estudio) | *305* |
| (D) Segunda depuración, después de validación por los jefes inmediatos en función de nombramiento, actividades desempeñadas y responsabilidades asignadas | *288* |
| (E) Tercera depuración, después de la confirmación uno-a-uno | *218* |
| **Muestra definida** | *218* |
| (F) Número total de participantes esperados esperada | *218* |
| (G) Instrumentos de evaluación respondidos y registrados en la plataforma | *202* |
| (H) Instrumentos de evaluación validados después de CI y CE | *170* |

**Fuente:** Elaboración propia (2016)

En esta fase (13) también se validaron las bases de datos por parte de los jefes inmediatos de los participantes potenciales, lo que agrega una mayor confiabilidad a al perfil de los participantes. La muestra definida fue de 218 trabajadores.

**Tabla 3.2. Criterios de inclusión y exclusión**

| Criterios de inclusión (CI) | Criterios de exclusión (CE) |
|---|---|
| ✓ **CI1**. Que el participante en el estudio sea trabajador de algunas de las áreas funcionales que se consideraron en el | × **CE1**. Si las personas no realizan actividades vinculadas a: mandos medios, supervisión o directivos**.** |
| diseño muestral<br>✓ **CI2.** Los participantes son de puestos que En virtud de que el estudio está dirigido hacia las personas que promueven la gestión del conocimiento y de la innovación al interior de las organizaciones, se eliminaron algunos puestos de la muestra, de la parte.<br>✓ **CI3.** Puestos que realizan labores de control, fiscalización, supervisión y toma de decisiones que tienen trascendencia en el rumbo de la institución. Son puestos que normalmente tendrían una influencia alta en la organización.<br>✓ **CI4.** Se dejaron los mandos medios y directivos de confianza. | nombramiento de personal de confianza o directivo, excluir.<br>✗ **CE2.** Sí la persona está activo o jubilado, si está jubilado entonces excluir.<br>✗ **CE3**. Los instrumentos de evaluación fueron respondidos después del plazo delimitado |

**Fuente:** Elaboración propia, a partir de las recomendaciones de diversos autores (Ali, Ali Babar, Chen, & Stol, 2010; Kitchenham & Charters, 2007; Moncayo & Anticona, n.d.; Zahedi, Shahin, & Ali Babar, 2016)

Para la (14) aplicación del instrumento se evaluaron distintas herramientas, entre ellas SurveyMonkey®, Google Survey® y Qualtrics®. Finalmente, se decidió utilizar Qaultrics®, ya que permite realizar la aplicación en línea, la exportación de resultados en Excel y en formato CSV (Comma separated value), de manera tal que permite un mejor análisis; el monitoreo, el reporteo del comportamiento de respuesta en tiempo real, con alta disponibilidad y multiplataforma.

Así el instrumento puede tener mayor solidez metodológica en su aplicación, análisis de datos e interpretación. Otras de las ventajas de la plataforma seleccionada, es que ofrece fácil acceso desde cualquier dispositivo, está basado en web, y el manejo de los datos de manera gráfica.

Consecuentemente, el instrumento de evaluación se aplicó a los participantes a través de Quatrics®, una herramienta que permite llegar, incrementa los niveles de alcance potenciales del presente estudio y por los atributos tecnológicos de seguridad, estabilidad y confiabilidad que incrementan la calidad del estudio, y por lo tanto de los resultados. Algunas de las ventajas que se observaron para seleccionar la herramienta se enlistan a continuación.

**Ventajas:**

1) Qualtrics® ofrece la posibilidad que durante la aplicación del cuestionario los ítems se presentan al participante de manera aleatoria, lo que podría incrementar la confiabilidad del instrumento de evaluación; es decir que en condiciones de aleatoriedad y en circunstancias distintas los participantes muestran consistencia con sus respuestas, como un elemento de validez concurrente (Malhotra, 2008).
2) Otra prestación que ofrece esta plataforma es que permite observar, incluso desde la prueba piloto, el tiempo promedio que le toma cada uno de los ítems y también los tiempos globales del instrumento de evaluación a los participantes. De esta manera el investigador tiene la posibilidad de realizar las adecuaciones necesarias a fin de fortalecer la claridad, consistencia, credibilidad, confiabilidad y validez del instrumento de evaluación y, consecuentemente, del estudio y de los resultados, en su totalidad.
3) En ese orden de ideas, la plataforma también cuenta con un primer filtro de discriminación de instrumentos aplicados que se aceptan, previo a la validación por parte del investigador, ya que la plataforma discrimina entre los cuestionarios contestados parcialmente y los realizados en su totalidad.
4) Finalmente, otro elemento adicional que apoya la elección de la plataforma es la capacidad de realizar el estudio en un periodo de tiempo limitado, ya que permite llegar al total de los participantes en tiempo real. Es de destacar que la mayoría de los participantes contestaron el instrumento de evaluación en los primeros días de aplicación. Ello podría responder a que los participantes son personas que normalmente, como se ha señalado en la sección de diseño muestral, realizan actividades directivas o asignadas a personal de confianza, con una fuerte relación en la toma de decisiones de la organización.

5) Consecuentemente, los datos depurados que arroja, facilitan el tratamiento estadístico y la inferencia que a partir de ellos permite al investigador observar el objeto de estudio y aproximarse a implicaciones prácticas fundamentadas en la validación empírica.

De esta manera, el instrumento de evaluación fue aplicado a través de la herramienta Qualtrics®, para ello se envió a los participantes, mediante un correo electrónico el enlace de acceso al instrumento de evaluación. Fue durante el periodo del 10 al 17 de noviembre del año 2016. El primer día, con corte a las 18:00 horas, se recibieron 110 respuestas, lo que equivale al *44%* de la participación, lo cual demuestra una ventaja enunciada en la evaluación de la plataforma utilizada, para la aplicación de este tipo de instrumentos.

Es importante mencionar que el enlace al instrumento de evaluación fue colocado en un enlace corto (goo.gl, *shortener)*. En las figuras 3.5, 3.6 y 3.7, podemos observar los datos de procedencia de los usuarios, así como el sistema operativo que usan, la hora a la que ingresan a responder el cuestionario, y el tipo de navegador que usan.

**Figura 3.5. Distribución de los sistemas operativos que utilizaron los participantes**

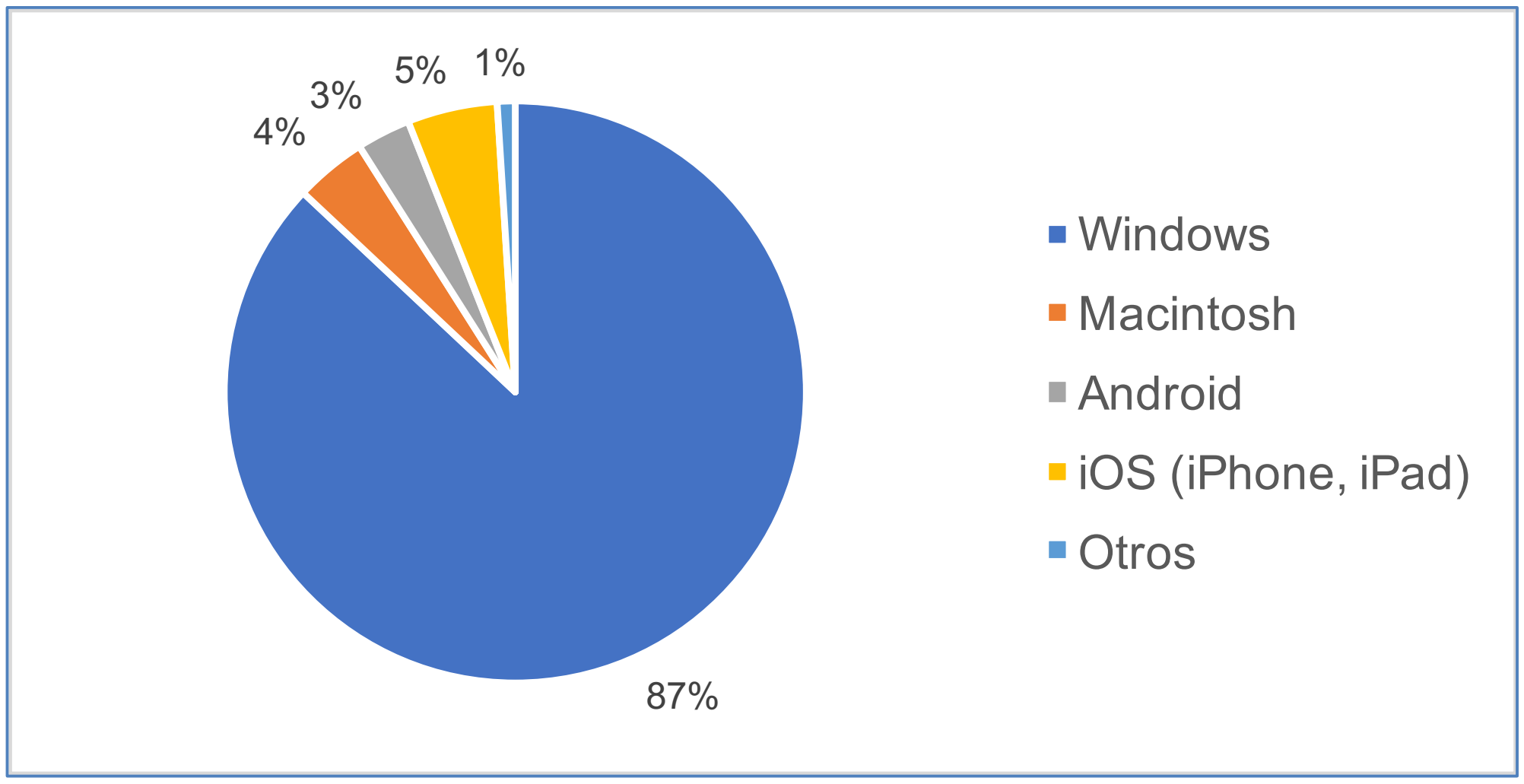

**Fuente:** Elaboración propia (2016).

**Figura 3.6. Distribución de los navegadores que utilizaron los participantes**

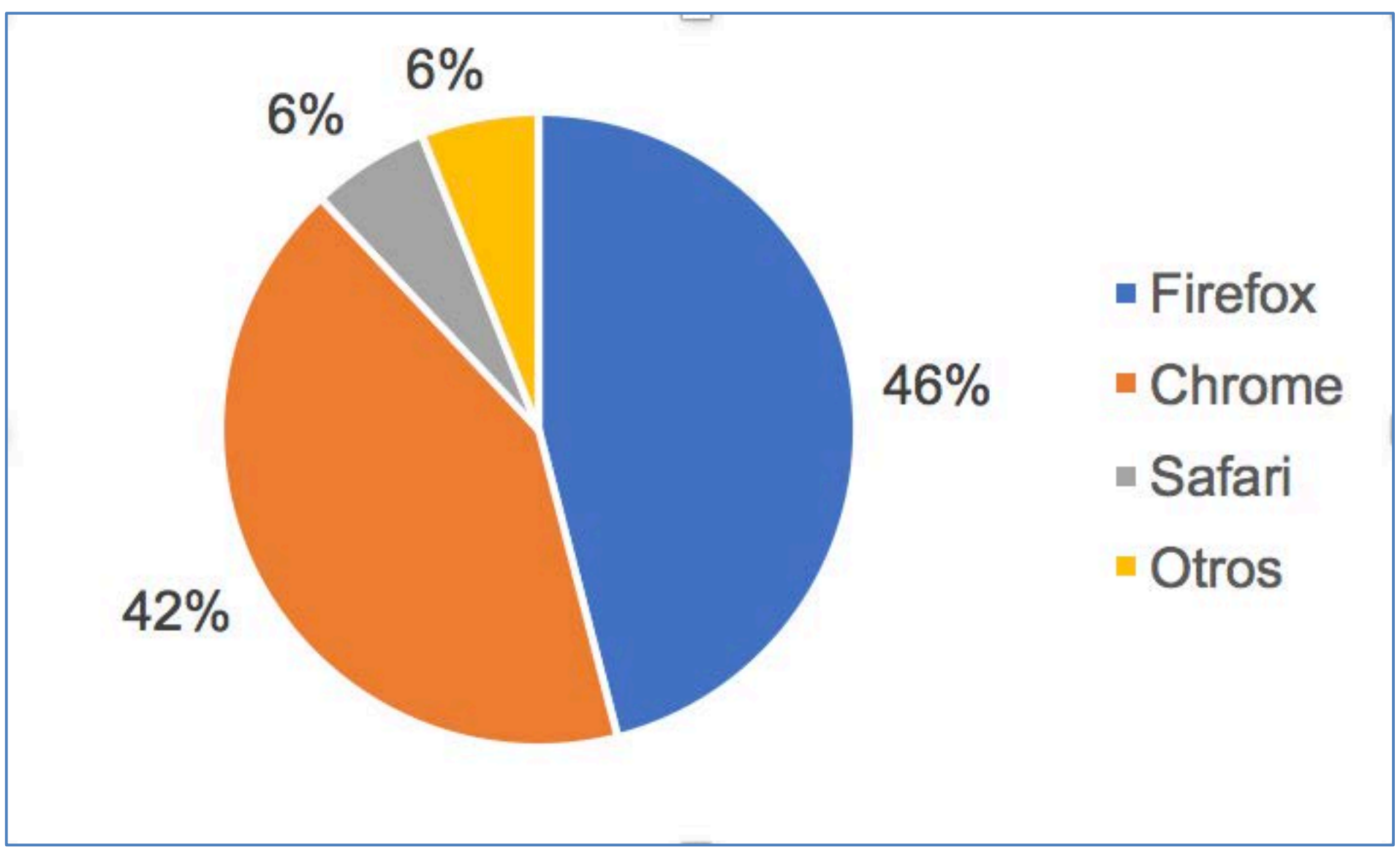

**Fuente:** Elaboración propia (2016).

**Figura 3.7. Información sobre los participantes obtenida de goo.gl**

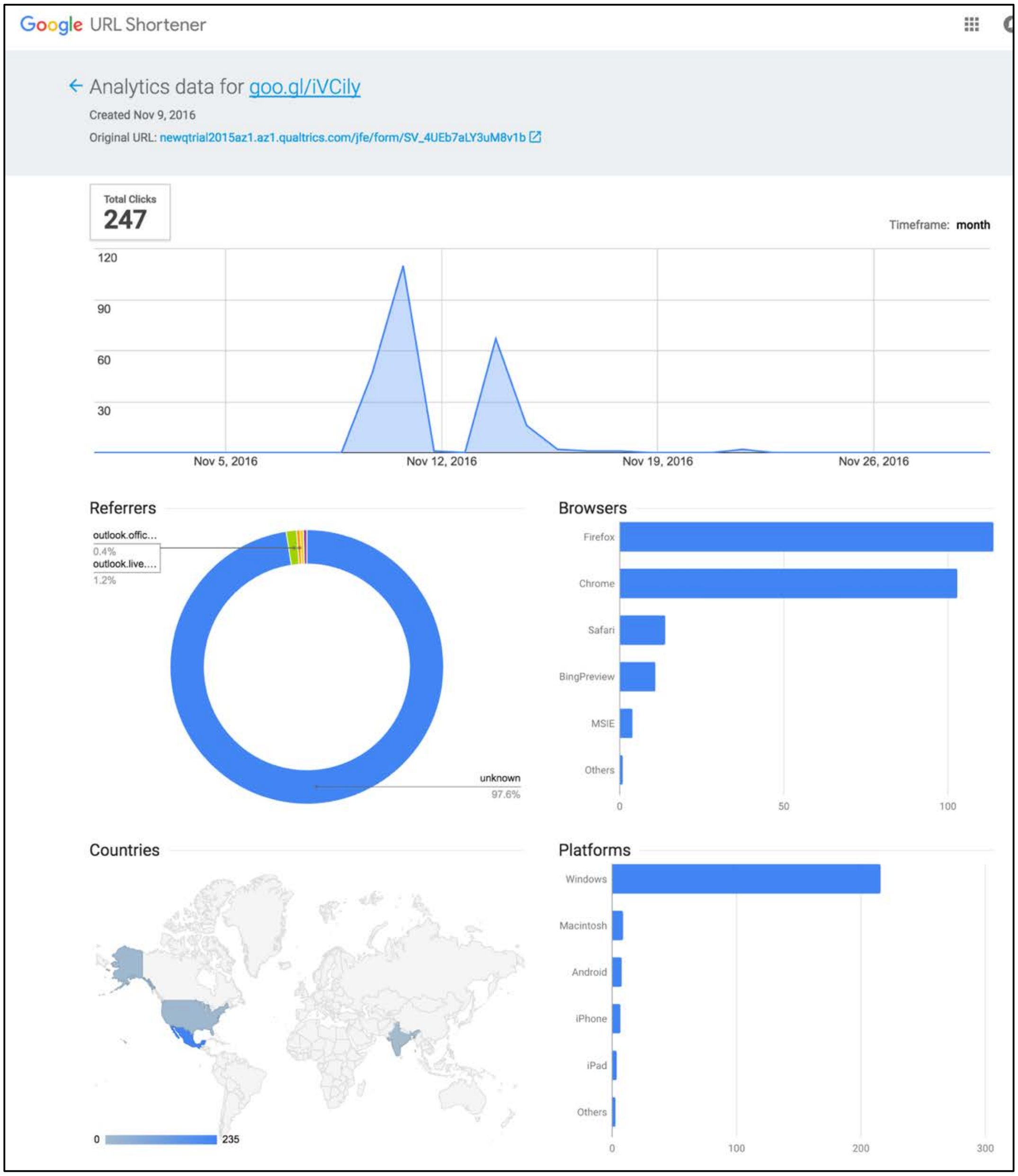

**Fuente:** goo.gle®

Toda vez que se terminó el plazo delimitado para responder al instrumento de evaluación (15), se obtuvieron todos los instrumentos de evaluación que habían sido respondidos por los participantes en la plataforma Qualtrics®. Se tuvo una participación de 202 trabajadores en el estudio (92.6% de la muestra definida)*;* enseguida*,* se realizaron etapas de depurado de la información obtenida, véase la tabla 3.1 y tabla 3.2.

De esta manera, se determinó utilizar *170 i*nstrumentos de evaluación, respondidos en su totalidad y considerados válidos, de acuerdo con las etapas de depuración definidas, y observadas en la literatura (Malhotra, 2008), en total se eliminaron 34 instrumentos de evaluación. Consecuentemente, se (16) sistematizó e integró una única base de datos en una carpeta digital, que facilitara su localización y tratamiento, los instrumentos se clasificaron con un código único en función de cuándo habían sido respondidos y se integraron, a manera de documentación del presente estudio.

Finalmente, (17) se recopiló toda la información en una matriz en Excel® para facilitar su análisis y posterior tratamiento estadístico, se realizaron pruebas de Alfa de Cronbach, para observa la confiabilidad de consistencia interna del estudio y se obtuvieron los siguientes resultados por factor:

**Tabla 3.3. Cálculo del Alpha Cronbach**

| Alpha Cronbach (Cho & Kim, 2015; Cronbach, 1951) | | | |
|---|---|---|---|
| **Valores A** | **Resultado** | **Absoluto** | **¿Satisfactoria?** (Malhotra, 2008, p. 285) |
| A_F1 | 0.926205 | *0.926* | Si |
| A_F2 | 0.932474 | *0.932* | Si |
| A_F3 | 0.920399 | *0.920* | Si |
| A_F4.1^ | 0.963760 | *0.964* | Si |
| A_F4.2^ | 0.946214792 | *0.946* | Si |

**Fuente**: Elaboración propia (2016).

**^Nota:** Para facilitar el análisis estadístico, y garantizar la validez y confiabilidad del instrumento de evaluación, el factor 4 se analizó en función de sus dos componentes: estrategia hacia la gestión del conocimiento y estrategia hacia la innovación.

***(18) Procesamiento de los datos***

En cuanto al análisis de las variables se procesó la información en el programa SPSS® versión 23, realizando un análisis factorial exploratorio para confirmar las cargas factoriales de cada ítem (McIntosh, Edwards, & Antonakis, 2014), y su pertenecía al factor y variable descritos previamente en el planteamiento teórico.

Adicionalmente, para analizar los datos obtenidos se utilizó SmartPLS® Versión 3.2.6, que permite realizar análisis de regresión entre los ítems del instrumento de evaluación y de los factores, estos resultados se presentan en el Apéndice 2. Además permite el modelado de educaciones estructurales (por mínimos cuadrados parciales, PLS) (Hair, Hult, Ringle, & Sarstedt, 2014; Zhang, 2012).

**(19) Presentación de los principales resultados**

La descripción de las características de los participantes en el presente estudio se muestra a continuación, caracterizada por una muestra poblacional que tiene 9 años de antigüedad promedio en la organización. La información sobre: (a) tipo de puesto, (b) unidad administrativa de pertenencia y (c) grado académico de la muestra, se observa en las figuras 3.8, 3.9 y 3.10.

Consideramos que se tuvo una alta participación con respecto al total esperado en las distintas unidades administrativas, lo que garantiza una representatividad de todas las áreas potencialmente elegibles para responder el instrumento de evaluación. Estas áreas como se ha señalado antes, se relacionan con la gestión de las tecnologías, compras, infraestructura, recursos financieros, recursos humanos, sistema de salud, y la alta dirección administrativa, de la organización.

En total los participantes respondieron en función del nombre de 12 puestos, que fueron clasificados de acuerdo al tabulador, funciones y responsabilidades que se desprenden de los manuales organizacionales. Para ello, además se utilizó el tabulador de confianza de la organización para clasificar los puestos de los participantes, quedando de la siguiente manera:

**Figura 3.8. Participación por tipo de puesto**

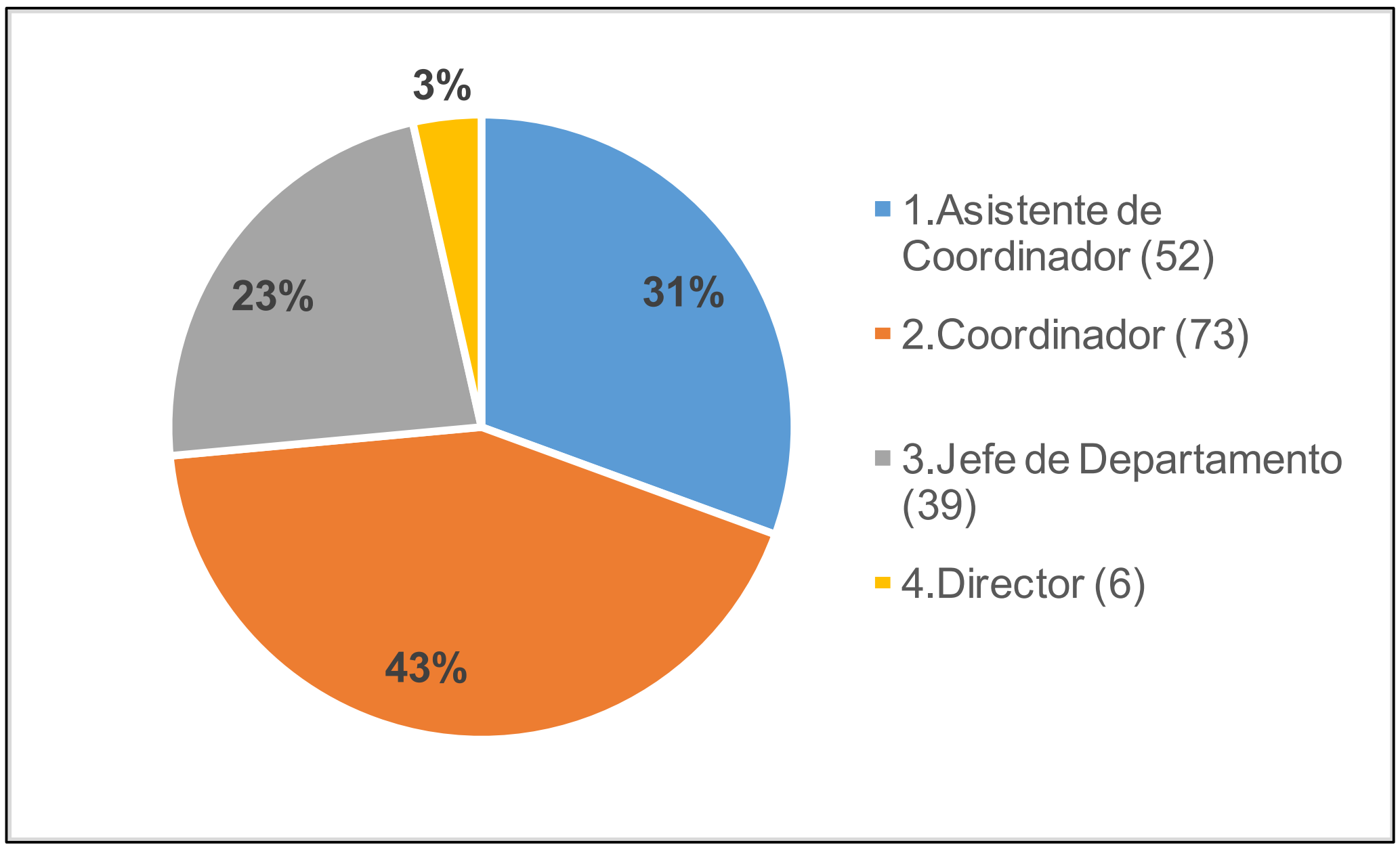

**Fuente**: Elaboración propia (2016)

**Figura 3.9. Participación por nivel de estudios**

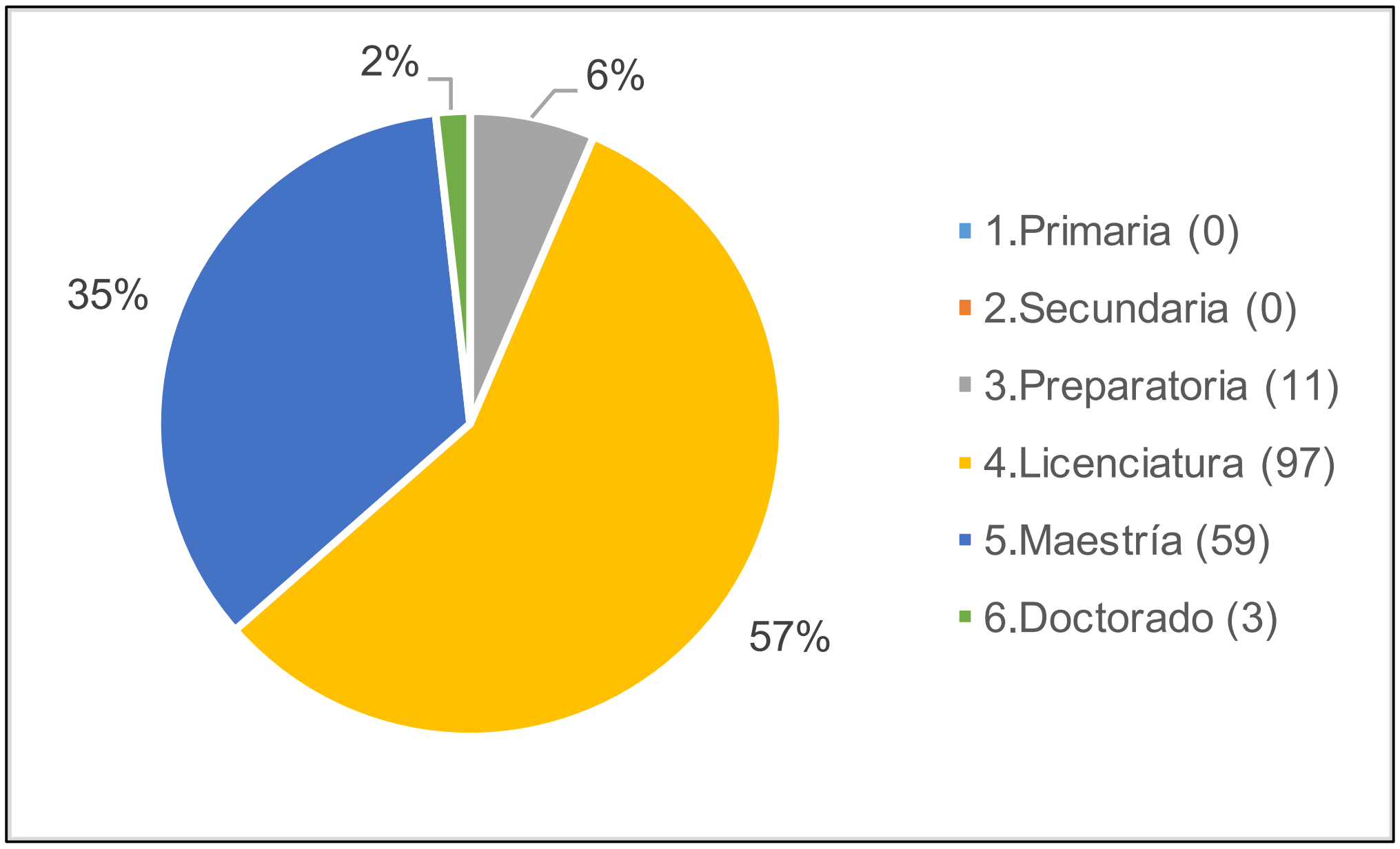

Fuente: Elaboración propia (2016)

**Figura 3.10. Participación por unidad administrativa**

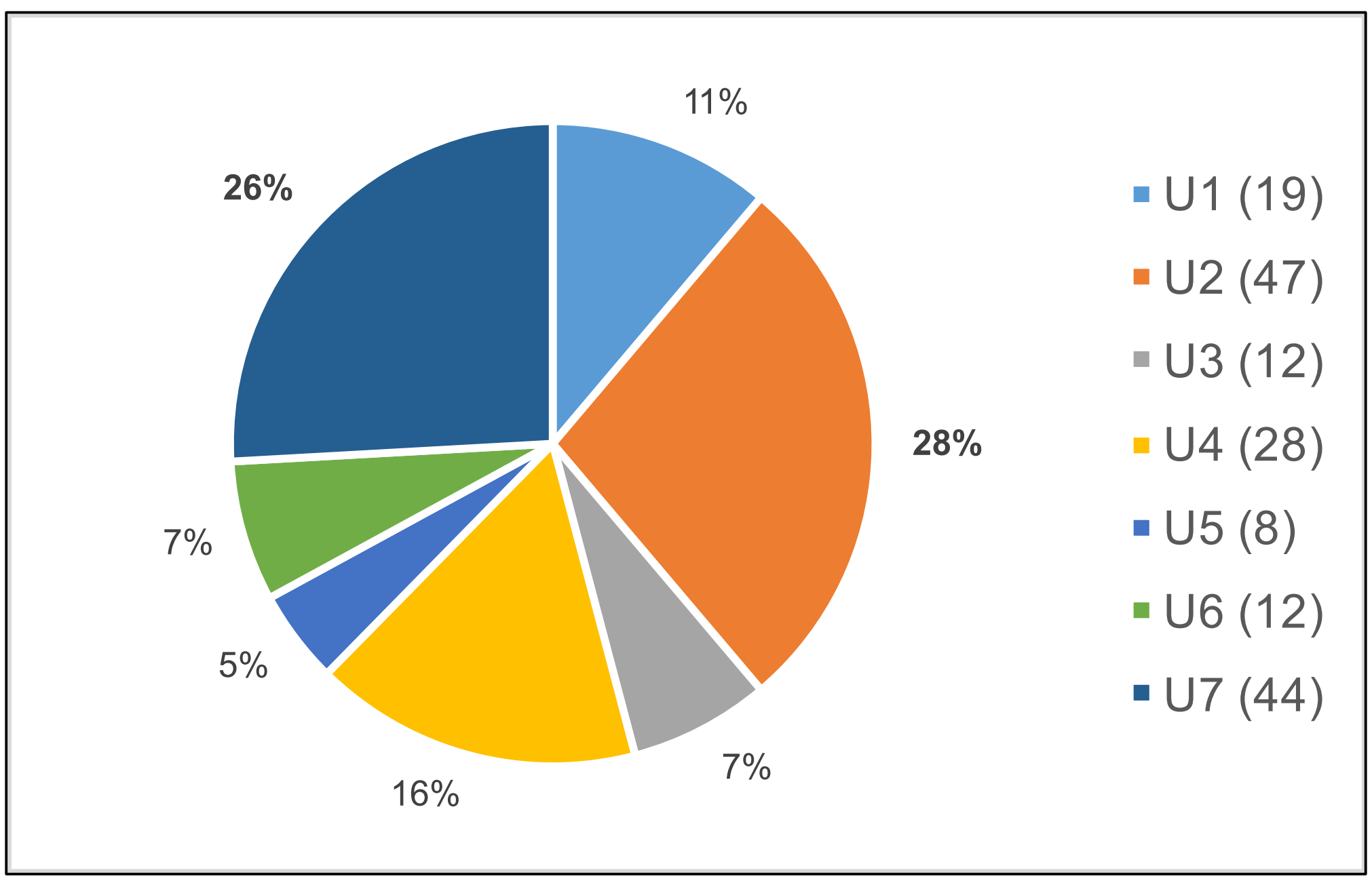

**Fuente**: Elaboración propia (2016)

a. **Análisis factorial**: Se puede confirmar que el instrumento de evaluación identificó los componentes principales del instrumento, que confirman el marco teórico utilizado y los cuatro factores evaluados en el modelo de investigación planteado.

En este análisis se mide la correlación entre los ítems y permite reducirlos agrupándolos en factores (reducción de factores) donde: sí, los ítems no estuvieran asociados el valor otorgado a estos se aproxima a cero (Morales-Vallejo, 2011).

Se realizó el análisis factorial exploratorio en el cual se identificaron los factores principales del instrumento (F1:F4), que confirman el marco teórico utilizado y los cuatro factores evaluados en el modelo planteado. Los ítems que conforman los factores del modelo propuesto se ven reflejados en el análisis factorial y confirman las teorías utilizadas de referente para generar el instrumento de evaluación. Derivado del análisis realizado con el software SPSS®, la matriz de los resultados del componente rotado se presenta en el **Apéndice 2.**

También se debe hacer mención que fueron realizados algunos ejercicios de confirmación de criterio utilizando distintas variables de control, y se observa que el criterio se confirma, los factores se conforman de acuerdo al modelo planteado.

b. Del **Análisis de correlaciones** entre los ítems del estudio, que se muestra en el **Apéndice 3,** se desprende que existen ocho ítems que se han encontrado en el estudio de correlaciones que confirman algunos supuestos de investigación (M. Allameh et al., 2011; Lowry & Gaskin, 2014; Sensuse, Cahyaningsih, & Wibowo, 2015), se ha decidido observar a los ítems que tienen una correlación mayor a 0.7, tal como se sugiere en la literatura académica (Jääskeläinen, 2009). La explicación de estas relaciones se describe en la tabla 3.4.

**Tabla 3.4. Estudio de correlaciones**

| Ítem | Valor de correlación | Explicación |
|---|---|---|
| 11.10 y 11.6 | 0.759 | Los sistemas deberán enfocarse en ser sencillos de usar, adecuados a las necesidades de la organización y orientados en mejorar la experiencia de usuario para favorecer la gestión del conocimiento y la innovación |
| 11.11 y 11.6 | 0.781 | Los sistemas que proveen información actualizada y que resultan útiles para la realización de las funciones de los colaboradores en la organización favorecen la gestión del conocimiento y la innovación |
| 12.7 y 12.3 | 0.754 | Cuando se valora el conocimiento producido por los colaboradores, así como cuando la organización los apoya en su formación profesional se favorece la gestión del conocimiento y la innovación |
| 12.9 y 12.3 | 0.771 | Se debe apoyar el desarrollo de competencias y la profesionalización de los colaboradores para incentivar la gestión del conocimiento y la innovación |
| 12.10 y 12.7 | 0.761 | Cuando se incentiva el uso y deseo de aprender a utilizar nuevas tecnologías y se valoran sus aportaciones, se favorece la gestión del conocimiento y la innovación |
| 13.38 y 13.35* | *0.790* | Cuando existe un firme compromiso de los directivos de la organización con la implementación de nuevos sistemas y tecnologías y una visión institucional clara sobre la aplicación del conocimiento y sus beneficios, se favorece la gestión del conocimiento y la innovación |
| 13.27 y 13.19 | 0.759 | Cuando la organización genera proyectos que se vinculan con otras organizaciones o sectores de la sociedad y la organización aprende continuamente de sus aciertos y errores se favorece la gestión del conocimiento y la innovación |
| 13.28 y 13.27 | 0.771 | Cuando existe una estrategia clara que guía a la organización en la producción de nuevos conocimientos y genera proyectos que se vinculan con otras organizaciones o sectores de la sociedad |
| | | se favorece la gestión del conocimiento y la innovación |

**Fuente**: Elaboración propia (2016),

**Describe el valor de correlación más alto entre ítems del presente estudio (Henseler et al., 2014)* — 

**c. De análisis estructural de ecuaciones,** se refiere al uso de una técnica de segunda generación para análisis de datos en investigaciones sociales. Este tipo de análisis es común cuando se trata de validar factores que se asocian comúnmente a fenómenos que no son observables a simple vista, y que por lo tanto es necesario la construcción de un modelo de interpretación de estas relaciones (Hair et al., 2014; Ringle, Wende, & Becker, 2005, 2015).

Para efectos de observar cómo los diversos factores propuestos en el modelo de investigación de esta tesis se relacionan e inciden en la gestión del conocimiento y la innovación, se realizó una modelación que contempla únicamente el nivel cuatro del estándar, por ser el nivel donde ocurre el fenómeno objeto de estudio de esta tesis. Se ha decidido solamente considerar los 47 ítems que corresponden a los niveles tres y cuatro del EMGCIT propuesto, tal y como se ilustra en la figura 3.11.

Consecuentemente, se puede observar en la figura 3.11, que el factor 4, ***estrategia*** es el que mayor peso tiene sobre la gestión del conocimiento y la innovación (0.968). Estos pesos distribuidos de esta manera demuestran que es necesario contar con una estrategia, como elemento de soporte y que es determinante para que ocurra tanto la gestión del conocimiento como la innovación en las organizaciones. En sentido, la utilización de herramientas de planeación estratégica que favorezcan el logro de los objetivos organizacionales y que fue objeto de estudio del doctorado, han permitido amalgamar todos los conceptos revisados, para converger en una propuesta que hoy se materializa en el estándar que se propone.

Resultados del modelo:

1. Es posible explicar la gestión del conocimiento y la innovación a partir de los factores propuestos y observados mediante el EMGCIT.
2. La estrategia resalta como el factor más significativo para la gestión del conocimiento y la innovación en el contexto del presente estudio.
3. Es posible que el factor humano tenga menos incidencia en los niveles superiores del estándar como consecuencia de la automatización de procesos y el uso de tecnologías en todos los niveles de la organización.

**Figura 3.11. Modelado de ecuaciones estructurales (PLS) para la propuesta de Estándar Mexicano de la Gestión del Conocimiento y la Innovación**

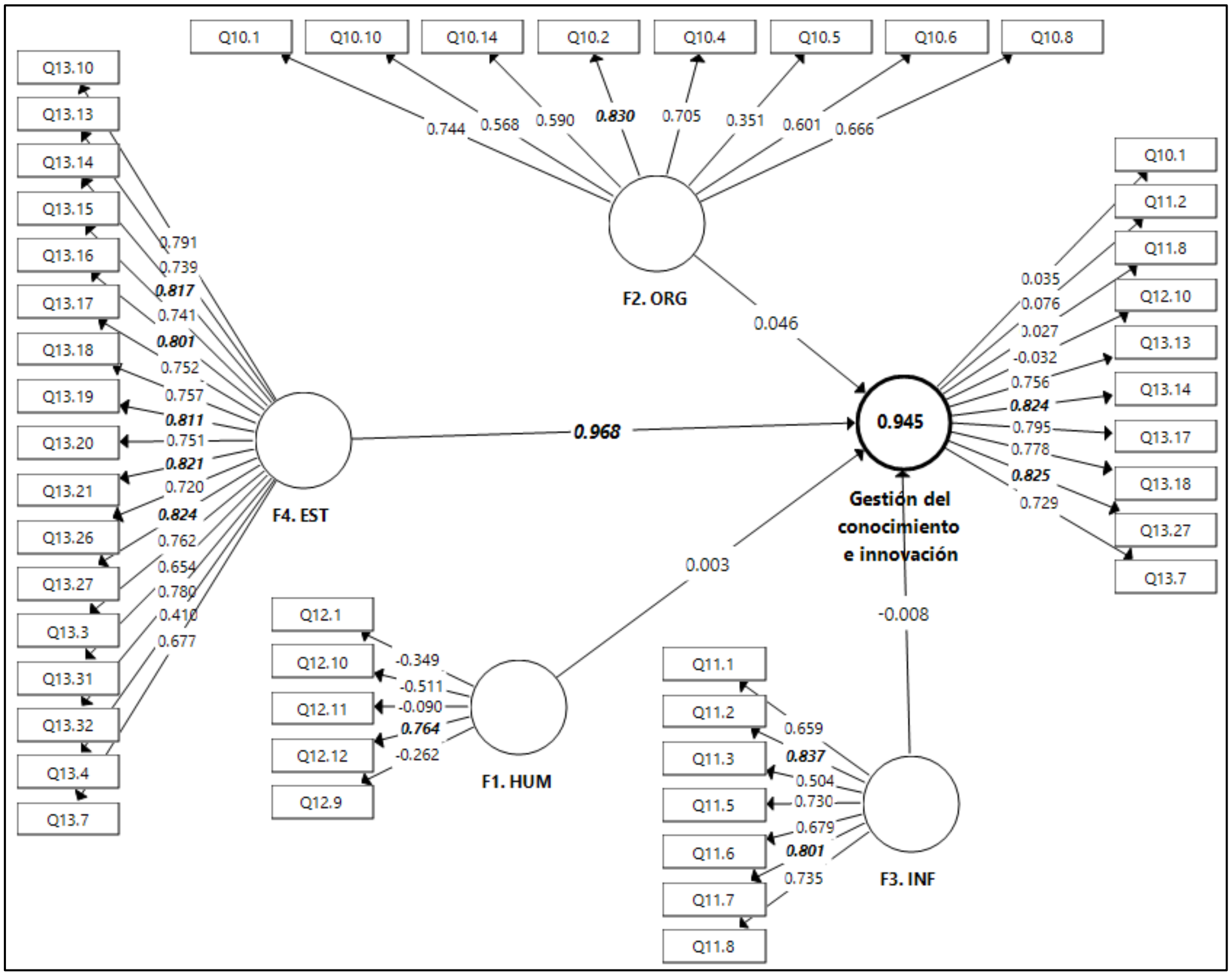

**Fuente**: Elaboración propia utilizando Smart PLS® (2016).

### d. Del análisis en función de los niveles propuestos en el estándar

Para llevar a cabo este análisis se ha decidido asumir la escala Likert (de siete puntos) asumiendo que el valor 7 corresponde al 100% en los niveles del EMGCIT. De esta manera y observando las características particulares de los participantes en este estudio, se han obtenido evaluaciones individuales del grado de aproximación a cada uno de los niveles propuestos en el estándar por parte de las áreas organizacionales involucradas.

**Tabla 3.5. Nivel de cumplimiento de las unidades organizacionales del Estándar Mexicano de Gestión del Conocimiento e Innovación Tecnológica**

| | X | U1 | U2 | U3 | U4 | U5 | U6 | U7 |
|---|---|---|---|---|---|---|---|---|
| **Nivel 1** | 73% | 75% | 70% | 73% | 78% | ***85%*** | 78% | 73% |
| **Nivel 2** | 71% | 71% | 65% | 70% | 73% | ***81%*** | 76% | 72% |
| **Nivel 3** | 71% | 73% | 67% | 70% | 77% | ***80%*** | 78% | 72% |
| **Nivel 4** | 71% | 69% | 66% | 66% | 73% | ***78%*** | 73% | 70% |
| *Promedio general* | 72% | 72% | 67% | 70% | 75% | ***81%*** | 76% | 72% |

Fuente: Elaboración propia (2016)

De este análisis se concluye que la unidad organizacional que tuvo mayor alcance en todos los niveles, fue la 5, ello podría deberse a la preparación y continuidad en la permanencia laboral de los integrantes de la unidad y a la alta demanda de calidad en el servicio que se ofrece, aunado a que esta unidad se caracteriza por tener sus procesos estandarizados, intensa producción del conocimiento, se fomenta el aprendizaje organizacional, así como la actualización constante del personal y su grado de preparación académica es alto. Por otra parte, en la Unidad 2 es necesario utilizar estrategias que gradualmente posibiliten ascender a niveles superiores en el estándar

En tal sentido, la figura 3.12, da cuenta del grado de alcance que tuvo cada una de las unidades organizacionales participantes con respecto al EMGCIT.

**Figura 3.12. Alcance de las unidades con respecto al EMGCIT**

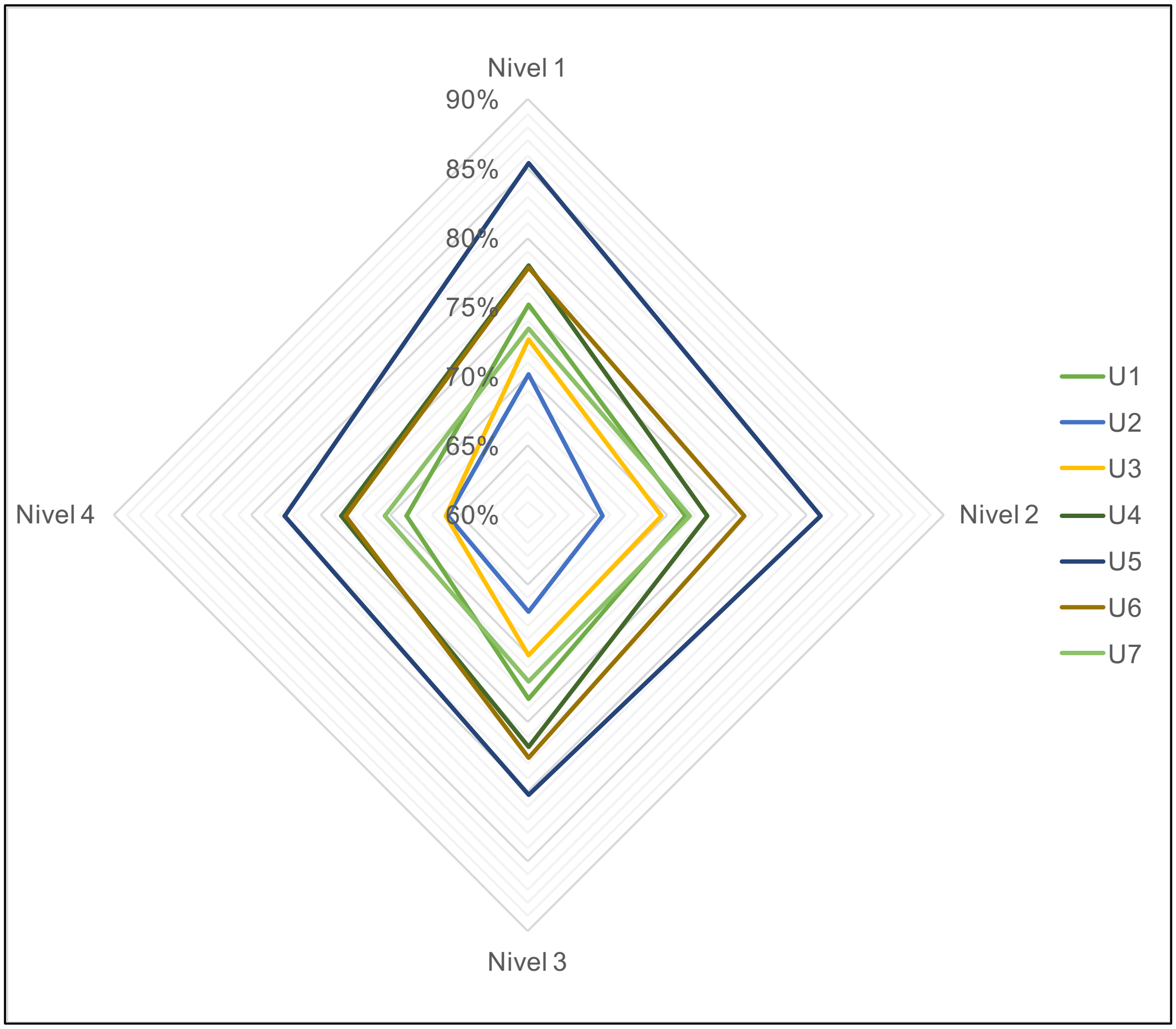

**Fuente**: Elaboración propia (2016)

# CAPÍTULO 4. DISCUSIÓN DE RESULTADOS, IMPLICACIONES PRÁCTICAS, LIMITACIONES Y FUTURAS LÍNEAS DE INVESTIGACIÓN

Para esta sección se ha definido, representar los resultados del estudio obtenido en función de los objetivos y supuestos de investigación. De esta manera se discuten los principales resultados obtenidos, sus implicaciones prácticas, así como las limitaciones y futuras líneas de investigación. Finalmente se presenta, a manera de resumen, las principales contribuciones de esta tesis.

## 4.1.Discusión de resultados

Se ha logrado el objetivo de investigación mediante la identificación de las variables que integrarán el estándar de referencia, a través de la construcción un instrumento válido y confiable elaborado ex profeso que puede ser aplicado en las organizaciones mexicanas. Así mismo se ha podido llevar a cabo la aplicación del estándar propuesto en una organización que se caracteriza por el uso intensivo del conocimiento y el desarrollo de la innovación.

A partir de esta tesis doctoral se ha podido demostrar que:

1. La definición de un Estándar de Gestión de Conocimiento e Innovación Tecnológica posibilita a las organizaciones incrementar sus capacidades para crear, adquirir, compartir, almacenar, incrementar y usar racionalmente el conocimiento para mejorar su eficiencia y productividad.
2. Consecuentemente la gestión del conocimiento favorece la innovación en las organizaciones y les permite ofrecer mejores productos y servicios.

3. Y finalmente en la medida en que se hace mejor gestión del conocimiento se facilita la innovación.

Para lograr los objetivos de esta investigación se desarrolló una propuesta metodológica que conjuga el análisis sistemático de la literatura, la elaboración de un instrumento de evaluación propio, la utilización de tecnologías, y herramientas de vanguardia para la investigación social, que permiten por un lado ofrecer una contribución a la literatura académica, y por otra parte validar empíricamente la propuesta, tal y como los asesores de este trabajo lo solicitaron a lo largo del desarrollo de esta investigación.

### 4.2. Implicaciones prácticas

A la par de la posible implementación del estándar, las organizaciones deben desarrollar estrategias para hacer uso de la gestión del conocimiento en beneficio de la innovación traducida en la creación de nuevos productos o servicios. Es por esto que la aplicación de metodologías como la denominada Stage-Gate® concebida por Cooper (R G Cooper, 1990; Robert G. Cooper, 1990) o bien la de Vijay (Govindarajan, 2012; Immelt, Govindarajan, & Trimble, 2009; Vijay K Jolly, 1997), que brindan a detalle los pasos y acciones para que, a manera de mapa de ruta de forma ágil y esbelta, se puedan lograr en menores plazos y con mayores tasas de éxito nuevos diseños de productos o servicios que mejoren o sustituyan a los existentes. En particular la de Cooper se viene ofreciendo asociada a un software como herramienta de apoyo que es comercializado incluso por empresas de la talla de SAP®.

De igual forma, es pertinente que las organizaciones clasifiquen a su personal, independientemente de las áreas en las que laboren o las tareas que desempeñen, en alguno de los niveles del Ciclo de Vida de Adopción de Tecnología de Moore (1999), en especial aquellas organizaciones que desarrollan tecnología de punta en cualquier industria o bien hacen uso, o producen y transfieren intensivamente conocimiento.

Una de las implicaciones de este trabajo es contar con un instrumento que, desarrollado en forma de estándar, de aplicación gradual y diferenciada, podría facilitar a las organizaciones mexicanas ubicarse dentro de su contexto de influencia y actuación en qué nivel específico se encuentran y qué estrategias y acciones deberían desplegar para potenciar sus resultados de creación de conocimiento y desarrollo de innovaciones, como se resume en la figura 4.1. De esta forma las organizaciones estarían en posibilidad de hacer un mejor uso de todos los recursos de que disponen para mejorar su productividad y competitividad.

**Figura 4.1. Resumen de los elementos que caracterizan los niveles del EMGCIT**

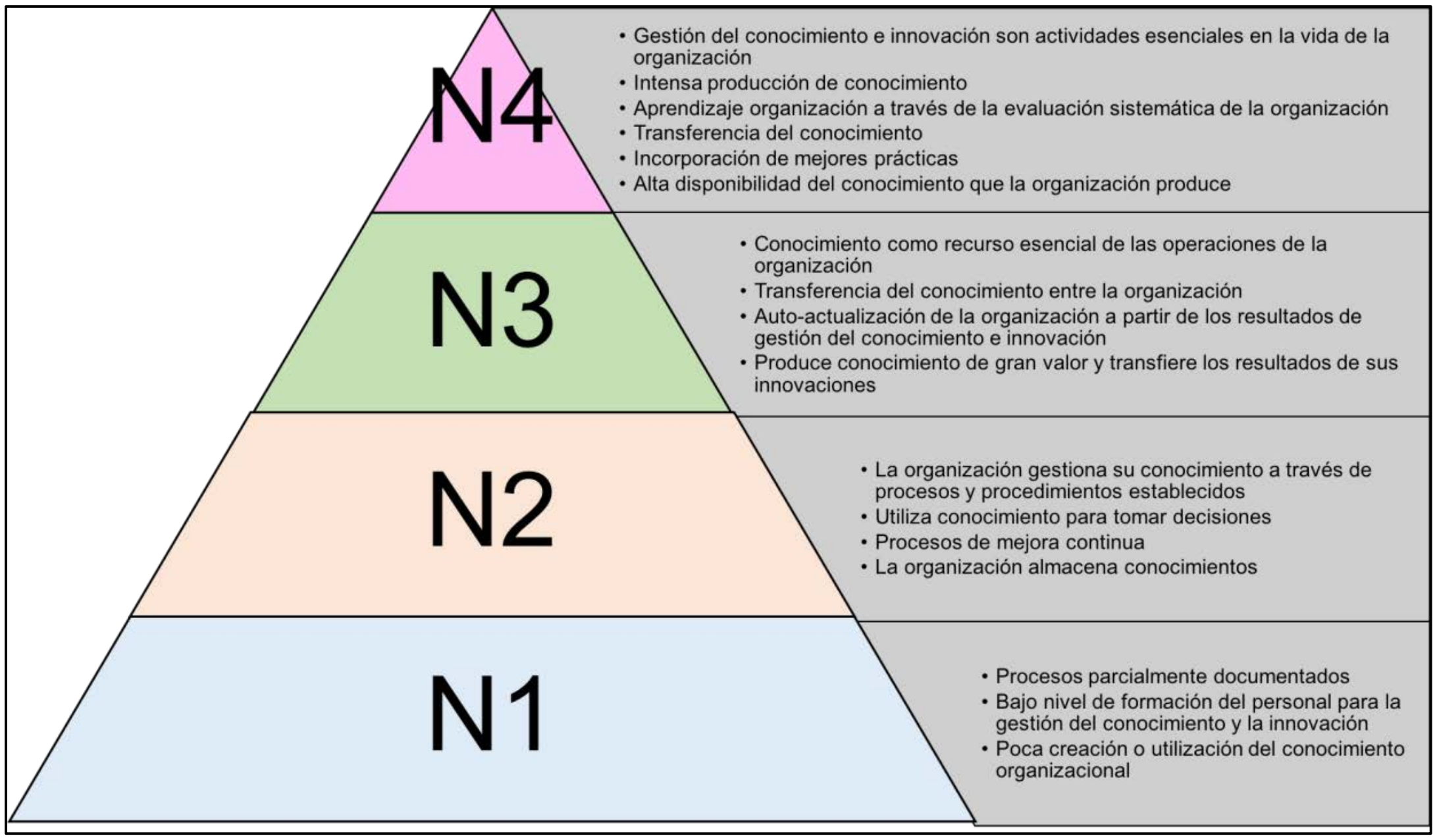

**Fuente**: Elaboración propia (2016)

## 4.3.Limitaciones

Una de las principales limitaciones del estudio fue cumplir con los tiempos propuestos en el calendario definido para la realización de esta tesis doctoral. Sin embargo, gracias al diseño de investigación y a las facilidades otorgadas por la institución en donde se realizó el estudio, fue posible lograr el objetivo que se presenta en este trabajo.

Otra limitación pudiera encontrarse en la aplicación del instrumento de evaluación en una sola organización. Sin embargo, dado que el mismo es susceptible de ponerse al alcance para que a futuro se pueda aplicar a más organizaciones.

Por último, los escasos referentes académicos hasta el momento encontrados en el mundo académico, impidieron realizar una contrastación de los elementos propuestos en este trabajo. Así como la falta de comprobación empírica de los modelos existentes.

## 4.4. Futuras líneas de investigación

Es posible realizar un estudio longitudinal para comprobar si los resultados se mantienen al alterar alguna de las variables (factores, componentes) del modelo de investigación propuesto y constatar su impacto en la gradualidad obtenida en el estándar.

Además, se podría aplicar el instrumento de evaluación en diversas organizaciones que tengan características similares a la organización estudiada en la presente tesis doctoral. Así mismo, en estudios posteriores se pudiera revisar y depurar el instrumento de evaluación para afinar la propuesta del estándar.

Finalmente, es posible realizar un análisis de la implicación del Factor Humano, con sus componentes, a cada uno de los niveles del EMGCIT, relacionándolo con la gestión del conocimiento y la innovación, en el entendido que éste comprende la parte exógena del modelo.

## 4.5 Conclusiones

Como todo trabajo de investigación se trata de una propuesta susceptible de mejoras; el mundo académico no tiene puntos finales. Ese es nuestro camino a recorrer todos los días. Sin duda, es necesario contribuir a la nivelación de las capacidades de gestión del conocimiento e innovación globales que permitan a nuestra nación equipararse con el escenario a nivel mundial, de tal suerte que se puedan observar y medir estos dos elementos como claves para mejorar la productividad y competitividad de las organizaciones.

Se requiere una mentalidad abierta hacia la gestión del conocimiento y la innovación en especial de las instituciones que se dedican a su creación, preservación, obtención, aplicación y diseminación. Por ello, es relevante generar políticas y estrategias idóneas para la implementación de un estándar en la materia. Organizaciones como la ADIAT, Secretaría del Trabajo, Secretarias de Desarrollo Económico, entre otras, podrían valorar el uso de un estándar como el propuesto en esta investigación, en sus organizaciones asociadas.

A medida que la gestión del conocimiento y la innovación son componentes estratégicos que se vuelven más globales, es necesario que las organizaciones tomen conciencia del alto valor que significa contar con personas que poseen conocimientos y experiencias en diversos ámbitos y cuya contribución – no solo monetaria- facilita el logro de los objetivos organizacionales.

Es necesario desarrollar programas permanentes de formación institucional para todos los integrantes de las organizaciones. Esto contribuye no sólo a la auto preservación del conocimiento sino a la motivación y crecimiento personal y profesional de sus integrantes. Generar “comunidades de práctica”, en todas sus manifestaciones –cursos, talleres, blogs, wikis, etc.- permite diseminar y obtener nuevo conocimiento y propiciar la innovación.

No hay receta mágica para crear ecosistemas de innovación perfectos, pero sí es posible contribuir a crear incentivos para que las organizaciones mejoren sus prácticas de gestión del conocimiento e innovación.

En la medida en la que las organizaciones valoran de manera consciente y proporcional cada uno de los componentes del estándar, generarán estrategias para dimensionar la contribución que tienen en el logro de los objetivos de la misma.

### 4.6 Principales contribuciones de la tesis

A continuación, se enlistan las principales contribuciones derivadas de la realización de esta tesis doctoral.

a. Se ha realizado una revisión sistemática de la literatura para construir una propuesta metodológica propia y original para el desarrollo de este trabajo que integra los principales referentes conceptuales identificados en la literatura, lo que ha permitido construir y validar el instrumento de evaluación del estándar propuesto.

b. Contar con un estándar mexicano que permita parametrizar y en su momento nivelar las características mínimas que deben compartir las organizaciones que quieran implementar iniciativas de gestión del conocimiento e innovación tecnológica.

c. Al ser un trabajo que aborda este tema en el contexto específico de una organización mexicana, permite avanzar hacia marcos teóricos y conceptuales más sólidos que puedan contribuir a materializar la gestión del conocimiento y la innovación tecnológica en las organizaciones mexicanas.

d. Esta tesis ofrece una herramienta que coadyuva a las organizaciones a ubicarse en el contexto competitivo internacional con referencia al tema de estudio.

## REFERENCIAS

Adams, R., Bessant, J., & Phelps, R. (2006). Innovation management measurement: A review. *International Journal of Management Reviews*, *8*(1), 21–47. http://doi.org/10.1111/j.1468-2370.2006.00119.x

Akhavan, P., Jafari, M., & Fathian, M. (2006). Critical success factors of knowledge management systems: a multi-case analysis. *European Business Review*, *18*(2), 97–113. http://doi.org/10.1108/09555340610651820

Al-Alawi, A. I., Al-Marzooqi, N. Y., & Mohammed, Y. F. (2007). Organizational culture and knowledge sharing: critical success factors. *Journal of Knowledge Management*, *11*(2), 22–42. http://doi.org/10.1108/13673270710738898

Al-Tit, A. A. (2015). The Mediating Role of Knowledge Management and the Moderating Part of Organizational Culture between HRM Practices and Organizational Performance. *International Business Research*, *9*(1), 43. http://doi.org/10.5539/ibr.v9n1p43

Alavi, M., Kayworth, T. R., & Leidner, D. E. (2006). An Empirical Examination of the Influence of Organizational Culture on Knowledge Management Practices. *Journal of Management Information Systems*, *22*(3), 191–224. http://doi.org/10.2753/MIS0742-1222220307

Aldas-Mnzano, J. (2013). Análisis Factorial Confirmatorio: Apuntes y ejercicios, 55.

Ali, M. S., Ali Babar, M., Chen, L., & Stol, K.-J. (2010). A systematic review of comparative evidence of aspect-oriented programming. *Information and Software Technology*. http://doi.org/10.1016/j.infsof.2010.05.003

Alkhuraiji, A., Liu, S., Oderanti, F. O., & Megicks, P. (2015). New structured knowledge network for strategic decision-making in IT innovative and implementable projects. *Journal of Business Research*. http://doi.org/10.1016/j.jbusres.2015.10.012

Allameh, M., Zamani, M., & Davoodi, S. M. R. (2011). The relationship between organizational culture and knowledge management. *Procedia Computer Science*, *3*, 1224–1236. http://doi.org/10.1016/j.procs.2010.12.197

Allameh, S. M., Zare, S. M., & Davoodi, S. M. R. (2011). Examining the impact of KM enablers on knowledge management processes. *Procedia Computer Science*, *3*, 1211–1223. http://doi.org/10.1016/j.procs.2010.12.196

Allen, R. H., & Sriram, R. D. (2000). The Role of Standards in Innovation. *Technological Forecasting and Social Change*, *64*(2–3), 171–181. http://doi.org/10.1016/S0040-1625(99)00104-3

Alreemy, Z., Chang, V., Walters, R., & Wills, G. (2016). Critical success factors (CSFs)

for information technology governance (ITG). *International Journal of Information Management*, *36*(6), 907–916. http://doi.org/10.1016/j.ijinfomgt.2016.05.017

Alsadhan, A. O., Zairi, M., & Keoy, K. H. A. (2008). From P Economy to K Economy: An empirical study on knowledge-based quality factors. *Total Quality Management & Business Excellence*, *19*(7–8), 807–825. http://doi.org/10.1080/14783360802159469

Altunok, T., & Cakmak, T. (2010). A technology readiness levels (TRLs) calculator software for systems engineering and technology management tool. *Advances in Engineering Software*, *41*(5), 769–778. http://doi.org/10.1016/j.advengsoft.2009.12.018

Anantatmula, V. S., & Kanungo, S. (2007). Modeling enablers for successful KM implementation. *Proceedings of the Annual Hawaii International Conference on System Sciences*. http://doi.org/10.1109/HICSS.2007.387

Andrés, H., Vallé, C., Peralta, G., Farioli, M., & Giacosa, L. (2016). Entrepreneurial and Innovative Practices in Public Institutions, 19–39. http://doi.org/10.1007/978-3-319-32091-5

Anthony, S. D., Johnson, M. W., & Sinfield, J. V. (2008). Institutionalizing Innovation. *MIT Sloan Management Review*, (49216).

Aranda, D. A., & Molina-Fernández, L. M. (2002). Determinants of innovation through a knowledge-based theory lens. *Industrial Management & Data Systems*, *102*(5/6), 289–296. http://doi.org/10.1108/02635570210428320

Argyris, C. (1977). Double loop learning in organizations. *Harvard Business Review*, 115–126.

Armson, G. (2008). How innovative is your culture? *Training & Development in Australia*, *54*(3), 20–23. Retrieved from http://search.ebscohost.com/login.aspx?direct=true&db=buh&AN=41563804&site=ehost-live

Arora, R. (2002). Implementing KM – a balanced score card approach. *Journal of Knowledge Management*, *6*(3), 240–249. http://doi.org/10.1108/13673270210434340

Atkinson, R. D., & Ezell, S. (2015). Principles for National Innovation Success. In *2015 Global Innovation Index* (pp. 89–98). INSEAD, the World Intellectual Property Organization, and Cornell University. Retrieved from http://www.wipo.int/edocs/pubdocs/en/wipo_pub_gii_2015-chapter4.pdf

Audretsch, D. B., Coad, A., & Segarra, A. (2014). Firm growth and innovation. *Small Business Economics*, *43*(4), 743–749. http://doi.org/10.1007/s11187-014-9560-x

Augustine, S. (2008). *Agile Project Management*. *Pace Pacing And Clinical*

*Electrophysiology* (Sixth Edit, Vol. 48). Elsevier. http://doi.org/10.1145/1101779.1101781

Badewi, A., & Shehab, E. (2016). The impact of organizational project benefits management governance on ERP project success: Neo-institutional theory perspective. *International Journal of Project Management*, *34*(3), 412–428. http://doi.org/10.1016/j.ijproman.2015.12.002

Balaid, A., Abd Rozan, M. Z., Hikmi, S. N., & Memon, J. (2016). Knowledge maps: A systematic literature review and directions for future research. *International Journal of Information Management*, *36*(3), 451–475. http://doi.org/10.1016/j.ijinfomgt.2016.02.005

Baller, S., Dutta, S., & Lanvin, B. (2016). *The Global Information Technology Report 2016. Innovating in the Digital Economy*. Geneva. Retrieved from http://www3.weforum.org/docs/GITR2016/GITR_2016_full report_final.pdf

Banco Mundial. (2016). Datos.

Bate, S. P., & Robert, G. (2002). Knowledge management and communities of practice in the private sector: lessons for modernizing the National Health Service in England and Wales. *Public Administration*, *80*(4), 643–663. http://doi.org/10.1111/1467-9299.00322

Batini, C., & Scannapieco, M. (2016). Methodologies for Information Quality Assessment and Improvement. In *Data and Information Quality* (pp. 353–402). Cham: Springer International Publishing. http://doi.org/10.1007/978-3-319-24106-7_12

Becerra-Fernandez, I., & Sabherwal, R. (2015). *Knowledge Management: Systems and Processes* (2nd ed.). New York: Routledge.

Belkahla, W., & Triki, A. (2011). Customer knowledge enabled innovation capability: proposing a measurement scale. *Journal of Knowledge Management*, *15*(4), 648–674. http://doi.org/10.1108/13673271111152009

Bindé, J., & Matsuura, K. (2005). *Towards Knowledge Societies. UNESCO world report* (Vol. [1]).

Birkinshaw, J. M., Hamel, G., & Mol, M. J. (2008a). Management innovation. *Academy of Management Review*, *33*(4), 825–845. http://doi.org/10.5465/AMR.2008.34421969

Birkinshaw, J. M., Hamel, G., & Mol, M. J. (2008b). Management Innovation. *Academy of Management Review*, *33*(4), 825–845. http://doi.org/10.5465/AMR.2008.34421969

Blind, K. (2013). *The Impact of Standardization and Standards on Innovation* (15 No. 13). *NESTA Compendium of Evidence on Innovation Policy Intervention*.

Manchester. Retrieved from http://www.innovation-policy.net/compendium/

Blind, K. (2016). Standardization and Standards as Research and Innovation Indicators: Current opportunities and future challenges. Ghent: OECD Blue Sky III. Retrieved from https://www.google.com.mx/url?sa=t&rct=j&q=&esrc=s&source=web&cd=1&cad=rja&uact=8&ved=0ahUKEwjF_s-ijYLRAhXh3YMKHR_nCXEQFgggMAA&url=https%3A%2F%2Fwww.oecd.org%2Fsti%2F049%2520-%2520BlueSky_Standards_Blind.pdf&usg=AFQjCNGnTZ1Tgb-HwgV6OyC1IoFN2011eA&sig2=fPG

Board, E., & Chen, P. (2013). *Knowledge Discovery, Knowledge Engineering and Knowledge Management*. (A. Fred, J. L. G. Dietz, K. Liu, & J. Filipe, Eds.) (Vol. 272). Berlin, Heidelberg: Springer Berlin Heidelberg. http://doi.org/10.1007/978-3-642-29764-9

Boland, R. J., Singh, J., Salipante, P., Aram, J. D., Fay, S. Y., & Kanawattanachai, P. (2001). Knowledge Representations and Knowledge Transfer. *Academy of Management Journal*, *44*(2), 393–417. http://doi.org/10.2307/3069463

Bolles, D., & Hubbard, D. G. (2007). What Is Standardization? In *The Power of Enterprise-Wide Project Management* (pp. 79–81). New York: AMACOM.

Bonilla, A., & Agencia Informativa Conacyt. (2016). México debe sumarse a la economía del conocimiento: Enrique Cabrero. Retrieved December 8, 2016, from http://conacytprensa.mx/index.php/sociedad/politica-cientifica/12357-mexico-debe-sumarse-a-la-economia-del-conocimiento

Bontis, N. (2001). Assessing knowledge assets: a review of the models used to measure intellectual capital. *International Journal of Management Reviews*, *3*(1), 41–60. http://doi.org/10.1111/1468-2370.00053

Bosilj Vukšić, V., & Pejić Bach, M. (2015). Background and scope of the special issue on "Innovations driven by knowledge management." *Baltic Journal of Management*, *10*(4). http://doi.org/10.1108/BJM-07-2015-0145

Bouthillier, F., & Shearer, K. (2002). Understanding knowledge management and information management: The need for an empirical perspective. *Information Research*. http://doi.org/10.1177/1524839907309867

Brand, A. (1998). Knowledge Management and Innovation at 3M. *Journal of Knowledge Management*, *2*(1), 17–22. http://doi.org/10.1108/EUM0000000004605

Brockmann, C., & Roztocki, N. (2015). Topics on Knowledge Management: An Empirical Insight into Articles Published in the International Journal of Knowledge Management. *2015 48th Hawaii International Conference on System Sciences*, 3834–3840. http://doi.org/10.1109/HICSS.2015.460

Burford, S., & Ferguson, S. (2011). The Adoption Of Knowledge Management Standards And Frameworks In The Australian Government Sector. *Journal of Knowledge Management Practice*, *12*(1). Retrieved from http://www.tlainc.com/articl249.htm

Cambridge University Press. (2016). Standardization. Retrieved January 1, 2015, from http://dictionary.cambridge.org/dictionary/english/standardization

Cano-Kollmann, M., Cantwell, J., Hannigan, T. J., Mudambi, R., & Song, J. (2016). Knowledge connectivity: An agenda for innovation research in international business. *Journal of International Business Studies*, *47*(3), 255–262. http://doi.org/10.1057/jibs.2016.8

Carmo, W. C. do, & Albuquerque, A. B. (2014). Project Management Suported by Business Process Management. In *2014 9th International Conference on the Quality of Information and Communications Technology* (pp. 236–241). IEEE. http://doi.org/10.1109/QUATIC.2014.39

Carneiro, A. (2000). How does knowledge management influence innovation and competitiveness? *Journal of Knowledge Management*, *4*(2), 87–98. http://doi.org/10.1108/13673270010372242

Chan, H. S., & Chow, K. W. (2007, December 1). Public Management Policy and Practice in Western China: Metapolicy, Tacit Knowledge, and Implications for Management Innovation Transfer. *The American Review of Public Administration*. http://doi.org/10.1177/0275074006297552

Chang Lee, K., Lee, S., & Kang, I. W. (2005). KMPI: measuring knowledge management performance. *Information & Management*, *42*(3), 469–482. http://doi.org/10.1016/j.im.2004.02.003

Chatwin, D. (2006). Knowledge management -- a guide. *inCite*, *27*, 8–8 1p. Retrieved from http://search.ebscohost.com/login.aspx?direct=true&db=cin20&AN=105924511&site=ehost-live

Chen, C.-J., Shih, H.-A., & Yang, S.-Y. (2009). The Role of Intellectual Capital in Knowledge Transfer. *IEEE Transactions on Engineering Management*, *56*(3), 402–411. http://doi.org/10.1109/TEM.2009.2023086

Chesbrough, H., Vanhaverbeke, W., & West, J. (2006). *Open Innovation: Researching a New Paradigm*. *Oxford*. Oxford University Press. Retrieved from http://books.google.com/books?hl=de&lr=&id=wBmA_ft_5lgC&pgis=1

Chick, S. E., Huchzermeier, A., & Netessine, S. (2014). Europe ' s Solution Factories. *Harvard Business Review*, (April), 111–114.

Cho, E., & Kim, S. (2015). Cronbach's Coefficient Alpha: Well Known but Poorly Understood. *Organizational Research Methods*, *18*(2), 207–230.

http://doi.org/10.1177/1094428114555994

Chong, C. W., & Chong, S. C. (2009). Knowledge management process effectiveness: measurement of preliminary knowledge management implementation. *Knowledge Management Research and Practice*. http://doi.org/10.1057/kmrp.2009.5

Choo, C. W. (1998). The Knowing Organization: How organizations use information to construct meaning, create knowledge, and make decisions. *International Journal of Information Management*, *16*(5), 329–340. http://doi.org/10.1093/acprof:oso/9780195176780.001.0001

Choo, C. W. (2006). *The Knowing Organization: How Organizations Use Information to Construct Meaning, Create Knowledge, and Make Decisions* (2nd ed.). New York: Oxford University Press.

Chutivongse, N., & Gerdsri, N. (2015). Proposed steps to analyze organizational characteristics and develop a roadmap for being an innovative organization. In *2015 Portland International Conference on Management of Engineering and Technology (PICMET)* (pp. 728–735). IEEE. http://doi.org/10.1109/PICMET.2015.7273263

Coff, R. W. (2003). The Emergent Knowledge-Based Theory of Competitive Advantage: An Evolutionary Approach to Integrating Economics and Management. *Managerial and Decision Economics*, *24*(4), 245–251. http://doi.org/10.1002/1127

Cohen, B. J. (1999). Fostering Innovation in a Large Human Services Bureaucracy. *Administration in Social Work*, *23*(2), 47–59. http://doi.org/10.1300/J147v23n02_04

Coleman, J. S. (1988). Social Capital in the Creation of Human Capital. *American Journal of Sociology*, *94*, S95–S120. http://doi.org/10.2307/2780243

CONACYT. (2016). Top 10 de científicos y centros de investigación mexicanos. Retrieved October 24, 2016, from http://www.conacytprensa.mx/index.php/sociedad/politica-cientifica/11169-top-10-de-cientificos-y-centros-de-investigacion-mexicanos

Cooper, R. G. (1983). The new product process: an empirically-based classification scheme. *R&D Management*, *13*(1), 1–13. http://doi.org/10.1111/j.1467-9310.1983.tb01124.x

Cooper, R. G. (1990). Stage-gate systems: A new tool for managing new products. *Business Horizons*, *33*(3), 44–54. http://doi.org/10.1016/0007-6813(90)90040-I

Cooper, R. G. (1990). Stage-Gate Systems: A New Tool for Managing New Products. *Business Horizons*, *33*(3), 44–54. http://doi.org/10.1016/0007-6813(90)90040-I

Corbetta, P. (2007). *Metodología y técnicas de investigación social*. Madrid: McGraw-Hill.

Cornell University, INSEAD, & WIPO. (2015). *The Global Innovation Index 2015: Effective Innovation Policies for Development*. (D. Soumitra, B. Lanvin, & S. Wunsch-Vincent, Eds.). Ithaca and Geneva: World Intellectual Property Organization. Retrieved from https://www.globalinnovationindex.org/userfiles/file/reportpdf/gii-full-report-2015-v6.pdf

Cornell University, INSEAD, & WIPO. (2016). *The Global Innovation Index 2016: Winning with Global Innovation*. (S. Dutta, B. Lanvin, & S. Wunsch-Vincent, Eds.). Ithaca, Fontainebleau, and Geneva: Cornell University, INSEAD, and the World Intellectual Property Organization. Retrieved from https://www.globalinnovationindex.org/gii-2016-report

Corsi, P., & Neau, E. (2015). *Innovation Capability Maturity Model*. London: ISTE Ltd.

Costa, V., & Monteiro, S. (2016). Key knowledge management processes for innovation: a systematic literature review. *VINE Journal of Information and Knowledge Management Systems*, *46*(3), null. http://doi.org/10.1108/VJIKMS-02-2015-0017

Cronbach, L. J. (1951). Coefficient alpha and the internal structure of tests. *Psychometrika*, *16*(3), 297–334. http://doi.org/10.1007/BF02310555

Cronbach, L. J., & Meehl, P. E. (1955a). Construct validity in psychological tests. *Psychological Bulletin*, *52*(4), 281–302. http://doi.org/10.1037/h0040957

Cronbach, L. J., & Meehl, P. E. (1955b). Construct validity in psychological tests. *Psychological Bulletin*, *52*(4), 281–302. http://doi.org/10.1037/h0040957

Daft, R. (2011). *Teoría y diseño organizacional* (10th ed.). México, DF: Cengage Learning Editores.

Dalkir, K. (2005). *Knowledge Management in Theory and Practice*. Boston, MA: Elsevier.

Darroch, J. (2005). Knowledge management, innovation and firm performance. *Journal of Knowledge Management*, *9*(3), 101–115. http://doi.org/10.1108/13673270510602809

Darroch, J., & McNaughton, R. (2002). Examining the link between knowledge management practices and types of innovation. *Journal of Intellectual Capital*, *3*(3), 210–222. http://doi.org/10.1108/14691930210435570

Davenport, T. H., & Prusak, L. (1998a). *Working Knowledge: How Organizations Manage what They Know*. Harvard Business Press.

Davenport, T. H., & Prusak, L. (1998b). Working Knowledge How Organization Manage What They Know. http://doi.org/10.1145/348772.348775

Dayan, R., & Evans, S. (2006). KM your way to CMMI. *Journal of Knowledge Management*, *10*(1), 69–80. http://doi.org/10.1108/13673270610650111

de Sordi, J. O., & Carvalho Azevedo, M. (2008). Análise de Competências Individuais e Organizacionais Associadas à Prática de Gestão do Conhecimento. *Revista Brasileira de Gestão de Negócios*, *10*(29), 391–407.

Dehghani, R., & Ramsin, R. (2015). Methodologies for developing knowledge management systems: an evaluation framework. *Journal of Knowledge Management*, *19*(4), 682–710. http://doi.org/10.1108/JKM-10-2014-0438

Den Hertog, P. (2000). Knowledge Intensive Business Services As Co-Producers of Innovation. *International Journal of Innovation Management*, *4*(4), 4–6. http://doi.org/10.1142/S136391960000024X

den Ouden, E. (2012). *Innovation Design*. *Innovation Design - Creating Value for People, Organizations and Society*. http://doi.org/10.1007/978-1-4471-2268-5

Department of Economic and Social Affairs, P. D. (2015). World Population Prospects 2015, 66. http://doi.org/Working Paper No. ESA/P/WP.241.

Diakoulakis, I. E., Georgopoulos, N. B., Koulouriotis, D. E., & Emiris, D. M. (2004). Towards a holistic knowledge management model. *Journal of Knowledge Management*, *8*(1), 32–46. http://doi.org/10.1108/13673270410523899

Ding, W., Liang, P., Tang, A., & van Vliet, H. (2014). Knowledge-based approaches in software documentation: A systematic literature review. *Information and Software Technology*, *56*(6), 545–567. http://doi.org/10.1016/j.infsof.2014.01.008

Dodgson, M., Gann, D., & Salter, A. (2008). *The Management of Technological Innovation: Strategy and Practice* (2nd ed.). Oxford, UK: Oxford University Press.

Donate, M. J., & Guadamillas, F. (2013). An empirical study on the relationships between knowledge management, knowledge-oriented human resource practices and innovation. *Knowledge Management Research & Practice*, *13*(2), 134–148. http://doi.org/10.1057/kmrp.2013.36

Donate, M. J., & Sánchez de Pablo, J. D. (2014). The role of knowledge-oriented leadership in knowledge management practices and innovation. *Journal of Business Research*, *68*(2), 360–370. http://doi.org/10.1016/j.jbusres.2014.06.022

Dorasamy, M., Raman, M., & Kaliannan, M. (2013). Knowledge management systems in support of disasters management: A two decade review. *Technological Forecasting and Social Change*, *80*(9), 1834–1853. http://doi.org/10.1016/j.techfore.2012.12.008

Drahos, P., & Maher, I. (2004). Innovation, competition, standards and intellectual property: policy perspectives from economics and law. *Information Economics and Policy*, *16*(1), 1–11. http://doi.org/10.1016/j.infoecopol.2003.09.001

Dugan, R. E., & Gabriel, K. J. (2013). "Special Forces" Innovation: How DARPA attacks problems. *Harvard Business Review*, (OCT).

Dutta, S. (2015). *The Global Innovation Index 2015. Stronger Innovation Linkages for*. http://doi.org/978-2-9522210-8-5

Dutta, S. (2016). *The Global Innovation Index 2015. Stronger Innovation Linkages for*. http://doi.org/978-2-9522210-8-5

Edenius, M., & Borgerson, J. (2003). To manage knowledge by intranet. *Journal of Knowledge Management*, *7*(5), 124–136. http://doi.org/10.1108/13673270310505430

Edison, H., Bin Ali, N., & Torkar, R. (2013). Towards innovation measurement in the software industry. *Journal of Systems and Software*, *86*(5), 1390–1407. http://doi.org/10.1016/j.jss.2013.01.013

Engineering, S., & Committee, S. (2013). *IEEE Standard — Adoption of ISO / IEC 20000-2 : 2012 , Information technology — Service management — Part 2 : Guidance on the application of service management systems IEEE Computer Society*.

Errasti, N., & Zabaleta, N. (2011). A review and conceptualisation of innovation models from the past three decades. *International Journal of Technology Management*, *55*(3/4), 194–200. http://doi.org/10.1504/IJTM.2011.041946

Esterhuizen, D., Schutte, C. S. L., & Du Toit, A. S. A. (2012). Knowledge creation processes as critical enablers for innovation. *International Journal of Information Management*, *32*(4), 354–364. http://doi.org/10.1016/j.ijinfomgt.2011.11.013

Eyring, M. J., & Johnson, M. W. (2005). Spotlight on business model innovation. *Harvard Business Review*, *89*(February), 88–95. Retrieved from http://www.mendeley.com/research/t-c-design-winning-op-y-o-r-p-os-t-t-c-op-y-o-r-p-os-t/

Farrell, J., & Saloner, G. (1985). Standardization, Compatibility, and Innovation. *Source: The RAND Journal of Economics Rand Journal of Economics*, *16*(1), 70–83. http://doi.org/10.2307/2555589

Farzin, M. R., Kahreh, M. S., Hesan, M., & Khalouei, A. (2014). A Survey of Critical Success Factors for Strategic Knowledge Management Implementation: Applications for Service Sector. *Procedia - Social and Behavioral Sciences*, *109*, 595–599. http://doi.org/10.1016/j.sbspro.2013.12.512

Fast-Berglund, Å., Bligård, L.-O., Åkerman, M., & Karlsson, M. (2014). Using the TRL-methodology to Design Supporting ICT-tools for Production Operators. *Procedia CIRP*, *17*, 726–731. http://doi.org/10.1016/j.procir.2014.02.039

Fateh Rad, M., Seyedesfahani, M. M., & Jalilvand, M. R. (2015). An effective

collaboration model between industry and university based on the theory of self organization. *Journal of Science and Technology Policy Management*, *6*(1), 2–24. http://doi.org/10.1108/JSTPM-08-2014-0035

Ferguson, S. (2006). AS 5037–2005: knowledge management blueprint for Australian organisations? *The Australian Library Journal*, *55*(3), 196–209. http://doi.org/10.1080/00049670.2006.10721852

Fine, C. H., & Hax, A. C. (1985). Manufacturing Strategy:A Methodology and an Illustration. *Interfaces*, *15*(6), 28–46. http://doi.org/10.1287/inte.15.6.28

Galbraith, J. R. (1982). Designing the Innovative Organization. *Organizational Design*, *10*(3), 5–25. http://doi.org/10.1016/0090-2616(82)90033-X

Galindo-Rueda, F., & Millot, V. (2015). *Measuring Design and its Role in Innovation* (OECD Science, Technology and Industry Working Papers No. 2015/01). http://doi.org/10.1787/5js7p6lj6zq6-en

Garlatti, A., & Massaro, M. (2016). Is KM declining? *The Electronic Journal of Knowledge Management*, *14*(1), 1–4.

Garlatti, A., Massaro, M., Dumay, J., & Zanin, L. (2014). Intellectual Capital and Knowledge Management within the public sector. A systematic literature review and future developments. In *International Conference on Intellectual Capital, Knowledge Management & Organizational Learning (2014)* (pp. 175–184). Academic Conferences Limited.

Gasik, S. (2011). A model of project knowledge management. *Project Management Journal*, *42*(3), 23–44. http://doi.org/10.1002/pmj.20239

Giudice, M. Del, & Peruta, M. R. Della. (2016). The impact of IT-based knowledge management systems on internal venturing and innovation: A structural equation modeling approach to corporate performance. *Journal of Knowledge Management*, *20*(3), 484–498. http://doi.org/10.1108/JKM-07-2015-0257

González, E. R. V., & Rodríguez, S. E. (2016). Knowledge and Technology Transfer Relationship between a Research Center and the Production Sector: CIMAT Case Study. *Latin American Business Review*, *17*(4), 271–288. http://doi.org/10.1080/10978526.2016.1232577

Gonzalez, R., Llopis, J., & Gasco, J. (2013). Innovation in public services: The case of Spanish local government. *Journal of Business Research*, *66*(10), 2024–2033. http://doi.org/10.1016/j.jbusres.2013.02.028

Govindarajan, V. (2012). A Reverse- Innovation Playbook. *Harvard Business Review*, *90*(April), 120–124. http://doi.org/10.1108/sd.2012.05628iaa.008

Greenhalgh, T., Robert, G., Macfarlane, F., Bate, P., & Kyriakidou, O. (2004). Diffusion of innovations in service organizations: systematic review and recommendations.

*The Milbank Quarterly*, *82*(4), 581–629. http://doi.org/10.1111/j.0887-378X.2004.00325.x

Hafeez, K., & Abdelmeguid, H. (2003). Dynamics of human resource and knowledge management. *Journal of the Operational Research Society*, *54*(2), 153–164. http://doi.org/10.1057/palgrave.jors.2601513

Hair, J. F. J., Hult, G. T. M., Ringle, C., & Sarstedt, M. (2014). *A Primer on Partial Least Squares Structural Equation Modeling (PLS-SEM)*. *Long Range Planning* (Vol. 46). http://doi.org/10.1016/j.lrp.2013.01.002

Hamann, P. M., Schiemann, F., Bellora, L., & Guenther, T. W. (2013). Exploring the Dimensions of Organizational Performance A Construct Validity Study. *Organizational Research Methods*, *16*(1), 67–87. http://doi.org/10.1177/1094428112470007

Hamel, G. (2006). The why, what, and how of management innovation. *Harvard Business Review*, *84*(2), 72–84, 163. http://doi.org/10.1016/0002-9610(92)90118-B

Handzic, M. (2011). Integrated socio-technical knowledge management model: an empirical evaluation. *Journal of Knowledge Management*, *15*(2), 198–211. http://doi.org/10.1108/13673271111119655

Hardy, B., & Ford, L. R. (2014). It's Not Me, It's You: Miscomprehension in Surveys. *Organizational Research Methods*, *17*(2), 138–162. http://doi.org/10.1177/1094428113520185

Haslinda, a., & Sarinah, a. (2009). A Review of Knowledge Management Models. *The Journal of International Social Research*, *2*(9), 187–198.

Heisig, P. (2009). Harmonisation of knowledge management – comparing 160 KM frameworks around the globe. *Journal of Knowledge Management*, *13*(4), 4–31. http://doi.org/10.1108/13673270910971798

Heisig, P., Suraj, O. A., Kianto, A., Kemboi, C., Perez Arrau, G., & Fathi Easa, N. (2016). Knowledge management and business performance: global experts' views on future research needs. *Journal of Knowledge Management*, *20*(6), 1169–1198. http://doi.org/10.1108/JKM-12-2015-0521

Henning, F. (2013). The impact of interoperability standards adoption on organisations in government information networks. In *Proceedings of the 7th International Conference on Theory and Practice of Electronic Governance - ICEGOV '13* (pp. 250–259). New York, New York, USA: ACM Press. http://doi.org/10.1145/2591888.2591934

Henseler, J., Ringle, C. M., & Sarstedt, M. (2014). A new criterion for assessing discriminant validity in variance-based structural equation modeling. *Journal of the Academy of Marketing Science*, 115–135. http://doi.org/10.1007/s11747-014-0403-

8

Hernandez-Munoz, L., Torane, M., Amini, A., & Vivekanandan-Dhukaram, A. (2015). A State-of-the-art Analysis of Innovation Models and Innovation Software Tools. In *PROCEEDINGS OF THE 10TH EUROPEAN CONFERENCE ON INNOVATION AND ENTREPRENEURSHIP (ECIE 2015)* (pp. 237–245).

Hernández Sampieri, R., Fernández-Collado, C., & Baptista Lucio, P. (2010). *Metodología de la investigación*. (J. Mares Chacón, Ed.), *México Trillas* (Quinta edi, Vol. 18). México, Df: McGRAW-HILL / INTERAMERICANA EDITORES, S.A. DE C.V.

Herrero, J. (2010). El Análisis Factorial Confirmatorio en el estudio de la Estructura y Estabilidad de los Instrumentos de Evaluación: Un ejemplo con el Cuestionario de Autoestima CA-14. *Psychosocial Intervention*, *19*(3), 289–300. http://doi.org/10.5093/in2010v19n3a9

Herzog, P. (2011). *Open and Closed Innovation* (2nd ed.). Wiesbaden: Gabler Verlag. http://doi.org/10.1007/978-3-8349-6165-5

Hislop, D. (2005). *Knowledge management in organizations: A critical introduction*. Oxford, UK: Oxford University Press.

Ho, C.-F., Hsieh, P.-H., & Hung, W.-H. (2014). Enablers and processes for effective knowledge management. *Industrial Management & Data Systems*, *114*, 734–754. http://doi.org/10.1108/IMDS-08-2013-0343

Holsapple, C. W., & Joshi, K. D. (2001). Organizational knowledge resources. *Decision Support Systems*, *31*(1), 39–54. http://doi.org/10.1016/S0167-9236(00)00118-4

Holsapple, C. W., & Luo, W. (1996). A framework for studying computer support of organizational infrastructure. *Information & Management*, *31*(1), 13–24. http://doi.org/10.1016/S0378-7206(96)01067-1

Hubbard, D. (2014). *How to measure anything*.

Hunter, L., Webster, E., & Wyatt, A. (2005). Measuring intangible capital: a review of current practice. *Australian Accounting Review*, *15*(2), 4–21. http://doi.org/10.1111/j.1835-2561.2005.tb00288.x

Ichijo, K., & Nonaka, I. (2007). *Knowledge creation and management: new challenges for managers*. *Oxford University Press*. New York: Oxford University Press.

IEEE. (2004). IEEE Guide Adoption of PMI Standard a Guide to the Project Management Body of Knowledge. New York: The Institute of Electrical and Electronics Engineers. http://doi.org/10.1109/IEEESTD.2004.94565

Immelt, J. R., Govindarajan, V., & Trimble, C. (2009). by Jeffrey R. Immelt, Vijay Govindarajan, and Chris Trimble 56. *Harvard Business Review*, (October), 56–65.

Imnc, N. M., & Projects-, T. T. (2008). Gesti6n de Ia Tecnologia- Proyectos tecnol6gicos - Requisites.

INEGI. (2015). ESTADÍSTICAS A PROPÓSITO DEL... DÍA MUNDIAL DEL INTERNET (17 DE MAYO). Aguascalientes, Ags: INEGI. Retrieved from http://www.inegi.org.mx/saladeprensa/aproposito/2015/internet0.pdf

Inkinen, H. (2016). Review of empirical research on knowledge management practices and firm performance. *Journal of Knowledge Management*, *20*(2), 230–257. http://doi.org/10.1108/JKM-09-2015-0336

Inkinen, H. T., Kianto, A., & Vanhala, M. (2015). Knowledge management practices and innovation performance in Finland. *Baltic Journal of Management*, *10*(4), 432–455. http://doi.org/10.1108/BJM-10-2014-0178

İzadi, A., Zarrabi, F., & Zarrabi, F. (2013). Firm-Level Innovation Models. *Procedia - Social and Behavioral Sciences*, *75*, 146–153. http://doi.org/10.1016/j.sbspro.2013.04.017

Jääskeläinen, A. (2009). Identifying a suitable approach for measuring and managing public service productivity. *Electronic Journal of Knowledge Management*, *7*(4), 447–458.

Jain, A. K., & Moreno, A. (2015). Organizational learning, knowledge management practices and firm's performance. *The Learning Organization*, *22*(1), 14–39. http://doi.org/10.1108/TLO-05-2013-0024

Jain, P. (2009). Knowledge Management In e-Government. *Journal of Knowledge Management Practice*, *10*(4), 1–11. Retrieved from http://www.tlainc.com/articl203.htm

Jakobs, K. (2006). Shaping user-side innovation through standardisation. *Technological Forecasting and Social Change*, *73*(1), 27–40. http://doi.org/10.1016/j.techfore.2005.06.007

Jennex, M. E., & Olfman, L. (2004). Assessing knowledge management success/effectiveness models. In IEEE (Ed.), *Proceedings of the 37th Hawaii International Conference on System Sciences*. Hawaii: IEEE. http://doi.org/10.1109/HICSS.2004.1265571

Jennex, M. E., & Smolnik, S. (2011). *Strategies for Knowledge Management Success: Exploring Organizational Efficacy*. (M. E. Jennex & S. Smolnik, Eds.). Hershey PA: IGI Global. http://doi.org/10.4018/978-1-60566-709-6

Jenny S. Z. Eriksson Lundström, Wiberg, M., Hrastinski, S., Edenius, M., & Ågerfalk, P. J. (2013). *Managing Open Innovation Technologies*. (J. S. Z. Eriksson Lundström, M. Wiberg, S. Hrastinski, M. Edenius, & P. J. Ågerfalk, Eds.). Berlin, Heidelberg: Springer Berlin Heidelberg. http://doi.org/10.1007/978-3-642-31650-0

Johannessen, J.-A. (2008). Organisational innovation as part of knowledge management. *International Journal of Information Management*, *28*(5), 403–412. http://doi.org/10.1016/j.ijinfomgt.2008.04.007

Johnson, G., Yip, G. S., & Hensmans, M. (2012). Achieving Successful Strategic Transformation. *Mit Sloan Management Review*, *53*(53308), 25–+.

Jolly, V. K. (1997). *Commercializing new technologies: getting from mind to market*. Boston: Harvard Business Press.

Jolly, V. K. (1997). Commercializing New Technologies - Getting from Mind to Market. *Innovation: Management, Policy & Practice*, *1*, 1–30. http://doi.org/10.5172/impp.1998.1.5-6.40

Jonsson, A. (2013). Beyond knowledge management – understanding how to share knowledge through logic and practice. *Knowledge Management Research & Practice*, *13*(1), 1–14. http://doi.org/10.1057/kmrp.2013.28

Karlsson, M. (2013). Who needs a standard for Innovation Management? Retrieved January 1, 2015, from http://conference.ispim.org/wp-content/uploads/sites/2/2013/11/ISPIM2013_Karlsson.pdf

Karr, A. F. (2016). Data Sharing and Access. *Annual Review of Statistics and Its Application*, *3*(1), 113–132. http://doi.org/10.1146/annurev-statistics-041715-033438

Kerlinger, F. (1975). *La Investigación del Comportamiento. Técnicas y Metodologías*. México, DF: Interamericana.

Kerzner, H. (2013). *Project management metrics, KPIs, and dashboards: a guide to measuring and monitoring project performance* (1st ed.). New York: John Wiley & Sons.

Khedhaouria, A., & Jamal, A. (2015). Sourcing knowledge for innovation: knowledge reuse and creation in project teams. *Journal of Knowledge Management*, *19*(5), 932–948. http://doi.org/10.1108/JKM-01-2015-0039

Kimble, C., de Vasconcelos, J. B., & Rocha, Á. (2016). Competence management in knowledge intensive organizations using consensual knowledge and ontologies. *Information Systems Frontiers*. http://doi.org/10.1007/s10796-016-9627-0

Kitchenham, B., & Charters, S. (2007). *Guidelines for performing Systematic Literature Reviews in Software Engineering*. UK. Retrieved from http://www.dur.ac.uk/ebse/resources/guidelines/Systematic-reviews-5-8.pdf

Klaus-Dieter, A., Frank, B., & Carsten, T. (2000). Knowledge Management for Building Learning Software Organizations. *Information Systems Frontiers*, *2*, 349. Retrieved from
http://proquest.umi.com/pqdweb?did=404921041&Fmt=7&clientId=18803&RQT=

309&VName=PQD

Kogut, B., & Zander, U. (1992). Knowledge of the Firm, Combinative Capabilities, and the Replication of Technology. *Organization Science*, *3*(3), 383–397. Retrieved from http://www.jstor.org/stable/2635279

Koh, C. E., Ryan, S., & Prybutok, V. R. (2005). Creating value through managing knowledge in an e-government to constituency (G2C) environment. *JOURNAL OF COMPUTER INFORMATION SYSTEMS*, *45*, 32–41.

Kör, B., & Maden, C. (2013). The Relationship between Knowledge Management and Innovation in Turkish Service and High-Tech Firms. *International Journal of Business and Social Science*, *4*(4), n/a. Retrieved from http://search.proquest.com/docview/1356021472/abstract/79B0B6A2DAFA4AB9PQ/7?accountid=17242%5Cnhttp://media.proquest.com/media/pq/classic/doc/2982471491/fmt/pi/rep/NONE?hl=knowledge,knowledge,management,management,turkeys,turkey,turkeys,turkey

Kotter, J. P. (2012). Accelerate. How the most innovative companies capitalize on today's rapid-fire strategic challenges and still make their numbers. *Harvard Business Review*, (nov.), 44–59. http://doi.org/10.1057/palgrave.crr.1540128

Kovačič, A. (2007). Process-based knowledge management: Towards e-government in Slovenia. *Management*, *12*(1), 45–64.

Kulkarni, U., Ravindran, S., & Freeze, R. (2007). A Knowledge Management Success Model: Theoretical Development and Empirical Validation. *J. Manage. Inf. Syst.*, *23*(3), 309–347. http://doi.org/10.2753/MIS0742-1222230311

Lassi A, L. (2010). Summary of the knowledge-creating company. *New York*. Oxford University Press.

Leal-Millan, A., Roldan, J. L., Leal-Rodriguez, A. L., & Ortega-Gutierrez, J. (2016). IT and relationship learning in networks as drivers of green innovation and customer capital: Evidence from the automobile sector. *Journal of Knowledge Management*, *20*(3), 444–464. http://doi.org/10.1108/JKM-05-2015-0203

Leal-Rodríguez, A., Leal-Millán, A., Roldán-Salgueiro, J. L., & Ortega-Gutiérrez, J. (2013). Knowledge management and the effectiveness of innovation outcomes: The role of cultural barriers. *Electronic Journal of Knowledge Management*, *11*(1), 62–71.

Ledesma, R., Ibañez, G., & Mora, P. (2002). Análisis de consistencia interna mediante Alfa de Cronbach: un programa basado en gráficos dinámicos. *Psico-USF*, *7*(2), 143–152. http://doi.org/10.1590/S1413-82712002000200003

Lee, C. S., & Wong, K. Y. (2015). Development and validation of knowledge management performance measurement constructs for small and medium enterprises. *Journal of Knowledge Management*, *19*(4), 711–734.

http://doi.org/10.1108/JKM-10-2014-0398

Lee, H., & Choi, B. (2003). Knowledge Management Enablers, Processes, and Organizational Performance: An Integrative View and Empirical Examination. *Journal of Management Information Systems*, *20*(1), 179–228. http://doi.org/10.1080/07421222.2003.11045756

Lester, E. I. A. (2014). Information Management. In *Project Management, Planning and Control* (pp. 359–364). Elsevier. http://doi.org/10.1016/B978-0-08-098324-0.00037-8

Levett, G. P., & Guenov, M. D. (2000). A methodology for knowledge management implementation. *Journal of Knowledge Management*, *4*(3), 258–270. http://doi.org/10.1108/13673270010350066

Liberona, D., & Ruiz, M. (2013). Análisis de la implementación de programas de gestión del conocimiento en las empresas chilenas. *Estudios Gerenciales*, *29*(127), 151–160. http://doi.org/10.1016/j.estger.2013.05.003

Little, T., & Deokar, A. V. (2012). Knowledge creation in the context of knowledge-intensive business processes. *Proceedings of the 2012 SIGBPS Workshop on Business Processes and Services SIGBPS 12*, 54–58. http://doi.org/10.1108/JKM-11-2015-0443

Lj Todorović, M., Č Petrović, D., Mihić, M. M., Lj Obradović, V., & Bushuyev, S. D. (2015). Project success analysis framework: A knowledge-based approach in project management. *International Journal of Project Management*, *33*, 772–783. http://doi.org/10.1016/j.ijproman.2014.10.009

Lowry, P. B., & Gaskin, J. (2014). Partial Least Squares (PLS) Structural Equation Modeling (SEM) for Building and Testing Behavioral Causal Theory: When to Choose It and How to Use It. *IEEE Transactions on Professional Communication*, *57*(2), 123–146. http://doi.org/10.1109/TPC.2014.2312452

Luna-Reyes, L. F., Gil-Garcia, J. R., & Cruz, C. B. (2007). Collaborative digital government in Mexico: Some lessons from federal Web-based interorganizational information integration initiatives. *Government Information Quarterly*, *24*(4), 808–826. http://doi.org/10.1016/j.giq.2007.04.003

Lundvall, B.-Å., & Nielsen, P. (2007). Knowledge management and innovation performance. *International Journal of Manpower*, *28*(3/4), 207–223. http://doi.org/10.1108/01437720710755218

Lustri, D., Miura, I., & Takahashi, S. (2007). Knowledge management model: practical application for competency development. *The Learning Organization*, *14*(2), 186–202. http://doi.org/10.1108/09696470710727023

Mageswari, U., Sivasubramanian, C., & Dath, S. (2015). Knowledge Management Enablers, Processes and Innovation in Small Manufacturing Firms: A Structural

Equation Modeling Approach. *IUP Journal of Knowledge Management*, *13*(1), 33–58.

Maier, R., & Schmidt, A. (2014). Explaining organizational knowledge creation with a knowledge maturing model. *Knowledge Management Research & Practice*, (November 2013), 1–21. http://doi.org/10.1057/kmrp.2013.56

Makani, J., & Marche, S. (2010). Towards a typology of knowledge-intensive organizations: determinant factors. *Knowledge Management Research & Practice*, *8*(3), 265–277. http://doi.org/10.1057/kmrp.2010.13

Malhotra, N. K. (2008). *Investigación de mercados* (5th ed.). México, DF: Pearson Educación.

Man, A.-P. De. (2008). Knowledge management and innovation in networks, 215. http://doi.org/10.4337/9781848443846

Mankins, J. C. (1995). Technology readiness levels. Retrieved January 1, 2015, from http://fellowships.teiemt.gr/wp-content/uploads/2016/01/trl.pdf

Mao, H., Liu, S., Zhang, J., & Deng, Z. (2016). Information technology resource, knowledge management capability, and competitive advantage: The moderating role of resource commitment. *International Journal of Information Management*, *36*(6), 1062–1074. http://doi.org/10.1016/j.ijinfomgt.2016.07.001

March, J. G. (1991). Exploration and Exploitation in Organizational Learning. *Organization Science*, *2*(1), 71–87. http://doi.org/10.1287/orsc.2.1.71

Marinova, D., & Phillimore, J. (2003). Models of Innovation. In *The International Handbook on Innovation* (pp. 44–53). Elsevier. http://doi.org/10.1016/B978-008044198-6/50005-X

Marsal-Llacuna, M.-L., & Wood-Hill, M. (2017). The Intelligenter method (III) for “smarter” standards development and standardisation instruments. *Computer Standards & Interfaces*, *50*, 142–152. http://doi.org/10.1016/j.csi.2016.09.010

Martínez Piva, J. M. (2008). *Generación y protección del conocimiento: propiedad intelectual, innovación y desarrollo económico*. México, DF: Naciones Unidas. Retrieved from http://www.cepal.org/cgi-bin/getProd.asp?xml=/publicaciones/sinsigla/xml/2/32772/P32772.xml&xsl=/mexico/tpl/p10f.xsl&base=/mexico/tpl/top-bottom.xsl

Mas-Machuca, M., & Martínez Costa, C. (2012). Exploring critical success factors of knowledge management projects in the consulting sector. *Total Quality Management & Business Excellence*, *23*(11–12), 1297–1313. http://doi.org/10.1080/14783363.2011.637778

Massa, S., & Testa, S. (2004). Innovation or imitation? *Benchmarking: An International Journal*, *11*(6), 610–620. http://doi.org/10.1108/14635770410566519

Massaro, M., Dumay, J., & Garlatti, A. (2015). Public sector knowledge management: a structured literature review. *Journal of Knowledge Management*, *19*(3), 530–558. http://doi.org/10.1108/JKM-11-2014-0466

Massingham, P. (2014a). An evaluation of knowledge management tools: Part 1-managing knowledge resources. *Journal of Knowledge Management*, *18*(6), 1075–1100. http://doi.org/10.1108/JKM-11-2013-0449

Massingham, P. (2014b). An evaluation of knowledge management tools: Part 2 – managing knowledge flows and enablers. *Journal of Knowledge Management*, *18*(6), 1101–1126. http://doi.org/10.1108/JKM-03-2014-0084

Massingham, P. R., & Massingham, R. K. (2014). Does knowledge management produce practical outcomes? *Journal of Knowledge Management*, *18*(2), 221–254. http://doi.org/10.1108/JKM-10-2013-0390

Matei, A., & Savulescu, C. (2014). Enhancing the capacity for innovation of public administration. An exploratory study on e-Governance, ICT, knowledge management in Romania. *Theoretical and Applied Economics*, *XXI*(11), 7–26.

Matzler, K. (2013). Business model innovation: coffee triumphs for Nespresso. *Journal of Business Strategy*, *34*(2), 30–37. http://doi.org/10.1108/02756661311310431

Maxwell, J. W. (1998). Minimum quality standards as a barrier to innovation. *Economics Letters*, *58*(3), 355–360. http://doi.org/10.1016/S0165-1765(97)00293-0

McAdam, R., & McCreedy, S. (1999). A critical review of knowledge management models. *The Learning Organization*, *6*(3), 91–101. http://doi.org/10.1108/09696479910270416

McIntosh, C. N., Edwards, J. R., & Antonakis, J. (2014). Reflections on Partial Least Squares Path Modeling. *Organizational Research Methods*, *17*(2), 210–251. http://doi.org/10.1177/1094428114529165

Merrill, J., Keeling, J., & Gebbie, K. (2009). Toward standardized, comparable public health systems data: A taxonomic description of essential public health work. *Health Services Research*, *44*(5 PART 2), 1818–1841. http://doi.org/10.1111/j.1475-6773.2009.01015.x

Michelman, P. (2016). Frontiers: Exploring the Digital Future of Management Frontiers: Exploring the Digital Future of Management 1 Introduction.

Milner, E. M. (2000). *Managing Information and Knowledge in the Public Sector*. Abingdon, UK: Taylor & Francis. http://doi.org/10.4324/9780203458631

Miniwatts Marketing Group. (2016). Internet World Stats. Retrieved May 1, 2016, from http://www.internetworldstats.com/stats.htm

Mir, M., Casadesús, M., & Petnji, L. H. (2016). The impact of standardized innovation management systems on innovation capability and business performance: An empirical study. *Journal of Engineering and Technology Management*, *41*, 26–44. http://doi.org/10.1016/j.jengtecman.2016.06.002

Mitre-Hernández, H. A., Mora-Soto, A., López-Portillo, H. P., & Lara-Alvarez, C. (2015). Strategies for fostering Knowledge Management Programs in Public Organizations. In A. Garlatti & M. Massaro (Eds.), *16th European Conference on Knowledge Management* (pp. 539–547). Reading: Academic Conferences and Publishing International Limited. Retrieved from http://arxiv.org/abs/1506.03828

Moffett, S., & McAdam, R. (2009). Knowledge management: a factor analysis of sector effects. *Journal of Knowledge Management*, *13*(3), 44–59. http://doi.org/10.1108/13673270910962860

Mojibi, T., Khojasteh, Y., & Khojasteh-Ghamari, Z. (2015). The Role of Infrastructure Factors in Knowledge Management Implementation. *Knowledge and Process Management*, *22*(1), 34–40. http://doi.org/10.1002/kpm.1459

Moncayo, M. B., & Anticona, M. T. (n.d.). A Systematic Review Based on Kitchengam ' s Criteria About use of Specific Models to Implement E-government Solutions, 75–80.

Moore, G. A. (1999). *Crossing the Chasm: Marketing and Selling High-Tech Products to Mainstream Customers*. *New York HarperBusiness* (Vol. Rev.). http://doi.org/10.1017/CBO9781107415324.004

Mora-Soto, A. (2011). *Marco Metodológico y Tecnológico para la Gestión del Conocimiento Organizativo que de Soporte al Despliegue de Buenas Prácticas de Ingeniería del Software*. Universidad Carlos III de Madrid.

Morales-Vallejo, P. (2011). El Análisis Factorial en la construcción e interpretación de tests, escalas y cuestionarios. *Universidad Pontificia Comillas*. Madrid: Universidad Pontificia Comillas. Retrieved from http://www.upcomillas.es/personal/peter/investigacion/AnalisisFactorial.pdf

Morten T.; Birkinshaw, H. and J. (2007). The innovation value chain. *Harvard Business Review*, (June), 127–148. http://doi.org/Article

Muench Dean. (1994). *The Sybase Development Framework*. Oakland, CA: Sybase Inc.

Murray, R., Caulier-Grice, J., & Mulgan, G. (2010). *The open book of social innovation*. London: Nesta and the Young Foundation. Retrieved from http://youngfoundation.org/publications/the-open-book-of-social-innovation/

Naghavi, M., Dastaviz, A. H., & Nezakati, H. (2013). Relationships among Critical Success Factors of Knowledge Management and Organizational Performance. *Journal of Applied Sciences*, *13*(5), 755–759. http://doi.org/10.3923/jas.2013.755.759

Nasa. (2014). Definition Of Technology Readiness Levels. *NASA Technology Readiness Level*, 9. Retrieved from http://www.nasa.gov/content/technology-readiness-level/#.U9-RylZaSP0

NASA. (2012). Technology Readiness Level. Retrieved October 1, 2015, from https://www.nasa.gov/directorates/heo/scan/engineering/technology/txt_accordion1.html

Ng, A. H. H., Yip, M. W., Din, S. B., & Bakar, N. A. (2012). Integrated Knowledge Management Strategy: A Preliminary Literature Review. *Procedia - Social and Behavioral Sciences*, *57*, 209–214. http://doi.org/10.1016/j.sbspro.2012.09.1176

Nonaka, I. (1991). The Knowledge-Creating Company. *Harvard Business Review*, *85*(7/8), 162–171. Retrieved from http://search.ebscohost.com/login.aspx?direct=true&db=bth&AN=25358848&lang=es&site=ehost-live

Nonaka, I. (1994). A Dynamic Theory of Organizational Knowledge Creation. *Organization Science*, *5*(1), 14–37. http://doi.org/10.1287/orsc.5.1.14

Nonaka, I. (2006). Creating Sustainable Competitive Advantage through Knowledge-Based Management.

Nonaka, I. (2008). *The Knowledge Creating Company*. Boston, MA: Harvard Business Press.

Nonaka, I., & D. Teece. (2001). *Managing Industrial Knowledge; Creation, Transfer and Utilization*. London: SAGE Publications.

Nonaka, I., & Takeuchi, H. (1999). *La organización creadora del conocimiento. Cómo las empresas japonesas crean la dinámica de la innovación*. México, DF: Oxford University Press.

O'Brien, E., Clifford, S., & Southern, M. (2010). *Knowledge Management for Process, Organizational and Marketing Innovation: Tools and Methods*. (E. O'Brien, S. Clifford, & M. Southern, Eds.). IGI Global. http://doi.org/10.4018/978-1-61520-829-6

OECD. (2000). *Knowledge Management in the Learning Society*. OECD Publishing. http://doi.org/10.1787/9789264181045-en

OECD. (2004). *Innovation in the Knowledge Economy*. Paris: OECD Publishing. http://doi.org/10.1787/9789264105621-en

OECD. (2010). *Measuring Innovation: A New Perspective*. *Healthcare executive* (Vol. 26). Paris: OECD Publishing.

OECD. (2014). *Measuring the Digital Economy*. OECD Publishing. http://doi.org/10.1787/9789264221796-en

OECD. (2015a). *Data-Driven Innovation: Big Data for Growth and Well-Being*. Paris: OECD Publishing. http://doi.org/10.1787/9789264229358-en

OECD. (2015b). *Frascati Manual 2015*. Paris: OECD Publishing. http://doi.org/10.1787/9789264239012-en

OECD. (2015c). OECD Innovation Strategy 2015 An Agenda for Policy Action. *OECD Reviews of Innovation Policy*, (June), 395–423. http://doi.org/10.1787/9789264039827-en

OECD. (2015d). *OECD Science, Technology and Industry Scoreboard 2015*. OECD Publishing. http://doi.org/10.1787/sti_scoreboard-2015-en

OECD. (2015e). *The innovation imperative*. Paris: OECD Publishing. http://doi.org/10.1067/mtc.2000.110491

OECD. (2015f). *The Innovation Imperative in the Public Sector*. OECD Publishing. http://doi.org/10.1787/9789264236561-en

OECD. (2016). Blue Sky Forum: Agenda. Retrieved November 16, 2016, from http://www.oecd.org/sti/blue-sky-2016-agenda.htm#ps1_d1

OECD. (2016). *Education at a Glance 2016*. OECD Publishing. http://doi.org/10.1787/eag-2016-en

OECD. (2016). *Governing Education in a ComplexWorld, Educational Research and Innovation*. (T. Burns & F. Köster, Eds.). Paris: OECD Publishing. http://doi.org/10.1787/9789264255364-en

OECD. (2016). *OECD Factbook 2015-2016*. Paris: OECD Publishing. http://doi.org/10.1787/factbook-2015-en

OECD. (2016). *OECD Science, Technology and Innovation Outlook 2016*. Paris: OECD Publishing. http://doi.org/10.1787/sti_in_outlook-2016-en

OECD, & Eurostat. (2005). *Oslo Manual: Guidelines for Collecting and Interpreting Innovation Data, 3rd Edition*. OECD Publishing. http://doi.org/10.1787/9789264013100-en

OECD Multilingual Summaries OECD Science , Technology and Industry Outlook Perspectivas de Ciencia , tecnología e industria de la OCDE 2012 Innovación en tiempos de crisis. (2012), 11–13.

Omar Sharifuddin Syed-Ikhsan, S., & Rowland, F. (2004). Knowledge management in a public organization: a study on the relationship between organizational elements and the performance of knowledge transfer. *Journal of Knowledge Management*, *8*(2), 95–111. http://doi.org/10.1108/13673270410529145

Pandey, K. N. (2016). Introduction. In *Studies in Systems, Decision and Control* (Vol. 60, pp. 1–70). New Delhi: Springer India. http://doi.org/10.1007/978-81-322-2785-

4_1

Paterson, P. (2013). Knowledge and innovation – how do they relate? Retrieved January 1, 2015, from https://innovation.govspace.gov.au/knowledge-and-innovation-how-do-they-relate

Pee, L. G., & Kankanhalli, A. (2008). Understanding the drivers, enablers, and performance of knowledge management in public organizations. In *2nd International Conference on Theory and Practice of Electronic Governance (ICEGOV 2008)* (pp. 439–466). Cairo: ACM. http://doi.org/10.1145/1509096.1509188

Perez Lopez-Portillo, H. (2016, September 9). *Knowledge management and measurement in Public Sector Organizations*. University of Guanajuato. Retrieved from http://arxiv.org/abs/1609.02995

Pérez López-Portillo, H., Romero Hidalgo, J. A., & Mora Martínez, E. O. (2016). Factores Previos para la Gestión del Conocimiento en la Administración Pública Costarricense. In *Administrar lo Público 3* (pp. 102–129). San José, Costa Rica: CICAP, Universidad de Costa Rica.

Peter, J. P. (1979). Reliability: A Review of Psychometric Basics and Recent Marketing Practices. *Journal of Marketing Research*, *16*(1), 6. http://doi.org/10.2307/3150868

Petty, R., & Guthrie, J. (2000). Intellectual capital literature review. *Journal of Intellectual Capital*, *1*(2), 155–176. http://doi.org/10.1108/14691930010348731

Phusavat, K., Anussornnitisarn, P., Helo, P., & Dwight, R. (2009). Performance measurement: roles and challenges. *Industrial Management & Data Systems*, *109*(5), 646–664. http://doi.org/10.1108/02635570910957632

Ping, Z. (2008). A Strategy for Knowledge Management in E-Government. In *2008 International Seminar on Business and Information Management* (pp. 222–225). IEEE. http://doi.org/10.1109/ISBIM.2008.30

Piraquive, F. N. D., García, V. H. M., & Crespo, R. G. (2015). Knowledge Management Model for Project Management. In L. Uden, M. Heričko, & I.-H. Ting (Eds.), *Knowledge Management in Organizations* (pp. 235–247). Springer International Publishing. http://doi.org/10.1007/978-3-319-21009-4_18

Pisano, G. P. (2015). You Need An Innovation Strategy. *Harvard Business Review*, *93*(6), 44–54. Retrieved from http://search.ebscohost.com/login.aspx?direct=true&db=buh&AN=102786227&site=eds-live

Plessis, M., & du Plessis, M. (2007). The role of knowledge management in innovation. *Journal of Knowledge Management*, *11*(4), 20–29. http://doi.org/10.1108/13673270710762684

Polyani, M. (1958). *Personal Knowledge*. London: Taylor & Francis Group.

Popadiuk, S., & Choo, C. W. (2006). Innovation and knowledge creation: How are these concepts related? *International Journal of Information Management*, *26*(4), 302–312. http://doi.org/10.1016/j.ijinfomgt.2006.03.011

Porter, M. E. (1990). The Competitive Advantage of Nations. *Harvard Business Review*, *68*, 73–93. http://doi.org/Article

Porter, M. E. (2008). The five forces that shape competitive strategy. *Harvard Buiness Review*, (January), 78–94.

Powell, W. W., & Snellman, K. (2004). The Knowledge Economy. *Annual Review of Sociology*, *30*(1), 199–220. http://doi.org/10.1146/annurev.soc.29.010202.100037

Prahalad, C. K., & Mashelkar, R. A. (2010). Innovation's Holy Grail. *Harvard Business Review*, (August), 132–142.

Project Management Institute. (2013). *A guide to the project management body of knowledge (PMBOK ® guide). Project Management Institute*. http://doi.org/10.1002/pmj.20125

Puron-Cid, G. (2014). Factors for a successful adoption of budgetary transparency innovations: A questionnaire report of an open government initiative in Mexico. *Government Information Quarterly*, *31*(SUPPL.1), S49–S62. http://doi.org/10.1016/j.giq.2014.01.007

Quast, L. (2012). Why Knowledge Management Is Important To The Success Of Your Company. Retrieved January 1, 2015, from http://www.forbes.com/sites/lisaquast/2012/08/20/why-knowledge-management-is-important-to-the-success-of-your-company/#54e0b05c5e1d

Quintane, E., Mitch Casselman, R., Sebastian Reiche, B., & Nylund, P. A. (2011). Innovation as a knowled-based outcome. *Journal of Knowledge Management*, *15*(6), 928–947. http://doi.org/10.1108/13673271111179299

Rachuri, S., Subrahmanian, E., Bouras, A., Fenves, S. J., Foufou, S., & Sriram, R. D. (2008). Information sharing and exchange in the context of product lifecycle management: Role of standards. *Computer-Aided Design*, *40*(7), 789–800. http://doi.org/10.1016/j.cad.2007.06.012

Ragab, M. a. F., & Arisha, A. (2013). Knowledge management and measurement: a critical review. *Journal of Knowledge Management*, *17*(6), 873–901. http://doi.org/10.1108/JKM-12-2012-0381

Razmerita, L., Phillips-Wren, G., & Jain, L. C. (Eds.). (2016). *Innovations in Knowledge Management* (Vol. 95). Berlin, Heidelberg: Springer Berlin Heidelberg. http://doi.org/10.1007/978-3-662-47827-1

Reinhardt, W., Schmidt, B., Sloep, P., & Drachsler, H. (2011). Knowledge Worker Roles and Actions — Results of Two Empirical Studies. *Knowledge and Process Management*, *18*(3), 150–174. http://doi.org/10.1002/kpm

Ringle, C. M., Wende, S., & Becker, J.-M. (2005). SmartPLS. Hamburg.

Ringle, C. M., Wende, S., & Becker, J.-M. (2015). SmartPLS 3. Boenningstedt, Germany: SmartPLS GmbH.

Rodríguez, R. (2013). Una propuesta de iniciativas estratégicas basadas en Gestión de Conocimiento para apoyar la innovación en las organizaciones, 14.

Romero Hidalgo, J. A., Pérez López-Portillo, H., & Rodríguez Carvajal, R. (2016). In press. Hacia un estándar de gestión del conocimiento e innovación. In *Memorias Congreso de Investigación Científica Multidisciplinaria (ISSN 2395-9711)*. Chihuahua: Instituto Tecnológico y de Estudios Superiores de Monterrey.

Rooney, D., Hearn, G., & Ninan, A. (2005). *Handbook on the Knowledge Economy*. (D. Rooney, G. Hearn, & A. Ninan, Eds.). Cheltenham, UK: Edward Elgar Publishing.

Ruiz Olabuénaga, J. I. (2012). *Metodología de la investigación cualitativa* (5th ed.). Bilbao: Universidad de Deusto.

Salleh, K., Chong, S. C., Syed Ahmad, S. N., & Syed Ikhsan, S. O. S. (2012). Learning and knowledge transfer performance among public sector accountants: an empirical survey. *Knowledge Management Research & Practice*, *10*(2), 164–174. http://doi.org/10.1057/kmrp.2011.46

Sánchez, L. E., & Morrison-Saunders, A. (2011). Learning about knowledge management for improving environmental impact assessment in a government agency: the Western Australian experience. *Journal of Environmental Management*, *92*(9), 2260–2271. http://doi.org/10.1016/j.jenvman.2011.04.010

Sarvary, M. (1999). Knowledge Management and Competition in the Consulting Industry. *California Management Review*, *41*(2), 95–107.

Sauser, B., Verma, D., Ramirez-Marquez, J., & Gove, R. (2006). From TRL to SRL: The concept of systems readiness levels. *Conference on Systems Engineering Research, Los Angeles, CA*, 1–10. Retrieved from http://www.boardmansauser.com/downloads/2005SauserRamirezVermaGoveCSER.pdf

Savvas, I., & Bassiliades, N. (2009). A process-oriented ontology-based knowledge management system for facilitating operational procedures in public administration. *Expert Systems with Applications*, *36*(3 PART 1), 4467–4478. http://doi.org/10.1016/j.eswa.2008.05.022

Schumpeter, J. A. (1954). *History of Economic Analysis*. *The Economic History Review* (Vol. 8). http://doi.org/10.2307/2591782

Schwab, K., & World Economic Forum. (2016). The Fourth Industrial Revolution: what it means, how to respond. Retrieved January 14, 2016, from https://www.weforum.org/agenda/2016/01/the-fourth-industrial-revolution-what-it-means-and-how-to-respond/

Scopus®. (2016). Analyze search results. Retrieved November 1, 2016, from https://www-scopus-com

Sensuse, D. I., Cahyaningsih, E., & Wibowo, W. C. (2015). Identifying Knowledge Management Process of Indonesian Government Human Capital Management Using Analytical Hierarchy Process and Pearson Correlation Analysis. *Procedia Computer Science*, *72*(81), 233–243. http://doi.org/10.1016/j.procs.2015.12.136

Shafique, F. (2015). Where is information society, it is lost in the knowledge society: Survival issues for developing countries in the knowledge society and need of knowledge management initiatives in the education sector. In *2015 Science and Information Conference (SAI)* (pp. 305–310). London, UK: IEEE. http://doi.org/10.1109/SAI.2015.7237160

Shareef, M. A., Kumar, V., Kumar, U., & Dwivedi, Y. K. (2011). E-Government Adoption Model (GAM): Differing service maturity levels. *Government Information Quarterly*, *28*(1), 17–35. http://doi.org/10.1016/j.giq.2010.05.006

Shepard, L. a. (1993a). Chapter 9: Evaluating Test Validity. *Review of Research in Education*, *19*(1), 405–450. http://doi.org/10.3102/0091732X019001405

Shepard, L. a. (1993b). Chapter 9: Evaluating Test Validity. *Review of Research in Education*, *19*(1), 405–450. http://doi.org/10.3102/0091732X019001405

Shin, D.-H., Kim, H., & Hwang, J. (2015). Standardization revisited: A critical literature review on standards and innovation. *Computer Standards & Interfaces*, *38*, 152–157. http://doi.org/10.1016/j.csi.2014.09.002

Siekmann, J. (2006). *Advances in Knowledge Acquisition and Management*. (A. Hoffmann, B. Kang, D. Richards, & S. Tsumoto, Eds.) (Vol. 4303). Berlin, Heidelberg: Springer Berlin Heidelberg. http://doi.org/10.1007/11961239

Society, S. S. E. S. C. of the I. C. (2007). *IEEE Standard Adoption of ISO/IEC 15939:2007—Systems and Software Engineering—Measurement Process*. http://doi.org/10.1109/IEEESTD.2009.4775910

Software Engineering Institute. (2010). CMMI for Development, Version 1.3. *Carnegie Mellon University*, (November), 482. http://doi.org/CMU/SEI-2010-TR-033 ESC-TR-2010-033

Sokhanvar, S., Matthews, J., & Yarlagadda, P. (2014). Importance of Knowledge Management processes in a project-based organization: A case study of research enterprise. *Procedia Engineering*, *97*, 1825–1830. http://doi.org/10.1016/j.proeng.2014.12.336

Soto-Acosta, P., & Cegarra-Navarro, J.-G. (2016). New ICTs for Knowledge Management in Organizations. *Journal of Knowledge Management*, *20*(3), 1–10. http://doi.org/10.1108/JKM-02-2016-0057

Spender, J.-C., & Grant, R. M. (1996). Knowledge and the firm: Overview. *Strategic Management Journal*, *17*(S2), 5–9. http://doi.org/10.1002/smj.4250171103

Spender, J.-C. J. (1996). Making knowledge the basis of a dynamic theory of the firm. *Strategic Management Journal*, *17*(S2), 45–62. http://doi.org/10.1002/smj.4250171106

Standard, I. (2009). INTERNATIONAL STANDARD ISO / IEC / IEEE Systems and software engineering — Life, *2009*.

Standards Australia. (2005). AS 5037 — 2005 Knowledge management - a guide. Retrieved from https://infostore.saiglobal.com/store/PreviewDoc.aspx?saleItemID=391611

Stehr, N., Adolf, M., & Mast, J. L. (2013). Knowledge Society, Knowledge-Based Economy, and Innovation. In E. G. Carayannis (Ed.), *Encyclopedia of Creativity, Invention, Innovation and Entrepreneurship* (pp. 1186–1191). New York, NY: Springer New York. http://doi.org/10.1007/978-1-4614-3858-8_440

Stošić, B., & Išljamović, S. (2010). TOWARDS INTEGRATIVE INNOVATION MODELS. In *Proceedings of the Multi-Conference on Innovative Developments in ICT* (pp. 219–222). SciTePress - Science and and Technology Publications. http://doi.org/10.5220/0003037902190222

Straub, J. (2015). In search of technology readiness level (TRL) 10. *Aerospace Science and Technology*, *46*, 312–320. http://doi.org/10.1016/j.ast.2015.07.007

Striukova, L., & Rayna, T. (2015). University-industry knowledge exchange. *European Journal of Innovation Management*, *18*(4), 471–492. http://doi.org/10.1108/EJIM-10-2013-0098

Strutt, J. E., Sharp, J. V., Terry, E., & Miles, R. (2006). A Capability maturity models for offshore organisational management. *Environment International*, *32*(8), 1094–1105. http://doi.org/10.1016/j.envint.2006.06.016

Sun, H. (2012). *Management of Technological Innovation in Developing and Developed Countries*. (H. Sun, Ed.). Rijeka: InTech. http://doi.org/10.5772/2313

Swan, J., Newell, S., Scarbrough, H., & Hislop, D. (1999). Knowledge management and innovation: networks and networking. *Journal of Knowledge Management*, *3*(4), 262–275. http://doi.org/10.1108/13673279910304014

Tabrizi, R. S., Ebrahimi, N., & Delpisheh, M. (2011). KM criteria and success of KM programs: an assessment on criteria from importance and effectiveness perspectives. *Procedia Computer Science*, *3*, 691–697.

http://doi.org/10.1016/j.procs.2010.12.115

Tan, J., Fischer, E., Mitchell, R., & Phan, P. (2009). At the Center of the Action : Innovation and Technology Strategy Research in t ..., *47*(3), 233–262.

Taylor, D., & Procter, M. (2008). *The literature review: a few tips on conducting it*. *University of Toronto*. Toronto.

Teece, D. J., Pisano, G., & Shuen, A. (1997). Dynamic capabilities and strategic management. *Strategic Management Journal*, *18*(7), 509–533. http://doi.org/10.1002/(SICI)1097-0266(199708)18:7<509::AID-SMJ882>3.0.CO;2-Z

Thomas, B., Miller, C., & Murphy, L. (2011). Innovation and Small Business. Volume 1. Ventus Publishing ApS. Retrieved from http://bookboon.com/en/innovation-and-small-business-volume-1-ebook

Thomas, N., & Vohra, N. (2015). Development of Network Measures for Knowledge Processes: A Relational Framework. *Knowledge and Process Management*, *22*(2), 126–139. http://doi.org/10.1002/kpm.1465

Tianyong Zhang. (2010). Application of knowledge management in public administration. In *2010 International Conference on Educational and Network Technology* (pp. 572–575). IEEE. http://doi.org/10.1109/ICENT.2010.5532103

Tidd, J. (2006). *A review of innovation models* (No. 1). *Imperial College London, Tanaka Business School*. London. Retrieved from www.emotools.com/static/upload/files/innovation_models.pdf

Tocan, M. C. (2012). Knowledge Based Strategies for Knowledge Based Organizations. *Journal of Knowledge Management, Economics and Information Technology*, *2*(6), 167–177. Retrieved from http://www.scientificpapers.org/wp-content/files/1324_Madalina_TOCAN_Knowledge_based_strategies_for_knowledge_based_organizations.pdf

Tyrrell, S. (2009). SPSS: Stats Practically Short and Simple. Ventus Publishing ApS. Retrieved from http://bookboon.com/en/stats-practically-short-and-simple-ebook

U.S. PATENT AND TRADEMARK OFFICE. (2016). U.S. Patent Statistics Chart 1963-2015. Retrieved May 1, 2016, from http://www.uspto.gov/web/offices/ac/ido/oeip/taf/us_stat.htm

Un Jan, A., & Contreras, V. (2016). Success model for knowledge management systems used by doctoral researchers. *Computers in Human Behavior*, *59*, 258–264. http://doi.org/10.1016/j.chb.2016.02.011

UNESCO, & Bindé, J. (2005). *Hacia las sociedades del conocimiento*. París: UNESCO. Retrieved from http://www.unesco.org/new/es/communication-and-information/resources/publications-and-communication-materials/publications/full-

list/towards-knowledge-societies-unesco-world-report/

United Nations. (2014). *United Nations E-Government Survey 2014. "E-Government for the Future We Want."* New York. Retrieved from http://unpan3.un.org/egovkb/Portals/egovkb/Documents/un/2014-Survey/E-Gov_Complete_Survey-2014.pdf

United Nations. (2015). *World economic situation and prospects*. New York.

van der Heiden, P., Pohl, C., Mansor, S., & van Genderen, J. (2016). Necessitated absorptive capacity and metaroutines in International Technology Transfer: A new model. *Journal of Engineering and Technology Management*, *41*, 65–78. http://doi.org/10.1016/j.jengtecman.2016.07.001

Van der Panne, G., Van Beers, C., & Kleinknecht, A. (2003). Success and failure of innovation: a literature review. *International Journal of Innovation Management*, *7*(3), 309–338.

Vasconcellos, E., Bruno, M. A. C., Campanario, M. de A., & Noffs, S. L. (2009). A new graphic format to facilitate the understanding of technological innovation models: the seesaw of competitiveness. *Technology Analysis & Strategic Management*, *21*(5), 565–582. http://doi.org/10.1080/09537320902969067

Venkatraman, & Grant, J. (1986). Construct measurement in organizational strategy research: a critique and proposal. *Academy of Management Review*, *11*(1), 71–87.

Vrakking, W. J. (1990). The innovative organization. *Long Range Planning*, *23*(2), 94–102. http://doi.org/10.1016/0024-6301(90)90204-H

VV.AA. (2005). *Knowledge management: Organizational and technological dimensions*. (J. Davis, E. Subrahmanian, & A. Westerberg, Eds.). New York: Physica Verlag Heidelberg.

VV AA. (2009). Best Practices for the Knowledge Society. Knowledge, Learning, Development and Technology for All. In M. D. Lytras, P. Ordóñez de Pablos, E. Damiani, D. Avison, A. Naeve, & D. G. Horner (Eds.), *SecondWorld Summit on the Knowledge Society, WSKS 2009*. Chania, Crete: Springer.

VV AA. (2015). *Working and Learning in Times of Uncertainty: Challenges to Adult, Professional and Vocational Education*. (S. Bohlinger, U. Haake, C. H. Jørgensen, H. Toiviainen, & A. Wallo, Eds.). Rotterdam: Sense Publishers. Retrieved from https://www.sensepublishers.com/media/2466-working-and-learning-in-times-of-uncertainty.pdf#page=112

Weill, P., & Woerner, S. L. (2015). Thriving in an Increasingly Digital Ecosystem. *MIT Sloan Management Review*, *56*(4), 27–34. Retrieved from http://mitsmr.com/1BkdvAq

Wiig, K. M. (1997a). Integrating intellectual capital and knowledge management. *Long*

*Range Planning*, *30*(3), 399–405. http://doi.org/10.1016/S0024-6301(97)90256-9

Wiig, K. M. (1997b). Knowledge Management: An Introduction and Perspective. *Journal of Knowledge Management*, *1*(1), 6–14. http://doi.org/10.1108/13673279710800682

Wiig, K. M. (2002). Knowledge management in public administration. *Journal of Knowledge Management*, *6*(3), 224–239. http://doi.org/10.1108/13673270210434331

Wilcox King, A., & Zeithaml, C. P. (2003). Measuring organizational knowledge: a conceptual and methodological framework. *Strategic Management Journal*, *24*(8), 763–772. http://doi.org/10.1002/smj.333

Williamson, P. J., & Yin, E. (2014). Accelerated Innovation: The New Challenge From China. *MIT Sloan Management Review*, *55*(4), 27–34. Retrieved from http://search.proquest.com.library.capella.edu/docview/1543709914?accountid=27965%5Cnhttp://wv9lq5ld3p.search.serialssolutions.com.library.capella.edu/?ctx_ver=Z39.88-2004&ctx_enc=info:ofi/enc:UTF-8&rfr_id=info:sid/ProQ:abiglobal&rft_val_fmt=info:ofi/fmt:k

Winter, R. (2013). *Organizational Change and Information Systems*. (P. Spagnoletti, Ed.) (Vol. 2). Berlin, Heidelberg: Springer Berlin Heidelberg. http://doi.org/10.1007/978-3-642-37228-5

Wong, K. Y. (2005). Critical success factors for implementing knowledge management in small and medium enterprises. *Industrial Management & Data Systems*, *105*(3), 261–279. http://doi.org/10.1108/02635570510590101

Wong, K. Y., & Aspinwall, E. (2005). An empirical study of the important factors for knowledge-management adoption in the SME sector. *Journal of Knowledge Management*, *9*(3), 64–82. http://doi.org/10.1108/13673270510602773

Woolf, K. (2010). Information Matters: Government's Strategy to Build Capability in Managing Its Knowledge and Information Assets. *Legal Information Management*, *10*(1), 47. http://doi.org/10.1017/S1472669610000162

World Economic Forum. (2016a). *Digital Media and Society Implications in a Hyperconnected Era*. Geneva.

World Economic Forum. (2016b). *The Future of Jobs. Employment, Skills and Workforce Strategy for the Fourth Industrial Revolution*. Geneva. Retrieved from http://www3.weforum.org/docs/WEF_Future_of_Jobs.pdf

Xia, E., Zhang, M., Zhu, H., & Jia, S. (2012). Research on the performance of innovation subjects in different innovation models. In *2012 International Conference on Management Science & Engineering 19th Annual Conference Proceedings* (pp. 1591–1596). IEEE. http://doi.org/10.1109/ICMSE.2012.6414385

Yamaguchi, T. (2014). Standardizing HR Practices Around the World. *Harvard Business Review*, *92*(9), 80–81. Retrieved from http://www.redi-bw.de/db/ebsco.php/search.ebscohost.com/login.aspx%3Fdirect%3Dtrue%26db%3Dbth%26AN%3D97509153%26site%3Dehost-live

Yang, T. M., & Maxwell, T. a. (2011). Information-sharing in public organizations: A literature review of interpersonal, intra-organizational and inter-organizational success factors. *Government Information Quarterly*, *28*(2), 164–175. http://doi.org/10.1016/j.giq.2010.06.008

Yoo, Y., Lyytinen, K., & Yang, H. (2005). The role of standards in innovation and diffusion of broadband mobile services: The case of South Korea. *The Journal of Strategic Information Systems*, *14*(3), 323–353. http://doi.org/10.1016/j.jsis.2005.07.007

Young, R., & Organization, A. P. (2010). *Knowledge management tools and techniques manual*. London.

Yousuf Al-Aama, A. (2014). Technology knowledge management (TKM) taxonomy. *VINE*, *44*(1), 2–21. http://doi.org/10.1108/VINE-12-2012-0052

Yuen, Y. (2007). Overview of Knowledge Management in the Public Sector. *7th Global Forum on Reinventing Goverment, Building Trust in Goverment*, 1–16. Retrieved from http://unpan1.un.org/intradoc/groups/public/documents/UNPAN/UNPAN026297.pdf

Zahedi, M., Shahin, M., & Ali Babar, M. (2016). A systematic review of knowledge sharing challenges and practices in global software development. *International Journal of Information Management*, *36*(6), 995–1019. http://doi.org/10.1016/j.ijinfomgt.2016.06.007

Zapata Cantu, L. E., & Mondragon, C. E. (2016). Knowledge management in Mexican NPOs: a comparative study in organizations with a local and national presence. *Journal of Knowledge Management*, *20*(1), 69–87. http://doi.org/10.1108/JKM-12-2014-0494

Zhang, D. (2012). Vocabulary and Grammar Knowledge in Second Language Reading Comprehension: A Structural Equation Modeling Study. *The Modern Language Journal*, *96*(4), 558–575. http://doi.org/10.1111/j.1540

Zheng, H., Chanaron, J., You, J., & Chen, X. (2009). Designing a key performance indicator system for technological innovation audit at firm's level: A framework and an empirical study. In *2009 IEEE International Conference on Industrial Engineering and Engineering Management* (pp. 1–5). IEEE. http://doi.org/10.1109/IEEM.2009.5373498

Zorrilla, S. (2009). *Introducción a la metodología de la investigación*. México, DF:

Ediciones Cal y Arena.

Zyngier, S., Burstein, F., & McKay, J. (2006). Governance of Strategies to Manage Organizational Knowledge. In *Case Studies in Knowledge Management* (Vol. 4, pp. 83–103). IGI Global. http://doi.org/10.4018/978-1-59140-351-7.ch006

## Apéndice 1. Instrumento de evaluación del Estándar Mexicano de Gestión del Conocimiento e Innovación Tecnológica

| Ítem | Factor | Componente | Redacción del ítem |
|---|---|---|---|
| Q10.1 | Organización | Metodologías | Ofrecemos metodologías de … |
| Q10.10 | Organización | Procesos y procedimientos | Realizo mis actividades de … |
| Q10.11 | Organización | Documentación | Utilizo formatos que … |
| Q10.12 | Organización | Documentación | Se encuentran documentados … |
| Q10.13 | Organización | Procesos y procedimientos | Existen procedimientos establecidos … |
| Q10.14 | Organización | Procesos y procedimientos | Los procedimientos son … |
| Q10.2 | Organización | Procesos y procedimientos | Tiene formalizados sus … |
| Q10.3 | Organización | Metodologías | Utilizamos metodologías … |
| Q10.4 | Organización | Metodologías | A partir de experiencias … |
| Q10.5 | Organización | Metodologías | Las actividades, los procesos … |
| Q10.6 | Organización | Documentación | El conocimiento está … |
| Q10.7 | Organización | Procesos y procedimientos | Existen criterios comunes… |
| Q0. | Organización | Metodologías | Resuelvo problemas con … |
| Q10.9 | Organización | Documentación | Conozco el manual de … |
| Q11.1 | Infraestructura | Sistemas | Desarrollamos sistemas … |
| Q11.10 | Infraestructura | Sistemas | Los sistemas con … |
| Q11.11 | Infraestructura | Reutilización de información | Los sistemas proveen … |
| Q11.12 | Infraestructura | Sistemas | Cuenta con sistemas … |
| Q11.13 | Infraestructura | Infraestructura | La red permite … |
| Q11.14 | Infraestructura | Infraestructura | El hardware y los … |
| Q11.15 | Infraestructura | Infraestructura | La red tiene un … |
| Q11.16 | Infraestructura | Infraestructura | Constantemente se renueva … |
| Q11.2 | Infraestructura | Sistemas | Compartimos nuestros sistemas … |
| Q11.3 | Infraestructura | Reutilización de información | Existen mecanismos para … |
| Q11.4 | Infraestructura | Reutilización de información | Utilizamos datos, información … |
| Q11.5 | Infraestructura | Sistemas | Cuento con sistemas … |
| Q11.6 | Infraestructura | Sistemas | Los sistemas informáticos … |
| Q11.7 | Infraestructura | Sistemas | Los sistemas están integrados … |
| Q11.8 | Infraestructura | Sistemas | Los sistemas son compatibles … |
| Q11.9 | Infraestructura | Sistemas | En esta organización … |
| Q12.1 | Humano | Cultura | Promovemos que se … |
| Q12.10 | Humano | Formación | Se incentiva el uso y … |
| Q12.11 | Humano | Colaboración | Transmito mis conocimientos … |
| Q12.12 | Humano | Colaboración | Resuelvo problemas con … |
| Q12.13 | Humano | Formación | Tengo las competencias … |
| Q12.14 | Humano | Formación | Las personas, con las … |
| Q12.15 | Humano | Cultura | Las personas tienen … |
| Q12.2 | Humano | Colaboración | El trabajo en equipo nos … |
| Q12.3 | Humano | Formación | Se apoya a los empleados … |
| Q12.4 | Humano | Formación | Las personas tienen la … |
| Q12.5 | Humano | Colaboración | Realizamos reuniones para … |
| Q12.6 | Humano | Formación | Conocemos nuestras brechas … |
| Q12.7 | Humano | Cultura | El conocimiento producido por … |
| Q12.8 | Humano | Formación | Sus miembros tienen … |
| Q12.9 | Humano | Formación | Se apoya el aprendizaje … |
| Q13.1 | Estrategia | Estrategia de innovación | Se implementan cambios … |
| Q13.10 | Estrategia | Estrategia de innovación | Realizamos transferencia … |
| Q13.11 | Estrategia | Estrategia de gestión del conocimiento | Existen políticas para … |
| Q13.12 | Estrategia | Estrategia de gestión del conocimiento | Se han obtenido … |
| Q13.13 | Estrategia | Estrategia de innovación | Obtenemos ingresos … |
| Q13.14 | Estrategia | Estrategia de gestión del conocimiento | Contamos con un … |
| Q13.15 | Estrategia | Estrategia de gestión del conocimiento | Se ha implementado un … |
| Q13.16 | Estrategia | Estrategia de gestión del conocimiento | Contamos con indicadores de … |
| Q13.17 | Estrategia | Estrategia de innovación | Se diseñan … |
| Q13.18 | Estrategia | Estrategia de innovación | Otras organizaciones solicitan … |
| Q13.19 | Estrategia | Estrategia de gestión del conocimiento | Considero que trabajo en … |
| Q13.2 | Estrategia | Estrategia de gestión del conocimiento | Las políticas y regulaciones … |
| Q13.20 | Estrategia | Estrategia de innovación | Ofrecemos productos o … |
| Q13.21 | Estrategia | Estrategia de gestión del conocimiento | Replicamos las … |
| Q13.22 | Estrategia | Estrategia de gestión del conocimiento | Existe un … |
| Q13.23 | Estrategia | Estrategia de innovación | Implementamos … |
| Q13.24 | Estrategia | Estrategia de gestión del conocimiento | Continuamente se … |
| Q13.25 | Estrategia | Estrategia de gestión del conocimiento | En la organización … |
| Q13.26 | Estrategia | Estrategia de gestión del conocimiento | La organización se … |
| Q13.27 | Estrategia | Estrategia de innovación | La organización genera … |
| Q13.28 | Estrategia | Estrategia de gestión del conocimiento | Existe una estrategia … |
| Q13.29 | Estrategia | Estrategia de gestión del conocimiento | Tengo acceso a … |
| Q13.3 | Estrategia | Estrategia de innovación | Existe apoyo … |
| Q13.30 | Estrategia | Estrategia de gestión del conocimiento | Los sistemas de gestión … |
| Q13.31 | Estrategia | Estrategia de gestión del conocimiento | Transferimos a … |
| Q13.32 | Estrategia | Estrategia de gestión del conocimiento | Se usa el conocimiento … |
| Q13.33 | Estrategia | Estrategia de gestión del conocimiento | El conocimiento y la … |
| Q13.34 | Estrategia | Estrategia de gestión del conocimiento | Los miembros de la … |
| Q13.35 | Estrategia | Estrategia de innovación | Existe un firme … |
| Q13.36 | Estrategia | Estrategia de gestión del conocimiento | Los objetivos de la … |
| Q13.37 | Estrategia | Estrategia de gestión del conocimiento | Existe un plan … |
| Q13.38 | Estrategia | Estrategia de gestión del conocimiento | En esta organización … |
| Q13.4 | Estrategia | Estrategia de gestión del conocimiento | Se usa el … |
| Q13.5 | Estrategia | Estrategia de gestión del conocimiento | Se documentan las … |
| Q13.6 | Estrategia | Estrategia de innovación | Ponemos a disponibilidad … |
| Q13.7 | Estrategia | Estrategia de innovación | Los sistemas de evaluación … |
| Q13.8 | Estrategia | Estrategia de innovación | Los líderes de la organización … |
| Q13.9 | Estrategia | Estrategia de innovación | Los líderes de la … |

Fuente: **Elaboración propia** (2016)

# Apéndice 2. Matriz de componente rotado

| Matriz de componente rotado[a] | | | | | | |
|---|---|---|---|---|---|---|
| | Componente | | | | | |
| | **1** | **2** | **3** | **4** | **5** | **6** |
| Q10.1 | | .617 | | | | |
| Q10.2 | | .600 | | | | .444 |
| Q10.3 | | .647 | | | | |
| Q10.4 | | .594 | | | | |
| Q10.5 | .437 | .737 | | | | |
| Q10.6 | | .708 | | | | |
| Q10.7 | | .661 | | | | |
| Q10.8 | | .708 | | | | |
| Q10.9 | | .717 | | | | |
| Q10.10 | | .629 | | | | |
| Q10.11 | | | | .626 | | |
| Q10.12 | | .717 | | | | |
| Q10.13 | | .730 | | | | |
| Q10.14 | | .685 | | | | |
| Q11.1 | | .452 | .490 | | | |
| Q11.2 | .454 | | .435 | | | .471 |
| Q11.3 | | .461 | .553 | | | |
| Q11.4 | | .444 | .545 | | | |
| Q11.5 | .400 | | .607 | | | |
| Q11.6 | .417 | | .659 | | | |
| Q11.7 | | | .643 | | | |
| Q11.8 | | | .616 | | | |
| Q11.9 | | .439 | .650 | | | |
| Q11.10 | | .466 | .605 | | | |
| Q11.11 | | | .669 | | | |
| Q11.12 | .421 | .442 | .628 | | | |
| Q11.13 | | | .699 | | | |
| Q11.14 | | | .734 | | | |
| Q11.15 | | | .719 | | | |
| Q11.16 | | | .671 | | | |
| Q12.1 | .484 | | | .406 | .485 | |
| Q12.2 | .483 | | | .571 | | |
| Q12.3 | .439 | | | | .620 | |
| Q12.4 | .505 | | .409 | .517 | | |
| Q12.5 | .472 | | | | .535 | |
| Q12.6 | .500 | | | | .481 | |
| Q12.7 | .524 | | | | .586 | |
| Q12.8 | .493 | | | .526 | | |
| Q12.9 | .458 | | | | .607 | |
| Q12.10 | .467 | | | | .627 | |
| Q12.11 | .455 | | | .539 | | |
| Q12.12 | .481 | | | .663 | | |
| Q12.13 | .484 | | | .607 | | |
| Q12.14 | .531 | | | .529 | | |
| Q12.15 | .440 | | | .630 | | |
| Q13.1 | .731 | | | | | |
| Q13.2 | .705 | | | | | |
| Q13.3 | .713 | | | | | |
| Q13.4 | .588 | | | | | |
| Q13.5 | .678 | | | | | |
| Q13.6 | .732 | | | | | |
| Q13.7 | .680 | | | | | |
| Q13.8 | .711 | | | | | |
| Q13.9 | .663 | | | .433 | | |
| Q13.10 | .716 | | | | | |
| Q13.11 | .694 | | | | | |
| Q13.12 | .740 | | | | | |
| Q13.13 | .665 | | | | | |
| Q13.14 | .724 | | | | | |
| Q13.15 | .726 | | | | | |
| Q13.16 | .703 | | | | | |
| Q13.17 | .696 | | | .423 | | |
| Q13.18 | .715 | | | | | |
| Q13.19 | .731 | | | | | |
| Q13.20 | .697 | | | | | |
| Q13.21 | .746 | | | | | |
| Q13.22 | .718 | | | | | |
| Q13.23 | .751 | | | | | |
| Q13.24 | .699 | | | | | |
| Q13.25 | .710 | | | | | |
| Q13.26 | .710 | | | | | |
| Q13.27 | .736 | | | | | |
| Q13.28 | .732 | | | | | |
| Q13.29 | .735 | | | | | |
| Q13.30 | .690 | | | | | |
| Q13.31 | .701 | | | | | |
| Q13.32 | .687 | | | | | |
| Q13.33 | .728 | | | | | |
| Q13.34 | .731 | | | | | |
| Q13.35 | .780 | | | | | |
| Q13.36 | .701 | | | .419 | | |
| Q13.37 | .752 | | | | | |
| Q13.38 | .716 | | | | | |

[1]Método de extracción: análisis de componentes principales.

[2]Método de rotación: Varimax con normalización Kaiser.

a. La rotación ha convergido en 9 iteraciones, por las recomendaciones en la literatura (Aldas-Mnzano, 2013; Tyrrell, 2009)

**Fuente:** Elaboración propia (2016)

# Apéndice 3. Estudio de correlaciones entre los ítems del instrumento de evaluación

| | Q10.1 | Q10.2 | Q10.3 | Q10.4 | Q10.5 | Q10.6 | Q10.7 | Q10.8 | Q10.9 | Q10.10 | Q10.11 | Q10.12 | Q10.13 | Q10.14 | Q11.1 | Q11.2 | Q11.3 | Q11.4 | Q11.5 | Q11.6 | Q11.7 | Q11.8 | Q11.9 | Q11.10 | Q11.11 | Q11.12 | Q11.13 | Q11.14 |
|---|---|---|---|---|---|---|---|---|---|---|---|---|---|---|---|---|---|---|---|---|---|---|---|---|---|---|---|---|
| **Q10.1** | 1 | | | | | | | | | | | | | | | | | | | | | | | | | | | |
| **Q10.2** | 0.588 | 1 | | | | | | | | | | | | | | | | | | | | | | | | | | |
| **Q10.3** | 0.496 | 0.439 | 1 | | | | | | | | | | | | | | | | | | | | | | | | | |
| **Q10.4** | 0.557 | 0.562 | 0.634 | 1 | | | | | | | | | | | | | | | | | | | | | | | | |
| **Q10.5** | 0.549 | 0.428 | 0.665 | 0.467 | 1 | | | | | | | | | | | | | | | | | | | | | | | |
| **Q10.6** | 0.601 | 0.504 | 0.579 | 0.499 | 0.714 | 1 | | | | | | | | | | | | | | | | | | | | | | |
| **Q10.7** | 0.526 | 0.42 | 0.595 | 0.504 | 0.659 | 0.589 | 1 | | | | | | | | | | | | | | | | | | | | | |
| **Q10.8** | 0.486 | 0.498 | 0.66 | 0.511 | 0.627 | 0.607 | 0.612 | 1 | | | | | | | | | | | | | | | | | | | | |
| **Q10.9** | 0.409 | 0.39 | 0.471 | 0.362 | 0.625 | 0.557 | 0.39 | 0.52 | 1 | | | | | | | | | | | | | | | | | | | |
| **Q10.10** | 0.519 | 0.497 | 0.651 | 0.544 | 0.684 | 0.623 | 0.585 | 0.594 | 0.514 | 1 | | | | | | | | | | | | | | | | | | |
| **Q10.11** | -0.015 | -0.015 | 0.081 | 0.077 | 0.012 | -0.025 | -0.006 | 0.043 | 0.171 | 0.063 | 1 | | | | | | | | | | | | | | | | | |
| **Q10.12** | 0.505 | 0.399 | 0.59 | 0.44 | 0.736 | 0.686 | 0.6 | 0.532 | 0.544 | 0.569 | -0.085 | 1 | | | | | | | | | | | | | | | | |
| **Q10.13** | 0.563 | 0.581 | 0.566 | 0.581 | 0.593 | 0.677 | 0.612 | 0.653 | 0.524 | 0.57 | -0.029 | 0.596 | 1 | | | | | | | | | | | | | | | |
| **Q10.14** | 0.583 | 0.391 | 0.486 | 0.486 | 0.67 | 0.61 | 0.608 | 0.524 | 0.553 | 0.59 | 0.041 | 0.657 | 0.614 | 1 | | | | | | | | | | | | | | |
| **Q11.1** | 0.512 | 0.374 | 0.462 | 0.463 | 0.475 | 0.432 | 0.468 | 0.333 | 0.388 | 0.364 | 0.084 | 0.446 | 0.547 | 0.57 | 1 | | | | | | | | | | | | | |
| **Q11.2** | 0.65 | 0.566 | 0.408 | 0.51 | 0.413 | 0.399 | 0.349 | 0.322 | 0.343 | 0.404 | 0.06 | 0.277 | 0.432 | 0.389 | 0.575 | 1 | | | | | | | | | | | | |
| **Q11.3** | 0.493 | 0.477 | 0.547 | 0.508 | 0.494 | 0.473 | 0.427 | 0.48 | 0.369 | 0.49 | -0.007 | 0.543 | 0.549 | 0.364 | 0.532 | 0.521 | 1 | | | | | | | | | | | |
| **Q11.4** | 0.461 | 0.462 | 0.523 | 0.52 | 0.483 | 0.416 | 0.487 | 0.351 | 0.342 | 0.43 | 0.111 | 0.515 | 0.493 | 0.462 | 0.568 | 0.455 | 0.663 | 1 | | | | | | | | | | |
| **Q11.5** | 0.35 | 0.268 | 0.511 | 0.541 | 0.435 | 0.488 | 0.459 | 0.454 | 0.423 | 0.502 | 0.153 | 0.406 | 0.452 | 0.437 | 0.45 | 0.41 | 0.518 | 0.523 | 1 | | | | | | | | | |
| **Q11.6** | 0.381 | 0.334 | 0.501 | 0.497 | 0.489 | 0.493 | 0.455 | 0.487 | 0.399 | 0.511 | 0.054 | 0.434 | 0.521 | 0.432 | 0.457 | 0.422 | 0.557 | 0.537 | 0.746 | 1 | | | | | | | | |
| **Q11.7** | 0.435 | 0.421 | 0.457 | 0.399 | 0.375 | 0.427 | 0.371 | 0.411 | 0.243 | 0.453 | -0.005 | 0.36 | 0.47 | 0.33 | 0.439 | 0.524 | 0.519 | 0.529 | 0.598 | 0.543 | 1 | | | | | | | |
| **Q11.8** | 0.468 | 0.455 | 0.38 | 0.437 | 0.356 | 0.405 | 0.39 | 0.342 | 0.257 | 0.419 | -0.028 | 0.306 | 0.462 | 0.326 | 0.513 | 0.552 | 0.522 | 0.472 | 0.477 | 0.522 | 0.589 | 1 | | | | | | |
| **Q11.9** | 0.445 | 0.336 | 0.438 | 0.45 | 0.531 | 0.561 | 0.481 | 0.433 | 0.402 | 0.537 | 0.041 | 0.522 | 0.532 | 0.417 | 0.478 | 0.434 | 0.558 | 0.554 | 0.599 | 0.627 | 0.6 | 0.523 | 1 | | | | | |
| **Q11.10** | 0.47 | 0.333 | 0.587 | 0.55 | 0.568 | 0.551 | 0.526 | 0.489 | 0.466 | 0.574 | 0.07 | 0.515 | 0.564 | 0.486 | 0.444 | 0.423 | 0.537 | 0.557 | 0.699 | 0.759 | 0.593 | 0.479 | 0.697 | 1 | | | | |
| **Q11.11** | 0.425 | 0.337 | 0.442 | 0.473 | 0.477 | 0.497 | 0.477 | 0.403 | 0.351 | 0.496 | 0.027 | 0.429 | 0.455 | 0.464 | 0.483 | 0.452 | 0.513 | 0.542 | 0.689 | 0.781 | 0.546 | 0.609 | 0.656 | 0.673 | 1 | | | |
| **Q11.12** | 0.447 | 0.376 | 0.435 | 0.418 | 0.522 | 0.527 | 0.476 | 0.433 | 0.444 | 0.442 | -0.054 | 0.551 | 0.521 | 0.443 | 0.489 | 0.474 | 0.536 | 0.525 | 0.542 | 0.626 | 0.513 | 0.576 | 0.626 | 0.616 | 0.597 | 1 | | |
| **Q11.13** | 0.272 | 0.174 | 0.34 | 0.307 | 0.367 | 0.371 | 0.312 | 0.364 | 0.268 | 0.405 | 0.121 | 0.339 | 0.381 | 0.316 | 0.396 | 0.298 | 0.548 | 0.447 | 0.571 | 0.605 | 0.531 | 0.4 | 0.584 | 0.522 | 0.589 | 0.454 | 1 | |
| **Q11.14** | 0.263 | 0.236 | 0.234 | 0.293 | 0.336 | 0.296 | 0.372 | 0.289 | 0.191 | 0.395 | 0.026 | 0.286 | 0.252 | 0.273 | 0.361 | 0.324 | 0.454 | 0.414 | 0.571 | 0.601 | 0.505 | 0.471 | 0.542 | 0.451 | 0.532 | 0.463 | 0.602 | 1 |
| **Q11.15** | 0.287 | 0.235 | 0.309 | 0.293 | 0.317 | 0.329 | 0.299 | 0.268 | 0.192 | 0.345 | 0.013 | 0.336 | 0.389 | 0.327 | 0.359 | 0.35 | 0.414 | 0.419 | 0.502 | 0.596 | 0.55 | 0.414 | 0.558 | 0.568 | 0.568 | 0.512 | 0.606 | 0.519 |
| **Q11.16** | 0.42 | 0.314 | 0.353 | 0.411 | 0.445 | 0.473 | 0.425 | 0.388 | 0.24 | 0.436 | 0.018 | 0.357 | 0.39 | 0.452 | 0.422 | 0.44 | 0.389 | 0.446 | 0.542 | 0.578 | 0.554 | 0.561 | 0.551 | 0.521 | 0.573 | 0.571 | 0.503 | 0.659 |
| **Q12.1** | 0.428 | 0.338 | 0.44 | 0.525 | 0.347 | 0.401 | 0.394 | 0.436 | 0.259 | 0.397 | 0.024 | 0.319 | 0.437 | 0.43 | 0.316 | 0.379 | 0.392 | 0.382 | 0.466 | 0.446 | 0.378 | 0.461 | 0.365 | 0.442 | 0.494 | 0.352 | 0.266 | 0.354 |
| **Q12.2** | 0.308 | 0.312 | 0.386 | 0.441 | 0.258 | 0.256 | 0.221 | 0.302 | 0.263 | 0.349 | 0.164 | 0.222 | 0.318 | 0.259 | 0.307 | 0.351 | 0.398 | 0.412 | 0.441 | 0.377 | 0.446 | 0.229 | 0.269 | 0.35 | 0.318 | 0.234 | 0.334 | 0.348 |
| **Q12.3** | 0.414 | 0.299 | 0.389 | 0.521 | 0.475 | 0.393 | 0.372 | 0.367 | 0.367 | 0.487 | 0.211 | 0.312 | 0.393 | 0.536 | 0.373 | 0.273 | 0.352 | 0.441 | 0.428 | 0.437 | 0.382 | 0.342 | 0.39 | 0.448 | 0.476 | 0.265 | 0.332 | 0.378 |
| **Q12.4** | 0.432 | 0.328 | 0.472 | 0.548 | 0.373 | 0.332 | 0.419 | 0.389 | 0.222 | 0.421 | 0.131 | 0.366 | 0.445 | 0.394 | 0.417 | 0.469 | 0.468 | 0.471 | 0.521 | 0.529 | 0.472 | 0.439 | 0.506 | 0.559 | 0.513 | 0.401 | 0.438 | 0.411 |
| **Q12.5** | 0.344 | 0.213 | 0.434 | 0.5 | 0.339 | 0.372 | 0.372 | 0.391 | 0.268 | 0.44 | 0.114 | 0.199 | 0.353 | 0.464 | 0.382 | 0.349 | 0.261 | 0.333 | 0.485 | 0.483 | 0.374 | 0.448 | 0.352 | 0.424 | 0.485 | 0.341 | 0.327 | 0.395 |
| **Q12.6** | 0.365 | 0.333 | 0.418 | 0.471 | 0.321 | 0.386 | 0.298 | 0.429 | 0.258 | 0.319 | 0.064 | 0.267 | 0.37 | 0.391 | 0.264 | 0.34 | 0.329 | 0.334 | 0.448 | 0.475 | 0.369 | 0.386 | 0.321 | 0.461 | 0.434 | 0.283 | 0.291 | 0.267 |
| **Q12.7** | 0.586 | 0.424 | 0.504 | 0.578 | 0.55 | 0.543 | 0.5 | 0.559 | 0.432 | 0.542 | 0.101 | 0.449 | 0.561 | 0.668 | 0.504 | 0.448 | 0.405 | 0.5 | 0.515 | 0.533 | 0.488 | 0.49 | 0.484 | 0.568 | 0.565 | 0.461 | 0.384 | 0.386 |
| **Q12.8** | 0.511 | 0.395 | 0.393 | 0.463 | 0.42 | 0.416 | 0.361 | 0.381 | 0.302 | 0.458 | 0.064 | 0.412 | 0.438 | 0.462 | 0.396 | 0.432 | 0.411 | 0.429 | 0.438 | 0.382 | 0.415 | 0.443 | 0.436 | 0.447 | 0.411 | 0.426 | 0.28 | 0.326 |
| **Q12.9** | 0.473 | 0.299 | 0.455 | 0.58 | 0.459 | 0.432 | 0.398 | 0.46 | 0.344 | 0.421 | 0.145 | 0.356 | 0.409 | 0.565 | 0.344 | 0.261 | 0.338 | 0.399 | 0.477 | 0.446 | 0.406 | 0.296 | 0.322 | 0.428 | 0.38 | 0.289 | 0.261 | 0.348 |
| **Q12.10** | 0.433 | 0.272 | 0.378 | 0.558 | 0.343 | 0.382 | 0.395 | 0.429 | 0.311 | 0.436 | 0.12 | 0.243 | 0.37 | 0.506 | 0.324 | 0.37 | 0.302 | 0.361 | 0.465 | 0.475 | 0.377 | 0.41 | 0.385 | 0.486 | 0.472 | 0.346 | 0.296 | 0.349 |
| **Q12.11** | 0.424 | 0.391 | 0.381 | 0.468 | 0.311 | 0.342 | 0.297 | 0.408 | 0.354 | 0.436 | 0.088 | 0.25 | 0.377 | 0.395 | 0.312 | 0.411 | 0.39 | 0.396 | 0.445 | 0.421 | 0.338 | 0.406 | 0.31 | 0.38 | 0.471 | 0.331 | 0.233 | 0.283 |
| **Q12.12** | 0.146 | 0.162 | 0.229 | 0.167 | 0.114 | 0.179 | 0.2 | 0.215 | 0.166 | 0.226 | 0.065 | 0.216 | 0.201 | 0.155 | 0.206 | 0.132 | 0.362 | 0.362 | 0.352 | 0.379 | 0.315 | 0.181 | 0.297 | 0.272 | 0.307 | 0.296 | 0.297 | 0.282 |
| **Q12.13** | 0.196 | 0.234 | 0.149 | 0.137 | 0.273 | 0.293 | 0.187 | 0.229 | 0.212 | 0.255 | 0.025 | 0.164 | 0.133 | 0.234 | 0.177 | 0.182 | 0.223 | 0.218 | 0.191 | 0.276 | 0.12 | 0.257 | 0.109 | 0.122 | 0.281 | 0.157 | 0.145 | 0.298 |
| **Q12.14** | 0.377 | 0.32 | 0.367 | 0.491 | 0.374 | 0.349 | 0.377 | 0.358 | 0.257 | 0.335 | -0.054 | 0.418 | 0.4 | 0.43 | 0.343 | 0.314 | 0.361 | 0.367 | 0.415 | 0.359 | 0.314 | 0.221 | 0.282 | 0.319 | 0.372 | 0.369 | 0.322 | 0.305 |
| **Q12.15** | 0.185 | 0.137 | 0.269 | 0.21 | 0.144 | 0.134 | 0.204 | 0.013 | 0.08 | 0.22 | 0.092 | 0.263 | 0.157 | 0.195 | 0.232 | 0.27 | 0.317 | 0.293 | 0.313 | 0.262 | 0.256 | 0.341 | 0.277 | 0.286 | 0.307 | 0.295 | 0.243 | 0.28 |
| **Q13.5** | -0.006 | 0.081 | 0.004 | 0.016 | -0.03 | 0.019 | 0.022 | 0 | 0.093 | 0.078 | 0.063 | 0.044 | 0.077 | 0.036 | 0.062 | 0.048 | 0.006 | -0.066 | 0.032 | 0.083 | -0.066 | -0.001 | -0.082 | -0.03 | 0.029 | 0.082 | 0.005 | 0.009 |
| **Q13.6** | 0.052 | 0.107 | 0.099 | 0.046 | 0.105 | 0.121 | -0.018 | 0.077 | 0.174 | 0.124 | 0.04 | 0.072 | 0.1 | 0.175 | 0.14 | 0.12 | 0.066 | 0.08 | 0.136 | 0.139 | 0.092 | 0.052 | -0.017 | 0.065 | 0.115 | 0.152 | 0.129 | 0.013 |
| **Q13.10** | -0.028 | -0.051 | -0.06 | -0.13 | -0.124 | -0.027 | -0.122 | -0.047 | -0.031 | -0.065 | -0.05 | -0.116 | -0.044 | -0.054 | -0.091 | 0.009 | -0.167 | -0.158 | -0.091 | -0.058 | -0.029 | -0.056 | -0.127 | -0.105 | -0.028 | -0.095 | -0.016 | -0.084 |
| **Q13.13** | 0.06 | 0.1 | 0.056 | -0.003 | -0.087 | 0.047 | 0.004 | 0.053 | 0.009 | -0.032 | 0.021 | -0.014 | 0.095 | 0.046 | 0.036 | 0.039 | -0.032 | 0.006 | 0.072 | 0.021 | 0.017 | 0.046 | -0.077 | -0.065 | -0.03 | 0.016 | -0.008 | -0.014 |
| **Q13.14** | -0.03 | -0.006 | 0.027 | 0.004 | -0.066 | -0.02 | 0.024 | 0.007 | -0.026 | 0.007 | -0.01 | -0.004 | -0.002 | -0.028 | -0.048 | 0.016 | -0.065 | 0.021 | 0.02 | 0.025 | 0.027 | -0.021 | -0.035 | 0.028 | 0.012 | 0.022 | 0.023 | -0.039 |
| **Q13.16** | 0.029 | 0.015 | 0.046 | -0.012 | -0.084 | -0.002 | -0.076 | -0.082 | -0.036 | -0.106 | -0.003 | -0.038 | 0.026 | -0.029 | -0.009 | 0.037 | -0.064 | -0.031 | 0.038 | -0.061 | 0.014 | 0.003 | -0.041 | -0.011 | -0.042 | 0.008 | -0.045 | -0.115 |
| **Q13.17** | -0.005 | 0.014 | -0.003 | -0.043 | -0.113 | -0.043 | -0.004 | 0.011 | -0.056 | -0.032 | -0.018 | -0.013 | 0.014 | 0.025 | -0.086 | 0.024 | -0.109 | -0.085 | -0.074 | -0.082 | -0.001 | -0.041 | -0.095 | -0.118 | -0.058 | -0.072 | -0.012 | -0.102 |
| **Q13.18** | 0.037 | 0.086 | 0.156 | 0.109 | -0.007 | 0.086 | 0.013 | 0.075 | -0.018 | 0.071 | 0.012 | 0.08 | 0.086 | 0.1 | 0.107 | 0.033 | 0.023 | 0.06 | 0.044 | 0.03 | 0.036 | 0.027 | -0.029 | 0.037 | 0.024 | 0.037 | -0.041 | -0.02 |
| **Q13.20** | 0.056 | 0.07 | 0.108 | 0.031 | 0.031 | 0.089 | 0.055 | 0.063 | 0.068 | 0.109 | 0.064 | 0.096 | 0.069 | 0.168 | 0.151 | 0.066 | -0.026 | 0.077 | 0.022 | -0.015 | 0.016 | -0.03 | -0.018 | 0.001 | 0.02 | -0.006 | 0.084 | -0.107 |
| **Q13.21** | -0.055 | -0.045 | 0.024 | -0.12 | -0.054 | -0.035 | -0.037 | -0.008 | -0.063 | 0.035 | 0.004 | -0.054 | -0.051 | 0.017 | -0.057 | -0.036 | -0.058 | -0.012 | 0.006 | -0.013 | 0.06 | -0.017 | -0.064 | -0.044 | -0.014 | 0.004 | 0.029 | -0.009 |
| **Q13.23** | 0.019 | 0.079 | 0.078 | -0.002 | -0.009 | 0.018 | 0.043 | 0.078 | 0.057 | 0.091 | 0.003 | -0.022 | -0.007 | 0.096 | -0.048 | 0.031 | -0.068 | 0.003 | 0.055 | 0.083 | 0.014 | 0.048 | -0.134 | -0.015 | 0.078 | -0.02 | -0.061 | -0.029 |
| **Q13.25** | -0.061 | -0.012 | 0.024 | -0.043 | -0.103 | -0.069 | -0.092 | -0.017 | -0.016 | -0.017 | 0.064 | -0.044 | -0.017 | -0.012 | -0.065 | -0.036 | -0.047 | -0.085 | -0.04 | -0.008 | -0.056 | -0.044 | -0.111 | -0.079 | -0.054 | 0.035 | -0.032 | -0.081 |
| **Q13.29** | -0.007 | 0.024 | 0.206 | 0.097 | 0.086 | 0.118 | 0.032 | 0.103 | 0.104 | 0.134 | -0.036 | 0.031 | 0.081 | 0.081 | 0.006 | -0.006 | -0.038 | 0.039 | 0.057 | 0.048 | 0.024 | -0.007 | -0.068 | 0.024 | 0.026 | -0.015 | 0.027 | -0.1 |
| **Q13.30** | 0.085 | 0.152 | 0.061 | 0.092 | 0.067 | 0.18 | 0.038 | 0.037 | 0.146 | 0.073 | -0.005 | 0.151 | 0.178 | 0.2 | 0.242 | 0.126 | 0.1 | 0.152 | 0.131 | 0.163 | 0.049 | 0.067 | 0.03 | 0.121 | 0.109 | 0.169 | 0.132 | -0.033 |
| **Q13.31** | 0.012 | 0.117 | 0.007 | 0.04 | -0.017 | 0.057 | 0.011 | -0.035 | 0.078 | 0.019 | 0.035 | 0.034 | 0.048 | 0.13 | 0.134 | 0.073 | -0.035 | 0.047 | 0.087 | 0.082 | 0.009 | -0.02 | -0.06 | 0.035 | 0.066 | 0.087 | 0.032 | -0.037 |
| **Q13.32** | -0.014 | -0.029 | 0.01 | -0.068 | -0.098 | -0.029 | -0.052 | 0.044 | -0.016 | -0.048 | 0.01 | -0.087 | -0.005 | 0.018 | 0.012 | 0.069 | -0.055 | -0.069 | 0.039 | -0.015 | -0.004 | -0.095 | -0.102 | -0.076 | -0.037 | 0.04 | 0.013 | -0.092 |
| **Q13.33** | -0.009 | 0.035 | 0.105 | 0.008 | -0.008 | 0.04 | 0.021 | -0.017 | 0.066 | 0.065 | -0.004 | -0.02 | 0.037 | 0.027 | 0.048 | -0.002 | -0.063 | 0 | 0.004 | 0.042 | 0.033 | 0.064 | -0.037 | 0.017 | 0.053 | -0.012 | -0.043 | -0.028 |
| **Q13.34** | -0.056 | -0.006 | 0.058 | -0.057 | -0.073 | -0.038 | -0.018 | -0.004 | 0.009 | 0.015 | 0.076 | -0.097 | -0.043 | -0.021 | -0.066 | -0.023 | -0.117 | -0.074 | -0.031 | -0.011 | -0.06 | -0.076 | -0.168 | -0.128 | -0.039 | -0.017 | -0.063 | -0.077 |
| **Q13.35** | -0.07 | -0.027 | 0.109 | -0.019 | -0.058 | -0.008 | -0.055 | 0.022 | 0 | -0.037 | -0.039 | -0.05 | -0.016 | -0.015 | 0.011 | -0.009 | -0.08 | 0.001 | 0.067 | 0.017 | 0.026 | -0.05 | -0.09 | -0.041 | 0.027 | 0.028 | -0.021 | -0.099 |
| **Q13.37** | 0.016 | 0.05 | 0.118 | -0.022 | 0.048 | 0.01 | 0.081 | 0.097 | 0.061 | 0.128 | 0.027 | 0.039 | 0.034 | 0.079 | -0.011 | 0.002 | -0.058 | -0.036 | 0.014 | 0.018 | 0.064 | -0.017 | -0.071 | 0.04 | 0.044 | -0.002 | -0.005 | -0.025 |
| **Q13.38** | -0.088 | -0.067 | -0.057 | -0.089 | -0.159 | -0.102 | -0.136 | -0.097 | -0.133 | -0.111 | -0.044 | -0.09 | -0.083 | -0.115 | -0.06 | -0.023 | -0.076 | -0.048 | -0.044 | -0.076 | -0.027 | -0.088 | -0.123 | -0.126 | -0.051 | -0.103 | -0.082 | -0.118 |
| **Q13.1** | -0.064 | 0.019 | 0.076 | 0.007 | -0.057 | -0.035 | 0.002 | 0.012 | -0.016 | -0.008 | 0.034 | -0.014 | -0.038 | -0.032 | -0.117 | -0.014 | -0.089 | -0.041 | 0.036 | -0.012 | -0.077 | -0.089 | -0.097 | -0.045 | 0.02 | 0 | -0.082 | -0.129 |
| **Q13.2** | -0.082 | 0.054 | -0.011 | -0.052 | -0.096 | -0.024 | -0.073 | 0.07 | -0.021 | -0.065 | 0.035 | -0.109 | 0.041 | 0.003 | -0.072 | -0.009 | -0.037 | -0.072 | 0.025 | 0.088 | 0.021 | -0.034 | -0.111 | -0.017 | 0.056 | -0.021 | 0.014 | -0.015 |
| **Q13.3** | 0.056 | 0.05 | 0.128 | 0.063 | -0.002 | 0.083 | 0.022 | 0.018 | 0.024 | 0.027 | 0 | -0.018 | 0.055 | 0.068 | 0.049 | 0.073 | -0.078 | 0 | 0.012 | -0.005 | 0.006 | 0 | -0.054 | 0.039 | 0.028 | -0.019 | -0.039 | -0.152 |
| **Q13.4** | 0.154 | 0.127 | 0.119 | 0.172 | 0.205 | 0.2 | 0.217 | 0.115 | 0.091 | 0.167 | -0.044 | 0.114 | 0.236 | 0.191 | 0.179 | 0.13 | 0.011 | 0.03 | 0.088 | 0.072 | 0.088 | 0.101 | 0.118 | 0.046 | 0.148 | 0.173 | -0.026 | -0.03 |
| **Q13.7** | -0.044 | -0.073 | 0.008 | -0.036 | -0.077 | -0.023 | 0.001 | -0.02 | -0.081 | -0.01 | 0.087 | -0.102 | -0.127 | -0.033 | -0.056 | 0.002 | -0.043 | -0.072 | 0.036 | 0.018 | 0.047 | -0.008 | -0.04 | -0.044 | 0.06 | -0.074 | 0.025 | 0.047 |
| **Q13.8** | 0.047 | 0.018 | 0.187 | 0.036 | 0.049 | 0.094 | 0.012 | 0.063 | 0.073 | -0.001 | -0.059 | 0.075 | 0.081 | 0.128 | 0.068 | 0.012 | -0.009 | 0.005 | 0.101 | 0.144 | 0.05 | 0.094 | -0.022 | 0.032 | 0.132 | 0.008 | 0.031 | -0.052 |
| **Q13.9** | -0.006 | 0.058 | 0.079 | 0.059 | -0.038 | 0.035 | 0.026 | 0.094 | -0.038 | -0.012 | 0.011 | -0.05 | 0.057 | 0.014 | -0.037 | 0.029 | -0.054 | -0.035 | 0.034 | 0.007 | 0.011 | -0.059 | -0.066 | -0.013 | 0.003 | -0.063 | -0.019 | -0.092 |
| **Q13.11** | -0.062 | -0.05 | -0.069 | -0.067 | -0.089 | -0.053 | -0.086 | -0.13 | -0.052 | -0.072 | -0.025 | -0.147 | -0.041 | -0.014 | -0.029 | -0.009 | -0.22 | -0.05 | -0.035 | -0.132 | -0.041 | -0.065 | -0.236 | -0.203 | -0.109 | -0.137 | -0.114 | -0.068 |
| **Q13.12** | -0.044 | 0.081 | 0.013 | -0.059 | -0.112 | 0.002 | -0.021 | 0.045 | 0.047 | -0.001 | 0.023 | -0.039 | 0.067 | -0.025 | -0.049 | 0 | -0.049 | -0.088 | 0.031 | 0.007 | 0.07 | 0.008 | -0.084 | -0.03 | 0.029 | -0.035 | -0.027 | -0.026 |
| **Q13.15** | 0.01 | 0.002 | 0.155 | 0.089 | 0.032 | 0.049 | 0.048 | 0.059 | 0.013 | 0.076 | -0.001 | 0.056 | 0.043 | 0.07 | 0.077 | 0.024 | -0.035 | 0.115 | 0.069 | 0.033 | 0.013 | 0.02 | -0.011 | -0.018 | 0.074 | 0.019 | 0.072 | -0.044 |
| **Q13.19** | -0.018 | 0.068 | 0.053 | -0.031 | -0.074 | 0.006 | 0.041 | 0.03 | 0.037 | -0.032 | 0.064 | -0.036 | 0.087 | 0.025 | -0.007 | 0.042 | -0.096 | -0.064 | 0.035 | 0.042 | 0.026 | 0.003 | -0.111 | -0.073 | 0.009 | -0.008 | -0.002 | -0.062 |
| **Q13.22** | -0.024 | 0.076 | 0.058 | 0.006 | -0.041 | 0.035 | 0.026 | 0.095 | 0.048 | 0.019 | 0.075 | -0.02 | 0.063 | 0.026 | -0.006 | 0.014 | -0.09 | -0.049 | 0.044 | 0.043 | -0.069 | -0.064 | -0.104 | -0.013 | 0.021 | 0.018 | -0.011 | -0.074 |
| **Q13.24** | -0.003 | 0.079 | 0.075 | -0.024 | -0.052 | 0.089 | 0.056 | 0.069 | 0.097 | 0.004 | 0.066 | -0.034 | 0.066 | 0.033 | 0.039 | 0.064 | -0.051 | -0.08 | 0.047 | 0.076 | -0.037 | 0.004 | -0.108 | -0.02 | 0.033 | 0.011 | -0.014 | -0.039 |
| **Q13.26** | -0.035 | -0.002 | 0.059 | -0.109 | -0.028 | -0.032 | -0.014 | 0.025 | -0.016 | 0.007 | 0.006 | -0.015 | -0.011 | 0.031 | -0.039 | -0.005 | -0.054 | 0.019 | 0.051 | 0.059 | 0.071 | 0.002 | -0.103 | 0.002 | 0.039 | -0.003 | -0.026 | -0.028 |
| **Q13.27** | 0.042 | 0.07 | 0.062 | 0.004 | -0.099 | -0.011 | 0.038 | 0.031 | -0.001 | -0.004 | 0.041 | -0.109 | 0.067 | 0.011 | 0.056 | 0.118 | -0.038 | -0.005 | 0.049 | 0.019 | 0.038 | 0.048 | -0.112 | -0.037 | 0.004 | 0.004 | -0.017 | -0.083 |
| **Q13.28** | -0.079 | 0.015 | 0.042 | -0.069 | -0.087 | -0.07 | 0.04 | 0.03 | -0.066 | 0.009 | 0.039 | -0.086 | -0.001 | -0.043 | -0.056 | 0.036 | 0.023 | -0.003 | 0.006 | 0.053 | 0.081 | 0.038 | -0.124 | -0.037 | 0.046 | 0.023 | 0.034 | -0.014 |
| **Q13.36** | -0.017 | 0.047 | 0.081 | 0.041 | -0.054 | 0.003 | 0.076 | 0.117 | -0.028 | 0.032 | -0.019 | 0.02 | 0.064 | 0.051 | -0.034 | -0.008 | -0.008 | 0.06 | 0.069 | 0.042 | 0.081 | 0.015 | -0.015 | -0.002 | 0.046 | 0.074 | 0.089 | -0.024 |

| Q11.15 | Q11.16 | Q12.1 | Q12.2 | Q12.3 | Q12.4 | Q12.5 | Q12.6 | Q12.7 | Q12.8 | Q12.9 | Q12.10 | Q12.11 | Q12.12 | Q12.13 | Q12.14 | Q12.15 | Q13.5 | Q13.6 | Q13.10 | Q13.13 | Q13.14 | Q13.16 | Q13.17 | Q13.18 | Q13.20 | Q13.21 | Q13.23 | Q13.25 |
|---|---|---|---|---|---|---|---|---|---|---|---|---|---|---|---|---|---|---|---|---|---|---|---|---|---|---|---|---|
| | | | | | | | | | | | | | | | | | | | | | | | | | | | | |
| | | | | | | | | | | | | | | | | | | | | | | | | | | | | |
| | | | | | | | | | | | | | | | | | | | | | | | | | | | | |
| | | | | | | | | | | | | | | | | | | | | | | | | | | | | |
| | | | | | | | | | | | | | | | | | | | | | | | | | | | | |
| | | | | | | | | | | | | | | | | | | | | | | | | | | | | |
| | | | | | | | | | | | | | | | | | | | | | | | | | | | | |
| | | | | | | | | | | | | | | | | | | | | | | | | | | | | |
| | | | | | | | | | | | | | | | | | | | | | | | | | | | | |
| | | | | | | | | | | | | | | | | | | | | | | | | | | | | |
| | | | | | | | | | | | | | | | | | | | | | | | | | | | | |
| | | | | | | | | | | | | | | | | | | | | | | | | | | | | |
| | | | | | | | | | | | | | | | | | | | | | | | | | | | | |
| | | | | | | | | | | | | | | | | | | | | | | | | | | | | |
| | | | | | | | | | | | | | | | | | | | | | | | | | | | | |
| | | | | | | | | | | | | | | | | | | | | | | | | | | | | |
| | | | | | | | | | | | | | | | | | | | | | | | | | | | | |
| | | | | | | | | | | | | | | | | | | | | | | | | | | | | |
| | | | | | | | | | | | | | | | | | | | | | | | | | | | | |
| | | | | | | | | | | | | | | | | | | | | | | | | | | | | |
| | | | | | | | | | | | | | | | | | | | | | | | | | | | | |
| | | | | | | | | | | | | | | | | | | | | | | | | | | | | |
| | | | | | | | | | | | | | | | | | | | | | | | | | | | | |
| | | | | | | | | | | | | | | | | | | | | | | | | | | | | |
| | | | | | | | | | | | | | | | | | | | | | | | | | | | | |
| | | | | | | | | | | | | | | | | | | | | | | | | | | | | |
| | | | | | | | | | | | | | | | | | | | | | | | | | | | | |
| | | | | | | | | | | | | | | | | | | | | | | | | | | | | |
| 1 | | | | | | | | | | | | | | | | | | | | | | | | | | | | |
| 0.492 | 1 | | | | | | | | | | | | | | | | | | | | | | | | | | | |
| 0.246 | 0.406 | 1 | | | | | | | | | | | | | | | | | | | | | | | | | | |
| 0.311 | 0.294 | 0.337 | 1 | | | | | | | | | | | | | | | | | | | | | | | | | |
| 0.29 | 0.432 | 0.565 | 0.48 | 1 | | | | | | | | | | | | | | | | | | | | | | | | |
| 0.439 | 0.412 | 0.515 | 0.506 | 0.466 | 1 | | | | | | | | | | | | | | | | | | | | | | | |
| 0.326 | 0.451 | 0.547 | 0.491 | 0.597 | 0.58 | 1 | | | | | | | | | | | | | | | | | | | | | | |
| 0.307 | 0.428 | 0.619 | 0.377 | 0.506 | 0.511 | 0.552 | 1 | | | | | | | | | | | | | | | | | | | | | |
| 0.346 | 0.561 | 0.649 | 0.462 | 0.754 | 0.57 | 0.648 | 0.618 | 1 | | | | | | | | | | | | | | | | | | | | |
| 0.281 | 0.383 | 0.48 | 0.437 | 0.454 | 0.688 | 0.492 | 0.346 | 0.556 | 1 | | | | | | | | | | | | | | | | | | | |
| 0.254 | 0.361 | 0.575 | 0.496 | 0.771 | 0.466 | 0.614 | 0.525 | 0.72 | 0.463 | 1 | | | | | | | | | | | | | | | | | | |
| 0.318 | 0.486 | 0.642 | 0.453 | 0.723 | 0.564 | 0.707 | 0.627 | 0.761 | 0.48 | 0.685 | 1 | | | | | | | | | | | | | | | | | |
| 0.278 | 0.268 | 0.603 | 0.328 | 0.513 | 0.368 | 0.456 | 0.479 | 0.507 | 0.335 | 0.517 | 0.504 | 1 | | | | | | | | | | | | | | | | |
| 0.329 | 0.139 | 0.141 | 0.318 | 0.118 | 0.275 | 0.146 | 0.133 | 0.134 | 0.339 | 0.124 | 0.145 | 0.267 | 1 | | | | | | | | | | | | | | | |
| 0.109 | 0.236 | 0.26 | 0.137 | 0.233 | 0.105 | 0.308 | 0.235 | 0.231 | 0.276 | 0.226 | 0.21 | 0.314 | 0.372 | 1 | | | | | | | | | | | | | | |
| 0.227 | 0.272 | 0.397 | 0.46 | 0.403 | 0.599 | 0.475 | 0.296 | 0.481 | 0.632 | 0.481 | 0.377 | 0.262 | 0.26 | 0.132 | 1 | | | | | | | | | | | | | |
| 0.282 | 0.213 | 0.262 | 0.312 | 0.199 | 0.444 | 0.3 | 0.148 | 0.191 | 0.512 | 0.128 | 0.279 | 0.112 | 0.366 | 0.137 | 0.312 | 1 | | | | | | | | | | | | |
| -0.08 | -0.129 | -0.026 | 0.058 | -0.004 | 0.043 | 0.135 | 0.01 | -0.016 | 0.025 | -0.028 | -0.019 | 0.036 | 0.05 | 0.034 | 0.084 | 0.082 | 1 | | | | | | | | | | | |
| 0.026 | 0.07 | -0.056 | 0.172 | 0.152 | 0.014 | 0.13 | 0.056 | 0.078 | 0.082 | 0.077 | 0.048 | 0.104 | 0.108 | 0.062 | 0.098 | 0.115 | 0.577 | 1 | | | | | | | | | | |
| -0.044 | -0.156 | -0.076 | -0.031 | -0.114 | -0.079 | 0.048 | -0.022 | -0.077 | -0.054 | -0.121 | -0.118 | -0.029 | -0.024 | 0.006 | -0.075 | 0.02 | 0.553 | 0.5 | 1 | | | | | | | | | |
| -0.061 | -0.048 | -0.016 | 0.088 | -0.073 | -0.012 | 0.125 | 0.053 | -0.015 | 0.026 | 0.029 | -0.055 | 0.064 | 0.047 | 0.107 | 0.07 | -0.023 | 0.555 | 0.513 | 0.542 | 1 | | | | | | | | |
| -0.053 | -0.066 | 0.032 | 0.095 | -0.026 | 0.088 | 0.136 | 0.03 | -0.023 | 0.048 | -0.011 | -0.004 | -0.041 | -0.008 | 0.008 | 0.068 | 0.15 | 0.621 | 0.544 | 0.651 | 0.488 | 1 | | | | | | | |
| -0.085 | -0.114 | -0.056 | 0.11 | -0.105 | -0.019 | 0.029 | -0.048 | -0.06 | 0.063 | -0.042 | -0.102 | -0.024 | -0.027 | -0.062 | 0.068 | 0.149 | 0.572 | 0.547 | 0.642 | 0.658 | 0.608 | 1 | | | | | | |
| -0.079 | -0.071 | 0.055 | 0.06 | -0.05 | 0.041 | 0.037 | 0.024 | -0.028 | 0.024 | -0.03 | -0.063 | -0.048 | -0.004 | -0.063 | 0.077 | 0.049 | 0.45 | 0.373 | 0.564 | 0.5 | 0.616 | 0.551 | 1 | | | | | |
| -0.04 | 0.015 | -0.066 | 0.146 | 0.06 | -0.016 | 0.12 | 0.01 | 0.004 | 0.013 | 0.05 | 0.003 | -0.008 | 0.005 | 0.04 | 0.002 | 0.109 | 0.432 | 0.587 | 0.547 | 0.547 | 0.551 | 0.572 | 0.526 | 1 | | | | |
| -0.048 | -0.072 | -0.054 | 0.085 | 0.006 | -0.017 | 0.048 | 0.011 | 0.052 | 0.084 | 0.034 | 0.032 | 0.011 | 0.056 | -0.006 | 0.082 | 0.049 | 0.474 | 0.497 | 0.502 | 0.507 | 0.61 | 0.571 | 0.51 | 0.56 | 1 | | | |
| -0.054 | -0.045 | -0.09 | 0.06 | -0.017 | -0.058 | 0.071 | -0.051 | -0.081 | 0 | 0.004 | -0.061 | -0.014 | -0.024 | -0.021 | 0.003 | 0.053 | 0.524 | 0.624 | 0.628 | 0.576 | 0.672 | 0.668 | 0.567 | 0.619 | 0.652 | 1 | | |
| -0.049 | -0.082 | 0.066 | 0.087 | 0.095 | -0.09 | 0.14 | 0.102 | 0.05 | -0.009 | 0.117 | 0.045 | 0.105 | 0.033 | 0.129 | 0 | 0.078 | 0.5 | 0.615 | 0.586 | 0.534 | 0.639 | 0.574 | 0.514 | 0.667 | 0.585 | 0.721 | 1 | |
| -0.115 | -0.077 | -0.071 | 0.058 | -0.062 | -0.064 | 0.026 | -0.049 | -0.085 | -0.1 | 0.001 | -0.071 | -0.061 | -0.035 | -0.109 | -0.013 | -0.013 | 0.536 | 0.542 | 0.54 | 0.497 | 0.58 | 0.608 | 0.575 | 0.663 | 0.496 | 0.665 | 0.668 | 1 |
| -0.066 | -0.065 | 0.067 | 0.101 | 0.064 | 0.012 | 0.145 | 0.051 | 0.052 | 0.04 | 0.113 | 0.076 | 0.081 | -0.054 | 0.029 | 0.053 | 0.026 | 0.492 | 0.486 | 0.589 | 0.438 | 0.604 | 0.527 | 0.589 | 0.563 | 0.642 | 0.611 | 0.609 | 0.447 |
| 0.008 | 0.007 | 0.002 | 0.093 | 0.057 | 0.077 | 0.104 | 0.029 | 0.076 | 0.079 | 0.078 | 0.026 | 0.09 | 0.031 | 0.037 | 0.12 | 0.063 | 0.615 | 0.698 | 0.421 | 0.532 | 0.527 | 0.489 | 0.384 | 0.504 | 0.512 | 0.466 | 0.425 | 0.426 |
| -0.027 | -0.054 | -0.049 | 0.075 | 0.026 | 0.062 | 0.109 | -0.012 | -0.017 | 0.063 | 0.091 | -0.03 | 0.038 | -0.001 | 0.043 | 0.112 | 0.071 | 0.65 | 0.67 | 0.405 | 0.477 | 0.538 | 0.563 | 0.422 | 0.462 | 0.495 | 0.451 | 0.449 | 0.463 |
| 0.023 | -0.109 | -0.132 | 0.085 | -0.094 | -0.058 | 0.046 | -0.075 | -0.057 | -0.012 | -0.047 | -0.091 | 0.026 | 0.108 | 0.064 | 0.069 | -0.037 | 0.563 | 0.546 | 0.598 | 0.548 | 0.573 | 0.665 | 0.559 | 0.525 | 0.57 | 0.63 | 0.604 | 0.609 |
| -0.08 | 0.015 | 0.042 | 0.097 | 0.034 | -0.055 | 0.107 | 0.008 | 0.013 | -0.024 | 0.041 | 0.022 | 0.023 | -0.012 | -0.032 | -0.064 | 0.093 | 0.444 | 0.497 | 0.488 | 0.401 | 0.526 | 0.478 | 0.469 | 0.582 | 0.6 | 0.585 | 0.66 | 0.556 |
| -0.142 | -0.114 | -0.131 | 0.054 | -0.068 | -0.088 | 0.077 | 0.002 | -0.06 | -0.104 | -0.062 | -0.034 | -0.035 | 0.034 | 0.064 | 0.012 | 0.044 | 0.647 | 0.482 | 0.619 | 0.51 | 0.607 | 0.601 | 0.5 | 0.523 | 0.599 | 0.66 | 0.628 | 0.603 |
| -0.043 | -0.112 | -0.069 | 0.096 | -0.042 | -0.026 | 0.134 | -0.025 | -0.034 | -0.018 | -0.001 | -0.035 | 0.002 | 0.066 | -0.01 | 0.095 | 0.123 | 0.568 | 0.586 | 0.614 | 0.465 | 0.623 | 0.608 | 0.615 | 0.523 | 0.582 | 0.607 | 0.598 | 0.548 |
| -0.061 | -0.068 | -0.013 | 0.065 | 0.045 | -0.046 | 0.108 | -0.044 | 0.022 | 0.006 | 0.045 | 0.02 | 0.034 | -0.063 | -0.022 | 0.009 | 0.015 | 0.526 | 0.53 | 0.518 | 0.5 | 0.566 | 0.552 | 0.539 | 0.598 | 0.669 | 0.672 | 0.685 | 0.573 |
| -0.07 | -0.158 | -0.114 | 0.061 | -0.168 | -0.013 | 0 | -0.052 | -0.155 | -0.041 | -0.111 | -0.174 | -0.054 | 0.035 | -0.021 | 0.02 | 0.03 | 0.584 | 0.471 | 0.609 | 0.471 | 0.673 | 0.61 | 0.689 | 0.496 | 0.569 | 0.606 | 0.537 | 0.524 |
| -0.082 | -0.216 | -0.006 | 0.015 | -0.092 | -0.024 | 0.002 | -0.038 | -0.079 | -0.002 | -0.086 | -0.108 | 0.012 | 0.139 | 0.004 | 0.03 | 0.139 | 0.605 | 0.567 | 0.617 | 0.478 | 0.656 | 0.652 | 0.52 | 0.547 | 0.556 | 0.539 | 0.602 | 0.583 |
| 0.028 | -0.074 | -0.021 | 0.095 | -0.008 | -0.016 | 0.106 | 0.03 | -0.028 | -0.029 | 0.034 | 0.005 | -0.01 | 0.092 | 0.045 | 0.072 | 0.099 | 0.435 | 0.506 | 0.566 | 0.538 | 0.505 | 0.51 | 0.521 | 0.568 | 0.459 | 0.586 | 0.598 | 0.548 |
| -0.028 | -0.102 | -0.035 | 0.032 | -0.007 | -0.034 | 0.095 | -0.009 | 0.018 | -0.022 | -0.005 | -0.051 | -0.045 | -0.065 | -0.067 | -0.025 | 0.035 | 0.493 | 0.503 | 0.612 | 0.499 | 0.624 | 0.541 | 0.557 | 0.666 | 0.608 | 0.512 | 0.611 | 0.563 |
| 0.01 | 0.044 | 0.056 | 0.025 | 0.118 | 0.066 | 0.184 | 0.043 | 0.155 | 0.1 | 0.118 | 0.125 | 0.003 | 0.039 | 0.102 | 0.204 | 0.042 | 0.339 | 0.256 | 0.213 | 0.295 | 0.286 | 0.273 | 0.299 | 0.265 | 0.371 | 0.224 | 0.285 | 0.207 |
| -0.043 | -0.021 | -0.063 | 0.133 | 0 | -0.012 | 0.066 | 0.055 | -0.055 | -0.07 | -0.005 | 0.011 | -0.016 | 0.116 | 0.075 | -0.053 | 0.092 | 0.345 | 0.558 | 0.575 | 0.46 | 0.506 | 0.488 | 0.499 | 0.568 | 0.42 | 0.576 | 0.569 | 0.467 |
| -0.014 | 0.006 | 0.063 | 0.06 | 0.118 | 0 | 0.194 | 0.125 | 0.112 | 0.054 | 0.108 | 0.06 | 0.03 | 0.025 | 0.103 | 0.092 | 0.072 | 0.392 | 0.497 | 0.474 | 0.528 | 0.497 | 0.484 | 0.551 | 0.55 | 0.459 | 0.573 | 0.589 | 0.451 |
| -0.015 | -0.104 | -0.012 | 0.118 | -0.031 | 0.084 | 0.109 | 0.059 | -0.049 | 0.103 | 0.064 | -0.022 | -0.02 | 0.146 | 0.105 | 0.15 | 0.089 | 0.353 | 0.399 | 0.447 | 0.403 | 0.431 | 0.511 | 0.559 | 0.487 | 0.4 | 0.442 | 0.419 | 0.484 |
| -0.118 | -0.135 | -0.102 | 0.018 | -0.044 | -0.132 | 0.101 | -0.07 | -0.093 | -0.044 | -0.05 | -0.131 | -0.046 | -0.116 | 0.095 | 0.016 | -0.028 | 0.432 | 0.442 | 0.681 | 0.527 | 0.562 | 0.548 | 0.533 | 0.461 | 0.471 | 0.581 | 0.569 | 0.451 |
| -0.044 | -0.101 | 0.044 | 0.064 | -0.008 | -0.046 | 0.129 | 0.042 | -0.051 | -0.011 | 0.036 | -0.014 | 0.049 | -0.017 | -0.036 | -0.013 | 0.097 | 0.603 | 0.577 | 0.597 | 0.544 | 0.686 | 0.601 | 0.562 | 0.561 | 0.589 | 0.675 | 0.701 | 0.608 |
| 0.023 | 0.003 | -0.037 | 0.126 | 0.072 | 0.007 | 0.154 | 0.098 | 0.057 | 0.023 | 0.014 | 0.017 | -0.002 | 0.029 | 0.003 | 0.038 | 0.11 | 0.402 | 0.525 | 0.551 | 0.431 | 0.585 | 0.448 | 0.638 | 0.586 | 0.576 | 0.546 | 0.552 | 0.429 |
| -0.017 | -0.153 | -0.024 | 0.068 | -0.054 | 0.004 | 0.12 | 0.04 | -0.028 | -0.026 | -0.024 | -0.051 | 0.031 | 0.017 | 0.07 | 0.078 | 0.067 | 0.687 | 0.561 | 0.708 | 0.639 | 0.651 | 0.673 | 0.566 | 0.487 | 0.541 | 0.61 | 0.612 | 0.599 |
| -0.107 | -0.148 | -0.066 | 0.115 | -0.019 | -0.047 | 0.102 | -0.031 | -0.037 | -0.017 | -0.01 | -0.048 | 0.049 | -0.034 | 0.028 | 0.051 | 0.034 | 0.621 | 0.626 | 0.597 | 0.66 | 0.659 | 0.676 | 0.589 | 0.588 | 0.585 | 0.627 | 0.624 | 0.564 |
| -0.131 | -0.079 | -0.012 | -0.025 | -0.116 | -0.091 | 0.104 | -0.011 | -0.063 | -0.033 | -0.034 | -0.084 | -0.03 | 0 | 0.022 | 0.009 | 0.051 | 0.635 | 0.54 | 0.632 | 0.535 | 0.55 | 0.558 | 0.467 | 0.498 | 0.513 | 0.548 | 0.582 | 0.589 |
| -0.078 | -0.03 | -0.016 | 0.123 | 0.018 | -0.046 | 0.121 | 0.016 | -0.008 | 0.061 | 0.027 | -0.022 | 0.024 | 0.088 | 0.04 | 0.049 | 0.104 | 0.453 | 0.625 | 0.542 | 0.569 | 0.591 | 0.554 | 0.46 | 0.568 | 0.463 | 0.717 | 0.705 | 0.602 |
| -0.115 | -0.119 | -0.024 | 0.071 | -0.078 | 0.02 | 0.063 | -0.027 | -0.018 | 0.02 | -0.02 | -0.063 | 0.009 | 0.038 | 0.05 | 0.04 | 0.059 | 0.627 | 0.554 | 0.615 | 0.576 | 0.689 | 0.608 | 0.602 | 0.488 | 0.581 | 0.638 | 0.636 | 0.612 |
| -0.057 | -0.065 | -0.055 | 0.077 | -0.035 | -0.021 | 0.092 | -0.005 | -0.039 | -0.062 | -0.017 | -0.021 | 0 | 0.04 | 0.05 | -0.016 | 0.08 | 0.598 | 0.543 | 0.58 | 0.532 | 0.686 | 0.591 | 0.515 | 0.564 | 0.537 | 0.743 | 0.703 | 0.69 |
| 0.027 | -0.026 | -0.025 | 0.114 | -0.045 | 0.065 | 0.117 | 0.053 | -0.003 | 0.081 | 0.07 | 0.017 | 0.046 | 0.116 | 0.023 | 0.159 | 0.127 | 0.478 | 0.529 | 0.553 | 0.588 | 0.547 | 0.531 | 0.558 | 0.539 | 0.511 | 0.604 | 0.507 | 0.534 |

| Q13.29 | Q13.30 | Q13.31 | Q13.32 | Q13.33 | Q13.34 | Q13.35 | Q13.37 | Q13.38 | Q13.1 | Q13.2 | Q13.3 | Q13.4 | Q13.7 | Q13.8 | Q13.9 | Q13.11 | Q13.12 | Q13.15 | Q13.19 | Q13.22 | Q13.24 | Q13.26 | Q13.27 | Q13.28 | Q13.36 |
|---|---|---|---|---|---|---|---|---|---|---|---|---|---|---|---|---|---|---|---|---|---|---|---|---|---|
| 1 | | | | | | | | | | | | | | | | | | | | | | | | | |
| 0.478 | 1 | | | | | | | | | | | | | | | | | | | | | | | | |
| 0.458 | 0.841 | 1 | | | | | | | | | | | | | | | | | | | | | | | |
| 0.52 | 0.43 | 0.457 | 1 | | | | | | | | | | | | | | | | | | | | | | |
| 0.657 | 0.385 | 0.399 | 0.491 | 1 | | | | | | | | | | | | | | | | | | | | | |
| 0.588 | 0.416 | 0.497 | 0.637 | 0.631 | 1 | | | | | | | | | | | | | | | | | | | | |
| 0.689 | 0.514 | 0.618 | 0.6 | 0.569 | 0.688 | 1 | | | | | | | | | | | | | | | | | | | |
| 0.649 | 0.401 | 0.461 | 0.57 | 0.744 | 0.648 | 0.616 | 1 | | | | | | | | | | | | | | | | | | |
| 0.582 | 0.514 | 0.579 | 0.583 | 0.486 | 0.638 | 0.79 | 0.541 | 1 | | | | | | | | | | | | | | | | | |
| 0.552 | 0.429 | 0.495 | 0.604 | 0.52 | 0.586 | 0.627 | 0.574 | 0.561 | 1 | | | | | | | | | | | | | | | | |
| 0.482 | 0.343 | 0.31 | 0.542 | 0.49 | 0.471 | 0.542 | 0.521 | 0.498 | 0.586 | 1 | | | | | | | | | | | | | | | |
| 0.686 | 0.466 | 0.453 | 0.532 | 0.587 | 0.535 | 0.615 | 0.622 | 0.546 | 0.641 | 0.487 | 1 | | | | | | | | | | | | | | |
| 0.397 | 0.302 | 0.347 | 0.272 | 0.356 | 0.386 | 0.396 | 0.392 | 0.321 | 0.339 | 0.279 | 0.331 | 1 | | | | | | | | | | | | | |
| 0.384 | 0.359 | 0.397 | 0.484 | 0.425 | 0.479 | 0.489 | 0.462 | 0.555 | 0.459 | 0.498 | 0.403 | 0.237 | 1 | | | | | | | | | | | | |
| 0.596 | 0.39 | 0.372 | 0.477 | 0.604 | 0.51 | 0.533 | 0.592 | 0.493 | 0.499 | 0.474 | 0.587 | 0.334 | 0.416 | 1 | | | | | | | | | | | |
| 0.505 | 0.368 | 0.469 | 0.498 | 0.376 | 0.412 | 0.578 | 0.38 | 0.531 | 0.528 | 0.517 | 0.469 | 0.337 | 0.484 | 0.332 | 1 | | | | | | | | | | |
| 0.626 | 0.377 | 0.439 | 0.533 | 0.468 | 0.603 | 0.628 | 0.536 | 0.563 | 0.483 | 0.433 | 0.556 | 0.33 | 0.403 | 0.502 | 0.393 | 1 | | | | | | | | | |
| 0.596 | 0.435 | 0.53 | 0.57 | 0.709 | 0.64 | 0.657 | 0.742 | 0.591 | 0.591 | 0.617 | 0.586 | 0.321 | 0.464 | 0.568 | 0.485 | 0.567 | 1 | | | | | | | | |
| 0.682 | 0.458 | 0.407 | 0.543 | 0.602 | 0.589 | 0.703 | 0.598 | 0.635 | 0.545 | 0.5 | 0.649 | 0.376 | 0.471 | 0.658 | 0.452 | 0.6 | 0.532 | 1 | | | | | | | |
| 0.502 | 0.492 | 0.54 | 0.621 | 0.466 | 0.67 | 0.676 | 0.608 | 0.654 | 0.646 | 0.609 | 0.531 | 0.34 | 0.524 | 0.489 | 0.505 | 0.588 | 0.636 | 0.505 | 1 | | | | | | |
| 0.566 | 0.6 | 0.651 | 0.596 | 0.442 | 0.61 | 0.639 | 0.572 | 0.594 | 0.593 | 0.538 | 0.54 | 0.352 | 0.523 | 0.442 | 0.521 | 0.591 | 0.634 | 0.537 | 0.684 | 1 | | | | | |
| 0.513 | 0.517 | 0.577 | 0.512 | 0.51 | 0.622 | 0.646 | 0.551 | 0.549 | 0.534 | 0.535 | 0.585 | 0.281 | 0.479 | 0.412 | 0.454 | 0.479 | 0.664 | 0.418 | 0.703 | 0.587 | 1 | | | | |
| 0.486 | 0.418 | 0.385 | 0.568 | 0.543 | 0.562 | 0.551 | 0.609 | 0.461 | 0.511 | 0.538 | 0.498 | 0.128 | 0.435 | 0.548 | 0.366 | 0.551 | 0.64 | 0.503 | 0.496 | 0.574 | 0.515 | 1 | | | |
| 0.601 | 0.569 | 0.569 | 0.671 | 0.502 | 0.688 | 0.691 | 0.58 | 0.691 | 0.616 | 0.508 | 0.613 | 0.345 | 0.475 | 0.503 | 0.508 | 0.591 | 0.625 | 0.581 | 0.759 | 0.671 | 0.629 | 0.546 | 1 | | |
| 0.478 | 0.429 | 0.462 | 0.653 | 0.576 | 0.719 | 0.587 | 0.647 | 0.591 | 0.602 | 0.633 | 0.488 | 0.293 | 0.551 | 0.506 | 0.435 | 0.511 | 0.698 | 0.456 | 0.709 | 0.614 | 0.585 | 0.647 | 0.771 | 1 | |
| 0.48 | 0.574 | 0.551 | 0.53 | 0.361 | 0.513 | 0.615 | 0.424 | 0.6 | 0.493 | 0.503 | 0.47 | 0.281 | 0.538 | 0.403 | 0.568 | 0.45 | 0.519 | 0.556 | 0.579 | 0.578 | 0.561 | 0.539 | 0.567 | 0.559 | 1 |

**Fuente**: Elaboración propia (2016*)*

****Nota**: El valor de Q.10.11, aparece en números negativos debido a que se trata de un ítem de valores inversos. En donde, la escala Likert se ha invertido.*

## Apéndice 4. Elementos mínimos de cumplimiento para cada nivel del Estándar Mexicano de Gestión del Conocimiento e Innovación Tecnológica

| Nivel | Algunos elementos presentes en el nivel |
|---|---|
| **Nivel básico** | 1. Mínima documentación de procesos y procedimientos<br>2. Criterios comunes para realizar actividades similares<br>3. Mecanismos para almacenar el conocimiento que se produce en la organización<br>4. La organización cuenta con sistemas fáciles de usar<br>5. Los trabajadores de la organización tienen las competencias necesarias para el desarrollo de sus funciones<br>6. Compromiso de los directivos de esta organización con la implementación de nuevos sistemas y tecnologías<br>7. Los miembros de la organización conocen sus objetivos, planes y/o políticas<br>8. Apoyo de los líderes de la organización a las ideas nuevas o innovadoras |
| **Nivel intermedio** | 1. Se utilizan metodologías para el desarrollo de las actividades<br>2. Las personas conocen y utilizan el manual de procedimientos de la organización<br>3. Existen procedimientos establecidos para integrar el conocimiento y la información que se produce en la organización<br>4. Se utilizan datos, información y/o conocimiento relevante del entorno para la toma de decisiones<br>5. Los sistemas cumplen con su propósito<br>6. El hardware y los dispositivos tecnológico, en las áreas de trabajo, están actualizados<br>7. Se realizan reuniones para resolver problemas y para proponer iniciativas<br>8. El conocimiento producido por los empleados es valorado por la organización<br>9. Se documentan las actividades innovadoras |
| **Nivel avanzado** | 1. La organización tiene formalizados sus procesos de transferencia de mejores prácticas<br>2. A partir de experiencias exitosas implementamos procesos estandarizados o metodologías propias de trabajo<br>3. El conocimiento está correctamente explicitado (en manuales, procedimientos, en la normatividad, políticas, entre otros)<br>4. La organización desarrolla sistemas propios que le permiten gestionar el conocimiento de la organización<br>5. Constantemente se renueva el equipo y la infraestructura |
| | tecnológica disponible de la organización<br>6. Resuelvo problemas con base en mi experiencia<br>7. La organización promueve que se compartan los conocimientos |
| **Nivel experto** | 1. La organización ofrece metodologías de diagnóstico a otras organizaciones a partir de las nuestras<br>2. Los sistemas de la organización son compatibles con los de otros departamentos e instituciones<br>3. En la organización se incentiva el uso y deseo de aprender a utilizar nuevas tecnologías<br>4. Se diseñan productos o servicios propios<br>5. Otras organizaciones solicitan acceso a nuestros sistemas de conocimiento para aplicarlo en la resolución de sus problemas<br>6. La organización genera proyectos que se vinculan con otras organizaciones o sectores de la sociedad<br>7. La organización integra elementos de transferencia de la innovación, para convertir en producto o servicios comercializables sus desarrollos, por ejemplo: patente licenciada o vendida, software desarrollado y licenciado o utilizado en la organización, asesoría y/o capacitación especializada, etc. |

**Fuente**: Elaboración propia (2016)